\documentclass{aa}  
\pdfoutput=1 
\usepackage{txfonts}
\usepackage{graphicx, graphics, rotating,amssymb,amsmath}
\usepackage{booktabs}
\usepackage{subfig}
\usepackage{float,capt-of}
\usepackage{natbib}
\usepackage{multirow}
\usepackage[english]{babel}
\usepackage{morefloats}
\usepackage{color}
\usepackage{sidecap}
\usepackage{epsfig,color}
\usepackage{wrapfig}
\usepackage{longtable}
\usepackage{caption}
\usepackage{url}
\newcommand\kms{km~s$^{-1}$}

\begin{document}
   \title{The final data release of ALLSMOG: a survey of CO in typical local low-$M_*$ star-forming galaxies}

   \author{C. Cicone \thanks{claudia.cicone@brera.inaf.it}
          \inst{1,2} 
          \and
         M. Bothwell  \inst{3,4}
         \and J. Wagg \inst{5}
         \and P. M{\o}ller \inst{6}
         \and C. De Breuck \inst{6}
         \and Z. Zhang \inst{7,6}
         \and S. Mart\'in \inst{8,9}
         \and R. Maiolino \inst{3,4}
          \and P. Severgnini \inst{1}
         \and M. Aravena \inst{10}
         \and F. Belfiore \inst{3,4}  
         \and D. Espada \inst{11,12}
          \and A. Fl\"utsch \inst{3,4}
         \and V. Impellizzeri \inst{9}
         \and Y. Peng \inst{13}
         \and M. A. Raj \inst{14}
         \and N. Ram\'irez-Olivencia \inst{15} 
         \and D. Riechers \inst{16}
         \and K. Schawinski \inst{2}
         }
      \institute{INAF-Osservatorio Astronomico di Brera, via Brera 28, 20121, Milan, Italy 
      \and Institute for Astronomy, Department of Physics, ETH Zurich, Wolfgang-Pauli-Strasse 27, CH-8093 Zurich, Switzerland 
      \and Cavendish Laboratory, University of Cambridge, 19 J. J. Thomson Ave, Cambridge CB3 0HE, UK 
      \and Kavli Institute for Cosmology, University of Cambridge, Madingley Road, Cambridge CB3 0HA, UK 
      \and Square Kilometre Array Organisation, Lower Withington, Cheshire, UK 
      \and European Southern Observatory, Karl-Schwarzschild-Strasse 2, 85748 Garching, Germany 
      \and Institute for Astronomy, University of Edinburgh, Royal Observatory, Blackford Hill, Edinburgh EH9 3HJ, UK 
      \and European Southern Observatory, Alonso de C\'ordova 3107, Vitacura, Santiago, Chile 
      \and Joint ALMA Observatory, Alonso de C\'ordova 3107, Vitacura, Santiago, Chile 
     	\and N\'ucleo de Astronom\'ia, Facultad de Ingenier\'ia, Universidad Diego Portales, Av. Ej\'ercito 441, Santiago, Chile 
 	\and National Astronomical Observatory of Japan (NAOJ), 2-21-1 Osawa, Mitaka, Tokyo 181-8588 
	\and Department of Astronomical Science, The Graduate University for Advanced Studies (SOKENDAI), 2-21-1 Osawa, Mitaka, Tokyo 181-8588, Japan 
      \and Kavli Institute for Astronomy and Astrophysics, Peking University, 100871 Beijing, PR China 
       \and Jodrell Bank Centre for Astrophysics, School of Physics \& Astronomy, The University of Manchester, Oxford Road, Manchester M13 9PL, UK 
      \and Instituto de Astrof\'isica de Andaluc\'ia (IAA-CSIC), Glorieta de la Astronom\'ia s/n, 18008 Granada, Spain 
      \and Department of Astronomy, Cornell University, 220 Space Sciences Building, Ithaca, NY 14853, USA 
      	}
	
   \date{Received 13-02-2017 / Accepted: 03-05-2017 }


\abstract{We present the final data release of the APEX low-redshift legacy survey for molecular gas (ALLSMOG), comprising CO(2-1) emission line observations of 88 nearby, low-mass ($10^{8.5}<M_{*} [M_{\odot}]<10^{10}$) star-forming galaxies carried out with the 230~GHz APEX-1 receiver on the APEX telescope. The main goal of ALLSMOG is to probe the molecular gas content of more typical and lower stellar mass galaxies than have been studied by previous CO surveys. We also present IRAM 30m observations of the CO(1-0) and CO(2-1) emission lines in nine galaxies aimed at increasing the $M_{*}<10^9~M_{\odot}$ sample size. In this paper we describe the observations, data reduction and analysis methods and we present the final CO spectra together with archival H{\sc i}~21cm line observations for the entire sample of $97$ galaxies. At the sensitivity limit of ALLSMOG, we register a total CO detection rate of 47\%. Galaxies with higher $M_*$, SFR, nebular extinction ($A_V$), gas-phase metallicity (O/H), and H{\sc i} gas mass have systematically higher CO detection rates. In particular, the parameter according to which CO detections and non-detections show the strongest statistical differences is the gas-phase metallicity, for any of the five metallicity calibrations examined in this work. We investigate scaling relations between the CO(1-0) line luminosity ($L^{\prime}_{\rm CO(1-0)}$) and galaxy-averaged properties using ALLSMOG and a sub-sample of COLD GASS for a total of 185 sources that probe the local main sequence (MS) of star-forming galaxies and its $\pm0.3$~dex intrinsic scatter from $M_* = 10^{8.5}~M_{\odot}$ to $M_* = 10^{11}~M_{\odot}$. $L^{\prime}_{\rm CO(1-0)}$ is most strongly correlated with the SFR, but the correlation with $M_*$ is closer to linear and almost comparably tight. The relation between $L^{\prime}_{\rm CO(1-0)}$ and metallicity is the steepest one, although deeper CO observations of galaxies with $A_V<0.5~mag$ may reveal an as much steep correlation with $A_V$. Our results suggest that star-forming galaxies across more than two orders of magnitude in $M_*$ obey similar scaling relations between CO luminosity and the galaxy properties examined in this work. Besides SFR, the CO luminosity is likely most fundamentally linked to $M_*$, although we note that stellar mass alone cannot explain all of the variation in CO emission observed as a function of O/H and $M_{\rm HI}$.}

 
   \maketitle


\section{Introduction}\label{sec:intro}

The cold phase of the interstellar medium (ISM), consisting of clouds of neutral and molecular gas dominated in mass by atomic and molecular hydrogen (H{\sc i} and H$_2$), has a central role in galaxy growth and evolution. Atomic gas, which can be traced through the H{\sc i}~21cm line, is widespread in galaxy disks but it generally avoids the central regions of massive spirals \citep{Tacconi+Young86,Walter+08}. H{\sc i} extends up to several tens of kpc and is the dominant ISM component at the periphery of the disk \citep{Giovanelli+Haynes88}. H{\sc i} filaments in galaxy disks are the cradle of molecular clouds, hence the atomic medium provides an important supply of fuel for galaxies \citep{Wong+Blitz02,Blitz+07}. Molecular gas is the ISM phase that is most directly linked to the star formation process (e.g. \citealt{Wong+Blitz02, Bigiel+08, Leroy+08}). In typical non-merging star-forming galaxies, massive stars form out of discrete, self-gravitating and virially bound molecular clouds denominated giant molecular clouds (GMCs, typical sizes of $\sim50$~pc, molecular gas masses of $M_{mol}>10^4~M_{\odot}$, see for example \cite{Kennicutt+Evans12}), which dominate the ISM mass within a few kpc from the galaxy centre \citep{Scoville+75,Solomon+79,Larson81,Dame+86, McKee+Ostriker07}. 

Carbon monoxide ($^{12}$CO, hereafter CO), the second most-abundant molecule after H$_2$, is the most convenient tracer of the bulk of molecular gas in large galaxy samples, thanks to its bright low-$J$ transitions at millimetre wavelengths that are easily excited down to gas temperatures of $T\sim10$~K \citep{Omont07,Carilli+Walter13}. A part from a few pioneering studies attempting to detect CO in `normal galaxies' \citep{Verter87,Solomon+Sage88,Young+95}, most past CO surveys have focussed on the very bright population of local starbursts and interacting galaxies, that is mostly luminous and ultra-luminous infrared galaxies ((U)LIRGs) \citep{Solomon+97,Downes+Solomon98,Wilson+08,Chung+09,Papadopoulos+12a}. Only more recently have there been successful efforts to trace CO emission in local massive ($M_*\gtrsim10^{10}~M_{\odot}$) galaxies characterised by a more quiescent star formation activity \citep{Leroy+09, Lisenfeld+11, Saintonge+11a, Saintonge+11b}. Hence, there is now a strong motivation to extend the dynamic range in galaxy properties probed by these previous CO studies and investigate the molecular gas content in typical star-forming galaxies with lower stellar masses of, $M_* \lesssim 10^{10}~M_{\odot}$. 

The stellar mass function of galaxies steepens towards lower masses, hence low-$M_*$ galaxies are the most numerous galaxy type in the Universe \citep{Bell+03, Weigel+16}. They are also interesting laboratories to study feedback mechanisms, being very susceptible to both internal (for example energy injection by star formation and AGN activity) and external (external ram pressure stripping, dynamical harassment) feedback due to their shallow potential well that cannot retain baryons \citep{Kormendy14}. 
The push to reach lower-$M_*$ sources with CO observations is also driven by the necessity of understanding better the high redshift Universe. In a hierarchical assembly scenario, massive dark matter halos form from the merging of small low-mass halos in earlier epochs, and so primitive dwarfs can be considered as the building blocks of present-day massive gas-rich galaxies \citep{Tosi03}. 
Primeval galaxies that formed in the early Universe are expected to share at least some of the properties of local low-$M_*$ metal-poor galaxies, as tentatively supported by recent millimetre and sub-millimetre observations at high-$z$ \citep{Riechers+14,Capak+15,Bouwens+16,Pavesi+16,Pentericci+16}.

Because of the mass-metallicity relation \citep{Tremonti+04}, the ISM of the majority of low-$M_*$ star-forming galaxies is low in metals, and these galaxies are notoriously difficult to detect in CO emission \citep{Elmegreen+80,Verter+Hodge95,Taylor+98}. For many years observational studies targeting CO in low-metallicity dwarfs have attempted to address the question of whether their low CO detection rate corresponds to an extremely low molecular gas content or if it is mainly driven by a very high $\alpha_{CO}$ - the CO-to-H$_2$ conversion factor \citep{Leroy+05,Schruba+12,Shi+15,Shi+16}. Recent observational efforts have begun to routinely detect CO in local compact blue dwarfs \citep{Hunt+15, Amorin+16}. However, these objects, despite being very metal poor, are characterised by vigorous and bursty star formation and so probe a population different from typical normal star-forming galaxies \citep{Amorin+16}. We note that CO observations of galaxies with stellar masses as low as $M_*\sim10^9~M_{\odot}$ have been recently conducted as part of the Herschel Reference Survey (HRS, \citealt{Boselli+14a}). However, most of the HRS low-$M_*$ sources belong to the Virgo cluster and are deficient in H{\sc i} because of ram pressure gas stripping and frequent dynamical harassment episodes \citep{Boselli+14c,Grossi+16}.

The APEX low-redshift legacy survey for molecular gas (ALLSMOG) is intended to enhance previous extragalactic CO line surveys by targeting low-$J$ CO line emission in normal star-forming galaxies with $8.5<\log(M_* [M_{\odot}])<10.0$.
In this stellar mass range, ALLSMOG probes quite uniformly the local main sequence (MS) of star-forming galaxies and its intrinsic scatter. The MS is   
a tight correlation between stellar mass ($M_*$) and star formation rate (SFR) followed by the majority of normal star-forming galaxies up to $z\sim4$ \citep{Noeske+07, Peng+10, Speagle+14, Whitaker+14, Schreiber+15, Renzini+Peng15}. 
We note that starburst galaxies such as ULIRGs, which are the targets of most previous CO surveys, are outliers of the MS and so are not representative of the local star-forming galaxy population. ALLSMOG constitutes a first step towards the construction of a galaxy sample that is highly representative of local actively star-forming galaxies. By significantly extending the dynamic range of galaxy parameters (such as $M_*$, metallicity and dust extinction) probed by existing CO surveys, ALLSMOG allows us to examine scaling relations between gas content and other galaxy-integrated properties in a statistically-sound way for typical star-forming galaxies. 

In this paper we present the final ALLSMOG survey data release, which comprises APEX CO(2-1) observations of 88 galaxies, including the 42 objects presented in the early data release published by \cite{Bothwell+14}, and an additional sample of nine galaxies observed in CO(1-0) and CO(2-1) line emission with the IRAM 30m telescope, for a total of 97 sources. In the initial part of the paper we describe in detail the sample selection (Section~\ref{sec:sample}), the observations, data reduction and analysis methodology (Section~\ref{sec:observations}), and the available ancillary optical and H{\sc i}~21cm data (Section~\ref{sec:ancillary}). In the second part of the paper we report on our initial results based on ALLSMOG observations. A sub-sample of the COLD GASS survey is used to complement the ALLSMOG dataset in the high-$M_*$ regime, and is selected as explained in Section~\ref{sec:coldgass}. The presentation of the results is organised as follows: in Section~\ref{sec:histo} we discuss the distribution of detections in ALLSMOG as a function of several galaxy properties, namely: M$_*$, SFR, SSFR (the specific SFR, SSFR$\equiv$SFR/M$_*$), redshift, $A_V$, five different calibrations of gas-phase metallicity, and H{\sc i} gas mass. We then move on in Section~\ref{sec:lco_vs_prop} to the investigation of the relations between the galaxy properties and the CO line luminosity, by using the full galaxy sample defined by the ALLSMOG survey and the sub-sample of star-forming galaxies in COLD GASS described in $\S$~\ref{sec:coldgass}. Finally, in Section~\ref{sec:discussion}, we discuss our results in the light of possible differences between low-$M_*$ and high-$M_*$ star-forming galaxies in the local Universe. We note that further analysis of the data will be presented in a number of additional papers in preparation by the team (Bothwell et al. in prep, Cicone et al. in prep). 

The ESO Phase 3 has been implemented for the APEX dataset of ALLSMOG, and the fully reduced APEX data products and catalogues are available for public download via the ESO science website. All data products including the IRAM~30m observations will be also made publicly available on the ALLSMOG survey website \footnote{\texttt{http://www.mrao.cam.ac.uk/ALLSMOG/}}. We note that for the purposes of the present data release, all observations including those presented by \cite{Bothwell+14} have been re-reduced and analysed in a consistent way. 

Throughout the paper, we adopt a standard $\Lambda$CDM cosmological model with $\rm H_0 = 67.3~km~s^{-1}~Mpc^{-1}$, $\Omega_{\Lambda}=0.685$, $\Omega_M = 0.315$ (Planck Collaboration 2014).

\section{The sample}\label{sec:sample}
\begin{figure*}[tbp]
\centering
    \includegraphics[clip=true,trim=8cm 3.5cm 1.5cm 3cm,width=0.5\textwidth,angle=180]{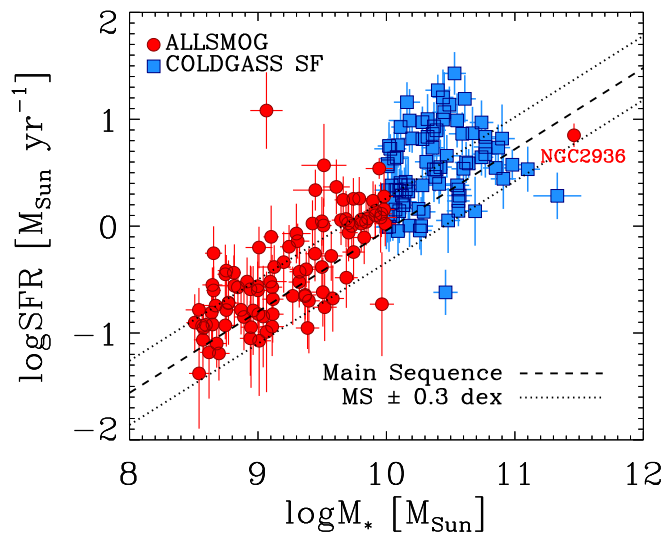}\\
       \vspace{0.2cm} 
     \caption{The ALLSMOG sample plotted in the M$_*$-SFR plane. The dashed line indicates the location of the local MS relation obtained by \cite{Renzini+Peng15}, and the dotted lines show the intrinsic scatter about the MS derived using optical surveys \citep{Peng+10}. The plot also displays a sub-sample of the COLD~GASS survey including only galaxies selected as `star-forming' according to their position on the classical BPT diagram as explained in $\S$~\ref{sec:coldgass}.}
   \label{fig:MS}
\end{figure*}

The criteria adopted to select the targets for ALLSMOG and their motivation were described by \cite{Bothwell+14}, hence we refer to that work for additional details on the sample selection, which we simply summarise in the following. 
The ALLSMOG sample is entirely drawn from the MPA-JHU catalogue\footnote{\texttt{http://www.mpa-garching.mpg.de/SDSS/}} of spectral measurements and galaxy parameters for the Sloan digital sky survey data release 7 (SDSS DR7, \cite{Abazajian+09}). The targets were selected according to the following criteria:
\begin{enumerate}
\item  Classified as `star-forming' according to their location on the $\log$([OIII]/H$\beta$) vs $\log$([NII]/H$\alpha$) diagram (the BPT diagram, \cite{BPT81}) using the division lines of \cite{Kauffmann+03} and the S/N criteria of \cite{Brinchmann+04}, hence excluding from this category galaxies with $S/N<3$ in one of the four emission lines of the BPT line-ratio diagram;
\item with stellar masses in the range, $10^{8.5}<M_* [M_{\odot}] < 10^{10}$. Targets are chosen at random to ensure a uniform sampling of this stellar mass range, similar to the COLD GASS survey \citep{Saintonge+11a}.
\item with redshift in the range, $0.01<z<0.03$;
\item at a declination, $\delta < 15$~deg;
\item with a gas-phase metallicity, $12+\log(O/H) \geq 8.5$, according to the calibration by \cite{Tremonti+04}. As already explained in \cite{Bothwell+14}, the cut on metallicity is intended to exclude sources with very high CO-to-H$_2$ conversion values for which a detection with APEX would be unfeasible. Because of the slightly higher sensitivity of the IRAM observations, this constraint was relaxed to include objects with metallicity down to $12+\log(O/H) \sim 8.3$ in the IRAM sample. 
\end{enumerate} 
In addition we checked that all targets had an existing archival H{\sc i}~21cm observation (although not all of them have a detection, see further discussion in $\S$~\ref{sec:HIdata}). 
We note that \cite{Bothwell+14} reported a slightly different threshold for the declination, that is $\delta<10$~deg. The reasons for the updated cut in declination are (i) the significant amount of APEX observing time obtained under good weather conditions (see also $\S$~\ref{sec:APEX_obs}), which allowed us to observe 26 sources at $\delta > 10$~deg despite their low elevation on the sky from the APEX site (down to $\sim30$~deg); and (ii) the inclusion of the northern sample observed with the IRAM~30m telescope, selected to have $10\leq \delta [{\rm deg}] <15$ (see also $\S~\ref{sec:IRAM30m_obs}$). 

During the APEX observations we realised that one of the sources, namely NGC2936, was extraordinarily brighter in its CO(2-1) emission compared to the others. Further investigation led us to discover that NGC2936 is actually an interacting object, with a stellar mass of $M_* \simeq 2.9 \times 10^{11}~M_{\odot}$ \citep{Xu+10}, a very different value from that reported in the MPA-JHU catalogue ($M_* \simeq 3.4\times10^{8}~M_{\odot}$), which is the reason why the source was accidentally included in our selection. We concluded that, due to the complex and extended morphology of this source in combination with the limited aperture of the SDSS spectroscopic fibre (3$\arcsec$), the MPA-JHU analysis of this galaxy may have mis-identified a bright region offset from the nucleus with the galaxy nucleus itself, hence biasing the estimates of the various galaxy parameters. Nevertheless, we decided to include NGC2936 in this paper as part of the ALLSMOG data release. However, we caution the reader that its physical properties may be affected by significant observational biases due to the limited apertures of the SDSS fibre and of the APEX beam compared to the extent of the source.

Figure~\ref{fig:MS} shows the position of the 97 ALLSMOG sources in the $M_*-{\rm SFR}$ plane with respect to the local MS of star-forming galaxies, where we use the following MS definition provided by \cite{Renzini+Peng15}:
\begin{equation}\label{eq:MSrelation}
\log {\rm SFR}_{MS} [M_{\odot}~yr^{-1}]  = 0.76~ \log M_* [M_{\odot}]- 7.64.  
\end{equation}
The $M_*$ and SFR of ALLSMOG sources are extracted from the MPA-JHU catalogue of SDSS DR7 observations (further information will be provided in $\S$~\ref{sec:ancillary}).
Figure~\ref{fig:MS} shows that, although the targets for ALLSMOG were not intentionally selected to lie on the MS (see the selection criteria listed above), the survey probes quite uniformly the local MS between $10^{8.5}<M_* [M_{\odot}]< 10^{10}$, in addition to a few objects that lie significantly above or below the MS relation (i.e. by more than a factor of two in SFR, which corresponds approximately to the MS intrinsic scatter derived by \cite{Peng+10}). 
Therefore, the ALLSMOG sample is highly representative of the bulk of the local low-M$_*$ star-forming galaxy population.
However, we note that our sample includes fewer outliers below the MS than above it, likely as a consequence of the S/N cut applied to the nebular emission lines used for the BPT classification. This potential bias, together with other possible selection effects introduced by our sample selection criteria (in particular by the cut in metallicity and by the BPT classification) will be discussed at the beginning of Section~\ref{sec:lco_vs_prop} and taken into consideration in the analysis of the statistical relations between galaxy properties and CO line luminosity presented in $\S$~\ref{sec:lco_sfr}-\ref{sec:lco_hi}.

\section{The ALLSMOG observations}\label{sec:observations}

ALLSMOG is an ESO Large Programme for the Atacama Pathfinder EXperiment (APEX, project no.: \texttt{E-192.A-0359}, principal investigator (PI): J. Wagg) targeting the CO(2-1) emission line (rest frequency, $\nu_{\rm CO(2-1)} = 230.538$~GHz) in 88 local, low-$M_*$ star-forming galaxies. The project was initially allocated $300$~hrs of ESO observing time over the course of four semesters, corresponding to $75$~hrs per semester throughout periods P92-P95 (October 2013 - September 2015). However, during P94 and P95 there was a slowdown in ALLSMOG observations, mainly due the installation of the visiting instrument Supercam in combination with better-than-average weather conditions - causing other programmes requiring more stringent precipitable water vapour (PWV) constraints to be prioritised. Because of the resulting $\sim50\%$ time loss for ALLSMOG during two semesters, the ESO observing programmes committee (OPC) granted a one-semester extension of the project, hence allowing us to complete the survey in P96 (March 2016). The final total APEX observing time dedicated to ALLSMOG amounts to 327~hrs, including the overheads due to setup and calibration but not accounting for possible additional time lost because of technical issues.

In 2014 a northern component of the ALLSMOG survey was approved at the IRAM 30m telescope (project code: 188-14, PI: S. Mart\'in), aimed at observing the CO(1-0) (rest frequency, $\nu_{\rm CO(1-0)} = 115.271$~GHz) and CO(2-1) emission lines in a sample of nine additional galaxies characterised by stellar masses, $M_*<10^9$~M$_{\odot}$. A total of $22$~hrs of observations were obtained with the IRAM 30m during two observing runs in November 2014 and May 2015.

\subsection{Observing strategy and survey goals}

We aimed to reach a line peak-to-rms ratio of $S/N\gtrsim3$ for the detections and an uniform rms for the non-detections, corresponding to rms = 0.8~mK (31.2~mJy) per $\delta \varv=50$~\kms  channels for the APEX 230~GHz observations and rms = 0.95~mK (5.7~mJy) per
$\delta \varv=50$~\kms channels for the IRAM 115~GHz observations. The rms was continuously checked during the observations through a baseline fitting in \texttt{CLASS}\footnote{Namely, the ``continuum and line analysis single-dish software'', part of the GILDAS software package.}. In case of a detection, we stopped integrating when a single-Gaussian fitting in \texttt{CLASS} on the baseline-subtracted spectrum would return a peak S/N above three. 

ALLSMOG is therefore intended to be a CO flux-limited survey and so, to first order, thanks to the narrow redshift distribution of the sample ($0.01<z<0.03$), a CO luminosity-limited survey. More specifically, for the APEX CO(2-1) observations, considering the mean redshift $\langle z \rangle _{\rm APEX} \simeq 0.02$, and assuming an average CO line width of $FWHM\sim160$~\kms (corresponding to the average CO line width of the detections, see Table~\ref{table:co_observations}), our rms goal allows us to detect at the three sigma level sources with a CO(2-1) line luminosity of $L_{\rm CO(2-1)}^{\prime} \gtrsim 4 \times 10^7$~K~\kms~pc$^2$ (the definition of $L^{\prime}_{\rm CO}$ is provided in $\S$~\ref{sec:lco_vs_prop}). For the IRAM observations, assuming the average redshift of the IRAM targets, $\langle z \rangle  _{\rm IRAM} = 0.026$, and an average CO line width of $FWHM\sim160$~\kms, our rms goal corresponds to a $3\sigma$ upper limit on the observable CO(1-0) luminosity of $L_{\rm CO(1-0)}^{\prime} \sim 4.8 \times 10^7$~K~\kms~pc$^2$. 

The upper limits on the CO luminosity may be translated into upper limits on the molecular gas mass, once an $\rm CO(2-1)/CO(1-0)$ luminosity ratio and a CO-to-H$_2$ conversion factor are assumed. If we assume $r_{21}=L^{\prime}_{\rm CO(2-1)}/L^{\prime}_{\rm CO(1-0)}=0.8$ (appropriate for nearby normal star-forming galaxies, further discussion in $\S$~\ref{sec:lco_vs_prop}), and a Milky-Way-type CO-to-H$_2$ conversion factor of $\alpha_{\rm CO}^{MW}\equiv M_{mol}/L^{\prime}_{\rm CO(1-0)} = 4.3$~M$_{\odot}$~(K~\kms~pc$^2$)$^{-1}$ \citep{Bolatto+13}, we obtain 3$\sigma$ upper limits on the measurable molecular gas mass of $M_{mol}\sim 2.2 \times 10^8~M_{\odot}$ and $M_{mol}\sim 2 \times 10^8~M_{\odot}$ for the APEX and IRAM observations, respectively.

\subsection{APEX observations (88 galaxies)}\label{sec:APEX_obs}

\begin{figure}[tbp]
\centering
    \includegraphics[clip=true,trim=0cm 0cm 0cm 0cm,angle=90,width=0.35\textwidth,angle=270]{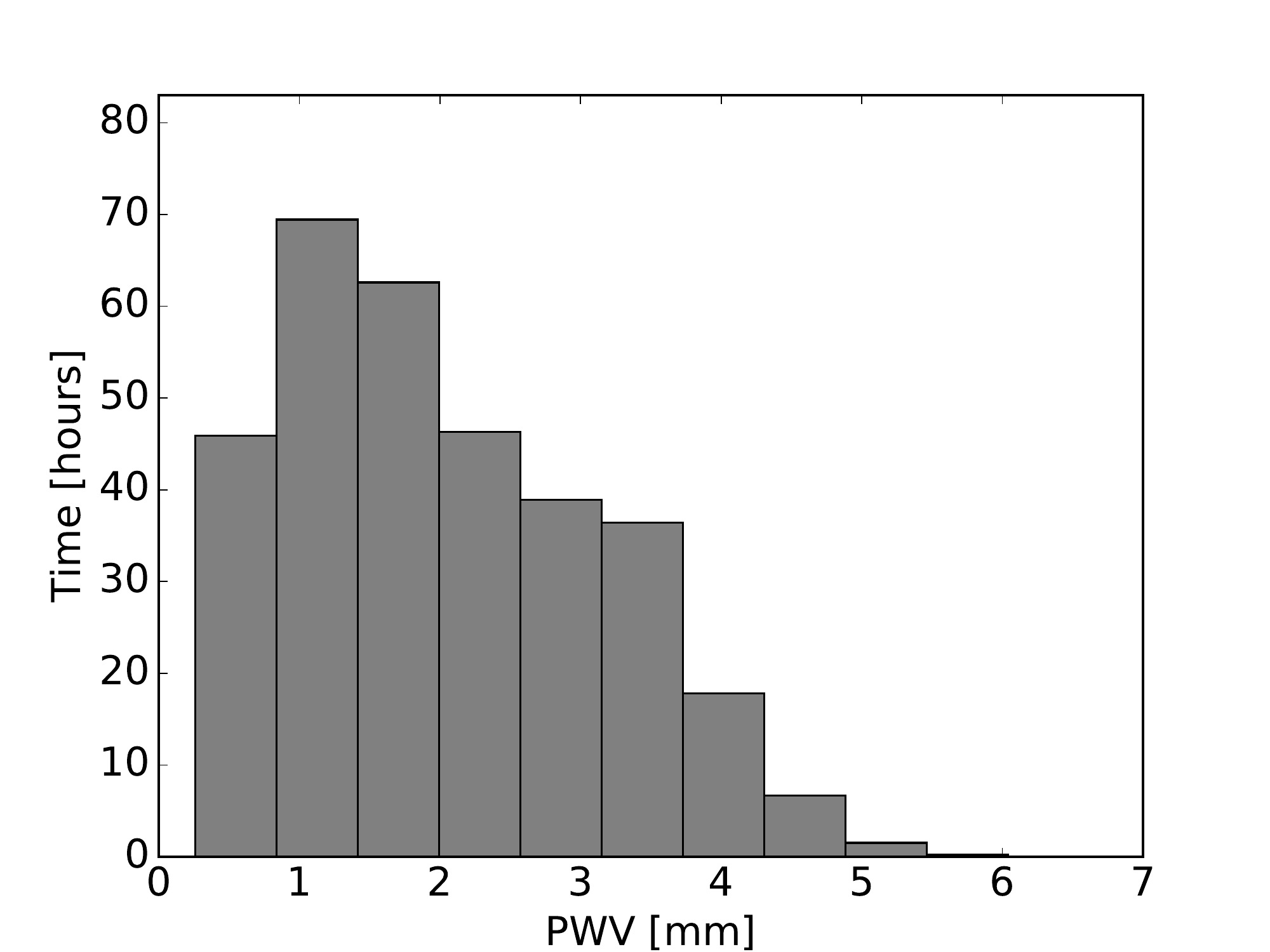}\quad    
      \caption{Distribution of precipitable water vapour conditions at the Chajnantor Plateau as measured by the APEX radiometer throughout the 327~hours of ALLSMOG observations.}
   \label{fig:PWV_stats}
\end{figure}

ALLSMOG was intended to exploit the conditions with higher atmospheric water column occurring at the Chajnantor Plateau, namely $PWV\geq 3$~mm, when the $PWV$ is prohibitive for observations at high frequencies in the sub-mm regime. However, as mentioned in $\S$~\ref{sec:sample}, the survey eventually benefited from a significant amount of `good weather' observing time. In Figure~\ref{fig:PWV_stats} we report the statistics of $PWV$ as measured by the APEX radiometer during the entire course of ALLSMOG observations ($\sim327$~hours), including the time used for setup and calibration. Fig.~\ref{fig:PWV_stats} shows that the bulk of the data were collected with $0.5<PWV~[\rm mm]<2.5$, hence demonstrating that, at the APEX telescope, even a `poor-weather' programme can profit of a significant amount of observing time under optimal weather conditions, provided the targets span a wide enough range in LST values to ease the scheduling of the observations.

The observations were performed with the APEX-1 single-pixel frontend that is part of the Swedish heterodyne facility instrument (SHeFI, \cite{Vassilev+08}).
APEX-1 offers a total bandwidth of $\Delta\nu=4$~GHz that, when the receiver is used in combination with the extended fast Fourier transform spectrometer (XFFTS2) spectral backend, is covered by two 2.5~GHz -wide spectrometer units overlapping by 1~GHz (see \cite{APECS_User_Manual_2016}). Each 2.5~GHz-wide spectral window is divided into 32,768 channels, resulting in a maximum spectral resolution of 76~kHz ($\sim0.1$~\kms~ in velocity units). The system temperature of the APEX-1 receiver in the frequency range of interest for ALLSMOG, $224\lesssim\nu\lesssim228$~GHz (LSB tuning), is $T_{sys}\sim 200-250~K$.

For each science target, we pointed the telescope at the galaxy's SDSS coordinates and tuned to the CO(2-1) frequency computed by using the optical redshift inferred from SDSS observations (further information in $\S$~\ref{sec:optdisk_par}). The sources were observed at a sky elevation between 30 and 80 deg. The typical ALLSMOG observing sequence included: (i) telescope focus correction in the $z$, $y$, $x$ directions, preferentially on a planet; (ii) telescope pointing correction using a bright source close to the target or at least at a matched elevation on sky; (iii) $1<N<10$ loops of a sequence including a minute-long calibration scan and a six-minutes long ON-OFF integration scan on the science target. Pointing was corrected every hour and focussing every 2-3 hours and after significant changes in atmospheric conditions (for example after sunset and dawn). The ON-OFF science observations were performed in the wobbler-switching symmetric mode with $60\arcsec$ chopping amplitude and a chopping rate of 1.5~Hz. The wobbler-switching mode produces much more stable spectral baselines than the position-switching mode.

Calibration was done by the online data calibrator program that automatically performs the atmospheric corrections and delivers calibrated data in $T_A^*$ units, corresponding to the antenna temperature corrected for atmospheric loss. We note that for the observations taken between March 2014 and June 2014 the $T_A^*$ values produced by the online calibrator had to be multiplied by a re-calibration factor ($\sim0.8-0.9$, depending on the exact observing date and on the frequency) to account for a small calibration error introduced by a hardware intervention to the SHeFI instrument in March 2014. The correction was applied at the data reduction stage following the instructions on the APEX website\footnote{\texttt{http://www.apex-telescope.org}}. The absolute flux calibration uncertainty of APEX-1 is typically $8\%$ \citep{Dumke+MacAuliffe10}, but it rises up to $\sim 12\%$ for the observations taken between March and June 2014.

\subsection{IRAM 30m observations (nine galaxies)}\label{sec:IRAM30m_obs}

The IRAM~30m observations were carried out on four different days: 26 Nov 2014 ($PWV\sim 2$~mm), 29 Nov 2014 ($PWV\sim6$~mm), 2 May 2015 ($PWV\sim 5$~mm), and 3 May 2015 ($PWV\sim4-8$~mm). Typical system temperatures were $T_{sys}\sim 150~K$ and $T_{sys}\sim 300~K$ respectively for the CO(1-0) and the CO(2-1) observations.

We used the eight mixer receiver (EMIR, \citealt{Carter+12}) E090 and E230 frontends tuned to 112.4~GHz and 224.8~GHz, respectively. This choice allowed us to observe all IRAM targets with the same tuning (the CO(1-0) line is placed within $112.40\pm 0.43$~GHz and the CO(2-1) line within $224.80\pm 0.85 $~GHz for all sources), hence zeroing the tuning overheads. 
The total instantaneous spectral bandwidth of EMIR is 8~GHz in dual polarisation. We used both the fast Fourier transform spectrometer (FTS) and the wide lineband multiple autocorrelator (WILMA) backends. FTS consists of two 4~GHz-wide spectrometer units (denominated upper inner (UI) and upper outer (UO)) covering the full 8~GHz IF bandwidth of EMIR at a spectral resolution of 0.195~MHz, which corresponds to $\sim0.5$~\kms~ in the 112.4 GHz band and to $\sim0.3$~\kms~ at 224.8 GHz. However, in practice, since the tuning frequencies were centred in the lower 4~GHz half of the USB, we worked only on the UI section of the FTS spectra. WILMA, which was used in parallel to double check the FTS results, has only one spectrometer (centred on the tuning frequency) and offers a bandwidth of 3.7~GHz at a 2~MHz spectral resolution, that is $\sim 5~$\kms~ and $\sim 3$~\kms~ at 112.4~GHz and 224.8~GHz respectively.

Focussing was checked every four hours on planets and pointing every two hours on nearby quasars. Typical pointing and focus corrections were respectively of $\sim2\arcsec$ and $\sim0.4~mm$. Science ON-OFF observations were performed in wobbler switching mode at a rate of 0.5~Hz and with an amplitude of $200\arcsec$. 
The output spectra obtained from the antenna are calibrated in the $T_A^*$ scale.

\subsection{Data reduction and analysis}\label{sec:datared}

We reduced the data using the \texttt{GILDAS/CLASS} software package, using a series of customised
scripts based on the statistics of the data. A similar data reduction method
was applied to both the APEX and IRAM 30m observations, and the procedure consists of
the following steps: 
\begin{enumerate}
\item For each APEX science observation subscan, we combined the two 2.5~GHz-wide spectral segments obtained from the two different spectrometer units into a single 4~GHz-wide spectrum, to cover the entire bandwidth offered by the APEX-1 receiver. To do this, we first assumed that the 1~GHz-wide overlap region receives the same signal in both units, and that the variation between them is only due to imperfect total power levels. We fitted with zero-order baselines the spectra from the two spectrometer units to match them in their overlap region, and finally averaged them together to produce a single combined spectrum for the subscan. This procedure allows us to obtain a very good alignment between the two segments of the spectrum, without creating artificial `breaks' at the edges of the central overlap region. This operation was unnecessary for the IRAM 30m FTS data, since we configured the CO lines to be placed at the centre of the UI unit and no obvious platforming effects were observed within this unit.
\item We checked each combined subscan spectrum by eye and discarded those with poor baselines showing standing waves, non-Gaussian noise patterns, or instrumental features placed close to the frequency of interest for the line detection. We rejected poor-quality subscans based on three different factors: (i) the presence of clear features or ripples visible by eye in the original high-resolution spectrum or in a rebinned (to $\delta \varv\sim10$~\kms) version of it; (ii) high (i.e. $\gg 1$) ratios
between the measured rms noise level and the theoretical value calculated using
the system temperature and integration time, and (iii) high values of the modified Allan variance factor, defined as below, which was modified from the original definition (calculated with time, see also Appendix A in \cite{Tan+11}): 
\begin{equation}\label{eq:Allan_factor}
D_{Allan} = \frac{\sigma_{rms, binned}}{\sigma_{rms, origin}}\sqrt{\frac{\delta\nu_{origin}}{\delta\nu_{binned}}},
\end{equation}
where $\sigma_{rms, origin}$ and $\sigma_{rms, binned}$ are the rms noise levels measured before and after rebinning, while $\delta\nu_{origin}$ and $\delta\nu_{binned}$ are the frequency resolutions before and after rebinning.
The Allan variance factor quantifies how much the rms noise level decreases after rebinning with respect to the theoretical value, which assumes a random white noise. In practice, this factor is significantly higher than unity when low frequency standing waves and/or $1/f$ noises dominate the baseline of the smoothed spectrum. This quality check of the individual subscans resulted in the flagging of $\sim17\%$ of the APEX data, and $<5\%$ of the IRAM~30m data.
\item At this point we collected all the subscan spectra belonging to observations targeting the same source (sometimes observed on different dates), fitted and subtracted a linear baseline individually from each of them by masking the central $\varv\in(-300, 300)$~\kms~ around the expected CO line and averaged them together to produce a single, high S/N and high resolution spectrum for each source.
\item We smoothed the high resolution spectrum to a spectral resolution of $\delta \varv = 50$~\kms~, performed a linear baseline fitting and subtraction and calculated the rms value per $\delta \varv = 50$~\kms~ channel. In this baseline fitting to the total integrated spectrum, the width of the central window masked from the fit was adjusted based on the width of the observed CO line. In case of a non-detection, we used a mask of $\varv\in(-300, 300)$~\kms. The final rms values are reported in Table~\ref{table:co_observations} for the entire ALLSMOG sample. For the non-detections, we show in Figure~\ref{fig:rms} the rms noise as a function of stellar mass and compare it with the survey rms goal. For most sources we managed to approach very closely or even surpass the rms goal, hence ensuring that the non-detections can be used to put informative upper limits on the molecular gas content in these objects.
\begin{figure}[tbp]
\centering
	\includegraphics[clip=true,trim=8cm 3cm 1.5cm 3cm,width=0.5\textwidth,angle=180]{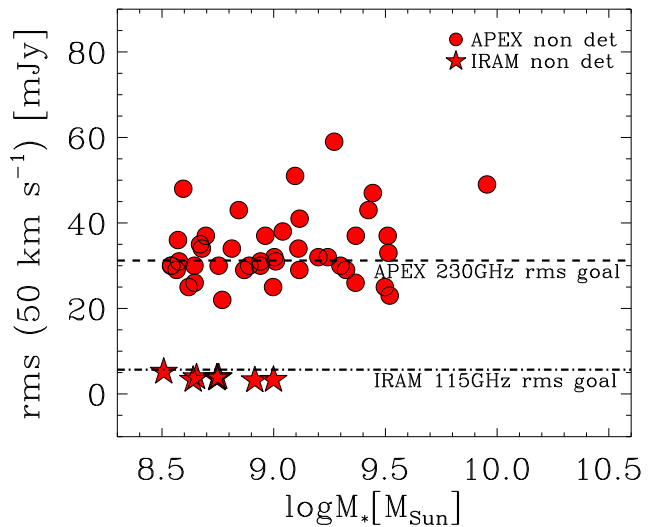}\\
      \caption{Distribution of the $1\sigma$ rms of ALLSMOG non-detections as a function of the targets' stellar masses. The rms is calculated over channels of $\delta \varv=50$~\kms. The survey rms goal for non-detections is indicated for both the APEX 230~GHz (dashed line) and the IRAM~30m~115~GHz (dot-dashed line) observations.}
   \label{fig:rms}
\end{figure}
\item We then produced the final spectrum to be used for the CO spectral analysis. For the majority of the galaxies (80/97) the final spectral analysis was performed on the same $\delta \varv = 50$~\kms~-binned spectrum used for rms calculation and obtained as described in Point 4. However for 15 (two) sources, characterised by particularly narrow (broad) CO emission, we produced a second, higher (lower) resolution spectrum to optimise the S/N of the detection. This final spectrum was imported into IDL, where we performed the remaining analysis.
\end{enumerate} 
Once imported in IDL, the CO spectra were converted from the $T_A^*$ to the flux density ($S_{\nu}$) scale using the following conversion factors: $S_{\nu}/T_A^* = 39$~Jy~K$^{-1}$ for the APEX CO(2-1) observations, $S_{\nu}/T_A^*= 5.98$~Jy~K$^{-1}$ and $S_{\nu}/T_A^*= 7.73$~Jy~K$^{-1}$ respectively for the IRAM~30m CO(1-0) and CO(2-1) observations\footnote{The corresponding conversion factors relating the main beam brightness temperature ($T_{mb} (K)$) to the flux density in Jansky can then be easily retrieved by applying the following relation: $T_{mb}=(F_{eff}/B_{eff})~T_A^*$, where $F_{eff}$ is the forward efficiency of the telescope and $B_{eff}$ is the main beam efficiency. By using the updated $B_{eff}$ and $F_{eff}$ values reported by \cite{Dumke+MacAuliffe10} we obtain $S_{\nu}/T_{mb} = 32$~Jy~K$^{-1}$ for the APEX-1 observations. The conversion factor we derive for IRAM is $S_{\nu}/T_{mb} = 4.96$~Jy~K$^{-1}$, which is of course the same at 115~GHz and 230~GHz. }. The IRAM conversion factors were computed by using the 2013 updated efficiencies reported on the IRAM website. We then performed the spectral analysis by fitting one or more Gaussian functions to the observed line profile using the function \texttt{mpfit} in IDL. In all galaxies but one (NGC2936) the CO profile is well fitted by a single Gaussian; in the case of NGC2936, two Gaussians are required to model the broad CO line emission. Table~\ref{table:co_observations} lists the best-fit parameters to the CO spectral profile for the entire ALLSMOG sample. 

For the sources where a line is only marginally detected (i.e. CO line peak signal-to noise ratio of $S/N\sim3$), we performed a sanity check by comparing the central velocity of the putative CO detection with that expected from the H{\sc i}~21cm spectrum. The H{\sc i}~21cm spectra are retrieved from public archival observations, further details will be provided in $\S$~\ref{sec:HIdata}. The motivation for this comparison is based on the notion that there is an excellent agreement between the central CO and HI velocities in local star-forming galaxies \citep{Braine+Combes93,Leroy+09,Schruba+11}.
For a few galaxies with a marginal detection\footnote{Namely, UGC00272, UGC09359, 2MASXJ1110+0411, CGCG078-021, PGC3127469, and VPC0873.}, we discarded the hypothesis of a CO detection because the central velocity of the putative CO line was found to be inconsistent with that of the H{\sc i}~21cm line. In the other cases, we confirmed the detection based on the consistency between the CO and H{\sc i} velocities and their similar spectral profiles. We acknowledge that a galaxy may display very different CO and H{\sc i} line profiles as a consequence of the different spatial distribution and extent of atomic and molecular gas. However, we still judge the comparison with the H{\sc i}~21cm spectrum a valuable first-order check to confirm marginal detections in CO, especially if dealing with single-dish observations that may be subject to baseline instabilities. The final CO detection rate in ALLSMOG is $\sim47$~\% (i.e. 46/97 detections). 

For non-detections, the 3~$\sigma$ upper limit on the total integrated line flux listed in Table~\ref{table:co_observations} were estimated as follows:
\begin{equation}\label{eq:COul}
\int S_{\rm CO}d\varv < 3 \sigma_{rms, channel} \sqrt{\delta \varv_{channel} \Delta \varv_{line}},
\end{equation}
where $\sigma_{rms, channel}$ is the rms noise per spectral channel (listed in Table~\ref{table:co_observations}), $\delta \varv_{channel}$ is the channel width ($=50~$\kms), and $\Delta \varv_{line}$ is the expected CO line width. We assumed $\Delta \varv_{line}$ equal to the average CO FWHM calculated using the CO detections, that is $\Delta \varv_{line}\sim \langle FWHM_{\rm CO}\rangle_{det} \sim 160$~\kms. 

In Appendix~\ref{sec:appendix_spectra} we show, for each galaxy in ALLSMOG, the SDSS {\it g r i} composite cutout optical image and the CO spectrum. The dashed grey circle placed over the SDSS thumbnail image shows the beam of our CO observations. At the frequency of the CO(2-1) line redshifted for ALLSMOG sources, the size of the APEX beam is $27\arcsec$ (FWHM), whereas the beamsize of the IRAM 30m dish is $22\arcsec$ for the CO(1-0) observations and $11\arcsec$ for the CO(2-1) observations.
In the sources with a CO detection, we overlay on the data shown in Figs.~\ref{fig:spectra1}-\ref{fig:spectra10} the best fit to the observed spectral profile. Below each CO spectrum we plot the corresponding H{\sc i}~21cm spectrum (see $\S$~\ref{sec:HIdata}), renormalised to an arbitrary scale for visualisation purposes.

\longtab{
\begin{longtable}{@{}llcccccl@{}}
\caption{Details of the CO observations}
\label{table:co_observations} \\
\hline
\hline
 ID			& Galaxy Name			& $t_{\rm ON}$ 	   	& rms (50 km s$^{-1}$)	   &  $\varv_{0}$  		& $\sigma_{\varv}$   	& $S_{\rm peak}$  & $\int S_{\rm CO} d\varv$	\\
		    &						& [min]			& [mJy]    				   &  [km s$^{-1}$]	& [km s$^{-1}$] & [mJy]		  & [Jy km s$^{-1}$]  \\
(1)			& (2)					& (3)		   	& (4)	   					& (5)			& (6)			& (7)			& (8)		 \\	
\hline
\endfirsthead
\multicolumn{8}{c}{\footnotesize Following from previous page}\\
\hline	
 ID			& Galaxy Name			& $t_{\rm ON}$ 	   	& rms (50 km s$^{-1}$)	   &  $\varv_{0}$  		& $\sigma_{\varv}$   	& $S_{\rm peak}$  & $\int S_{\rm CO} d\varv$	\\
		    &						& [min]			& [mJy]    &  [km s$^{-1}$]	& [km s$^{-1}$] & [mJy]		  & [Jy km s$^{-1}$]  \\
(1)			& (2)					& (3)		   	& (4)	   & (5)			& (6)			& (7)			& (8)		 \\	
\hline
\endhead
\hline
\multicolumn{8}{c}{\footnotesize Follows on next page}\\
\endfoot
\hline
\endlastfoot
\hline
\multicolumn{8}{c}{APEX CO(2-1) observations}\\
\hline
1  &	UGC11631	& 80   &    35 	&   -31 $\pm$14 	&    71  $\pm$  14 	&  170 $\pm$   30 	&    30 $\pm$ 8    		\\ 
2  &	UGC00272 &    	113  &    26 	&    --- 		     	&   --- 				&   --- 				&    $<$7   		 	\\ 
3  &	IC0159		 & 	46    &    40 	&   -12 $\pm$12 	&    67  $\pm$  12 	&  210 $\pm$   30 	&    35 $\pm$ 8   		\\ 
4  &	KUG0200-101   &     30    &    41 	&     --- 			&     --- 				&    --- 				&    $<$11    			\\ 
5  &	NGC1234	&     118   &    25 	&   -32 $\pm$10 	&    49  $\pm$  10 	&  130 $\pm$   20 	&    16 $\pm$4    		\\ 
6  &	UGC06838 &     41    &     24 	&  -100 $\pm$ 30 &    190  $\pm$  30 	&   75 $\pm$   11 		&    36 $\pm$8    		\\ 
7  &	UGC09359 &   	28    &    	37 	&    --- 			&     --- 				&    --- 				&    $<$10  			\\ 
8  &	2MASXJ2235-0845 & 70   &    28 	&     0 $\pm$ 6 	&    45  $\pm$ 6 		&  250 $\pm$   30 	&    28$\pm$5    		\\   
9  &	NGC5414			   &     19    &    72 &     11 $\pm$    7 &    47  $\pm$  7 &  450 $\pm$  60 &    52 $\pm$    11    		\\ 
10 &	NGC5405			   &     14   &    75 &     13 $\pm$    3 &    14  $\pm$   3 &  610 $\pm$  130 &    21 $\pm$     7    \\ 
11 &	UGC10005		   &    57    &    32 &     11 $\pm$   10 &    24  $\pm$  10 &  110 $\pm$   40 &     6 $\pm$     4    \\ 
12 &	MCG+00-25-005	   &     10    &    70 &     4 $\pm$   13 &    71  $\pm$  13 &  330 $\pm$   50 &    60 $\pm$    15    \\ 
13 &	UGC02004		   &     14    &   92 	&   -28 $\pm$   13 &    90  $\pm$  13 &  530 $\pm$   60 &   120 $\pm$    20    \\ 
14 &	UGC02529		   &     36    &    43 &     --- 			&     --- 			&    --- 				&    $<$11     \\ 
15 &	MCG+00-29-013      &     26    &    51 &    -8 $\pm$   12 &    71  $\pm$  12 &  260 $\pm$   40 &    47 $\pm$    11    \\ 
16 &	UGC06329           	   &     47    &    33 &   -20 $\pm$   18 &    48  $\pm$  18 &  100 $\pm$   30 &    12 $\pm$     6    \\ 
17 &	UGC05648               &     36    &    43 &     -10 $\pm$   30 &   120  $\pm$  30 &  140 $\pm$   30 &    42 $\pm$    12    \\ 
18 &	IC0605             	   &     28    &    46 &     3 $\pm$   14 	&    67  $\pm$  14 &  210 $\pm$   40 &    36 $\pm$    10    \\ 
19 &	2MASXJ0910+0752 &     14    &    58 &     0 $\pm$   18 &    80  $\pm$  19 &  210 $\pm$   40 &    42 $\pm$    13    \\ 
20 &	2MASXJ0939+0624 &     31    &    35 &     0 $\pm$   20 &   120  $\pm$  20 &  120 $\pm$   20 &    34 $\pm$     9    \\ 
21 &	2MASXJ0955+0632 &     30    &    35 &    19 $\pm$   18 &    68  $\pm$  18 &  130 $\pm$   30 &    22 $\pm$     7    \\ 	
22 &	2MASXJ1014+0748 &     25    &    37 &     --- 			&     --- 			&    --- 				&    $<$10  		  \\ 
23 &	2MASXJ1011+0746 &    46    	 &    34 &    20 $\pm$   20 &   100  $\pm$  20 &  120 $\pm$   20 &    32 $\pm$    9    \\ 
24 &	PGC031905             &     13    &    55 &    -14 $\pm$   9 &    30  $\pm$  11 	&  240 $\pm$   70 &    18 $\pm$    8    \\  
25 &	PGC031382             &     17    &    49 &     --- 			&     ---				&    -- 				&    $<$13   \\  
26 &	2MASXJ1110+0411  &     47    &    33 &     --- 			&     --- 			&    --- 				&    $<$9   \\ 
27 &	2MASXJ0855+0345  &     45   &    41 &   140 $\pm$ 30 &   110  $\pm$  30 &  100 $\pm$   20 &    27 $\pm$    10    \\ 
28 &	2MASXJ0839+0349   &     12   &    71 &    20 $\pm$   30 &   90  $\pm$  30 &  190 $\pm$   50 &    45 $\pm$    17    \\ 
29 &	UGC04977           		&    50    &    29 &   -14 $\pm$   16 &   98  $\pm$  16 &  138 $\pm$   19 &    34 $\pm$     7    \\ 
30 &	2MASXJ0846+0230    &     19    &   46 &   -12 $\pm$   10 &    37  $\pm$  9 &  230 $\pm$   50 &    21 $\pm$     7    \\ 
31 &	SDSSJ1328-0202      &     23    &    47 &   -28 $\pm$   17 &    51  $\pm$  17 &  150 $\pm$  40 &    19 $\pm$     8    \\ 
32 &	MCG+00-34-038      &     9    &    65 &    16 $\pm$   13 &    81  $\pm$  13 &  350 $\pm$   50 &    70 $\pm$    15    \\ 
33 &	UGC08526           &     22    &    42 &    -4 $\pm$   10 &    63  $\pm$  10 &  250 $\pm$   30 &    40 $\pm$     8    \\ 
34 &	CGCG017-017        &    55    &    33 &    34 $\pm$   14 &    84  $\pm$  14 &  170 $\pm$   20 &    35 $\pm$     8  	\\ 
35 &	NGC2936 (narrow)	       	&     11    	&    72 		&   216 $\pm$    4 &   34  $\pm$   4 &  810 $\pm$   80 &   69 $\pm$    11   	\\ 
     & 	NGC2936 (broad)	      		&          	&    		&   102 $\pm$    15 &   193  $\pm$  12 &  580 $\pm$   40 &   280 $\pm$    30   	\\ 
36 &	SDSSJ0937+0927     &    51    &    30 &     --- &     --- &   --- &     $<$8  	\\ 
37 &	SDSSJ0950+1118    	&   83	  &   31  &    --- &    ---  &   ---  &    $<$8 	\\ 
38 &	IC3069            			&    41 	&   25  &     --- &    ---  &   ---  &    $<$7    \\ 
39 &	SDSSJ1112+0931    	&    42 	&   30  &     --- &     ---  &   ---  &    $<$8  \\ 
40 &	SDSSJ0805+0659    	&   66 	&   25  &     --- &   ---  &   ---  &    $<$7    \\ 
41 &	SDSSJ1104+0507    	&   51 	&   34  &     --- &    ---  &   ---  &   $<$9  \\ 
42 &	CGCG050-042       	&   72 	&   30  &     --- &    ---  &   ---  &    $<$8    \\ 
43 &	2MASXJ0858+0345   	&    43 	&   27  &     4 $\pm$   14 &   66   $\pm$ 14  & 120  $\pm$  20  &   20  $\pm$    6    \\ 
44 &	SDSSJ1008+1428    	&    26 	&   43  &     --- &    ---  &   ---  &   $<$11  \\ 
45 &	SDSSJ0954+0458    	&    34 	&   26  &     --- &    ---  &   ---  &    $<$7   \\ 
46 &	SDSSJ1213+1056	&    30 	&   48  &     --- &    ---  &   ---  &   $<$13    \\ 
47 &	SDSSJ0945+0515	 &   22 	&   30  &     --- &    ---  &   ---  &   $<$8   \\ 
48 &	SDSSJ1122+1316	  	&    25 	&   37  &     --- &    ---  &   ---  &   $<$10   \\ 
49 &	SDSSJ1403+1003	 &   51	&   30  &     --- &    ---  &   ---  &    $<$8     \\ 
50 &	CGCG063-006		 &   54 	&   31  &     --- &    ---  &   ---  &    $<$8   \\ 
51 &	2MASXJ0843+1303	  &  65 	&   22  &     --- &    ---  &   ---  &    $<$6      \\ 
52 &	SDSSJ0920+0759	  &  80 &   29  &     --- &    ---  &   ---  &    $<$8   \\ 
53 &	CGCG080-042		  &   51 &   37  &    --- &    ---  &   --- &    $<$10  \\ 
54 &	CGCG078-021		  &   42 &   25  &     --- &    ---  &   ---  &    $<$7   \\ 
55 &	UGC04567		  	 &    29 &   33  &    10 $\pm$   40 &  100   $\pm$ 40  &  70  $\pm$  20  &   16  $\pm$    8    \\ 
56 &	SDSSJ1624+1251	  &    25 &   30  &   -60 $\pm$   50 &  100   $\pm$ 50  &  50  $\pm$  20  &   12  $\pm$    8    \\ 
57 &	CGCG058-066		  &   71 &   27  &     9 $\pm$   16 &   60   $\pm$ 16  & 100  $\pm$  20  &   15  $\pm$    5    \\ 
58 &	CGCG061-003		  &   93 &   29  &     --- &    ---  &   ---  &    $<$8   \\ 
59 &	CGCG051-037		  &  109 &   34 &     --- &    ---  &   ---  &   $<$9    \\ 
60 &	SDSSJ0854+0418	  &   62 &   25  &     -1 $\pm$  13 &   32  $\pm$ 9  &  100  $\pm$  20  &    8  $\pm$    3    \\ 
61 &	CGCG059-031		  &    21 &   59  &     --- &    ---  &   ---  &   $<$16   \\ 
62 &	2MASXJ0806+1249	  &   68 &   30  &     0 $\pm$    5 &   11   $\pm$  5  & 150  $\pm$  60  &    4  $\pm$   3    \\ 
63 &	VIIIZW039		  		 &   66 	&   26  &   -13 $\pm$   7 &   43   $\pm$  7  & 180  $\pm$  30  &   20  $\pm$    4    \\ 
64 &	2MASXJ0941+1056	 &    46	 &   33  &    21 $\pm$   13 &   31   $\pm$ 13  & 110  $\pm$  40  &    9  $\pm$   5    \\ 
65 &	SDSSJ0943+0356	  &    25 	&   38  &     --- &    ---  &   ---  &   $<$10   \\ 
66 &	CGCG035-063		  &   84 	&   47  &     --- &    ---  &   ---  &   $<$13    \\ 
67 &	CGCG036-048		  &   95 	&   32  &     --- &    ---  &   ---  &    $<$9    \\ 
68 &	SDSSJ1625+1142	  	&   61 	     &   51  &     --- &    ---  &   ---  &   $<$14   \\ 
69 &	SDSSJ1028+0424	 &    31     &   32  &     --- &    ---  &   ---  &    $<$9     \\ 
70 &	UGC06011		  		&   54 	&   29  &     --- &    ---  &   --- &    $<$8  \\ 
71 &	PGC051743 		  	&    20 	&   49  &    18 $\pm$ 7 &   38   $\pm$  7  & 350  $\pm$  60  &   33  $\pm$    9    \\   
72 &	2MASXJ1437+0500   	&    22 	&   43  &   -40 $\pm$ 20 &   80   $\pm$ 20  &  80  $\pm$  20  &   17 $\pm$   6    \\	
73 &	SDSSJ0944+0400	 &    37 &   32  &    24 $\pm$  9 &   16   $\pm$  6  &  130  $\pm$  40  &    5  $\pm$    3    \\ 
74 &	2MASXJ1048+1201	 &    31 &   32  &     --- &    ---  &   ---  &    $<$9     \\ 
75 &	SDSSJ1110+1345	  	&    41 	&   29  &     --- &    ---  &   ---  &    $<$8     \\ 
76 &	PGC012214		  	&    38 	&   30  &    --- &    ---  &   ---  &    $<$8   \\ 
77 &	PGC003530		  	&    45 	&   34  &     --- &    ---  &   ---  &    $<$9    \\ 
78 &	PGC073268		  	&   118 &   27  &    60 $\pm$   30 &   70   $\pm$ 30  &  70  $\pm$  20  &   12  $\pm$    6    \\ 
79 &	PGC1452135		  	&   77 	&   30  &     --- &    ---  &   ---  &    $<$8  \\ 
80 &	PGC3127469		  	&   67 	&   36  &     --- &    ---  &   ---  &   $<$10    \\ 
81 &	UGC00317		  	&   81 	&   30  &     3 $\pm$   17 &   31   $\pm$ 12  &  90  $\pm$  30  &    7  $\pm$    4    \\ 
82 &	SDSSJ0011+1428    	&    33 	&   35  &     --- &    ---  &   ---  &    $<$9     \\ 
83 &	PGC1455779		  	&   68 	&   23  &     --- &    ---  &   ---  &    $<$6   \\
84 &	PGC000010		  	&    20 	&   62  &     0 $\pm$   6 &   25   $\pm$  6  & 380  $\pm$  80  &   24  $\pm$    8    \\ 
85 &	PGC1446233		  	&    41 	&   34  &   -16 $\pm$  19 &  51   $\pm$ 19  &  100  $\pm$  30  &   13  $\pm$    6    \\ 
86 &	2MASXJ0020+1413   	&   87 	&   29  &    -8 $\pm$    7 &   25   $\pm$  7  & 180  $\pm$  40  &   11  $\pm$    4    \\ 
87 &	PGC1464874		  	&   74 	&   31  &     --- &   ---  &   ---  &    $<$8  \\ 
88 &    IC1706			  		&    12 	&   87  &   -38 $\pm$   17 &   33   $\pm$ 15  & 210  $\pm$  80  &   17  $\pm$   11    \\ 
\hline                            
\multicolumn{8}{c}{IRAM CO(1-0) observations}\\
\hline
89 & 	SDSSJ0944+1116	    & 72  &  3.3 &   --- &   ---  &  --- & $<$0.6 \\ 
90 & 	SDSSJ1049+1108	    & 90  &  3.2 &   --- &   ---  &  --- & $<$0.6 \\ 
91 & 	2MASXJ1336+1552  &  39  &  5.5 &  66 $\pm$ 8 &  33 $\pm$ 8  & 32 $\pm$ 7 & 2.6 $\pm$   0.8	\\ 
92 & 	SDSSJ1032+1227    & 72  &  3.9 &   --- &   ---  &  --- & $<$0.7 	\\ 
93 & 	SDSSJ1100+1207	    & 75  &  3.6 &   --- &   ---  &  --- & $<$0.7	\\ 
94 & 	SDSSJ1207+1200   & 73  &  3.9 &   --- &   ---  &  --- & $<$0.7	\\ 
95 & 	KUG1147+149		   & 84  &  3.9 &   --- &   --- &  --- & $<$0.7 \\ 
96 & 	SDSSJ1320+1524   & 83  &  5.2 &   --- &   ---  &  --- & $<$1.0	\\ 
97 & 	VPC0873			& 72  &  3.3 &  --- &   ---  &  --- & $<$0.6	\\ 
\hline                            
\multicolumn{8}{c}{IRAM CO(2-1) observations}\\
\hline
91 & 	2MASXJ1336+1552		&  39  &  11.4 &   68 $\pm$ 8 &  36 $\pm$ 8  & 62 $\pm$ 12 & 5.6 $\pm$ 1.7	\\ 
\hline
\end{longtable}  

\tablefoot{Col. (1): ALLSMOG ID. Col (2): Galaxy name. Col (3): On-source time. For the IRAM~30m CO(1-0) observations the `human' on source time is half the value listed in the table, because the EMIR receiver allows simultaneous dual-polarisation observations that were averaged together to produce the final spectrum. Col (4): 1$\sigma$ spectral rms calculated in channels of $\delta \varv = 50$~\kms. Col. (5) Central velocity of the CO emission line with respect to the optical redshift (inferred from SDSS observations) as derived through a single-Gaussian fitting. Col. (6): Velocity dispersion of the observed CO line derived through a single-Gaussian fitting. Col. (7): Amplitude of the Gaussian function fitted to the observed CO line profile. Col. (8): Total velocity-integrated flux of the Gaussian function fitted to the observed CO line profile. For the non-detections, we list the $3\sigma$-upper limit on the total integrated CO flux derived using Eq~\ref{eq:COul}. }

}

\subsection{Aperture correction}\label{sec:ap_cor}

The CO integrated fluxes measured by Gaussian fitting to the observed line profile (listed in Table~\ref{table:co_observations}) were corrected to account for possible flux losses due to any potential CO emission lying outside the telescope primary beam. We used the same technique as in \cite{Bothwell+14}, which consists in estimating the CO covering fraction of the APEX and IRAM beams (at the frequencies of interest for our observations), that is the ratio between the observed CO flux and the true total CO flux. This is done by integrating over the beam area an exponential disk model for CO emission, by taking into account the inclination of the galaxy optical disk ($i$, see $\S$~\ref{sec:optdisk_par}). The model uses the apparent optical diameter of the galaxy quantified by $d_{25}$ ($\S$~\ref{sec:optdisk_par}) and the relation between the CO exponential disk scaling length and $d_{25}$ found by \cite{Leroy+09} using resolved CO observations. We refer to \cite{Bothwell+14} for a detailed explanation of the method. The resulting CO coverage fractions for ALLSMOG galaxies are typically very close to unity (the median value is $0.98$), implying very small aperture corrections, and they are listed in Table~\ref{table:co_luminosity}. Table~\ref{table:co_luminosity} lists also the resulting CO fluxes corrected for beam coverage ($\int S_{\rm CO} d\varv~^{cor}$). The errors on $\int S_{\rm CO} d\varv~^{cor}$ were estimated by propagating the observational uncertainty on $\int S_{\rm CO} d\varv~$ and the uncertainty on the CO coverage fraction.

\longtab{
\begin{longtable}{@{}llclc@{}}
\caption{CO luminosity values}
\label{table:co_luminosity} \\
\hline
\hline
 ID			& Galaxy Name			& Beam coverage	& $\int S_{\rm CO} d\varv~^{cor}$ 	& $L_{\rm CO}^{'}$ 			\\
		    &						& (fraction)	& [Jy km s$^{-1}$] 			&  [$10^8$ K km s$^{-1}$ pc$^2$]	\\
(1)			& (2)					& (3)		   	& (4)		 				& (5)						\\	
\hline
\endfirsthead
\multicolumn{5}{c}{\footnotesize Following from previous page}\\
\hline	
 ID			& Galaxy Name			& Beam coverage	& $\int S_{\rm CO} d\varv~^{cor}$	& $L_{\rm CO}^{'}$ 			\\
		    &						& (fraction)	& [Jy km s$^{-1}$] 			&  [$10^8$ K km s$^{-1}$ pc$^2$]	\\
(1)			& (2)					& (3)		   	& (4)		 				& (5)						\\	
\hline
\endhead
\hline
\multicolumn{5}{c}{\footnotesize Follows on next page}\\
\endfoot
\hline
\endlastfoot
\hline
\multicolumn{5}{c}{APEX CO(2-1) Observations}\\
\hline
1  &	UGC11631		    &  0.90  $\pm$   0.06 &		  33  	 $\pm$     9   &     0.8  	$\pm$  0.2			 \\		  
2  &	UGC00272	    &  0.80  $\pm$   0.08 &		 $<$9    			   &     $<$0.18               		\\    
3  &	IC0159			    &  0.77  $\pm$   0.08 &		  45  	 $\pm$    11   &     0.9  	$\pm$  0.2               \\     
4  &	KUG0200-101	    &  0.96  $\pm$   0.03 &		  $<$11 			   &     $<$0.23  	              		 \\     
5  &	NGC1234		    &  0.72  $\pm$   0.09 &		  22  	 $\pm$     6   &     0.42  $\pm$  0.12               \\     
6  &	UGC06838	    &  0.95  $\pm$   0.04 &		  38  	 $\pm$     9   &     0.77  $\pm$  0.17               \\     
7  &	UGC09359	    &  0.93  $\pm$   0.04 &		  $<$11  	   	   	   &     $<$0.24  	              			\\  
8  &	2MASXJ2235-0845     &  0.97  $\pm$   0.03 &		  29  	 $\pm$     5   &    	 2.0  $\pm$  0.4               \\   
9  &	NGC5414			    &  0.86  $\pm$   0.07 &		  60  	 $\pm$    14   &     1.5  $\pm$  0.3              \\   
10 &	NGC5405			    &  0.84  $\pm$   0.09 &		  25  	 $\pm$     9   &     1.6  $\pm$  0.6               \\   
11 &	UGC10005		    &  0.90  $\pm$   0.06 &		   7  	 $\pm$     4   &     0.13  $\pm$  0.09               \\   
12 &	MCG+00-25-005	    &  0.93  $\pm$   0.05 &		  64  	 $\pm$    16   &     2.1  $\pm$  0.5               \\   
13 &	UGC02004		    &  0.94  $\pm$   0.04 &		 128  	 $\pm$    22   &     7.5  $\pm$  1.3               \\   
14 &	UGC02529		    &  0.981  $\pm$   0.019 &		  $<$12  	           &     $<$0.9  		                \\   
15 &	MCG+00-29-013       &  0.95  $\pm$   0.04 &		  49  	 $\pm$    12   &     3.9  $\pm$  0.9               \\   
16 &	UGC06329               &  0.88  $\pm$   0.07 &		   14  	 $\pm$     7   &     1.0  $\pm$  0.5               \\   
17 &	UGC05648               &  0.85  $\pm$   0.07 &		  49  	 $\pm$    15   &     3.2  $\pm$  0.9               \\   
18 &	IC0605              	   &  0.91  $\pm$   0.06 &		  40  	 $\pm$    11   &     2.3  $\pm$  0.6               \\   
19 &	2MASXJ0910+0752  &  0.97  $\pm$   0.02 &		  43  	 $\pm$    13   &     4.3  $\pm$  1.3               \\   
20 &	2MASXJ0939+0624     &  0.993  $\pm$   0.009 &		  34  	 $\pm$     9   &     2.6  $\pm$  0.7               \\   
21 &	2MASXJ0955+0632     &  0.988  $\pm$   0.014 &		  22  	 $\pm$     7   &     1.6  $\pm$  0.5              \\   
22 &	2MASXJ1014+0748     &  0.98  $\pm$   0.02 &		  $<$10  	           &     $<$1.0  	              \\        
23 &	2MASXJ1011+0746     &  0.96  $\pm$   0.03 &		  33  	 $\pm$    9   &     2.8  $\pm$  0.8               \\   
24 &	PGC031905           &  0.82  $\pm$   0.08 &		  22  	 $\pm$    10   &     1.1  $\pm$  0.5               \\   
25 &	PGC031382           &  0.94  $\pm$   0.04 &		  $<$14  	           &     $<$1.3  	                  \\    
26 &	2MASXJ1110+0411     &  0.993  $\pm$   0.010 &		  $<$9                &     $<$0.9         \\      
27 &	2MASXJ0855+0345     &  0.983  $\pm$   0.017 &		  27  	 $\pm$    10   &     2.6  $\pm$  1.0               \\    
28 &	2MASXJ0839+0349     &  0.989  $\pm$   0.013 &		  46  	 $\pm$    17   &     4.0  $\pm$  1.5               \\    
29 &	UGC04977            &  0.93  $\pm$   0.05 &		  37  	 $\pm$     8   &     3.5  $\pm$  0.7               \\    
30 &	2MASXJ0846+0230     &  0.992  $\pm$   0.011 &		  21  	 $\pm$     7   &     2.1  $\pm$  0.7               \\    
31 &	SDSSJ1328-0202      &  0.94  $\pm$   0.04 &		  20  	 $\pm$     9   &     0.38  $\pm$  0.16               \\  
32 &	MCG+00-34-038       &  0.94  $\pm$   0.04 &		  74  	 $\pm$    16   &     4.1  $\pm$  0.9               \\    
33 &	UGC08526            &  0.87  $\pm$   0.07 &		  46  	 $\pm$    10   &     0.90  $\pm$  0.19               \\    
34 &	CGCG017-017         &  0.80  $\pm$   0.08 &		  44  	 $\pm$    11   &     1.1  $\pm$  0.3               \\    
35 &	NGC2936 	        	&  1.012  $\pm$   0.017 &		 350  	 $\pm$    30   &    	24  $\pm$  2               \\    
36 &	SDSSJ0937+0927      &  0.95  $\pm$   0.03 &		   $<$8  			   &     $<$0.52  	                \\       
37 &	SDSSJ0950+1118    	&  0.998  $\pm$   0.005 &		   $<$8   			   &     $<$0.36      \\        
38 &	IC3069            			&  0.986  $\pm$   0.015 &		   $<$7  	 		   &     $<$0.31             \\      
39 &	SDSSJ1112+0931    	&  0.96  $\pm$   0.04 &		   $<$8                &     $<$0.44             \\      
40 &	SDSSJ0805+0659    	&  0.995  $\pm$   0.007 &		   $<$7                &     $<$0.19             \\      
41 &	SDSSJ1104+0507    	&  0.9999  $\pm$   0.0014 &		   $<$9               &     $<$0.34                \\       
42 &	CGCG050-042       	&  0.98  $\pm$   0.02 &		   $<$8                &     $<$0.12                \\       
43 &	2MASXJ0858+0345   	&  0.991  $\pm$   0.013 &		  20  	 $\pm$     6   &     1.8  $\pm$  0.5              \\    
44 &	SDSSJ1008+1428    	&  0.996  $\pm$   0.006 &		  $<$12    			   &     $<$0.57              \\      
45 &	SDSSJ0954+0458    	&  0.9998  $\pm$   0.0012 &		  $<$7   			   &     $<$0.25              \\      
46 &	SDSSJ1213+1056	  	&  0.9998 $\pm$   0.0014 &		  $<$13                &     $<$0.6            \\         
47 &	SDSSJ0945+0515	  	&  1.0001  $\pm$   0.0011 &		  $<$8                &     $<$0.44                  \\      
48 &	SDSSJ1122+1316	  	&  0.998  $\pm$   0.004 &		  $<$10                &     $<$0.51                \\         
49 &	SDSSJ1403+1003	  	&  0.991  $\pm$   0.011 &		   $<$8  	           &     $<$0.18  	            \\         
50 &	CGCG063-006		  	&  0.91  $\pm$   0.05 &		   $<$9  	           &     $<$0.27  	                \\   
51 &	2MASXJ0843+1303	  	&  0.997  $\pm$   0.004 &		   $<$6  	           &     $<$0.14  	              \\    
52 &	SDSSJ0920+0759	  	&  0.997  $\pm$   0.004 &		   $<$8                &     $<$0.43                 \\      
53 &	CGCG080-042		  	&  0.95  $\pm$   0.04 &		  $<$10                &     $<$0.37            \\        
54 &	CGCG078-021		  	&  0.96  $\pm$   0.03 &		   $<$7                &     $<$0.32             \\         
55 &	UGC04567		  		&  0.95  $\pm$   0.04 &		  17  	 $\pm$     8   &     0.38  $\pm$  0.19              \\   
56 &	SDSSJ1624+1251	  	&  0.97  $\pm$   0.02 &		  12  	 $\pm$     8   &     0.4  $\pm$  0.3              \\     
57 &	CGCG058-066		  	&  0.94  $\pm$   0.04 &		  16  	 $\pm$     5   &     0.46  $\pm$  0.15              \\   
58 &	CGCG061-003		  	&  0.95  $\pm$   0.04 &		   $<$8  	           &     $<$0.18  	               \\         
59 &	CGCG051-037		  	&  0.95  $\pm$   0.04 &		  $<$10  	           &     $<$0.36                \\         
60 &	SDSSJ0854+0418	  	&  0.98  $\pm$   0.02 &		   8  	 $\pm$     3   &     0.6  $\pm$  0.2              \\       
61 &	CGCG059-031		  	&  0.996  $\pm$   0.006 &		  $<$16  	           &     $<$0.46               \\         
62 &	2MASXJ0806+1249	  	&  0.995  $\pm$   0.008 &		   4  	 $\pm$     3   &     0.19  $\pm$  0.11              \\     
63 &	VIIIZW039		  			&  0.996  $\pm$   0.007 &		  20  	 $\pm$     4   &     0.83  $\pm$  0.17              \\     
64 &	2MASXJ0941+1056	  	&  1.000  $\pm$   0.014 &		   9  	 $\pm$     5   &     0.4  $\pm$  0.2              \\     
65 &	SDSSJ0943+0356	  	&  0.995  $\pm$   0.007 &		  $<$10                &     $<$0.48              \\ 
66 &	CGCG035-063		  	&  0.988  $\pm$   0.015 &		  $<$13  			   &     $<$0.6             \\  
67 &	CGCG036-048		  	&  0.991  $\pm$   0.011 &		   $<$9  			   &     $<$0.20             \\  
68 &	SDSSJ1625+1142	  		&  0.989  $\pm$   0.012 &		  $<$14  			   &     $<$0.44            \\   
69 &	SDSSJ1028+0424	  	&  0.989  $\pm$   0.013 &		   $<$9  			   &     $<$0.55            \\   
70 &	UGC06011		  			&  0.95  $\pm$   0.03 &		  $<$8  			   &     $<$0.34         \\       
71 &	PGC051743 		  		&  0.97  $\pm$   0.03 &		  34  	 $\pm$     9   &     3.0  $\pm$  0.8              \\   
72 &	2MASXJ1437+0500   		&  0.98  $\pm$   0.02 &		  17  	 $\pm$    6   &     1.4  $\pm$  0.5              \\   
73 &	SDSSJ0944+0400	  	&  1.0000  $\pm$   0.0019 &     5 	 $\pm$     3   &     0.25  $\pm$  0.12              \\ 
74 &	2MASXJ1048+1201	  	&  0.994  $\pm$   0.010 &		   $<$9  	           &     $<$0.50             \\     
75 &	SDSSJ1110+1345	  		&  0.993  $\pm$   0.009 &		   $<$8  	           &     $<$0.20            \\     
76 &	PGC012214		  		&  0.97  $\pm$   0.02 &		   $<$8  	           &     $<$0.50            \\     
77 &	PGC003530		  		&  0.998  $\pm$   0.003 &		   $<$9  	           &     $<$0.35              \\    
78 &	PGC073268		  		&  0.97  $\pm$   0.02 &		  12  	 $\pm$     6   &     0.5  $\pm$  0.2              \\   
79 &	PGC1452135		  		&  0.980  $\pm$   0.02 &		   $<$8  	           &     $<$0.25              \\     
80 &	PGC3127469		  		&  0.992  $\pm$   0.011 &		   $<$10  	           &     $<$0.38               \\     
81 &	UGC00317		  		&  0.96  $\pm$   0.04 &		   7  	 $\pm$     4   &     0.29  $\pm$  0.16              \\     
82 &	SDSSJ0011+1428    		&  0.9998  $\pm$   0.0013 &		   $<$9  	           &     $<$0.34  	           \\      
83 &	PGC1455779		  		&  0.98  $\pm$   0.02 &		   $<$6  	           &     $<$0.44              \\      
84 &	PGC000010		  		&  0.97  $\pm$   0.03 &		  25  	 $\pm$     8   &     1.7  $\pm$  0.6              \\  
85 &	PGC1446233		  		&  0.989  $\pm$   0.013 &		  13  	 $\pm$     6   &     0.9  $\pm$  0.4              \\  
86 &	2MASXJ0020+1413   		&  0.992  $\pm$   0.010 &		  11  	 $\pm$     4   &     0.43  $\pm$  0.16              \\       
87 &	PGC1464874		  		&  0.98  $\pm$   0.02 &		   $<$8  	           &     $<$0.24               \\         
88 &    IC1706			  			&  0.96  $\pm$   0.04 &		  18  	 $\pm$    11   &     1.0  $\pm$  0.7              \\      
\hline                            
\multicolumn{5}{c}{IRAM CO(1-0) Observations}		\\
\hline
89 & 	SDSSJ0944+1116	    &  0.90 $\pm$   0.06  &   $<$0.7   			&  $<$0.26 		\\
90 & 	SDSSJ1049+1108		&  0.98 $\pm$   0.02  &   $<$0.6            &  $<$0.21           \\
91 & 	2MASXJ1336+1552		&  0.93 $\pm$   0.05  &   2.8 $\pm$  0.9    &  0.9 $\pm$ 0.3          \\
92 & 	SDSSJ1032+1227		&  1.000 $\pm$   0.006  &   $<$0.7            & $<$0.27         \\
93 & 	SDSSJ1100+1207		&  0.991 $\pm$   0.012  &   $<$0.7            &  $<$0.22         \\
94 & 	SDSSJ1207+1200		&  0.984 $\pm$   0.017  &   $<$0.7            &  $<$0.31       \\
95 & 	KUG1147+149			&  0.995 $\pm$   0.007  &   $<$0.7            &  $<$0.17       \\
96 & 	SDSSJ1320+1524		&  0.998 $\pm$   0.004  &   $<$1.0            & $<$0.23        \\
97 & 	VPC0873				&  0.992 $\pm$   0.013  &   $<$0.6            &  $<$0.17         \\
\hline                            
\multicolumn{5}{c}{IRAM CO(2-1) Observations}		\\
\hline
91 & 	2MASXJ1336+1552		&  0.70 $\pm$   0.10  &   8 $\pm$  2    &  0.7 $\pm$ 0.2          \\  
\end{longtable}

\tablefoot{Col. (1): ALLSMOG ID. Col. (2): Galaxy name. Col. (3): Fraction of the expected total CO flux recovered by the beam of the single-dish telescope, calculated as explained in $\S$~\ref{sec:ap_cor} following the same method as \cite{Bothwell+14}. Col. (4): Total velocity-integrated CO flux corrected for beam coverage. Col. (5): CO line luminosity calculated from the aperture-corrected CO flux listed in Col. (4) by using Eq~\ref{eq:lco_def}.}
}

\section{Ancillary data}\label{sec:ancillary}

\longtab{
\begin{longtable}{@{}llcccccc@{}}
\caption{\label{table:sample_prop_1} Physical properties of the ALLSMOG galaxy sample}\\
\hline
\hline	
 ID			& Galaxy Name			& RA 	  	   & DEC  	    		& $z_{opt}$ 			& $D_L$			& $i$	 			  	& $d25$   \\
		         &						& (J2000)	   &		(J2000)	&       		& [Mpc]			& [deg] 				& [arcsec] \\
(1)			& (2)					& (3)		   & (4)				& (5)			& (6)			& (7)					& (8)			\\	
\hline
\endfirsthead
\multicolumn{8}{c}{\footnotesize Following from previous page}\\
\hline
 ID			& Galaxy Name			& RA 	  	   & DEC   	    		& $z_{opt}$			& $D_L$			& $i$		 			  	& $d25$   \\
		    &						& (J2000)	   & (J2000)			&       		& [Mpc]			& [deg] 				& [arcsec] \\	
(1)			& (2)					& (3)		   & (4)				& (5)			& (6)			& (7)					& (8)			\\			
\hline
\endhead
\hline
\multicolumn{8}{c}{\footnotesize Follows on next page}\\
\endfoot
\hline
\endlastfoot
\hline
\multicolumn{8}{c}{APEX Sample}\\
\hline
1  &	UGC11631		   & 20 47 59.9      &   -00 10 48    &  0.0141 &  63.6 &  90.0 &    59.7   \\
2  &	UGC00272		   & 00 27 49.7      &	 -01 11 60    &  0.0130 &  58.5 &  70.7 &    80.6   \\
3  &	IC0159			   & 01 46 25.1      &	 -08 38 12    &  0.0131 &  58.8 &  59.6 &    81.1   \\
4  &	KUG0200-101		   & 02 03 16.6      &	 -09 53 26    &  0.0130 &  58.3 &  68.3 &    39.0   \\
5  &	NGC1234			   & 03 09 39.1      &	 -07 50 46    &  0.0125 &  56.1 &  60.1 &    94.7   \\
6  &	UGC06838		   & 11 51 56.2      &	 -02 38 33    &  0.0129 &  58.2 &  68.0 &    42.5   \\
7  &	UGC09359		   & 14 33 15.8      &	 -01 08 24    &  0.0138 &  62.0 &  90.0 &    50.6   \\
8  &	2MASXJ2235-0845    & 22 35 07.0      &	 -08 45 55      &  0.0238 & 108.1 &  54.7 &    34.4   \\
9  &	NGC5414			   & 14 02 03.5      &   +09 55 46	&  0.0141 &  63.6 &  54.4 &    58.4   \\
10 &	NGC5405			   & 14 01 09.5      &   +07 42 08	&  0.0230 & 104.3 &  23.7 &    49.6   \\
11 &	UGC10005		   & 15 45 14.4      &   +00 46 20	&  0.0128 &  57.7 &  42.5 &    46.8   \\
12 &	MCG+00-25-005	   & 09 36 35.4      &   +01 07 00 	    &  0.0164 &  74.1 &  50.6 &    43.0   \\
13 &	UGC02004		   & 02 31 59.6      &   +00 54 36	    &  0.0218 &  98.9 &  51.3 &    42.5   \\
14 &	UGC02529		   & 03 05 29.56     &  -00 22 55     &  0.0249 & 112.9 &  51.7 &     30.0   \\
15 &	MCG+00-29-013      & 11 18 49.6      &   +00 37 10      &  0.0254 & 115.3 &  33.6 &    34.6   \\
16 &	UGC06329           & 11 18 56.2      &   +00 10 34      &  0.0249 & 113.1 &  38.5 &    48.2   \\
17 &	UGC05648           & 10 26 08.8      &   +04 22 22      &  0.0228 & 103.5 &  75.9 &    71.0   \\
18 &	IC0605             & 10 22 24.1      &   +01 11 54    &  0.0216 &  97.8 &  46.0 &    46.3   \\
19 &	2MASXJ0910+0752    & 09 10 58.8      &   +07 52 19    &  0.0285 & 129.6 &  56.5 &    34.1   \\
20 &	2MASXJ0939+0624    & 09 39 35.3      &   +06 24 52      &  0.0249 & 113.0 &  60.6 &    25.3   \\
21 &	2MASXJ0955+0632    & 09 55 09.8      &   +06 32 57    &  0.0240 & 109.0 &  51.6 &    27.1   \\
22 &	2MASXJ1014+0748    & 10 14 58.9      &   +07 48 03	    &  0.0283 & 128.8 &  62.7 &    33.2   \\
23 &	2MASXJ1011+0746    & 10 11 09.4      &   +07 46 49      &  0.0262 & 119.1 &  73.8 &    41.8   \\
24 &	PGC031905          & 10 52 28.9      &   +07 54 15      &  0.0206 &  93.0 &  38.2 &    59.9   \\  
25 &	PGC031382          & 10 35 42.3      &   +05 36 58      &  0.0272 & 123.7 &  60.3 &    44.6   \\  
26 &	2MASXJ1110+0411    & 11 10 37.4      &   +04 11 29      &  0.0290 & 131.8 &  54.1 &    24.8   \\
27 &	2MASXJ0855+0345    & 08 55 56.0      &   +03 45 30      &  0.0276 & 125.4 &  68.6 &    32.1   \\
28 &	2MASXJ0839+0349    & 08 39 39.2      &   +03 49 43      &  0.0267 & 121.2 &  49.6 &    26.3   \\
29 &	UGC04977           & 09 21 59.6     &   +03 22 43   &  0.0279 & 126.9 &  70.8 &    50.4   \\
30 &	2MASXJ0846+0230    & 08 46 54.0      &   +02 30 05      &  0.0282 & 128.3 &  37.3 &    23.2   \\
31 &	SDSSJ1328-0202     & 13 28 46.9      &  -02 02 28       &  0.0124 &  55.9 &  37.1 &    36.6   \\
32 &	MCG+00-34-038      & 13 29 50.4      &  -01 25 45       &  0.0213 &  96.3 &  49.1 &    40.2   \\
33 &	UGC08526           & 13 32 55.1      &  -01 09 34       &  0.0127 &  57.1 &  42.6 &    51.4   \\ 
34 &	CGCG017-017        & 13 32 50.2      &  -03 04 58       &  0.0142 &  64.1 &  90.0 &    86.5   \\
35 &	NGC2936          & 09 37 45.0      &   +02 45 34      &  0.0240 & 108.7 &  34.9 &     6.3   \\
36 &	SDSSJ0937+0927     & 09 37 09.0      &   +09 27 51      &  0.0225 & 101.8 &  67.4 &    42.3   \\
37 &	SDSSJ0950+1118    	& 09 50 11.2      &   +11 18 30     & 0.0188  &  85.1 &  22.9 &    17.5   \\
38 &	IC3069            	& 12 15 19.9      &   +10 09 39     & 0.0195  &  88.2 &  67.7 &    30.3    \\
39 &	SDSSJ1112+0931    	& 11 12 50.2      &   +09 31 39     & 0.0209  &  94.4 &  42.4 &    35.2    \\
40 &	SDSSJ0805+0659    	& 08 05 37.7      &	+06 59 35       &  0.0153 &  68.7 &  45.5 &    21.6    \\
41 &	SDSSJ1104+0507    	& 11 04 14.6      &	+05 07 37       &  0.0176 &  79.3 &  38.2 &    14.6    \\
42 &	CGCG050-042       	& 15 36 46.6      &	+07 50 01       &  0.0111 &  50.0 &  34.9 &    26.5    \\
43 &	2MASXJ0858+0345   	& 08 58 05.3      &	+03 45 23       &  0.0269 & 122.2 &  27.5 &    22.7    \\
44 &	SDSSJ1008+1428    	& 10 08 35.8      &	+14 28 19       &  0.0201 &  91.1 &  52.0 &    21.8    \\
45 &	SDSSJ0954+0458    	& 09 54 42.1      &	+04 58 32       &  0.0173 &  78.1 &  41.5 &    14.6    \\
46 &	SDSSJ1213+1056	  	& 12 13 59.2      &	+10 56 40	    &  0.0196 &  88.8 &  38.3 &    14.8    \\
47 &	SDSSJ0945+0515	  	& 09 45 58.0      &	+05 15 05	    &  0.0213 &  96.4 &  38.7 &    13.6    \\
48 &	SDSSJ1122+1316	  	& 11 22 31.0      &	+13 16 02	    &  0.0205 &  92.5 &  36.0 &    18.2    \\
49 &	SDSSJ1403+1003	  	& 14 03 17.8      &	+10 03 40	    &  0.0134 &  60.3 &  63.6 &    26.9    \\
50 &	CGCG063-006		  	& 09 32 51.9      &	+12 17 13	    &  0.0156 &  70.5 &  50.0 &    46.6    \\
51 &	2MASXJ0843+1303	  	& 08 43 35.1      &	+13 03 47	    &  0.0138 &  61.9 &  44.5 &    19.5    \\
52 &	SDSSJ0920+0759	  	& 09 20 43.0      &	+07 59 26	    &  0.0213 &  96.5 &  65.4 &    21.6    \\
53 &	CGCG080-042		  	& 16 29 44.1      &	+11 50 50	    &  0.0172 &  77.4 &  81.5 &    45.0    \\
54 &	CGCG078-021		  	& 15 35 45.5      &	+08 52 10	    &  0.0196 &  88.5 &  52.4 &    35.7    \\
55 &	UGC04567		  	& 08 44 42.7      &	+09 48 03	    &  0.0136 &  61.4 &  90.0 &    46.6    \\
56 &	SDSSJ1624+1251	  	& 16 24 14.8      &	+12 51 21	    &  0.0164 &  73.9 &  49.1 &    31.4    \\
57 &	CGCG058-066		  	& 07 54 17.1      &	+14 16 23	    &  0.0153 &  69.0 &  53.0 &    42.5    \\
58 &	CGCG061-003		  	& 08 44 32.4      &	+09 09 57	    &  0.0136 &  61.4 &  72.4 &    44.1    \\
59 &	CGCG051-037		  	& 16 05 24.0      &	+07 04 41	    &  0.0176 &  79.5 &  58.3 &    41.2    \\
60 &	SDSSJ0854+0418	  	& 08 54 44.7      &	+04 18 07	    &  0.0237 & 107.5 &  47.9 &    30.7    \\
61 &	CGCG059-031		  	& 08 03 50.2      &	+10 32 55	    &  0.0154 &  69.4 &  67.2 &    23.9    \\
62 &	2MASXJ0806+1249	  	& 08 06 16.2      &	+12 49 41	    &  0.0190 &  85.8 &  29.3 &    20.4    \\
63 &	VIIIZW039		  	& 09 09 52.1      &	+09 47 37	    &  0.0183 &  82.7 &  43.0 &    21.1    \\
64 &	2MASXJ0941+1056	  	& 09 41 01.2      &	+10 56 42	    &  0.0185 &  83.6 &  17.6 &    22.1    \\
65 &	SDSSJ0943+0356	  	& 09 43 08.6      &	+03 56 25	    &  0.0196 &  88.7 &  54.1 &    23.0    \\
66 &	CGCG035-063		  	& 09 49 00.7      &	+04 18 11	    &  0.0204 &  92.3 &  37.4 &    25.1    \\
67 &	CGCG036-048		  	& 10 10 37.2      &	+05 09 02	    &  0.0137 &  61.7 &  62.0 &    26.8    \\
68 &	SDSSJ1625+1142	  	& 16 25 24.7      &	+11 42 49	    &  0.0163 &  73.3 &  67.1 &    28.5    \\
69 &	SDSSJ1028+0424	  	& 10 28 15.8      &	+04 24 23	    &  0.0229 & 103.8 &  52.0 &    26.3    \\
70 &	UGC06011		  	& 10 53 20.8      &	-00 36 23	    &  0.0185 &  83.5 &  90.0 &    45.0    \\
71 &	PGC051743 		  	& 14 29 06.0      &	+07 50 46 	    &  0.0269 & 122.2 &  17.2 &    27.9    \\   
72 &	2MASXJ1437+0500   	& 14 37 50.7      &	+05 00 41       &  0.0252 & 114.3 &  66.1 &    33.7    \\	
73 &	SDSSJ0944+0400	  	& 09 44 34.5      &	+04 00 06	    &  0.0197 &  88.9 &  62.2 &    18.4    \\
74 &	2MASXJ1048+1201	  	& 10 48 15.6      &	+12 01 28	    &  0.0219 &  99.1 &  17.5 &    20.5    \\
75 &	SDSSJ1110+1345	  	& 11 10 29.6      &	+13 45 58	    &  0.0144 &  64.9 &  53.1 &    23.9    \\
76 &	PGC012214		  	& 03 17 20.0      &	-00 04 35       &  0.0222 & 100.6 &  72.8 &    37.0    \\
77 &	PGC003530		  	& 00 59 04.1      &	+01 00 04       &  0.0178 &  80.5 &  55.5 &    19.6    \\
78 &	PGC073268		  	& 00 10 25.5      &	+14 17 23       &  0.0182 &  82.1 &  52.1 &    32.2    \\
79 &	PGC1452135		  	& 02 00 44.0      &	+14 12 38       &  0.0160 &  72.0 &  71.6 &    34.1    \\
80 &	PGC3127469		  	& 00 33 44.5      &	+14 24 29       &  0.0179 &  80.8 &  35.9 &    22.7    \\
81 &	UGC00317		  	& 00 31 43.3      &	+00 54 03       &  0.0179 &  80.7 &  21.0 &    30.7    \\
82 &	SDSSJ0011+1428    	& 00 11 43.2      &	+14 28 01       &  0.0174 &  78.5 &  40.4 &    14.9    \\
83 &	PGC1455779		  	& 02 02 23.5      &	+14 20 55       &  0.0240 & 108.8 &  78.5 &    36.4    \\
84 &	PGC000010		  	& 00 00 07.8      &	-00 02 26       &  0.0237 & 107.3 &  47.5 &    33.5    \\
85 &	PGC1446233		  	& 01 29 49.2      &	+13 59 26       &  0.0233 & 105.4 &  71.8 &    29.7    \\
86 &	2MASXJ0020+1413   	& 00 20 48.6      &	+14 13 28       &  0.0178 &  80.5 &  40.0 &    22.9    \\
87 &	PGC1464874		  	& 02 01 04.1      &	+14 42 03       &  0.0152 &  68.6 &  76.5 &    35.3    \\
88 &    IC1706			  	& 01 27 31.0      &	+14 49 11       &  0.0216 &  97.9 &  18.6 &    30.1    \\
\hline
\multicolumn{8}{c}{IRAM Sample}\\
\hline
89 & 	SDSSJ0944+1116	     & 09 44 30.3	  &	+11 16 44	    &  0.0277 &  126.0 &  52.3 &    39.9 \\
90 & 	SDSSJ1049+1108		& 10 49 11.2	  &	+11 08 57	    &  0.0267 &  121.3 &  41.5 &    22.6 \\
91 & 	2MASXJ1336+1552		& 13 36 29.9	  &	+15 52 50	    &  0.0260 &  118.1 &  27.9 &    28.7 \\
92 & 	SDSSJ1032+1227		& 10 32 59.3	  &	+12 27 43	    &  0.0275 &  125.0 &  15.2 &    13.4 \\
93 & 	SDSSJ1100+1207		& 11 00 49.1	  &	+12 07 47	    &  0.0258 &  117.2 &  36.3 &    18.4 \\
94 & 	SDSSJ1207+1200		& 12 07 35.8	  &	+12 00 56	    &  0.0293 &  133.4 &  65.0 &    24.6 \\
95 & 	KUG1147+149			& 11 50 09.7	  &	+14 39 18	    &  0.0216 &   97.8 &  63.2 &    18.6 \\
96 & 	SDSSJ1320+1524		& 13 20 10.1	  &	+15 24 19	    &  0.0219 &   99.2 &  49.0 &    14.9 \\
97 & 	VPC0873				& 12 32 34.1	  &	+14 34 46	    &  0.0239 &  108.4 &  15.2 &    16.8 \\
\hline
\end{longtable}  

\tablefoot{Col. (1): ALLSMOG ID. Col (2): Galaxy name. Col. (3): The right ascension in the J2000.0 epoch. Col. (4): The declination in the J2000.0 epoch.
Col (5): Optical spectroscopic redshift extracted from the MPA-JHU catalogue (further details are given in $\S$~\ref{sec:optdisk_par}). Col. (6): Luminosity distance according to the adopted Cosmology. Col (7): Inclination of the optical disk. Col (8): Optical diameter as defined by the 25th magnitude B-band isophote. Both $i$ and $d_{25}$ are drawn from the Hyperleda database.}
}                                                 

\longtab{
\begin{longtable}{@{}llccccccc@{}}
\caption{Physical properties of the ALLSMOG galaxy sample - Part II}
\label{table:sample_prop_2} \\
\hline
\hline
 ID			& Galaxy Name			& $\log$M$_{*}$ 	   	& $\log$SFR   	    			& 			\multicolumn{5}{c}{Gas-phase metallicity (12+log(O/H))}				\\
		    &						& [M$_{\odot}$]	& [M$_{\odot}$~yr$^{-1}$]		& MPA-JHU     & N2 PP04		& N2 M13		& O3N2 PP04 & O3N2 M13\\	
(1)			& (2)					& (3)		   	& (4)						& (5)			& (6)			& (7)			& (8)		  & (9)			\\		
\hline
\endfirsthead
\multicolumn{9}{c}{\footnotesize Following from previous page}\\
\hline	
 ID			& Galaxy Name			& $\log$M$_{*}$	   	& $\log$SFR   	    			& \multicolumn{5}{c}{Gas-phase metallicity (12+log(O/H))}  \\
		    &						& [M$_{\odot}$]	& [M$_{\odot}$~yr$^{-1}$]	& MPA-JHU     & N2 PP04		& N2 M13 		& O3N2 PP04 & O3N2 M13 \\
(1)			& (2)					& (3)		   	& (4)						& (5)			& (6)			& (7)			& (8)		  & (9)			\\	
\hline
\endhead
\hline
\multicolumn{9}{c}{\footnotesize Follows on next page}\\
\endfoot
\hline
\endlastfoot
\hline
\multicolumn{9}{c}{APEX Sample}\\
\hline
1  &	UGC11631		   &    9.57$\pm$0.09 &  -0.28$\pm$0.23 &  8.84 & 8.54 & 8.45 & 8.59 & 8.44 \\
2  &	UGC00272		   &    9.37$\pm$0.07 &  -0.65$\pm$0.29 &  8.67 & 8.60 & 8.50 & 8.60 & 8.45 \\
3  &	IC0159			   &    9.71$\pm$0.07 &  -0.06$\pm$0.27 &  8.87 & 8.60 & 8.50 & 8.64 & 8.47 \\
4  &	KUG0200-101		   &    9.12$\pm$0.08 &  -0.68$\pm$0.21 &  8.61 & 8.41 & 8.34 & 8.37 & 8.29 \\
5  &	NGC1234			   &    9.58$\pm$0.09 &  -0.68$\pm$0.31 &  8.84 & 8.64 & 8.53 & 8.71 & 8.52 \\
6  &	UGC06838		   &    9.32$\pm$0.08 &  -0.52$\pm$0.24 &  8.87 & 8.60 & 8.50 & 8.68 & 8.50 \\
7  &	UGC09359		   &    8.96$\pm$0.09 &  -0.79$\pm$0.42 &  8.71 & 8.43 & 8.36 & 8.45 & 8.34 \\
8  &	2MASXJ2235-0845    &    9.79$\pm$0.08 &   0.26$\pm$0.20 &  9.01 & 8.62 & 8.52 & 8.69 & 8.51 \\
9  &	NGC5414			   &    9.74$\pm$0.07 &   0.26$\pm$0.19 &  8.92 & 8.57 & 8.47 & 8.57 & 8.42 \\
10 &	NGC5405			   &    9.98$\pm$0.07 &   0.28$\pm$0.27 &  9.03 & 8.61 & 8.51 & 8.79 & 8.57 \\
11 &	UGC10005		   &    9.07$\pm$0.13 &   1.08$\pm$0.36 &  8.79 & 8.66 & 8.55 & 8.67 & 8.49 \\
12 &	MCG+00-25-005	   &    9.70$\pm$0.08 &   0.03$\pm$0.22 &  9.01 & 8.64 & 8.53 & 8.75 & 8.54 \\
13 &	UGC02004		   &    9.61$\pm$0.10 &   0.37$\pm$0.26 &  9.15 & 8.68 & 8.56 & 8.87 & 8.63 \\
14 &	UGC02529		   &    9.42$\pm$0.05 &   0.02$\pm$0.22 &  8.70 & 8.48 & 8.40 & 8.48 & 8.36 \\
15 &	MCG+00-29-013      &    9.98$\pm$0.08 &   0.16$\pm$0.23 &  9.07 & 8.65 & 8.54 & 8.79 & 8.57 \\
16 &	UGC06329           &    9.80$\pm$0.07 &   0.02$\pm$0.26 &  8.98 & 8.63 & 8.52 & 8.72 & 8.53 \\
17 &	UGC05648           &    9.99$\pm$0.09 &   0.02$\pm$0.21 &  9.09 & 8.62 & 8.52 & 8.80 & 8.58 \\
18 &	IC0605             &    9.90$\pm$0.10 &   0.24$\pm$0.18 &  8.97 & 8.64 & 8.53 & 8.64 & 8.47 \\
19 &	2MASXJ0910+0752    &    9.86$\pm$0.08 &   0.07$\pm$0.21 &  9.11 & 8.66 & 8.55 & 8.82 & 8.60 \\
20 &	2MASXJ0939+0624    &    9.45$\pm$0.17 &   0.34$\pm$0.19 &  8.71 & 8.46 & 8.39 & 8.40 & 8.31 \\
21 &	2MASXJ0955+0632    &    9.50$\pm$0.07 &   0.04$\pm$0.23 &  8.91 & 8.56 & 8.47 & 8.61 & 8.45 \\
22 &	2MASXJ1014+0748    &    9.51$\pm$0.06 &   0.01$\pm$0.25 &  8.84 & 8.59 & 8.49 & 8.61 & 8.45 \\
23 &	2MASXJ1011+0746    &    9.96$\pm$0.08 &   0.12$\pm$0.22 &  8.86 & 8.59 & 8.49 & 8.63 & 8.47 \\
24 &	PGC031905          &    9.65$\pm$0.07 &   0.06$\pm$0.23 &  8.89 & 8.56 & 8.47 & 8.57 & 8.43 \\  
25 &	PGC031382          &    9.95$\pm$0.12 &   0.08$\pm$0.33 &  8.86 & 8.57 & 8.47 & 8.62 & 8.46 \\  
26 &	2MASXJ1110+0411    &    9.51$\pm$0.18 &   0.57$\pm$0.39 &  8.80 & 8.47 & 8.40 & 8.42 & 8.33 \\
27 &	2MASXJ0855+0345    &    9.73$\pm$0.08 &   0.00$\pm$0.23 &  8.81 & 8.56 & 8.47 & 8.59 & 8.44 \\
28 &	2MASXJ0839+0349    &    9.94$\pm$0.11 &   0.54$\pm$0.10 &  9.09 & 8.66 & 8.55 & 8.70 & 8.51 \\
29 &	UGC04977           &    9.91$\pm$0.08 &   0.14$\pm$0.25 &  8.95 & 8.60 & 8.50 & 8.73 & 8.53 \\
30 &	2MASXJ0846+0230    &    9.84$\pm$0.08 &   0.04$\pm$0.22 &  8.99 & 8.64 & 8.53 & 8.72 & 8.53 \\
31 &	SDSSJ1328-0202     &    8.82$\pm$0.13 &  -0.56$\pm$0.39 &  9.00 & 8.58 & 8.48 & 8.62 & 8.46 \\
32 &	MCG+00-34-038      &    9.91$\pm$0.08 &   0.10$\pm$0.22 &  9.09 & 8.63 & 8.53 & 8.80 & 8.58 \\
33 &	UGC08526           &    9.69$\pm$0.09 &  -0.48$\pm$0.28 &  9.02 & 8.64 & 8.54 & 8.81 & 8.58 \\ 
34 &	CGCG017-017        &    9.96$\pm$0.09 &  -0.73$\pm$0.49 &  - & 8.69 & 8.58 & 8.73 & 8.53 \\
35 &	NGC2936         	 &   11.46$\pm$0.00 &   0.85$\pm$0.11 &  9.17 & 8.64 & 8.53 & 8.88 & 8.63 \\
36 &	SDSSJ0937+0927     &    8.75$\pm$0.06 &  -0.79$\pm$0.34 &  8.64 & 8.52 & 8.44 & 8.50 & 8.38 \\
37 &	SDSSJ0950+1118    	&   8.58$\pm$0.08 & -0.95$\pm$0.23 &  8.61 & 8.44 & 8.37 & 8.42 & 8.33 \\
38 &	IC3069            	&   9.00$\pm$0.08 & -0.56$\pm$0.23 &  8.60 & 8.41 & 8.35 & 8.40 & 8.31  \\
39 &	SDSSJ1112+0931    	&   8.89$\pm$0.08 & -0.85$\pm$0.31 &  8.61 & 8.53 & 8.44 & 8.62 & 8.46  \\
40 &	SDSSJ0805+0659    	&   8.62$\pm$0.07 & -1.18$\pm$0.43 &  8.46 & 8.44 & 8.37 & 8.42 & 8.32  \\
41 &	SDSSJ1104+0507    	&   8.68$\pm$0.08 & -1.10$\pm$0.23 &  8.53 & 8.41 & 8.34 & 8.38 & 8.30  \\
42 &	CGCG050-042       	&   8.54$\pm$0.08 & -1.38$\pm$0.52 &  8.55 & 8.43 & 8.36 & 8.42 & 8.32  \\
43 &	2MASXJ0858+0345   	&   9.80$\pm$0.09 &  0.06$\pm$0.14 &  9.09 & 8.66 & 8.55 & 8.66 & 8.49  \\
44 &	SDSSJ1008+1428    	&   8.84$\pm$0.07 & -0.57$\pm$0.18 &  8.59 & 8.38 & 8.32 & 8.32 & 8.26  \\
45 &	SDSSJ0954+0458    	&   8.65$\pm$0.09 & -0.92$\pm$0.23 &  8.59 & 8.44 & 8.37 & 8.41 & 8.32  \\
46 &	SDSSJ1213+1056	  	&   8.60$\pm$0.08 & -0.92$\pm$0.17 &  8.62 & 8.39 & 8.33 & 8.35 & 8.28  \\
47 &	SDSSJ0945+0515	  	&   8.65$\pm$0.08 & -0.55$\pm$0.16 &  8.67 & 8.42 & 8.35 & 8.36 & 8.29  \\
48 &	SDSSJ1122+1316	  	&   8.70$\pm$0.08 & -1.19$\pm$0.25 &  8.67 & 8.43 & 8.36 & 8.42 & 8.33  \\
49 &	SDSSJ1403+1003	  	&   8.54$\pm$0.10 & -0.78$\pm$0.21 &  8.58 & 8.24 & 8.21 & 8.19 & 8.17  \\
50 &	CGCG063-006		  	&   8.94$\pm$0.06 & -0.59$\pm$0.22 &  8.54 & 8.42 & 8.35 & 8.37 & 8.29  \\
51 &	2MASXJ0843+1303	  	&   8.77$\pm$0.09 & -0.72$\pm$0.24 &  8.55 & 8.41 & 8.34 & 8.27 & 8.23  \\
52 &	SDSSJ0920+0759	  	&   8.87$\pm$0.08 & -0.78$\pm$0.23 &  8.58 & 8.47 & 8.39 & 8.48 & 8.36  \\
53 &	CGCG080-042		  	&   9.37$\pm$0.09 & -0.64$\pm$0.23 &  8.65 & 8.43 & 8.36 & 8.36 & 8.29  \\
54 &	CGCG078-021		  	&   9.50$\pm$0.09 & -0.38$\pm$0.25 &  8.99 & 8.65 & 8.54 & 8.77 & 8.56  \\
55 &	UGC04567		  	&   9.40$\pm$0.09 & -0.70$\pm$0.49 &  8.73 & 8.60 & 8.50 & 8.64 & 8.48  \\
56 &	SDSSJ1624+1251	  	&   9.07$\pm$0.07 & -0.99$\pm$0.56 &  8.54 & 8.54 & 8.45 & 8.57 & 8.42  \\
57 &	CGCG058-066		  	&   9.10$\pm$0.06 & -0.10$\pm$0.29 &  8.64 & 8.52 & 8.43 & 8.51 & 8.38  \\
58 &	CGCG061-003		  	&   9.11$\pm$0.08 & -0.82$\pm$0.27 &  8.70 & 8.51 & 8.43 & 8.52 & 8.39  \\
59 &	CGCG051-037		  	&   9.11$\pm$0.08 & -0.57$\pm$0.30 &  8.60 & 8.47 & 8.39 & 8.46 & 8.36  \\
60 &	SDSSJ0854+0418	  	&   9.01$\pm$0.10 & -0.20$\pm$0.19 &  8.72 & 8.48 & 8.41 & 8.45 & 8.34  \\
61 &	CGCG059-031		  	&   9.27$\pm$0.08 & -0.65$\pm$0.27 &  8.76 & 8.49 & 8.41 & 8.49 & 8.37  \\
62 &	2MASXJ0806+1249	  	&   9.13$\pm$0.13 & -0.38$\pm$0.20 &  8.69 & 8.45 & 8.38 & 8.38 & 8.30  \\
63 &	VIIIZW039		  	&   9.38$\pm$0.13 & -0.40$\pm$0.17 &  8.97 & 8.59 & 8.49 & 8.66 & 8.49  \\
64 &	2MASXJ0941+1056	  	&   9.39$\pm$0.09 & -0.95$\pm$0.20 &  8.84 & 8.64 & 8.53 & 8.71 & 8.52  \\
65 &	SDSSJ0943+0356	  	&   9.04$\pm$0.07 & -0.85$\pm$0.47 &  8.76 & 8.53 & 8.44 & 8.55 & 8.41  \\
66 &	CGCG035-063		  	&   9.44$\pm$0.07 & -0.26$\pm$0.26 &  8.73 & 8.59 & 8.49 & 8.59 & 8.44  \\
67 &	CGCG036-048		  	&   9.24$\pm$0.07 & -0.19$\pm$0.20 &  8.64 & 8.43 & 8.37 & 8.37 & 8.29  \\
68 &	SDSSJ1625+1142	  	&   9.10$\pm$0.13 & -0.52$\pm$0.20 &  8.67 & 8.41 & 8.35 & 8.38 & 8.30  \\
69 &	SDSSJ1028+0424	  	&   9.20$\pm$0.06 & -0.34$\pm$0.29 &  8.59 & 8.47 & 8.40 & 8.43 & 8.33  \\
70 &	UGC06011		  	&   9.32$\pm$0.08 & -0.43$\pm$0.28 &  8.59 & 8.57 & 8.47 & 8.56 & 8.42  \\
71 &	PGC051743 		  	&   9.66$\pm$0.07 &  0.25$\pm$0.22 &  9.01 & 8.62 & 8.51 & 8.67 & 8.49  \\   
72 &	2MASXJ1437+0500   	&   9.69$\pm$0.07 &  0.07$\pm$0.26 &  8.87 & 8.61 & 8.51 & 8.67 & 8.49  \\	
73 &	SDSSJ0944+0400	  	&   8.77$\pm$0.06 & -0.72$\pm$0.40 &  8.60 & 8.35 & 8.30 & 8.32 & 8.26  \\
74 &	2MASXJ1048+1201	  	&   9.00$\pm$0.08 & -0.82$\pm$0.23 &  8.69 & 8.58 & 8.48 & 8.56 & 8.42  \\
75 &	SDSSJ1110+1345	  	&   8.57$\pm$0.08 & -0.93$\pm$0.15 &  8.60 & 8.36 & 8.30 & 8.29 & 8.24  \\
76 &	PGC012214		  	&   9.30$\pm$0.11 & -0.07$\pm$0.28 &  8.62 & 8.41 & 8.35 & 8.38 & 8.30  \\
77 &	PGC003530		  	&   8.81$\pm$0.06 & -0.44$\pm$0.16 &  8.59 & 8.40 & 8.34 & 8.30 & 8.24  \\
78 &	PGC073268		  	&   9.50$\pm$0.09 & -0.62$\pm$0.32 &  8.76 & 8.66 & 8.55 & 8.64 & 8.48  \\
79 &	PGC1452135		  	&   8.94$\pm$0.08 & -1.05$\pm$0.35 &  8.54 & 8.49 & 8.41 & 8.48 & 8.37  \\
80 &	PGC3127469		  	&   8.57$\pm$0.07 & -1.06$\pm$0.30 &  8.58 & 8.42 & 8.35 & 8.38 & 8.30  \\
81 &	UGC00317		  	&   8.95$\pm$0.08 & -0.95$\pm$0.57 &  8.60 & 8.59 & 8.49 & 8.55 & 8.41  \\
82 &	SDSSJ0011+1428    	&   8.67$\pm$0.12 & -0.74$\pm$0.17 &  8.70 & 8.52 & 8.43 & 8.48 & 8.37  \\
83 &	PGC1455779		  	&   9.52$\pm$0.09 & -0.76$\pm$0.32 &  8.65 & 8.61 & 8.51 & 8.61 & 8.45  \\
84 &	PGC000010		  	&   9.83$\pm$0.08 & -0.11$\pm$0.25 &  8.94 & 8.67 & 8.56 & 8.67 & 8.49  \\
85 &	PGC1446233		  	&   9.11$\pm$0.09 & -0.94$\pm$0.28 &  8.53 & 8.45 & 8.38 & 8.43 & 8.34  \\
86 &	2MASXJ0020+1413   	&   9.31$\pm$0.09 & -0.14$\pm$0.16 &  8.79 & 8.50 & 8.42 & 8.43 & 8.34  \\
87 &	PGC1464874		  	&   9.01$\pm$0.09 & -1.07$\pm$0.52 &  8.62 & 8.43 & 8.36 & 8.38 & 8.30  \\
88 &    IC1706			  	&   9.74$\pm$0.09 & -0.24$\pm$0.26 &  8.96 & 8.65 & 8.54 & 8.68 & 8.50  \\
\hline                            
\multicolumn{8}{c}{IRAM Sample}\\
\hline
89 & 	SDSSJ0944+1116	    & 9.00$\pm$0.06 & -0.60$\pm$0.34 & 8.49 & 8.47 & 8.39 & 8.47 & 8.36 \\
90 & 	SDSSJ1049+1108		& 8.92$\pm$0.06 & -0.52$\pm$0.23 & 8.68 & 8.47 & 8.40 & 8.46 & 8.35 \\
91 & 	2MASXJ1336+1552		& 8.65$\pm$0.07 & -0.25$\pm$0.25 & 8.77 & 8.44 & 8.37 & 8.43 & 8.33 \\
92 & 	SDSSJ1032+1227		& 8.75$\pm$0.07 & -0.42$\pm$0.15 & 8.61 & 8.39 & 8.33 & 8.33 & 8.27 \\
93 & 	SDSSJ1100+1207		& 8.75$\pm$0.07 & -0.93$\pm$0.13 & 8.73 & 8.49 & 8.41 & 8.48 & 8.36 \\
94 & 	SDSSJ1207+1200		& 8.75$\pm$0.06 & -0.45$\pm$0.28 & 8.40 & 8.36 & 8.30 & 8.31 & 8.25 \\
95 & 	KUG1147+149			& 8.65$\pm$0.06 & -0.60$\pm$0.23 & 8.28 & 8.32 & 8.28 & 8.31 & 8.25 \\
96 & 	SDSSJ1320+1524		& 8.51$\pm$0.07 & -0.90$\pm$0.27 & 8.37 & 8.37 & 8.32 & 8.35 & 8.28 \\
97 & 	VPC0873				& 8.64$\pm$0.06 & -0.81$\pm$0.25 & 8.26 & 8.33 & 8.28 & 8.34 & 8.27 \\
\hline
\end{longtable}          

\tablefoot{Col. (1): ALLSMOG ID. Col (2): Galaxy name. Col (3): Stellar mass estimated from optical SDSS observations. The table lists the median value of the PDF for $\log M_*$ provided in the MPA-JHU catalogue. Its associated error is: $0.5\cdot (P84-P16)$, where P16 and P84 are respectively the 16th and the 84th percentile values of the PDF. Col (4): Star formation rate estimated from optical SDSS observations. The value listed in the table is the median value of the PDF for $\log SFR$ corrected for the SDSS fibre aperture provided in the MPA-JHU catalogue, and its associated error is $0.5\cdot (P84-P16)$. Col (5): Gas-phase metallicity derived using the calibration of \cite{Tremonti+04}. Similar to $M_*$ and SFR, the value reported in the table is the median of the PDF for $12+\log(O/H)$ provided in the MPA-JHU catalogue and the associated error is $0.5\cdot (P84-P16)$. Col (6): Gas-phase metallicity calculated using the N2 calibration provided of \cite{PP04}. Col (7): Gas-phase metallicity calculated using the N2 calibration proposed of \cite{Marino+13}. Col (8): Gas-phase metallicity calculated using the O3N2 calibration of \cite{PP04}. Col (9): Gas-phase metallicity calculated using the O3N2 calibration of \cite{Marino+13}. Further information on the quantities listed in this table is provided in $\S$~\ref{sec:ancillary}.}
}

\subsection{Physical parameters derived from optical observations}\label{sec:optdisk_par}

Table~\ref{table:sample_prop_1} lists for each ALLSMOG galaxy the SDSS coordinates, the optical redshift ($z_{opt}$), luminosity distance ($D_L$), disk inclination ($i$) and optical diameter as defined by the 25th magnitude B-band isophote ($d_{25}$). The redshift was extracted from the MPA-JHU catalogue, and corresponds to the systemic redshift of the galaxy inferred from a combination of stellar absorption lines and nebular emission lines at optical wavelengths \citep{Bolton+12}. The optical diameter and disk inclination were drawn from the Hyperleda database\footnote{\texttt{http://leda.univ-lyon1.fr/}} \citep{Makarov+14}. 

Table~\ref{table:sample_prop_2} lists the stellar mass and SFR values from the MPA-JHU catalogue, where $M_*$ is computed from a fit to the spectral energy distribution (SED) obtained using SDSS broad-band photometry \citep{Brinchmann+04, Salim+07}, and the SFR is based on the H$\alpha$ intrinsic line luminosity following the method of \cite{Brinchmann+04}. We note that we used the aperture-corrected SFRs, which should take into account any flux falling outside the $3\arcsec$-aperture of the SDSS spectroscopic fibres. In Table~\ref{table:sample_prop_2} we report the median values of the predicted probability density function (PDF) for $\log {\rm SFR}$ and $\log M_*$ provided in the MPA-JHU catalogue, and their associated errors estimated using the 16th ($P16$) and 84th ($P84$) percentile values of the corresponding PDFs, that is $0.5*(P84-P16)$. For the reasons discussed in $\S$~\ref{sec:sample}, the $M_*$ of NGC2936 is taken from \cite{Xu+10}. 

\subsubsection{The nebular visual extinction $A_V$}\label{sec:av_deriv}

The visual extinction ($A_V$) of the ionised interstellar medium was computed from the Balmer decrement (i.e. the observed H$\alpha$/H$\beta$ flux ratio) measured using the SDSS optical spectra probing the central 3$\arcsec$ region (corresponding to the size of the SDSS fibre). 
We note that we have used the far-IR maps produced by \cite{Schlegel+98} and \cite{Schlafly+11} to check for effects due to foreground extinction by dust in the Milky Way at the position of ALLSMOG galaxies and the resulting $E(B-V)$ values are typically very small, $E(B-V)<0.05$ in nearly all cases. For this reason, we will ignore Galactic extinction in our $A_V$ computation.


We calculated $A_V$ using two different attenuation curve models, the one proposed by \cite{CCM89} (hereafter CCM89) and the one of \cite{Calzetti+00} (hereafter simply indicated as ``Calzetti''). 
In the CCM89 case\footnote{The CCM89 attenuation curve values were computed using the `'absorption law calculator'' by Doug Welch available at \texttt{http://dogwood.physics.mcmaster.ca/Acurve.html}}, we use $R_V=3.1$ and the relation between $A_V$ and the observed Balmer decrement becomes: 
\begin{equation}\label{eq:ebv_ccm89}
A_V^{\rm CCM89}  = 7.16 \log \Biggl[\frac{(F_{\rm H\alpha}/F_{\rm H\beta})_{obs}}{2.86}\Biggr],
\end{equation}
where 2.86 is the theoretical H$\alpha$/H$\beta$ flux ratio expected in absence of dust absorption and for typical HII region conditions \citep{Osterbrock89}. By adopting the attenuation curve by \cite{Calzetti+00} (see also \cite{Dominguez+13}), we have instead $R_V=4.05$ and:
\begin{equation}\label{eq:ebv_calz}
A_V^{\rm Calzetti} = 7.98 \log \Biggl[\frac{(F_{\rm H\alpha}/F_{\rm H\beta})_{obs}}{2.86}\Biggl].
\end{equation}
The resulting $A_V$ values obtained for ALLSMOG galaxies range between $0 < A_V [mag] < 2.5$, with most objects displaying $A_V<1.5~mag$, independently of the attenuation curve model chosen (i.e. CCM89 or Calzetti). However the uncertainties on $A_V$ are quite large, on average $\Delta A_V \sim 0.5~mag$, because of the large measurement errors associated to the observed H$\alpha$/H$\beta$ flux ratio, which was computed using the H$\alpha$ and H$\beta$ flux values provided in the MPA-JHU catalogue.

\subsubsection{Gas-phase metallicities}\label{sec:metallicity}

The metal content of the ISM (gas-phase metallicity or simply metallicity in the remainder of the paper) is commonly quantified by the Oxygen abundance computed in units of $12+\log(O/H)$. In these units, the Oxygen abundance in the solar photosphere is $12+\log(O/H)=8.69\pm0.05$ \citep{Asplund+09}. Throughout this paper we will use the terms `metallicity' and `Oxygen abundance (or O/H)' as synonyms.

 In star-forming galaxies, the chemical composition of the ISM can be studied through nebular emission lines at optical wavelengths whose relative strengths are sensitive to the ionic abundances in HII regions, although they critically depend on a number of additional physical properties of the ionised gas and in particular on the electron temperature ($T_e$). When $T_e$ cannot be constrained directly using faint auroral lines, metallicities are estimated by adopting ad-hoc empirically- or theoretically-calibrated relations between a strong nebular line ratio of choice (`strong-line' methods) and the chemical (Oxygen) abundance. 

The literature offers plenty of different photoionisation model-based and empirical $T_e$-based metallicity calibrations, which often provide highly discrepant results by up to 0.7~dex even when applied to the same dataset, and no consensus has been reached so far on which one approximates best the true metallicity of galaxies \citep{Kewley+Ellison08}. It is therefore important to keep in mind that any estimate of the gas-phase metallicity in a galaxy cannot be decoupled from the methodology adopted to obtain it. 

In this work we explore five different strong-line diagnostics for determining the gas-phase metallicity of ALLSMOG galaxies (Cols. 5-9 in Table~\ref{table:sample_prop_2}):
\begin{itemize}
\item The \cite{Tremonti+04} calibration, which is based on photoionisation models and is adopted in the MPA-JHU catalogues. For each galaxy, we list in Table~\ref{table:sample_prop_2} the median of the probability distribution of metallicities derived with the \cite{Tremonti+04} method.
\item The \cite{PP04} (hereafter PP04) calibrations of the [NII]$\lambda$6584/H$\alpha$ (`N2') and ([OIII]$\lambda$5007/H$\beta$)/([NII]$\lambda$6584/H$\alpha$) (`O3N2') flux ratios, based mainly on observations of extragalactic HII regions, but model predictions were also used by \cite{PP04} to complement the data in the 
high-metallicity regime ($12+\log(O/H)\geq 9.0$);
\item The revisited empirical calibrations of the N2 and O3N2 indices recently proposed by \cite{Marino+13} (hereafter M13), which are based on observations of a large number of extragalactic HII regions, including recent data from the CALIFA survey \footnote{Calar Alto Legacy Integral Field spectroscopy Area survey.}.
\end{itemize}
There is a conceptual difference between the indirect approach adopted by \cite{Tremonti+04} and the direct PP04 and M13 methods. \cite{Tremonti+04} estimate the metallicity of a galaxy by simultaneously fitting all strong nebular emission lines with photoionisation models to generate a probability distribution of metallicities. PP04 and M13 instead first determine the chemical abundance of a sample of HII regions by using the direct method, by directly measuring $T_e$, $n_e$ and the ionic abundances, and then fit the relationship between the $T_e$-based metal abundances and the observed strong line ratios to extrapolate a general law that can be applied to calibrate the metallicity in other sources for which a direct determination of $T_e$ is unfeasible. All these methods suffer from several biases, which have been extensively discussed in the literature \citep{Kewley+Ellison08}. 
We note that the nebular N2 and O3N2 strong-line diagnostics and, in particular, the PP04 calibration of the O3N2 index, have so far shown the best agreement with the stellar metal abundances across a wide metallicity range, $8.1\lesssim12+\log(O/H)\lesssim9$ \citep{Bresolin+16}.


\subsection{HI Data}\label{sec:HIdata}

\longtab{
\begin{longtable}{@{}llccll@{}}
\caption{HI gas masses}
\label{table:HI_parameters} \\
\hline
\hline
 ID			& Galaxy Name			& $\log M_{\rm HI}$ 					& 	$\log M_{\rm HI}^{\rm corr}$	&   Telescope 	&  Reference \\
		    	&						& [$M_{\odot}$]						&	[$M_{\odot}$]				&    			&  		\\
(1)			& (2)					& (3)								&		(4)					  	&   	(5) 	&  (6) 	\\	
\hline
\endfirsthead
\multicolumn{6}{c}{\footnotesize Following from previous page}\\
\hline
 ID			& Galaxy Name			& $\log M_{\rm HI}$ 					& 	$\log M_{\rm HI}^{\rm corr}$  &   Telescope 	&  Reference	\\
		    &						& [$M_{\odot}$]						&	[$M_{\odot}$]			      &    			&  		\\
(1)			& (2)					& (3)								&		(4)					  &   	(5) 	&  (6) \\	
\hline
\endhead
\hline
\multicolumn{6}{c}{\footnotesize Follows on next page}\\
\endfoot
\hline
\endlastfoot
\hline
\multicolumn{6}{c}{APEX Targets}	\\
\hline
1  &	UGC11631		    &      9.70        $\pm$     0.02       &       9.78     $\pm$     0.02    	 & 	GBT		 &  \cite{Masters+14}		 \\	      
2  &	UGC00272		    &      9.83        $\pm$      0.05      &        9.95    $\pm$     0.05    	 & 	Arecibo	 & 	\cite{Springob+05}  \\         
3  &	IC0159			    &      9.53        $\pm$      0.07      &        9.56    $\pm$     0.07    	 & 	Nan{\c c}ay	 & 	\cite{Theureau+05}  \\      
4  &	KUG0200-101		    &     10.07        $\pm$      0.04      &       10.11    $\pm$     0.04    	 & 	Parkes	 & \cite{Barnes+01}$^\dag$		  \\                    
5  &	NGC1234			    &      9.74        $\pm$      0.05      &        9.77    $\pm$     0.05    	 & 	Nan{\c c}ay	 & 	 \cite{Theureau+05} \\                    
6  &	UGC06838		    &      9.55        $\pm$      0.09      &        9.59    $\pm$     0.09    	 & 	Nan{\c c}ay	 & 	 \cite{Theureau+05}	  \\                    
7  &	UGC09359		    &      9.68        $\pm$      0.07      &        9.80    $\pm$     0.07    	 & 	Arecibo	 & 	\cite{Springob+05}  \\                    
8  &	2MASXJ2235-0845     &      9.78        $\pm$      0.15      &        9.79    $\pm$     0.15    	 & 	Nan{\c c}ay	 & 	 \cite{Theureau+05}		  \\                    
9  &	NGC5414			    &      9.832       $\pm$      0.005     &        9.843   $\pm$     0.005   	 & 	Arecibo	 & 	\cite{Haynes+11}   \\                    
10 &	NGC5405			    &     10.096        $\pm$     0.007     &       10.096    $\pm$     0.007  	 & 	Arecibo	 & 	\cite{Haynes+11}   \\                    
11 &	UGC10005		    &      9.70        $\pm$      0.06      &        9.70    $\pm$     0.06    	 & 	NRAO~91m	 & 	\cite{Hewitt+83} \\                    
12 &	MCG+00-25-005	    &      9.80        $\pm$      0.13      &        9.82    $\pm$     0.13    	 & 	Parkes	 & 	\cite{Barnes+01}$^\dag$  \\                    
13 &	UGC02004		    &      9.93        $\pm$      0.05      &        9.96    $\pm$     0.05    	 & 	Arecibo	 & 	\cite{Springob+05}	  \\                    
14 &	UGC02529		    &     10.01        $\pm$      0.11      &       10.07    $\pm$     0.11    	 & 	Parkes	 & 	\cite{Barnes+01}$^\dag$	  \\                    
15 &	MCG+00-29-013       &         $\leq$9.69        			&     $\leq$9.69       			   	 & 	Parkes	 & 	\cite{Barnes+01}$^\dag$		 \\                      
16 &	UGC06329            &    10.15         $\pm$     0.05       &       10.16    $\pm$     0.05    	 & 	Arecibo	 & 	\cite{Lewis87} \\                          
17 &	UGC05648            &    10.131        $\pm$     0.009      &      10.181  $\pm$     0.008     	 & 	Arecibo	 & 	\cite{Haynes+11}  	\\                          
18 &	IC0605              &    10.00         $\pm$     0.07       &       10.01    $\pm$     0.07    	 & 	Arecibo	 & 	\cite{Paturel+03}  \\                          
19 &	2MASXJ0910+0752     &         $\leq$ 9.06        		    &     		$\leq$9.06       	   	 & 	Arecibo	 & 	\cite{Haynes+11}$^\dag$		\\                    
20 &	2MASXJ0939+0624     &  	 9.80          $\pm$  	0.02        &    	9.83     $\pm$     0.02    	 & 	Arecibo	 & 	\cite{Haynes+11} \\                          
21 &	2MASXJ0955+0632     &         $\leq$8.91         		    &     		$\leq$8.91      		 & 	Arecibo	 & 	\cite{Haynes+11}$^\dag$		\\                
22 &	2MASXJ1014+0748     &       9.73        $\pm$    0.02       &     	9.77   $\pm$     0.02  		 & 	Arecibo	 & 	\cite{Haynes+11}	   \\                        
23 &	2MASXJ1011+0746     &       9.82        $\pm$    0.02       &     	9.88   $\pm$     0.02  		 & 	Arecibo	 & 	\cite{Haynes+11}		   \\                        
24 &	PGC031905           &       9.619       $\pm$    0.013      &     	9.631  $\pm$     0.012 		 & 	Arecibo	 & 	\cite{Haynes+11}		    \\                        
25 &	PGC031382           &       9.94        $\pm$    0.02       &     	9.98   $\pm$     0.02  		 & 	Arecibo	 & 	\cite{Haynes+11}	   \\                        
26 &	2MASXJ1110+0411     &       9.72        $\pm$    0.03       &     	9.75   $\pm$     0.18  		 & 	Arecibo	 & 	\cite{Haynes+11} 		   \\                        
27 &	2MASXJ0855+0345     &          			$\leq$ 9.87        	&     		$\leq$ 9.92		   		 & 	Parkes	 & 	\cite{Barnes+01}$^\dag$		\\               
28 &	2MASXJ0839+0349     &          			$\leq$ 9.92         &         	$\leq$ 9.94          	 & 	Parkes	 & 	\cite{Barnes+01}$^\dag$	   \\              
29 &	UGC04977            &          			$\leq$ 9.84         &        	$\leq$ 9.88          	 & 	Parkes	 & 	\cite{Barnes+01}$^\dag$	   \\              
30 &	2MASXJ0846+0230     &     10.16       $\pm$    0.15         &     10.16      $\pm$     0.15  	 &	Parkes	 & 	\cite{Barnes+01}$^\dag$ \\                       
31 &	SDSSJ1328-0202      &      9.35       $\pm$    0.13         &      9.37      $\pm$     0.12  	 & 	Parkes	 & 	\cite{Barnes+01}$^\dag$	   \\                       
32 &	MCG+00-34-038       &      9.98       $\pm$    0.12         &     10.00      $\pm$     0.12  	 & 	Nan{\c c}ay   & 	\cite{Theureau+05}		   \\                       
33 &	UGC08526            &      9.86       $\pm$    0.09         &      9.88      $\pm$     0.09  	 & 	Parkes	 & 	\cite{Barnes+01}$^\dag$		   \\                       
34 &	CGCG017-017         &      9.86       $\pm$    0.05         &      9.94      $\pm$     0.06  	 & 	Nan{\c c}ay	 & 	\cite{Springob+05}   \\                       
35 &	NGC2936 	        &          			$\leq$9.95      	&  			$\leq$10.04	     	 	 & 	Parkes	 & 	\cite{Barnes+01}$^\dag$   \\               
36 &	SDSSJ0937+0927      &       9.784      $\pm$     0.012      &      9.831     $\pm$    0.012  	 & 	Arecibo	 & 	\cite{Haynes+11} 	  \\                         
37 &	SDSSJ0950+1118    	&       8.79      $\pm$     0.06         &      8.79     $\pm$    0.06   	 & 	Arecibo	 & 	\cite{Haynes+11}		 \\                         
38 &	IC3069            	&       9.32      $\pm$     0.03         &      9.37     $\pm$    0.03   	 & 	Arecibo	 & 	\cite{Haynes+11}		 \\                         
39 &	SDSSJ1112+0931    	&       9.35      $\pm$     0.02         &      9.37     $\pm$    0.02   	 & 	Arecibo	 & 	\cite{Haynes+11} 		 \\                         
40 &	SDSSJ0805+0659    	&       9.18      $\pm$     0.02         &      9.20     $\pm$    0.02   	 & 	Arecibo	 & 	\cite{Haynes+11} 		 \\                         
41 &	SDSSJ1104+0507    	&       8.93      $\pm$     0.04         &      8.95     $\pm$    0.04   	 & 	Arecibo	 & 	\cite{Haynes+11}		 \\                         
42 &	CGCG050-042       	&       9.03      $\pm$     0.02         &      9.04     $\pm$    0.02   	 & 	Arecibo	 & 	\cite{Haynes+11} 		 \\                         
43 &	2MASXJ0858+0345   	&           	$\leq$9.78	 	    	 &     		$\leq$9.79           	 & 	Parkes	 & 	\cite{Barnes+01}$^\dag$ 		 \\                 
44 &	SDSSJ1008+1428    	&      9.33        $\pm$    0.03        &     9.37     $\pm$   0.03      	 & 	Arecibo	 & 	\cite{Haynes+11} 		\\                         
45 &	SDSSJ0954+0458    	&      8.79        $\pm$    0.06        &     8.81     $\pm$   0.06      	 & 	Arecibo	 & 	\cite{Haynes+11}		\\                         
46 &	SDSSJ1213+1056	  	&      8.91        $\pm$    0.07        &     8.92     $\pm$   0.07      	 & 	Arecibo	 & 	\cite{Haynes+11}		\\                         
47 &	SDSSJ0945+0515	  	&      8.99        $\pm$    0.05        &     9.00     $\pm$   0.05      	 & 	Arecibo	 & 	\cite{Haynes+11} 		\\                         
48 &	SDSSJ1122+1316	  	&      9.02        $\pm$    0.04        &     9.03     $\pm$   0.04      	 & 	Arecibo	 & 	\cite{Haynes+11}		\\                         
49 &	SDSSJ1403+1003	  	&      8.70        $\pm$    0.04        &     8.74     $\pm$   0.04      	 & 	Arecibo	 & 	\cite{Haynes+11} 		\\                         
50 &	CGCG063-006		  	&      9.517        $\pm$    0.011      &     9.540     $\pm$   0.010    	 & 	Arecibo	 & 	\cite{Haynes+11} 		  \\                         
51 &	2MASXJ0843+1303	  	&      8.83        $\pm$    0.03        &     8.84     $\pm$   0.03      	 & 	Arecibo	 & 	\cite{Haynes+11}		\\                         
52 &	SDSSJ0920+0759	  	&      9.05        $\pm$    0.04       &      9.09    $\pm$    0.04      	 & 	Arecibo	 & 	\cite{Haynes+11}		\\                         
53 &	CGCG080-042		  	&      9.49        $\pm$    0.02       &      9.59    $\pm$    0.02      	 & 	Arecibo	 & 	\cite{Haynes+11}		\\                         
54 &	CGCG078-021		  	&      9.35        $\pm$    0.03       &      9.38    $\pm$    0.03      	 & 	Arecibo	 & 	\cite{Haynes+11}		\\                         
55 &	UGC04567		  	&      9.591        $\pm$    0.009     &      9.669    $\pm$    0.009    	 & 	Arecibo	 & 	\cite{Haynes+11}		  \\                         
56 &	SDSSJ1624+1251	  	&      9.43        $\pm$    0.02       &      9.46    $\pm$    0.02      	 & 	Arecibo	 & 	\cite{Haynes+11}		\\                         
57 &	CGCG058-066		  	&      9.38        $\pm$    0.02       &      9.40    $\pm$    0.02      	 & 	Arecibo	 & 	\cite{Haynes+11} 		\\                         
58 &	CGCG061-003		  	&      9.35        $\pm$    0.02       &      9.43    $\pm$    0.02      	 & 	Arecibo	 & 	\cite{Haynes+11}		\\                         
59 &	CGCG051-037		  	&      9.36        $\pm$    0.02       &      9.39    $\pm$    0.02      	 & 	Arecibo	 & 	\cite{Haynes+11} 		\\                         
60 &	SDSSJ0854+0418	  	&      9.42        $\pm$    0.03       &      9.45    $\pm$    0.03      	 & 	Arecibo	 & 	\cite{Haynes+11}		\\                         
61 &	CGCG059-031		  	&      9.13        $\pm$    0.04       &      9.19    $\pm$    0.04      	 & 	Arecibo	 & 	\cite{Haynes+11}		\\                         
62 &	2MASXJ0806+1249	  	&      9.05        $\pm$    0.03       &      9.07    $\pm$    0.03      	 & 	Arecibo	 & 	\cite{Haynes+11} 		\\                         
63 &	VIIIZW039		  	&      8.92        $\pm$    0.07       &      8.93    $\pm$    0.06      	 & 	Arecibo	 & 	\cite{Haynes+11}		\\                         
64 &	2MASXJ0941+1056	  	&      9.39        $\pm$    0.02       &      9.39    $\pm$    0.02      	 & 	Arecibo	 & 	\cite{Haynes+11}		\\                         
65 &	SDSSJ0943+0356	  	&      9.35        $\pm$    0.03       &      9.37    $\pm$    0.03      	 & 	Arecibo	 & 	\cite{Haynes+11} 		\\                         
66 &	CGCG035-063		  	&      9.676        $\pm$    0.014     &      9.681    $\pm$    0.014    	 & 	Arecibo	 & 	\cite{Haynes+11} 		  \\                         
67 &	CGCG036-048		  	&      9.10        $\pm$    0.02       &      9.10    $\pm$    0.02      	 & 	Arecibo	 & 	\cite{Haynes+11}		\\                         
68 &	SDSSJ1625+1142	  	&      9.01        $\pm$    0.04       &      9.07    $\pm$    0.04      	 & 	Arecibo	 & 	\cite{Haynes+11}		\\                         
69 &	SDSSJ1028+0424	  	&      9.55        $\pm$    0.02       &      9.60    $\pm$    0.02      	 & 	Arecibo	 & 	\cite{Haynes+11}		\\                         
70 &	UGC06011		  	&      9.99        $\pm$    0.11       &     10.03    $\pm$    0.11 		 & 	Parkes	 & 	\cite{Barnes+01}$^\dag$ 	     \\                         
71 &	PGC051743 		  	&      9.44        $\pm$    0.03       &      9.45    $\pm$    0.03 		 & 	Arecibo	 & 	\cite{Haynes+11} 	     \\                         
72 &	2MASXJ1437+0500   	&      9.69        $\pm$    0.02       &      9.74    $\pm$    0.02 		 & 	Arecibo	 & 	\cite{Haynes+11} 	     \\                         
73 &	SDSSJ0944+0400	  	&      9.55        $\pm$    0.02       &      9.58    $\pm$    0.02 		 & 	Arecibo	 & 	\cite{Haynes+11} 	     \\                         
74 &	2MASXJ1048+1201	  	&      9.24        $\pm$    0.03       &      9.27    $\pm$    0.03 		 & 	Arecibo	 & 	\cite{Haynes+11}	     \\                         
75 &	SDSSJ1110+1345	  	&         $\leq$9.22        		   &      $\leq$9.27      				 & 	Parkes	 & 	\cite{Barnes+01}$^\dag$  		 \\                
76 &	PGC012214		  	&        9.92     $\pm$     0.08      &    9.98    $\pm$     0.08   		 & 	Nan{\c c}ay	 &  \cite{Paturel+03}	    \\                          
77 &	PGC003530		  	&        9.78     $\pm$     0.10      &    9.80    $\pm$     0.09   		 & 	Parkes	 & 	\cite{Barnes+01}$^\dag$	    \\                          
78 &	PGC073268		  	&        9.24     $\pm$     0.04      &    9.27    $\pm$     0.04   		 & 	Arecibo	 & 	\cite{Haynes+11}    \\                          
79 &	PGC1452135		  	&        9.16     $\pm$     0.03      &    9.22    $\pm$     0.03   		 & 	Arecibo	 & 	\cite{Haynes+11} 	    \\                          
80 &	PGC3127469		  	&        9.08     $\pm$     0.03      &    9.09    $\pm$     0.03   		 & 	Arecibo	 & 	\cite{Haynes+11} 	    \\                          
81 &	UGC00317		  	&        9.2     $\pm$     0.2        &   9.2    $\pm$     0.2      		 & 	Parkes	 & 	\cite{Barnes+01}$^\dag$	 \\                          
82 &	SDSSJ0011+1428    	&        8.95     $\pm$     0.05      &    8.97    $\pm$     0.05   		 & 	Arecibo	 & 	\cite{Haynes+11} 	  	\\                             
83 &	PGC1455779		  	&        9.50     $\pm$     0.03      &    9.56    $\pm$     0.03   		 & 	Arecibo	 & 	\cite{Haynes+11}  	    \\                          
84 &	PGC000010		  	&        9.7     $\pm$     0.2        &    9.7    $\pm$     0.2     		 & 	Parkes	 & 	\cite{Barnes+01}$^\dag$  \\                          
85 &	PGC1446233		  	&        9.30     $\pm$     0.09      &    9.36    $\pm$     0.10   		 & 	Arecibo	 & 	\cite{Haynes+11}$^\dag$    \\                          
86 &	2MASXJ0020+1413   	&        9.05     $\pm$     0.03      &    9.06    $\pm$     0.03   		 & 	Arecibo	 & 	\cite{Haynes+11}	    \\                          
87 &	PGC1464874		  	&        9.633     $\pm$     0.010    &    9.70    $\pm$     0.010  		 & 	Arecibo	 & 	\cite{Haynes+11} 	     \\                          
88 &    IC1706			  	&        9.30     $\pm$     0.03      &    9.30    $\pm$     0.03   		 & 	Arecibo	 & 	\cite{Haynes+11}	    \\                          
\hline                            
\multicolumn{6}{c}{IRAM Targets}		\\
\hline
89 & 	SDSSJ0944+1116	    &  	    9.44      $\pm$     0.04       &    9.46     $\pm$      0.04  		   & 	Arecibo	 & 	\cite{Haynes+11}  \\	        
90 & 	SDSSJ1049+1108		&       9.49      $\pm$    0.04        &   9.53      $\pm$     0.04   			& 	Arecibo	 & 	\cite{Haynes+11} 		  \\          
91 & 	2MASXJ1336+1552		&       9.41      $\pm$    0.03        &   9.42      $\pm$     0.03   			& 	Arecibo	 & 	\cite{Haynes+11} 		  \\          
92 & 	SDSSJ1032+1227		&          		$\leq$9.55        		&     $\leq$9.55      		  			& 	Arecibo	 & 	\cite{Haynes+11}$^\dag$ 			\\            
93 & 	SDSSJ1100+1207		&      9.24       $\pm$    0.14       &     9.27     $\pm$   0.14     			& 	Arecibo	 & 	\cite{Haynes+11}$^\dag$ 		\\            
94 & 	SDSSJ1207+1200		&      9.50       $\pm$    0.09       &     9.53     $\pm$   0.09     			& 	Arecibo	 & 	\cite{Haynes+11}$^\dag$		\\            
95 & 	KUG1147+149			&      9.52       $\pm$    0.02       &     9.58     $\pm$   0.02     			& 	Arecibo	 & 	\cite{Haynes+11}		\\            
96 & 	SDSSJ1320+1524		&      8.85       $\pm$    0.18       &     8.87     $\pm$   0.18     			& 	Arecibo	 & 	\cite{Haynes+11}$^\dag$	 		\\            
97 & 	VPC0873				&      9.677       $\pm$    0.015       &     9.680     $\pm$   0.015     		& 	Arecibo	 & 	\cite{Haynes+11}		\\            
\hline                                                                                                                  
\end{longtable}     
$^\dag$ For these sources we have re-performed the H{\sc i}~21cm spectral analysis on the original data and 
used the H{\sc i} flux (or upper limit on it) measured by us to calculate $M_{\rm HI}$ (further details in \S~\ref{sec:HIdata}).

\tablefoot{Col. (1): ALLSMOG ID. Col (2): Galaxy name. Col (3): H{\sc i} gas mass computed from the integrated H{\sc i}~21cm emission line flux following Eq~\ref{eq:HI_mass}. Col (4): H{\sc i} gas mass corrected for self-absorption as explained in $\S$~\ref{sec:HIdata}. Col (5): Radio telescope used for the H{\sc i}~21cm observations. Col (6): Reference for the H{\sc i}~21cm observations used in this work. }
}

\subsubsection{H{\sc i}~21cm spectra}\label{sec:HI21cm}

As mentioned in $\S$~\ref{sec:sample}, all galaxies included in ALLSMOG have at least one publicly-available observation of the spin-flip 
neutral hydrogen transition at a rest-frame wavelength $\lambda=21.1$~cm ($\nu=1420.406$ ~MHz), although not 
all are detected in H{\sc i}. There are 10 ALLSMOG galaxies without a H{\sc i} detection, corresponding to 11\% of the total ALLSMOG sample. We note however that, unlike our APEX CO(2-1) and IRAM CO(1-0) observations, 
the sensitivity of the ancillary H{\sc i} data available for the ALLSMOG sources is not uniform throughout the sample, 
with different galaxies observed with different telescopes. Hence, within our sample, a non detection in H{\sc i} does not necessarily correspond to a smaller reservoir of H{\sc i} gas compared to a H{\sc i} detection.

Most ALLSMOG galaxies are part of two large surveys of H{\sc i} in local galaxies, such as
ALFALFA \citep{Haynes+11} and HIPASS \citep{Barnes+01}, conducted respectively with the 
Arecibo 305m and the Parkes 64m radio telescope. For the remaining objects we use H{\sc i} data acquired by the Nan{\c c}ay or Green Bank radio observatories. 
The references for the H{\sc i} observations of ALLSMOG galaxies that we employed in this work are listed in Table~\ref{table:HI_parameters}.

The H{\sc i}~21cm spectra are shown in Figs.~\ref{fig:spectra1}-\ref{fig:spectra10}, aligned in velocity space to the corresponding CO spectra to allow a direct comparison between the CO and H{\sc i} line profiles. For display purposes, the y-axes in these plots show the CO spectral units only, and the H{\sc i} spectra were renormalised to an arbitrary scale. All H{\sc i}~21cm spectra shown in this work are publicly available, and they can be retrieved either from the NED spectral database\footnote{\url{https://ned.ipac.caltech.edu/forms/SearchSpectra.html}} or from
the HIPASS website\footnote{\url{http://www.atnf.csiro.au/research/multibeam/release/}}.

\subsubsection{H{\sc i} gas masses and upper limits}

The H{\sc i} gas masses listed in Table~\ref{table:HI_parameters} were computed from the integrated H{\sc i}~21cm line fluxes following \cite{Catinella+10}:
\begin{equation}\label{eq:HI_mass}
M_{\rm HI} [M_{\odot}] = \frac{2.356 \times 10^5}{(1+z)} D_L [{\rm Mpc}]^2 \int S_{\rm HI}~d\varv~[{\rm Jy~km~s^{-1}}] 
\end{equation}
where $z$ and $D_L$ are the redshift and luminosity distance of the source (Table~\ref{table:sample_prop_1}), and $\int S_{\rm HI} d\varv$ is the velocity-integrated H{\sc i} emission line flux. In most cases we adopted the flux values reported in the corresponding catalogues referenced in Table~\ref{table:HI_parameters}, 
after having checked that these are consistent with the H{\sc i} line detections shown in Figs.~\ref{fig:spectra1}-\ref{fig:spectra10}. For 24 galaxies (indicated by a `$\dag$' next to the corresponding 
reference in Table~\ref{table:HI_parameters}), including the ten objects in which there is not a clear H{\sc i} line detection, we performed our own H{\sc i}~21cm spectral analysis. 

For the 10 non-detections in H{\sc i}, we estimated 3$\sigma$ upper limits on the total integrated line flux as follows:
\begin{equation}\label{eq:HI_upperlimit}
\int S_{\rm HI}~d\varv<3\sigma_{rms, channel}\sqrt{\delta \varv_{channel}\Delta \varv_{line} }, 
\end{equation}
where $\sigma_{rms, channel}$ is the rms noise per velocity channel, $\delta \varv_{channel}$ is the channel width, and $\Delta \varv_{line}$ is the expected H{\sc i} line width. For eight of these galaxies we have a CO detection from ALLSMOG observations, and so we assumed $\Delta \varv_{line}$ to be equal to the FWHM of the CO line, whereas in the other two objects that are undetected in both CO and H{\sc i} we inserted in Eq.~\ref{eq:HI_upperlimit} $\Delta \varv_{line}\sim160$~\kms, which is the average CO line FWHM measured in the ALLSMOG sample. 
In the other 14 sources showing an H{\sc i} detection we fitted the observed H{\sc i} line profile with Gaussian functions: among these, only two galaxies (UGC~08526 and PGC~1446233) display double-horn H{\sc i} profiles that required the use of two Gaussians instead of one.

The H{\sc i} gas masses calculated using Eq.~\ref{eq:HI_mass} were then corrected by a multiplicative factor to take into account 
self-absorption of H{\sc i} affecting the densest regions of galaxy disks. 
Such correction factor increases with disk inclination and is estimated to be $\sim (a/b)^{0.12}$ following \cite{Giovanelli+94} 
and \cite{Springob+05}, where $a$ and $b$ are the optical major and minor axes. We took the $a$ and $b$ values from the NASA/IPAC Extragalactic Database (NED). The final H{\sc i} gas 
mass values corrected for self-absorption are
given by $M_{\rm HI}^{corr}= (a/b)^{0.12}~M_{\rm HI}$. Table~\ref{table:HI_parameters} lists both the $M_{\rm HI}$ and the $M_{\rm HI}^{corr}$ values for ALLSMOG galaxies.


\section{Literature CO samples used in this work}\label{sec:coldgass}

With the aim of extending the dynamic range of galaxy properties probed by our study, we supplement the ALLSMOG dataset with CO observations of local (massive) galaxies available in the literature. The CO legacy database for the GASS survey (COLD GASS, \cite{Saintonge+11a}) is the optimal starting point to construct a sample that is well suited to complement ALLSMOG in the high-$M_*$ regime. We make use of the third data release (DR3) of the COLD GASS catalogue \footnote{\texttt{http://wwwmpa.mpa-garching.mpg.de/COLD\_GASS/}}, which contains IRAM~30m observations of the CO(1-0) emission line in 366 local galaxies with stellar masses, $M_*\gtrsim10^{10}~M_{\odot}$. The COLD GASS galaxies are randomly drawn from a parent sample that fulfils the following criteria: (i) overlap with: the SDSS spectroscopic survey, the ALFALFA survey and the projected footprint of the GALEX medium imaging survey; (ii) redshift, $0.025<z<0.05$; and (iii) stellar mass, $10^{10}<M_* [M_{\odot}] < 10^{11.5}$. For further information on the COLD GASS project and the relevant publicly available catalogues we refer to the works by \cite{Saintonge+11a, Saintonge+11b, Saintonge+12} and \cite{Catinella+10}. 

With a target selection based solely on stellar mass, COLD GASS includes many star-forming galaxies, located both on and above the local MS on the $M_*-SFR$ diagram, as well as more quiescent or passive galaxies significantly below the MS. Furthermore, since the COLD GASS selection criteria do not explicitly exclude AGNs, the survey includes also AGN host galaxies. In order to construct a sample suited to complement ALLSMOG, we applied to the COLD GASS catalogue the same BPT selection cut that we used to select the ALLSMOG targets (Point 1 in $\S$~\ref{sec:sample}), that is we included only galaxies classified as `star-forming' according to the division lines of \cite{Kauffmann+03} and the S/N criteria of \cite{Brinchmann+04}. This method leaves us with 88 COLD GASS sources, whose location on the $M_*-{\rm SFR}$ plane is shown in Fig.~\ref{fig:MS}. The final galaxy sample, including ALLSMOG and the sub-sample of 88 COLD GASS sources, covers quite uniformly the local MS locus and its $\pm0.3$~dex scatter from $M_*=10^{8.5}~M_{\odot}$ to $M_*\sim10^{11}~M_{\odot}$. The region of the $M_*- {\rm SFR}$ parameter space up to $\sim 0.5$~dex above the local MS is also probed quite evenly across the full stellar mass range, with a slight excess of `starburst' galaxies at $M_*>10^{10}~M_{\odot}$.

Since the COLD GASS DR3 catalogue lists CO(1-0) fluxes only for the sources detected in CO, we estimated 3$\sigma$ upper limits for the non-detections by inserting into Eq.~\ref{eq:COul} the $\sigma_{rms, channel}$ values provided in the catalogue, measured on channel widths of $\delta \varv_{channel}=21.57$~\kms, and by assuming an average CO line width of $\Delta \varv_{line}\sim300$~\kms.
For the detections we used the fluxes corrected for aperture effects provided by the COLD GASS team, while for the non-detections we estimated the beam corrections similar to ALLSMOG galaxies following the method described in \cite{Bothwell+14} (see also $\S$~\ref{sec:ap_cor}). We note that only two out of the 88 COLD GASS sources included in our analysis are not detected in CO(1-0), because most CO non-detections in COLD GASS are red (and low-SSFR) galaxies that are cut out by our BPT selection and S/N requirements on the optical lines. 

The physical parameters of COLD GASS galaxies used in this work are either directly extracted from the DR3 catalogue or computed by us following the same procedures as for the ALLSMOG sample described in $\S$~\ref{sec:ancillary}. More specifically, $z$, SFR, $M_*$, $d_{25}$ and $M_{\rm HI}$ are all provided in the COLD GASS DR3 catalogue\footnote{Unfortunately the catalogue does not provide upper limits for non-detections in H{\sc i} (i.e. 23 out of the 88 galaxies considered in our analysis).} and the disk inclination is drawn from the Hyperleda database as for the ALLSMOG sample ($\S$~\ref{sec:optdisk_par}). The gas-phase metallicity and the nebular visual extinction were calculated by us analogously to the ALLSMOG sample by exploiting the SDSS spectroscopic data products in the MPA-JHU catalogue ($\S$~\ref{sec:metallicity} and $\S$~\ref{sec:av_deriv}).



\section{Distribution of detections in ALLSMOG}\label{sec:histo}

In this section (Figs~\ref{fig:histo} to \ref{fig:histo_HI}) we present histogram plots displaying the distribution of the ALLSMOG detections and non-detections as a function of several galaxy physical parameters: $M_*$, SFR, SSFR, $z$, $A_V$, 12+$\log(O/H)$ (five different calibrations), and $M_{\rm HI}$ (see $\S$~\ref{sec:ancillary} for the details on the derivation of these parameters). At the bottom of each histogram we report the corresponding detection fraction per bin, defined as the ratio between the number of detections and the number of sources observed in that bin ($N_{det}/N_{obs}$). We use the two-sample Kolmogorov-Smirnov (KS) test to check if the distributions of detections and non-detections as a function of each galaxy parameter of interest are statistically different (see \cite{Press+92} for a description of the algorithm). The results of the K-S tests are reported in Table~\ref{table:KS_test}. We note that the K-S statistics is computed on the full unbinned dataset and so the K-S test results are completely independent of the histogram binning.

\subsection{As a function of $M_*$, SFR, SSFR and $z$}\label{sec:histo_1}

\begin{figure*}[tbp]
\centering
\includegraphics[clip=true,trim=4.5cm 4cm 0.8cm 2cm,width=0.37\textwidth,angle=180]{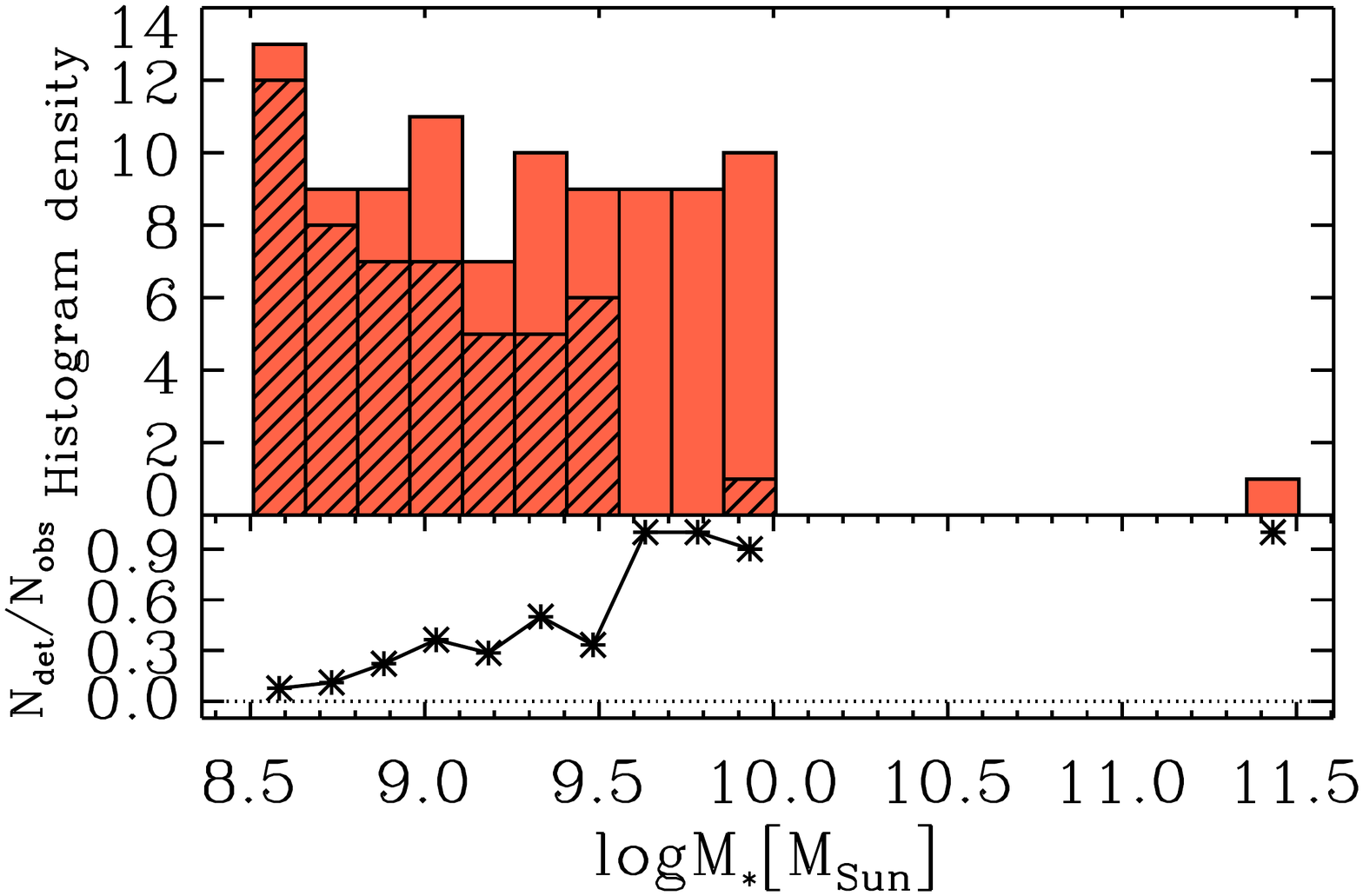}\quad
    \includegraphics[clip=true,trim=4.5cm 4cm 0.8cm 2cm,width=0.37\textwidth,angle=180]{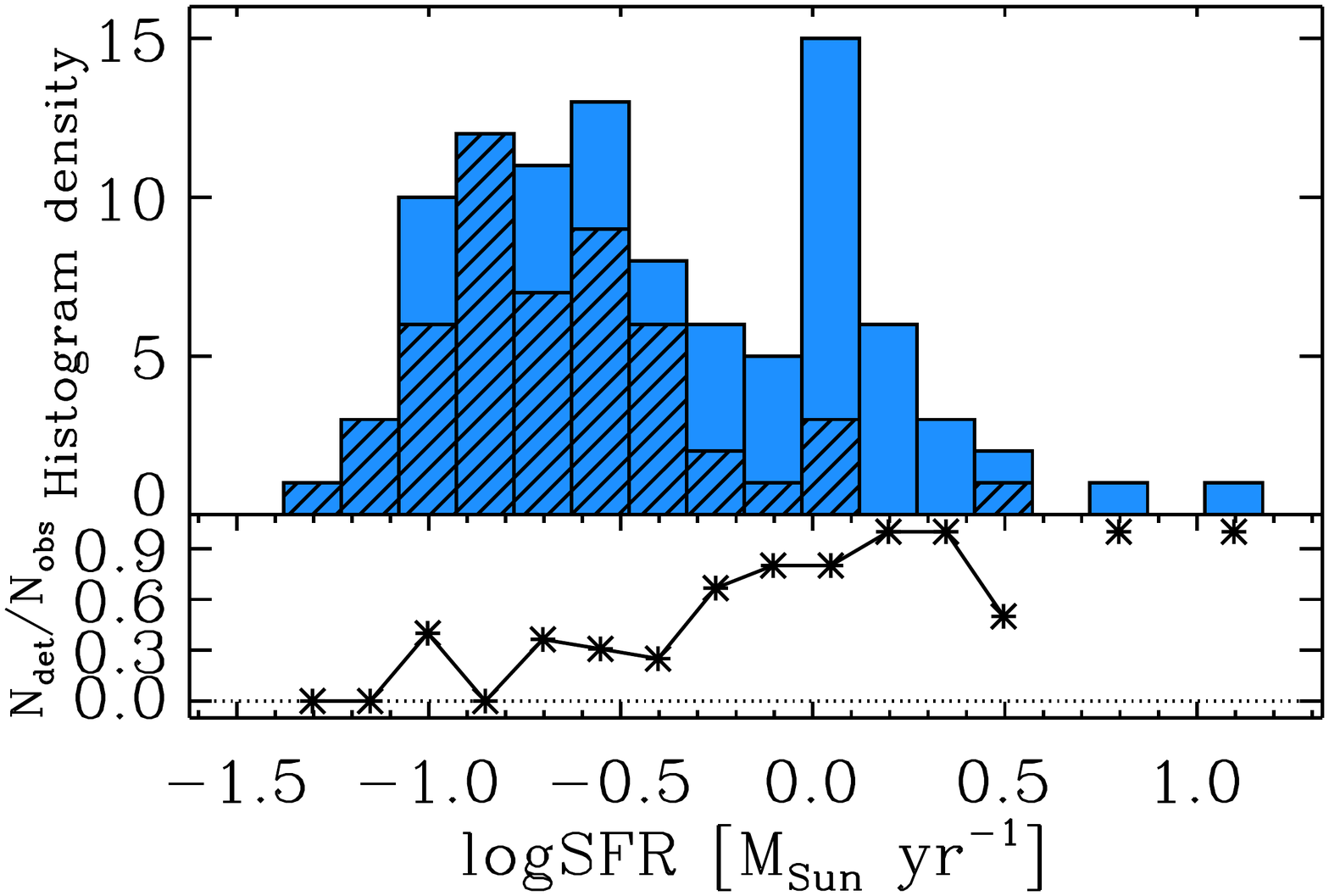}\\ 
    \vspace{0.1cm} 
        \includegraphics[clip=true,trim=4.5cm 4cm 0.8cm 2cm,width=0.37\textwidth,angle=180]{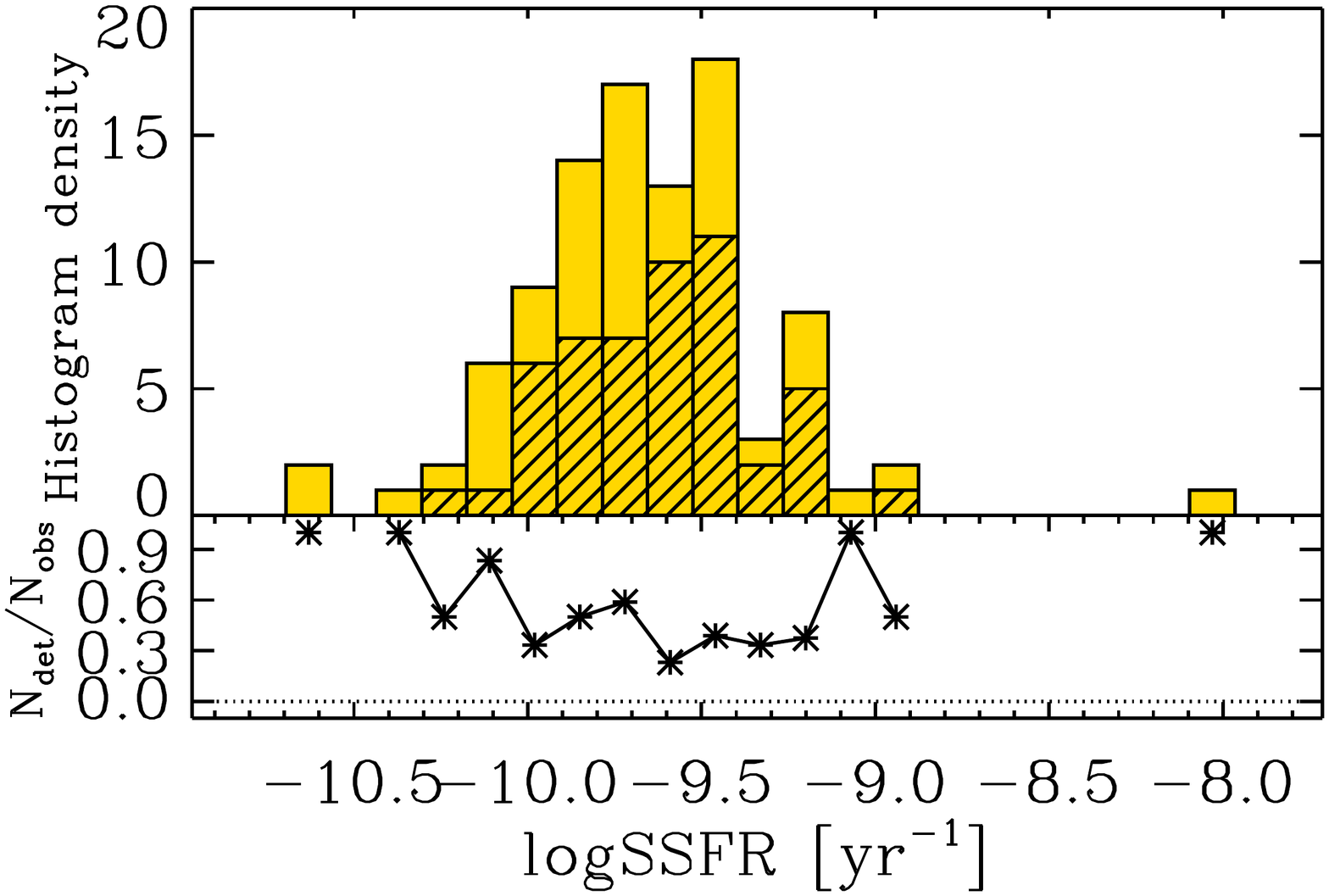}\quad 
    \includegraphics[clip=true,trim=4.5cm 4cm 0.8cm 2cm,width=0.37\textwidth,angle=180]{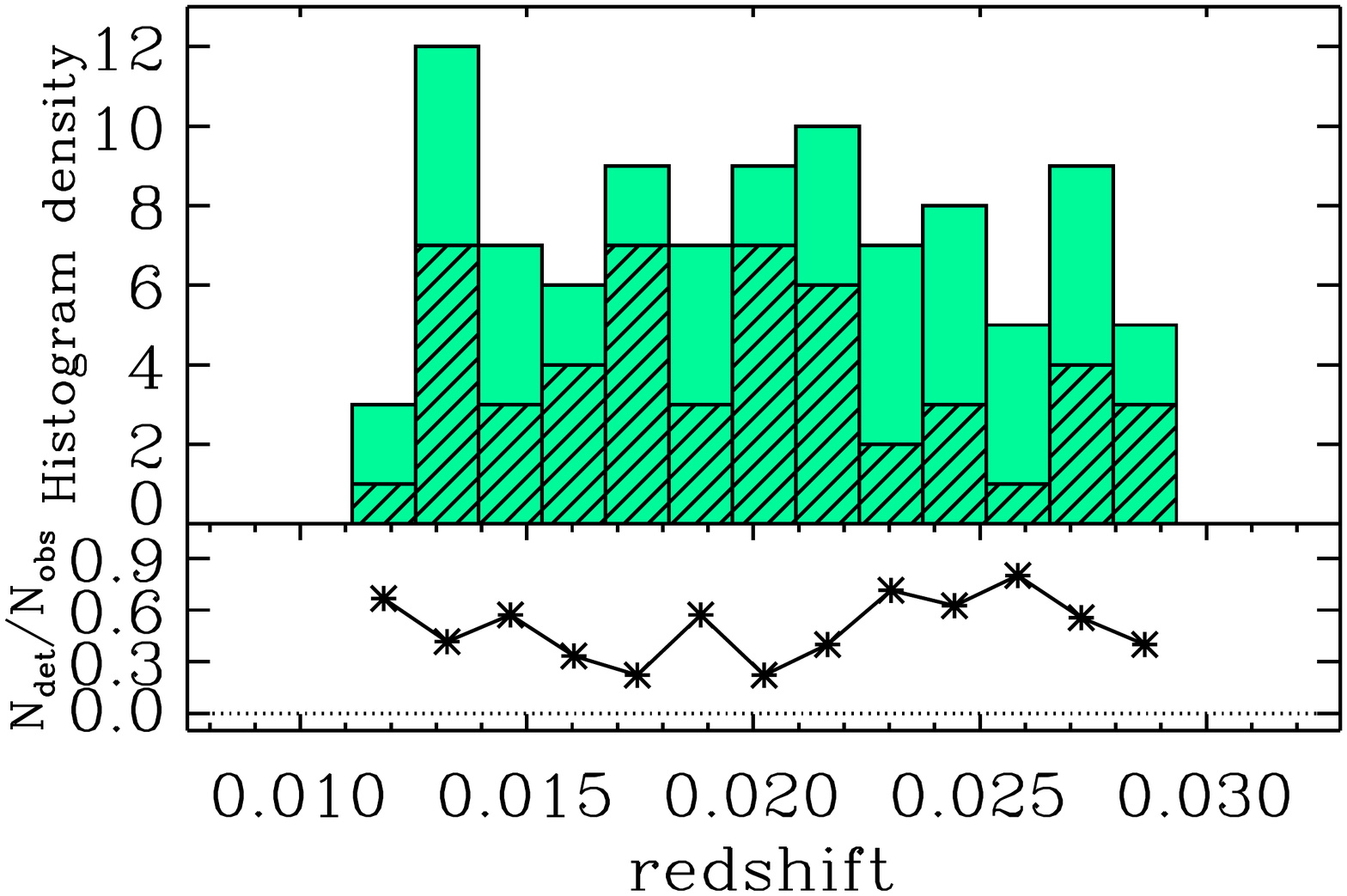}\\ 
    \vspace{0.2cm} 
     \caption{Distribution of ALLSMOG galaxies as a function of $M_*$ ({\it upper-left}), SFR ({\it upper-right}), SSFR ({\it bottom-left}) and redshift ({\it bottom-right}). The full sample is represented by the coloured histogram, and the hatched histogram trace the non-detections. At the bottom of each histogram we show the corresponding detection fraction per bin, computed as the ratio between the number of sources detected in CO and the number of sources observed in that bin.}
   \label{fig:histo}
\end{figure*}

The distribution of ALLSMOG detections and non-detections as a function of $M_*$, SFR, SSFR and redshift is displayed in Fig.~\ref{fig:histo}. From the histogram in the top-left panel of Fig.~\ref{fig:histo} it can be inferred that that more than half (30/51, i.e. $\sim59$~\%) of the non-detections are low-mass galaxies with $M_*<10^9~M_{\odot}$, and that almost all of them (47/51, i.e. $\sim92$~\%) have $M_*<10^{9.5}~M_{\odot}$. The detection fraction, reported at the bottom of the histogram, is indeed strongly dependent on $M_*$. In particular, we evidence a mild rise in the fraction of CO detections from $M_*=10^{8.5}~M_{\odot}$ to $M_*=10^{9.5}~M_{\odot}$, followed by an abrupt increase above this stellar mass value. The total detection rate at $M_*>10^{9.5}~M_{\odot}$ is $\simeq88$~\% (29/33 detections). Not surprisingly, the two-sample K-S test, whose results are summarised in Table~\ref{table:KS_test}, returns a very low probability associated to the null hypothesis that the two samples come from the same distribution in $M_*$.

The distribution in SFR, shown at the top-right of Fig.~\ref{fig:histo}, follows loosely the distribution in $M_*$, with the non-detections gathering towards lower SFRs. This is expected because, in our sample, SFR and $M_*$ are correlated, as demonstrated by Fig.~\ref{fig:MS}. Also in this case the K-S test returns a negligible probability that the two samples of detections and non-detections come from the same SFR distribution (Table~\ref{table:KS_test}).

The bottom panels of Fig.~\ref{fig:histo} show that in ALLSMOG, the detectability of CO does not depend significantly on the SSFR or redshift. The K-S test indicates that detections and non-detections are consistent with following the same distribution both in terms of SSFR and redshift (Table~\ref{table:KS_test}). The absence of an SSFR bias may be explained by the narrow SSFR range probed by ALLSMOG, which is short of big outliers of the MS relation (Figure~\ref{fig:MS}). Similarly, the lack of a redshift bias, which was already noted by \cite{Bothwell+14}, may be attributed to the narrow redshift range spanned by the survey.

\subsection{As a function of $A_V$}\label{sec:histo_av}

    \begin{figure*}[tbp]
\centering
    \includegraphics[clip=true,trim=4.5cm 4cm 0.8cm 2cm,width=0.37\textwidth,angle=180]{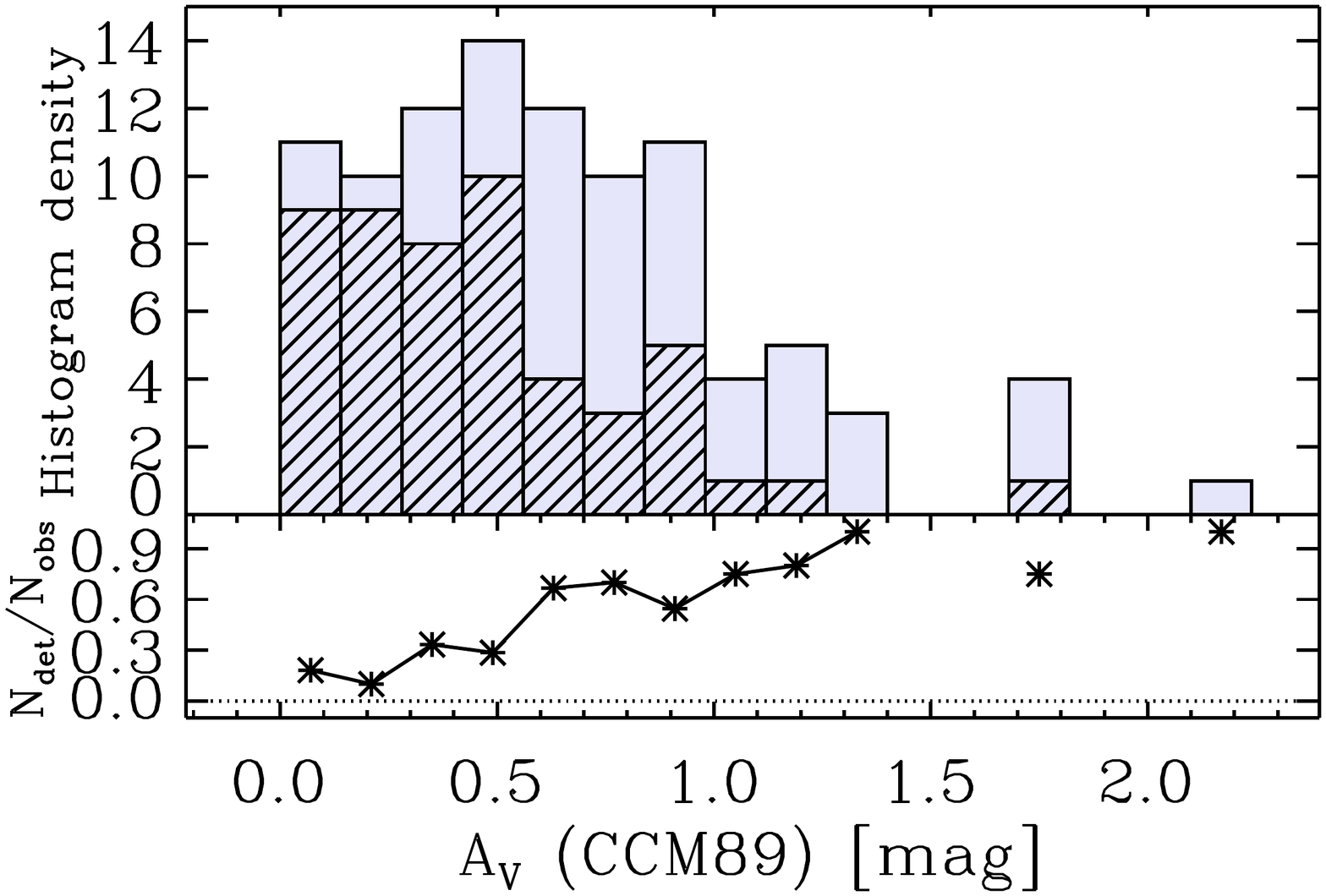}\quad 
   \includegraphics[clip=true,trim=4.5cm 4cm 0.8cm 2cm,width=0.37\textwidth,angle=180]{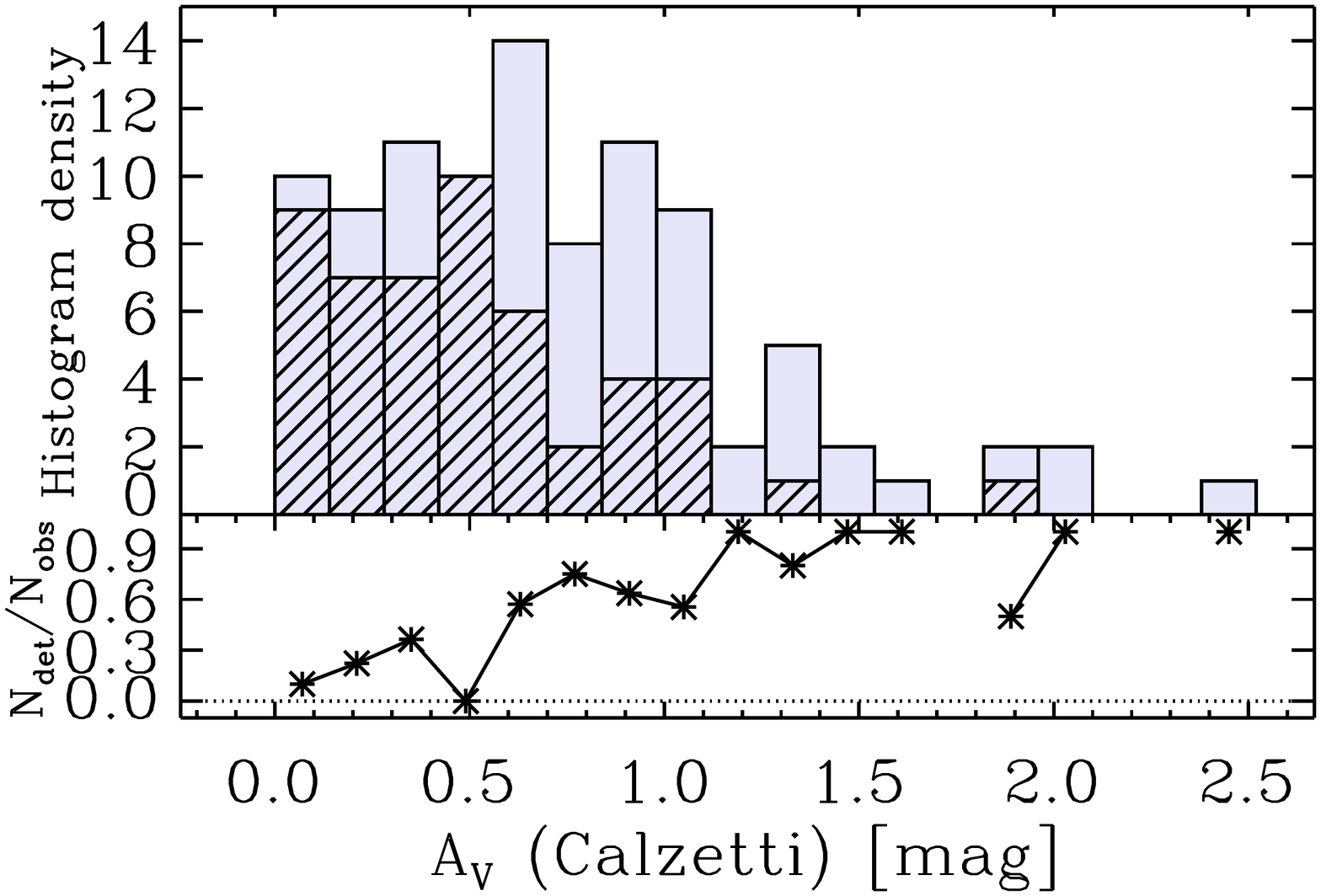}\\ 
  \vspace{0.2cm} 
     \caption{Distribution of ALLSMOG galaxies as a function of the $A_V$ computed using the CCM89 ({\it left}) and the Calzetti ({\it right}) models (see $\S$~\ref{sec:av_deriv}). The full sample is represented by the coloured histogram, and the hatched histogram traces the non-detections. The detection fraction per bin is shown at the bottom of each histogram.}
   \label{fig:histo_extinction}
\end{figure*}

The visual extinction of a galaxy depends on the quantity and properties of dust in the interstellar clouds intersecting our line of sight. The dust mass of a cloud is highly correlated with its molecular gas mass, because where dust is present, the medium is effectively shielded from UV radiation and so H$_2$ and other molecules can survive. The protection offered by dust is particularly important for CO, which, being less able to self-shield than $H_2$, can be photo-dissociated easily in the presence of an ambient UV radiation field. Therefore, the dust content of the ISM can strongly affect the ability of CO to trace molecular clouds, hence the $\alpha_{\rm CO}$. This idea is the starting point of theoretical studies that attempt to provide a prescription (empirically calibrated on the Milky Way) for $\alpha_{\rm CO}$ as a function of the median visual extinction of the gas averaged over several lines of sight (indicated with $\langle A_V \rangle$) \citep{Wolfire+10, Glover+MacLow11}. Estimating $\langle A_V \rangle$ for external galaxies is very difficult, but $\langle A_V \rangle$ is often assumed to be directly proportional to the metallicity of the ISM, and so the relation between $\alpha_{\rm CO}$ and $\langle A_V \rangle$ provided by the aforementioned models can be easily converted into a relation between $\alpha_{\rm CO}$ and gas-phase metallicity (the so-called ``metallicity-dependent $\alpha_{\rm CO}$'' recipes, see \cite{Bolatto+13}).

For ALLSMOG galaxies, we can investigate CO properties as a function of both the nebular $A_V$ and the gas-phase metallicity, which we estimated from the SDSS optical spectra as described in $\S$~\ref{sec:av_deriv} and $\S$~\ref{sec:metallicity}. We acknowledge that the $A_V$ computed using the optical nebular lines samples only one line-of-sight, and so for most galaxies it is likely to be lower than $\langle A_V \rangle$, which is the quantity used by theoretical models \citep{Wolfire+10, Glover+MacLow11}.  

Figure~\ref{fig:histo_extinction} shows the distribution of nebular $A_V$ values for the ALLSMOG sample, where, similar to Fig.~\ref{fig:histo}, the sources not detected in CO are indicated by the hatched histogram. We note that, regardless of the model used (CCM89 or Calzetti), the total ALLSMOG sample follows a roughly flat distribution in $A_V$ between $0<A_V [mag]\lesssim1$, reminiscent of the distribution in $M_*$ (which is flat by construction, see $\S$~\ref{sec:sample}), with an additional tail extending up to $A_V \sim 2.5~mag$ that includes $\sim20$\% of the sources. The detection fraction (displayed at the bottom of the histograms) shows a steady increase with $A_V$.  The K-S test (Table~\ref{table:KS_test}) rules out that detections and non-detections share the same distribution in $A_V$, and indeed the detections are on average higher in $A_V$ than the non-detections ($\langle A_V\rangle_{det}= 0.95$, $\langle A_V \rangle_{undet}= 0.51$).

\subsection{As a function of metallicity}

\begin{figure*}[tbp]
\centering
    \includegraphics[clip=true,trim=4.5cm 4cm 0.8cm 2cm,width=0.37\textwidth,angle=180]{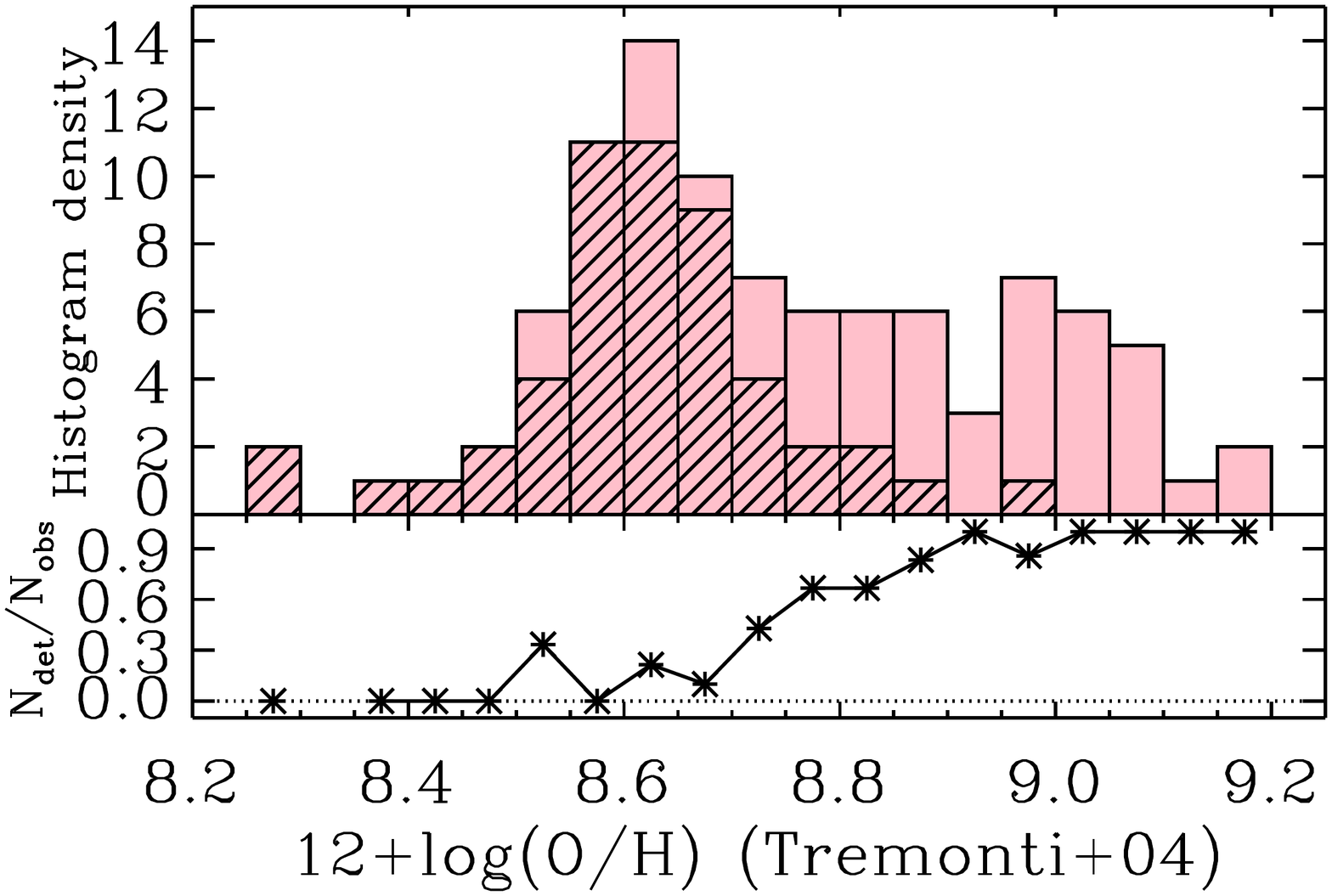}\quad  
    \includegraphics[clip=true,trim=4.5cm 4cm 0.8cm 2cm,width=0.37\textwidth,angle=180]{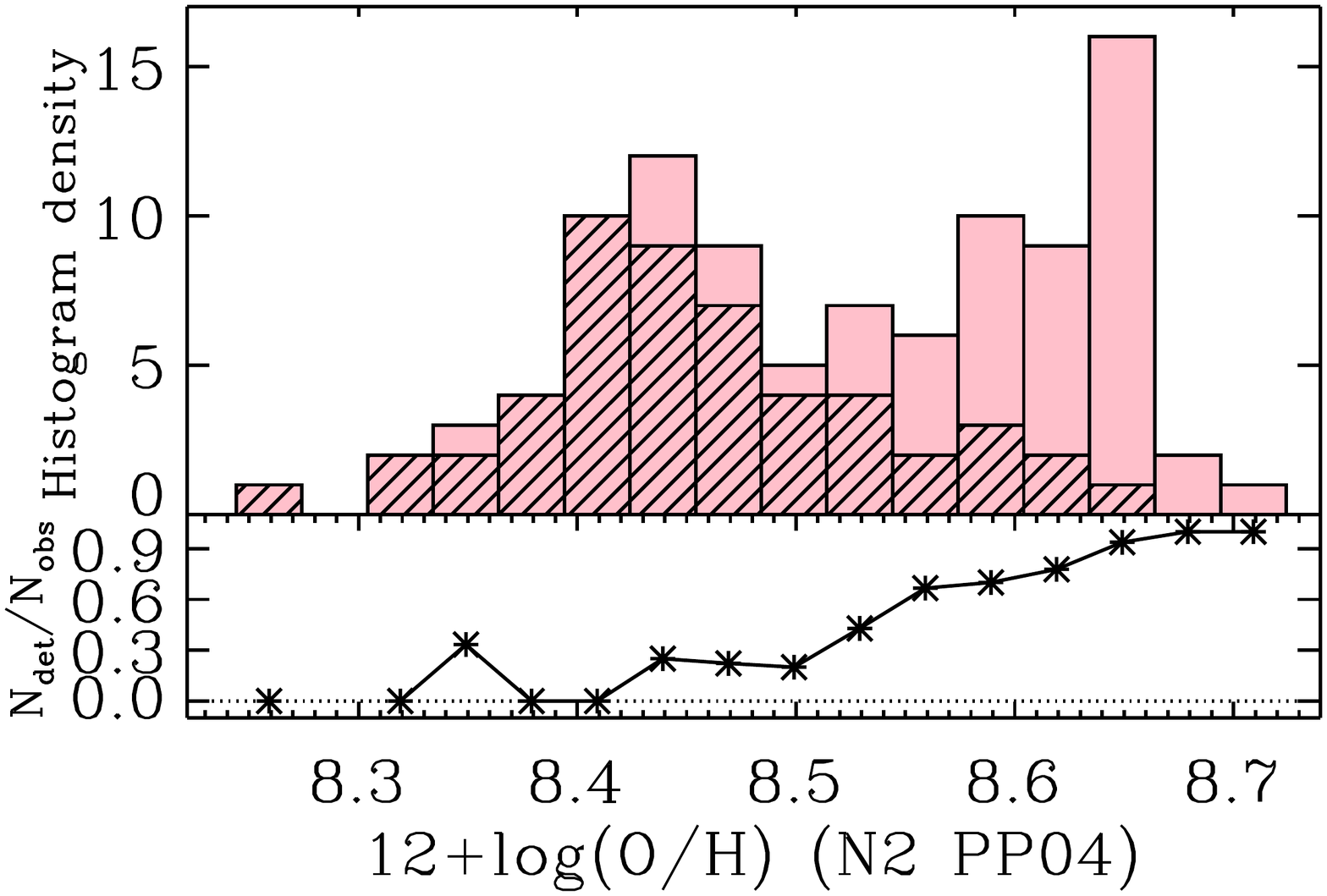}\\ 
    \vspace{0.1cm} 
    \includegraphics[clip=true,trim=4.5cm 4cm 0.8cm 2cm,width=0.37\textwidth,angle=180]{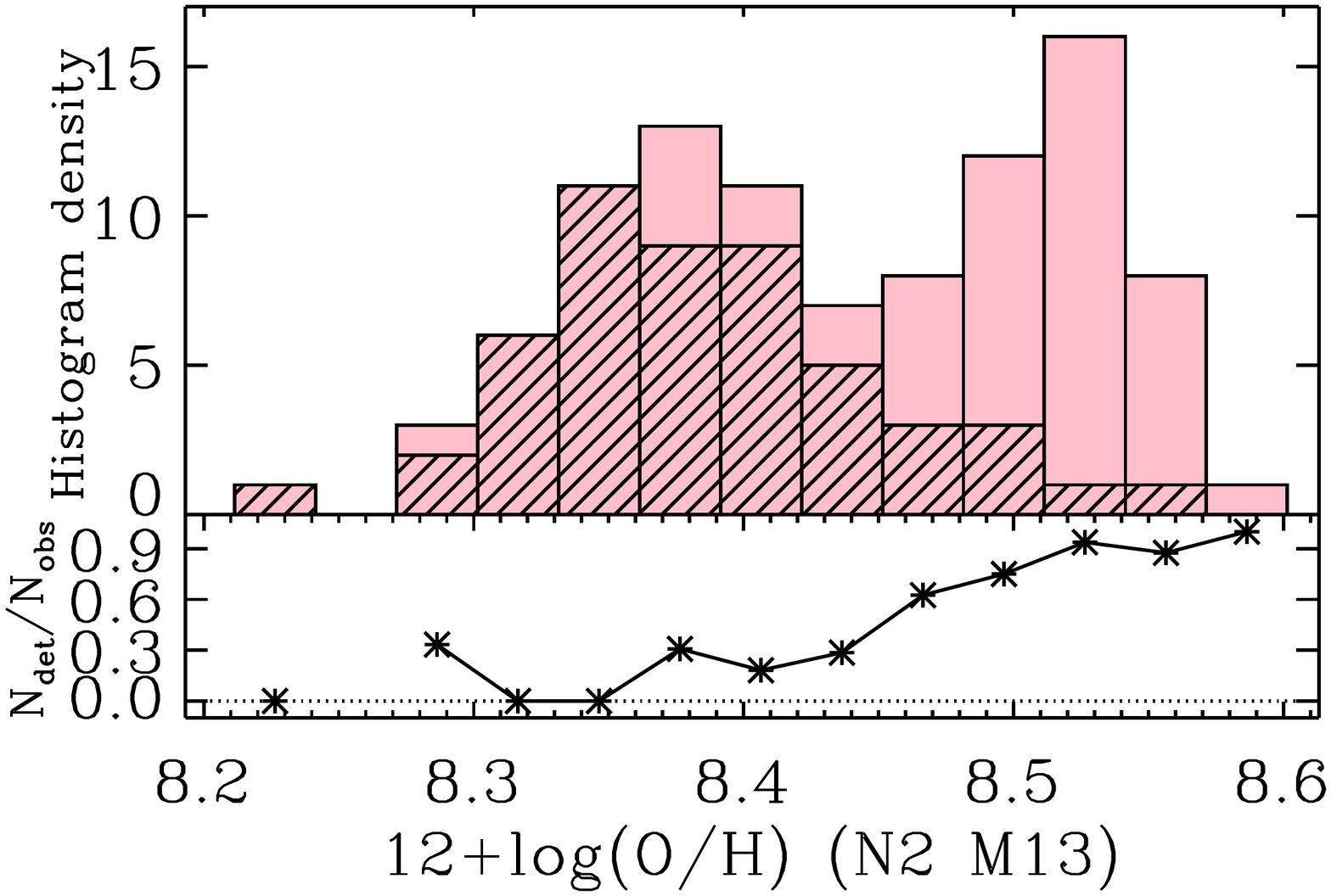}\quad 
    \includegraphics[clip=true,trim=4.5cm 4cm 0.8cm 2cm,width=0.37\textwidth,angle=180]{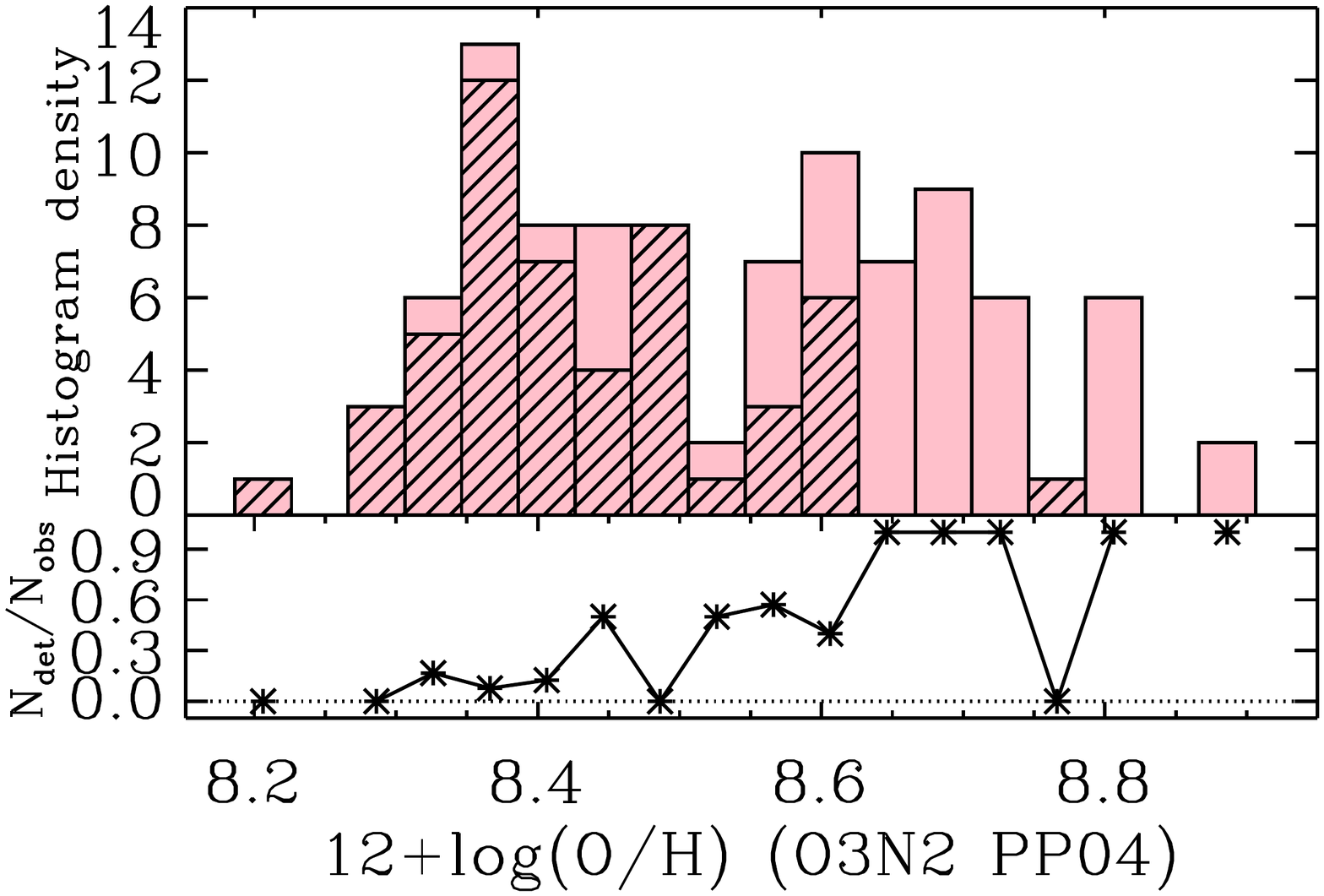}\\ 
    \vspace{0.1cm} 
    \includegraphics[clip=true,trim=4.5cm 4cm 0.8cm 2cm,width=0.37\textwidth,angle=180]{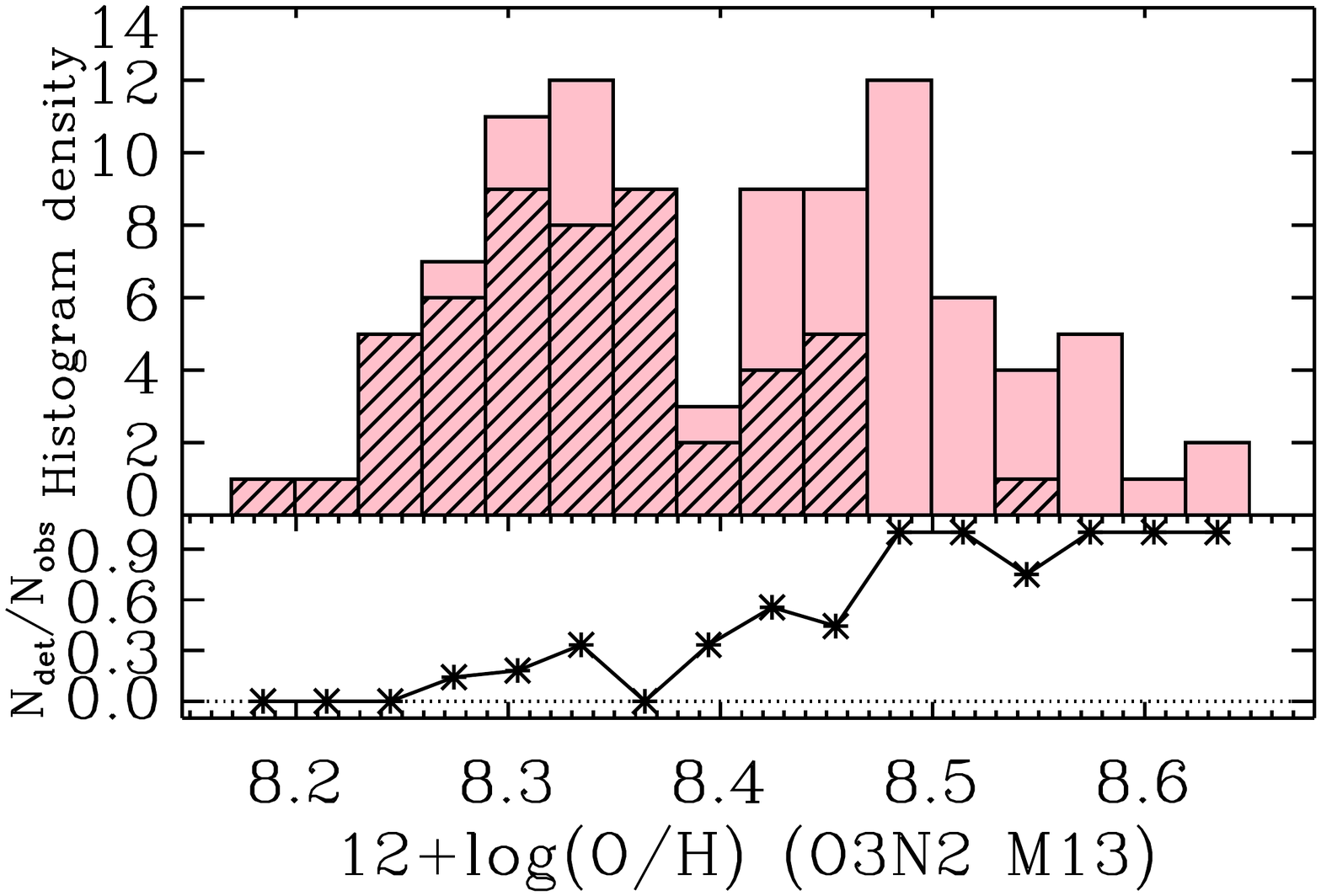}\\ 
     \vspace{0.2cm} 
     \caption{Distribution of ALLSMOG galaxies as a function of the gas-phase metallicity computed using five different strong-line calibrations (details in $\S$~\ref{sec:metallicity}). As in Fig~\ref{fig:histo} and \ref{fig:histo_extinction}, the full sample is represented by the coloured histogram, and the hatched histogram traces the non-detections. The detection fraction per bin is shown at the bottom of the histograms.}
   \label{fig:histo_met}
\end{figure*}

The distribution of gas-phase metallicities within the ALLSMOG sample is shown in Fig.~\ref{fig:histo_met} for each of the five metallicity calibrations adopted in this work ($\S$~\ref{sec:metallicity}). It is evident from Fig.~\ref{fig:histo_met} that the theoretical calibration by \cite{Tremonti+04} produces metallicities that are systematically offset to higher values than those obtained using empirical calibrations based on the $T_e$ method, as it was already noted by previous studies (e.g. \citealt{Bresolin+04}). Furthermore, we note that the distributions of non-detections peak at different Oxygen abundance values for different metallicity calibrations.

Figure~\ref{fig:histo_met} demonstrates that, for a flux-limited survey and for a sample of star-forming galaxies with $M_*<10^{10}~M_{\odot}$ such as ALLSMOG, the gas-phase metallicity is highly predictive of the CO detectability. For each metallicity calibration method explored in this paper, the detection fraction is found to increase with Oxygen abundance. However, the exact 12+$\log(O/H)$ value above which the number of CO detections exceeds the number of non-detections (i.e. $N_{det}/N_{obs}>0.5$) varies strongly depending on the calibration adopted. This value is the highest for the \cite{Tremonti+04} calibration ($12+\log(O/H) \sim 8.7$), it is slightly lower for the empirical calibrations by \cite{PP04} ($12+\log(O/H) \sim 8.5$), and even lower for the calibrations proposed by \cite{Marino+13} ($12+\log(O/H) \sim 8.4$). The K-S test confirms that the metallicity distributions of detections and non-detections are statistically different (Table~\ref{table:KS_test}).

\subsection{As a function of H{\sc i} gas mass}

\begin{figure}[tbp]
\centering
    \includegraphics[clip=true,trim=4.5cm 4cm 0.8cm 2cm,width=0.37\textwidth,angle=180]{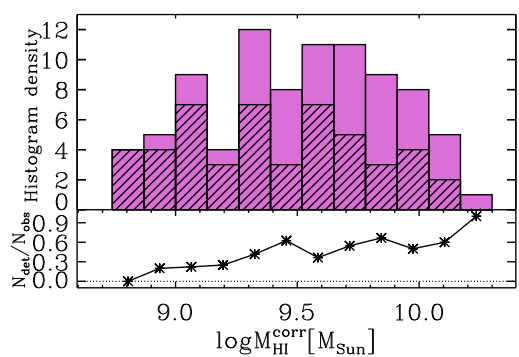}\\  
      \caption{Distribution of the ALLSMOG sample as a function of the H{\sc i} gas mass. As in Figs~\ref{fig:histo}-\ref{fig:histo_met}, the coloured histogram indicates the full sample, and the hatched histogram traces the non-detections. The detection fraction per $M_{\rm HI}$ bin is shown at the bottom of the histogram. Galaxies for which we have only an upper limit on $M_{\rm HI}$ are not included in this plot (see $\S$~\ref{sec:HIdata} and Table~\ref{table:HI_parameters}). We use the $M_{\rm HI}$ values corrected for self-absorption.}
   \label{fig:histo_HI}
\end{figure}

Lastly, in Fig~\ref{fig:histo_HI} we show the $M_{\rm HI}$ distribution of ALLSMOG galaxies, with CO non-detections identified by the hatched histogram similar to Figs.~\ref{fig:histo}-~\ref{fig:histo_met}. We note that, although we included in Fig.~\ref{fig:histo_HI} only the galaxies with an archival H{\sc i} detection, we do not expect the results to be affected by significant biases, because the H{\sc i} coverage of the ALLSMOG sample is not uniform in sensitivity. In fact, the sources without a H{\sc i} line detection do not necessarily correspond to lower $M_{\rm HI}$ values compared to the rest of the sample, as indicated by the upper limits on $M_{\rm HI}$ that we have estimated for the H{\sc i} non detections (Table~\ref{table:HI_parameters}).

Figure~\ref{fig:histo_HI} shows that the ALLSMOG galaxies span 1.5 orders of magnitude in $M_{\rm HI}$, from $M_{\rm HI}=10^{8.7}~M_{\odot}$ to $M_{\rm HI}=10^{10.2}~M_{\odot}$, with a median of $M_{\rm HI}=10^{9.5}~M_{\odot}$. The detectability of CO, as shown by the ratio of detected to observed sources in each bin reported at the bottom of the histogram, appears to depend on $M_{\rm HI}$. This is also confirmed by the KS test (Table~\ref{table:KS_test}).

\begin{table}
\caption{Results of the two-sample Kolmogorov--Smirnov test performed on the distribution of detections and non-detections as a function of the galaxy physical parameters explored in this work:} 
\label{table:KS_test}
\centering
\begin{tabular}{lcc}
\hline
\hline
  & $D$$^{\dag}$ & p-value$^{\ddag}$  \\
\hline
$\log M_{*}~[M_{\odot}]$ 			& 0.60 & 1.75E-8 \\
$\log SFR~[M_{\odot}~yr^{-1}]$ 	& 0.56 & 1.82E-7 \\
$\log SSFR~[yr^{-1}]$ 			& 0.21 & 0.19 \\
 redshift 							& 0.22 &  0.16 \\
$A_V~[mag]$ (CCM89)  			& 0.51 & 4.1E-6 \\
$A_V~[mag]$ (Calzetti)  			& 0.51 & 4.1E-6  \\
$12 + \log (O/H)$ (Tremonti+04) 	& 0.70 & 1.9E-11 \\
$12 + \log (O/H)$ N2 PP04 		& 0.67 & 2.2E-10 \\
$12 + \log (O/H)$ N2 M13 		& 0.67 & 2.2E-10  \\
$12 + \log (O/H)$ O3N2 PP04 	& 0.67 & 2.5E-10 \\
$12 + \log (O/H)$ O3N2 M13 		& 0.67 & 2.5E-10  \\
$\log M_{\rm HI}^{corr}$ 				& 0.30 & 0.034  \\
\hline

\end{tabular}
\tablefoot{$^{\dag}$ Value of the K-S statistic. $^{\ddag}$ The associated probability of the null hypothesis, that is the hypothesis according to which the two samples of detections and non-detections come from the same distribution. We reject the null hypothesis if the p-values $<0.05$.}
\end{table}


\section{CO line luminosity as a function of galaxy properties}\label{sec:lco_vs_prop}

We calculated the CO line luminosities from the integrated line fluxes corrected for aperture effects ($\S$~\ref{sec:ap_cor}) following the definition of \cite{Solomon+97}:
\begin{equation}\label{eq:lco_def}
L^{\prime}_{\rm CO} [{\rm K~km~s^{-1}~pc^2}] = 3.25 \times 10^7 \frac{D_L^2}{\nu_{obs}^{2} (1+z)^{3}} \int S_{\rm CO}~d\varv,
\end{equation}
where $D_L$ is expressed in units of Mpc, $\nu_{obs}$ is the observed frequency of the CO line in GHz and the velocity-integrated line flux is given in units of Jy~km~s$^{-1}$. The CO line luminosity defined in Eq~\ref{eq:lco_def} corresponds to the product of the velocity-integrated source brightness temperature $T_b$ (i.e. the surface brightness of the CO line) and the source area. Sometimes $L^{\prime}_{\rm CO}$ is referred to as ``brightness temperature CO luminosity'' in order to distinguish it from the CO line luminosity expressed in units of $L_{\sun}$, which is usually indicated as $L_{\rm CO}$. The $L_{\rm CO(2-1)}^{'}$ and $L_{\rm CO(1-0)}^{'}$ of ALLSMOG galaxies obtained respectively from the APEX CO(2-1) and IRAM CO(1-0) observations are listed in Table~\ref{table:co_luminosity}.

We recall that it is the luminosity in the lowest energy CO(1-0) transition that is typically used to estimate the total molecular gas mass through the assumption of an $\alpha_{CO}$ \citep{Solomon+97}. Therefore, for the 88 sources observed with APEX, we need to convert the $L^{\prime}_{\rm CO(2-1)}$ values into $L^{\prime}_{\rm CO(1-0)}$ estimates. We define $r_{21}$ as:
\begin{equation}\label{eq:r21_def}
r_{21} = L^{\prime}_{\rm CO(2-1)}/L^{\prime}_{\rm CO(1-0)}.
\end{equation}
Following on from the definition of $L^{\prime}_{\rm CO}$, $r_{21}$ corresponds to the ratio of the CO(2-1) and CO(1-0) intrinsic brightness temperatures averaged over the source \citep{Solomon+97}. We assume $r_{21} = 0.8 \pm 0.2$, consistently with integrated ($r_{21}\sim 0.8-0.9$, e.g. \citealt{Braine+Combes93}) and resolved observations of local star-forming spirals ($r_{21}\sim 0.6-1.0$, \citealt{Leroy+09}) as well as with recent results on $z\sim1.5-2.0$ star-forming disks ($r_{21}\sim 0.7-0.8$, \citealt{Aravena+10,Aravena+14}). Furthermore, we note that in the only ALLSMOG source for which we have a detection in both the CO(2-1) and CO(1-0) lines, 2MASXJ1336+1552 (observed with the IRAM~30m telescope), we derive $r_{21} = 0.8 \pm 0.3$, after correcting for the different beams of the IRAM~30m telescope at 115~GHz and 230~GHz (Table~\ref{table:co_luminosity}). 
Assuming local thermodynamic equilibrium (LTE) and optically thick gas, $r_{21}\sim0.8$ corresponds to a low beam-averaged excitation temperature of $T_{ex} \sim 10~K$ \citep{Eckart+90, Braine+Combes92, Leroy+09}, whereas $T_{ex}>20~K$ would yield $r_{21}\sim0.9-1$ under these conditions.

In the following we will address the relations between $L^{\prime}_{\rm CO(1-0)}$ and other integrated galaxy properties, namely: $M_*$, SFR, SSFR, $A_V$, 12+$\log(O/H)$, and $M_{\rm HI}$. The investigation will be carried out on the sample of 185 local star-forming galaxies whose distribution in the $M_*-SFR$ parameter space is shown in Fig.~\ref{fig:MS}, which comprises the ALLSMOG survey and the star-forming sub-sample of the COLD GASS survey defined in $\S$~\ref{sec:coldgass}. This galaxy sample is at the same time consistent in terms of selection criteria, because the same criteria adopted for ALLSMOG ($\S$~\ref{sec:sample}) were applied to COLD GASS to select the sub-sample used in this study ($\S$~\ref{sec:coldgass}), and quite broad in terms of $M_*$, SFR and ISM properties. By using a statistically-sound sample that is highly representative of the local star-forming galaxy population, we aim to identify the global galaxy parameters that are the best predictors of the CO luminosity in typical star-forming galaxies. Furthermore, we will exploit the broad dynamic range in $M_*$ probed to explore possible differences between the scaling relations defined by low-$M_*$ and high-$M_*$ star-forming galaxies.

Certainly, despite our efforts, the following analysis will not be completely free from selection effects and observational biases. In particular, we note:
\begin{itemize}
\item The cut on the gas-phase metallicity applied to the ALLSMOG sample selection ($\S$~\ref{sec:sample}), despite being low enough to include the majority of the SDSS galaxies with $M_*>10^{8.5}~M_{\odot}$ (see explanation in \cite{Bothwell+14}), inevitably biases our lowest-$M_*$ sample\footnote{From the mass-metallicity relation by \cite{Tremonti+04}, we expect this potential bias to be only valid for $\log M_{*} [M_{\odot}] <9.0$.} against the most metal poor sources.
\item The S/N cut on the nebular emission lines used for the BPT diagram has hindered the selection of galaxies with faint nebular lines and with a low-level star formation activity. The consequence of this cut can be appreciated in Fig.~\ref{fig:MS}, where it appears that, while the MS and its intrinsic $\pm 0.3$~dex scatter are probed quite uniformly at all stellar masses considered in our analysis, the outliers below the MS are fewer than the ones above the MS. This means that, although we are probing the bulk of the local star-forming galaxy population (from its high-SSFR tail up to at least one standard deviation below the ridge of the peak in the SFR-M$_*$ distribution as identified by \cite{Renzini+Peng15}), the low-SSFR end of the distribution is not well represented in our sample. For the purpose of the present study, this selection effect against faint star-forming galaxies is not particularly worrying and does not affect our results, as long as our conclusions are not extrapolated to galaxy populations other than the actively star-forming galaxy population.
\item The BPT method developed by \cite{BPT81} is widely used to classify galaxies based on their dominant source of photoionisation, separating star-forming galaxies from AGN hosts or galaxies with other sources of photoionisation such as evolved stellar populations and shocks. However, the BPT selection is known to fail in the case of `optically elusive' AGNs whose radiative output is low compared to that of the star formation in the host galaxy and so it does not dominate galaxy-averaged nebular emission line ratios (e.g. \citealt{Severgnini+03,Caccianiga+07}). 
\item There are intrinsic differences in the ALLSMOG and COLD GASS sample selections that we could not account for, namely, their different - but partially overlapping - redshift distributions ($\S$~\ref{sec:coldgass}), and the uniform and deeper HI coverage of the COLD GASS sample which was selected from the GASS survey \citep{Catinella+10, Saintonge+11a}. 
\item Finally, it is worth mentioning that the different observation methods (i.e. CO(2-1) vs CO(1-0), APEX vs IRAM) and data reduction methods may also introduce some biases in the final results, especially for what concerns the comparison between the relations defined by the low-$M_*$ galaxy sample ($M_*<10^{10}~M_{\odot}$, e.g. ALLSMOG), and the $M_*>10^{10}~M_{\odot}$ sample from COLD GASS.
\end{itemize}
Most of the statistical relations between CO luminosity and global galaxy properties that we will examine in the next sections have been already investigated in a number of previous studies, such as the dependency of $L^{\prime}_{\rm CO(1-0)}$ (or $M_{mol}$, if a constant $\alpha_{CO}$ is applied) on $M_*$ (as traced by the K-band Luminosity, e.g. \citealt{Leroy+05, Lisenfeld+11, Young+11, Amorin+16}), SFR (as traced by the FIR luminosity e.g. \citealt{Sanders+Mirabel85,Tacconi+Young87,Verter88,Solomon+Sage88, Young+Scoville91,Solomon+97,Gao+Solomon04b,Leroy+05,Chung+09,Lisenfeld+11,Schruba+11,Hunt+15,Amorin+16}), gas-phase metallicity (e.g. \citealt{Schruba+12,Amorin+16,Kepley+16}), and M$_{\rm HI}$ (e.g. \citealt{Stark+86,Verter88,Leroy+05,Lisenfeld+11,Saintonge+11a,Amorin+16}). Our results will be discussed in the context of these previous works, although we stress that, as mentioned in $\S$~\ref{sec:intro}, these previous analyses did not have the advantage of a wide dynamic range in galaxy properties (SFR, M$_*$, O/H) and a homogeneous CO dataset.

In order to quantify the strengths of possible correlations, and to determine the best fits to the data and their intrinsic scatters, we apply a Bayesian linear regression analysis following the approach of \cite{Kelly07}. This method is suited to deal with upper limits (non-detections in the $y$ variable) as well as with detected data with measurement errors in both $x$ and $y$ variables. The output of the regression analysis (performed using the IDL routine \texttt{linmix\_err} provided by \cite{Kelly07}) is a posterior probability distribution for each of the model fit parameters. In fact, the output parameter distributions are random draws from the posterior probability distributions obtained by using a Markov Chain Monte Carlo (MCMC) method. For the purposes of this study, for each parameter of interest, we use the median and a robust estimate of the standard deviation of the output probability distribution respectively as best-fit value and associated uncertainty. In $\S$~\ref{sec:lco_sfr} - $\S$~\ref{sec:lco_hi} we show the relationships between $L^{\prime}_{\rm CO(1-0)}$ and several galaxy physical parameters: $M_*$, SFR, SSFR, $A_V$, gas-phase metallicity (using five different calibrations), and $M_{\rm HI}$. The results of the linear regression analyses performed on these relationships is summarised in Table~\ref{table:regression_summary}. We note that for each relation, we carry out three different regression analyses: (i) on the total sample, (ii) on the low-$M_*$ sub-sample defined by $M_* < 10^{10}~M_{\odot}$, and (iii) on the massive sub-sample, defined by $M_* \geq 10^{10}~M_{\odot}$. In the following, wherever not specified, we refer to the results of the Bayesian regression analyses on the relations defined by the total sample. 

\subsection{$L_{\rm CO}^{\prime}$ vs $M_*$, SFR, SSFR}\label{sec:lco_sfr}

\begin{figure*}[tbp]
\centering
    \includegraphics[clip=true,trim=9cm 4cm 2.cm 3cm,width=0.32\textwidth,angle=180]{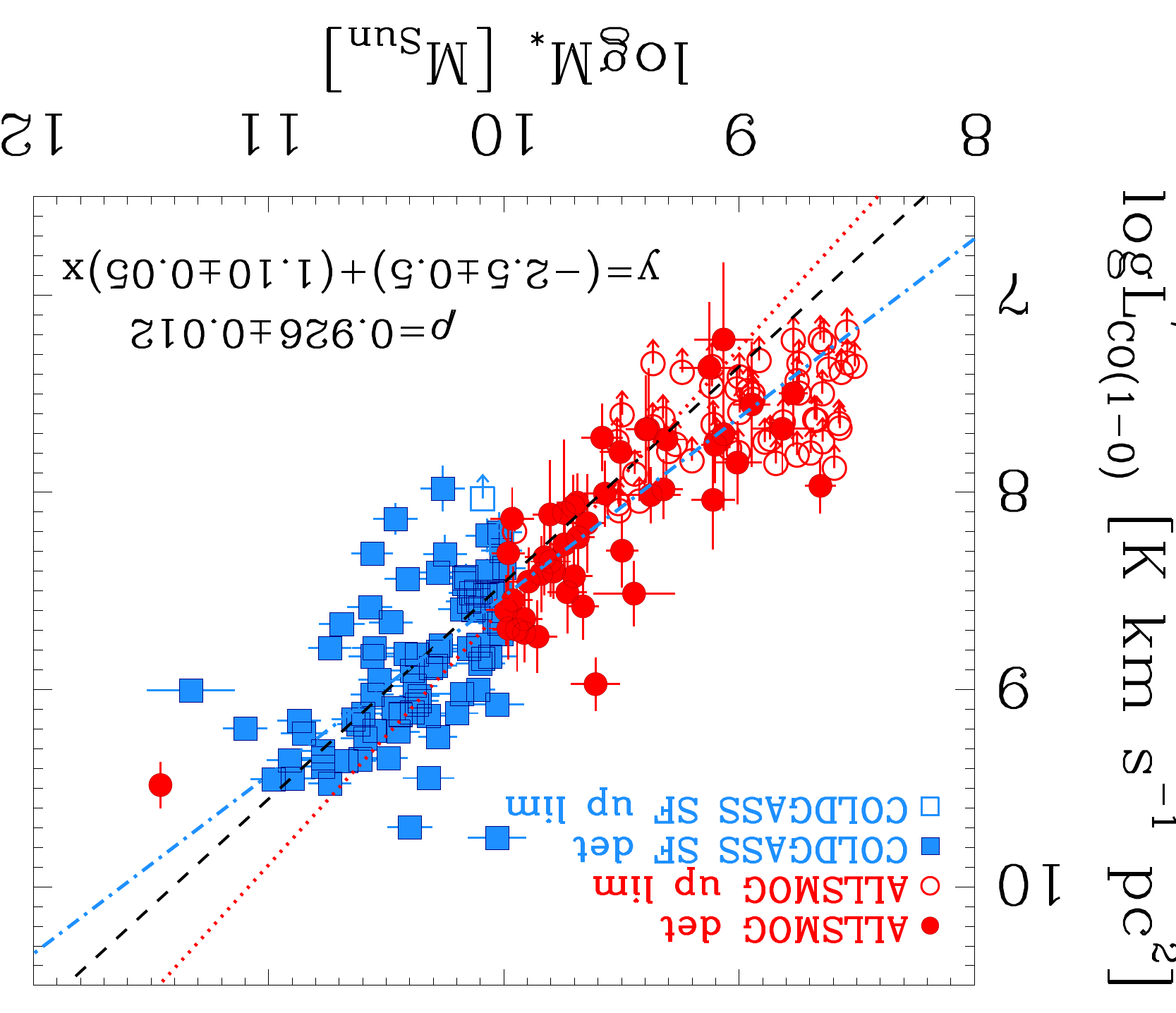}\quad 
    \includegraphics[clip=true,trim=9cm 4cm 2.cm 3cm,width=0.32\textwidth,angle=180]{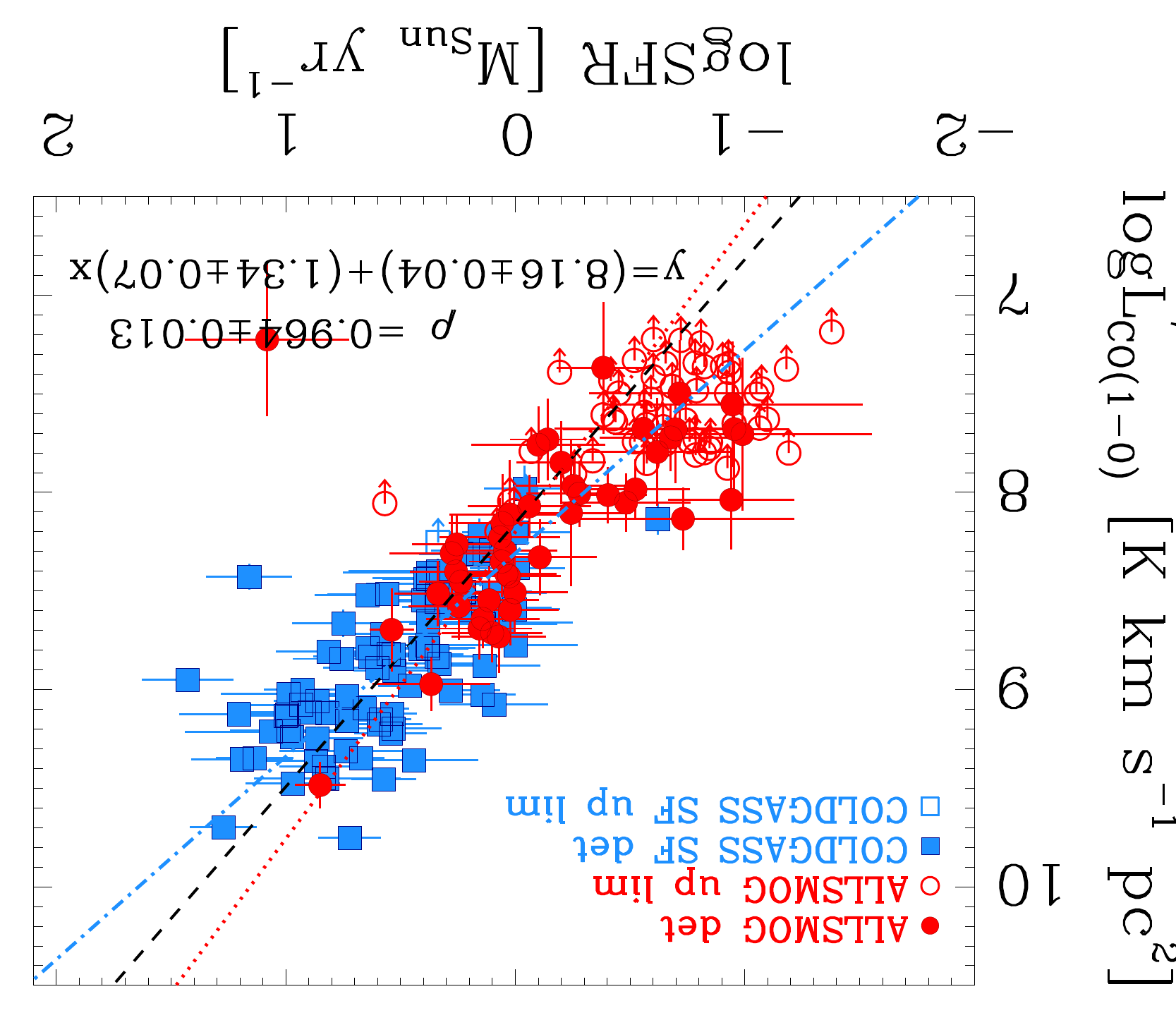}\quad 
   \includegraphics[clip=true,trim=9cm 4cm 2.cm 3cm,width=0.32\textwidth,angle=180]{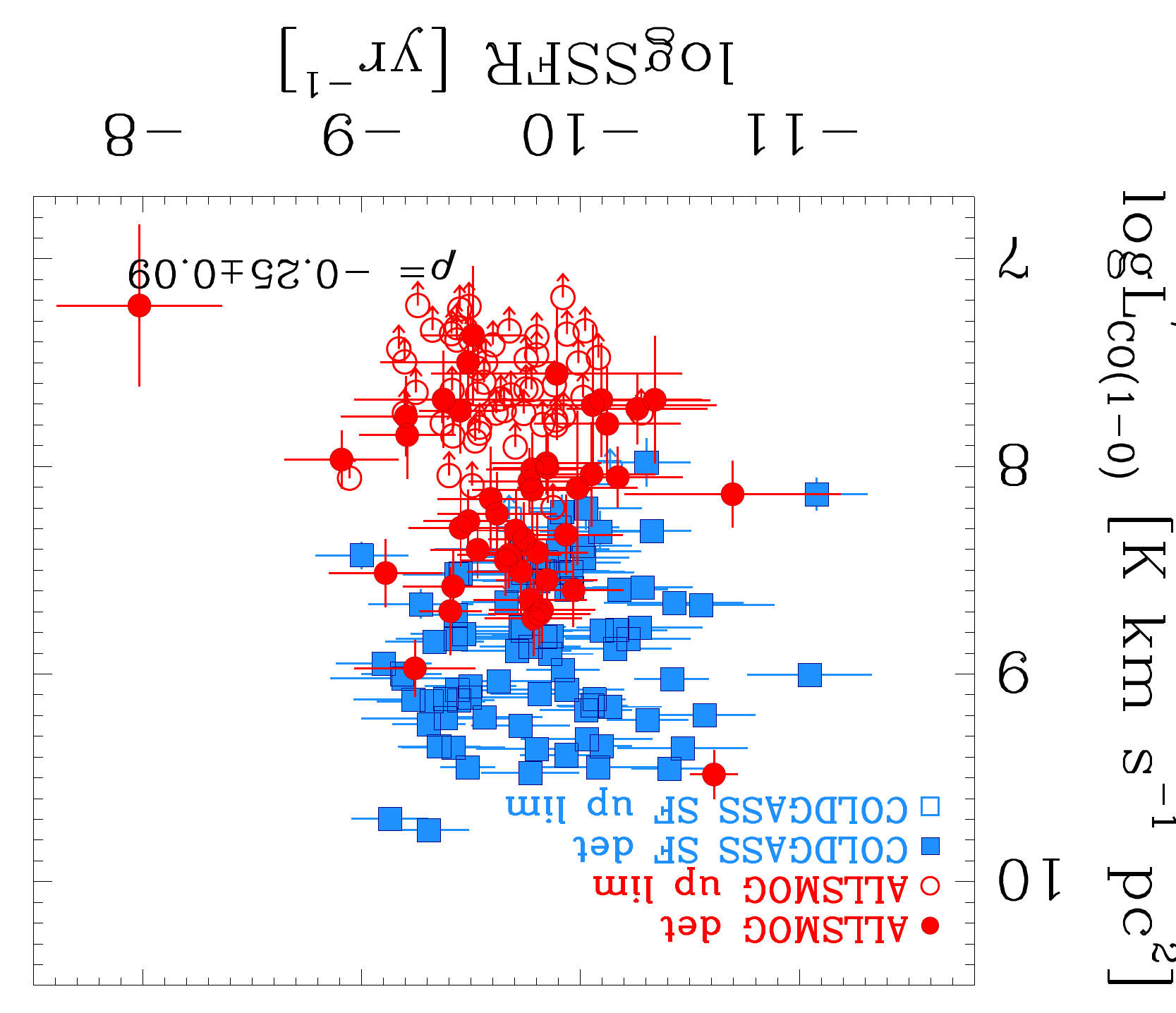}\\ 
     \vspace{0.2cm} 
     \caption{CO(1-0) line luminosity as a function of M$_*$ ({\it left panel}), SFR ({\it central panel}), and SSFR ({\it right panel}) in log-log scale for a sample of local star-forming galaxies defined by the ALLSMOG survey and by the sub-sample of COLD GASS selected in $\S$~\ref{sec:coldgass}. For the ALLSMOG sources observed only in their CO(2-1) transition (i.e. the APEX sample), $L^{\prime}_{\rm CO(1-0)}$ has been computed from $L^{\prime}_{\rm CO(2-1)}$ by assuming $r_{21} \equiv L^{\prime}_{\rm CO(2-1)}/L^{\prime}_{\rm CO(1-0)} = 0.8 \pm 0.2$, consistent with previous observations of normal, low-$M_*$ star-forming galaxies, and the uncertainty on $r_{21}$ is included in the error-bars. The empty symbols with downward arrows indicate the $3\sigma$ upper limits on $L^{\prime}_{\rm CO(1-0)}$ estimated for the sources not detected in CO. The black dashed line indicates the best fit relation obtained from a Bayesian linear regression analysis conducted on the total sample,
while the red dot and blue dot-dashed lines correspond to the best fit models obtained separately for the two samples of low-$M_*$ ($\log M_* [M_{\odot}] < 10.0$) and high-$M_*$ ($\log M_* [M_{\odot}] \geq 10.0$) galaxies, respectively. The best-fit regression parameters and correlation coefficient resulting from the analysis of the total sample are reported on the plots, and the full parameter list can be found in Table~\ref{table:regression_summary}.
We do not plot the best-fit to the $L^{\prime}_{\rm CO(1-0)}$ vs SSFR relation because the (anti-)correlation between these two quantities is weak, but we still report the best fit parameters in Table~\ref{table:regression_summary}.}
   \label{fig:lco_plots_1}
\end{figure*}

Figure~\ref{fig:lco_plots_1} shows $L^{\prime}_{\rm CO(1-0)}$ as a function of $M_*$, SFR and SSFR (in log-log scale). The Bayesian regression analysis (whose results are reported in Table~\ref{table:regression_summary}) demonstrates that both $M_*$ and SFR are strongly correlated with $L^{\prime}_{\rm CO(1-0)}$ with small intrinsic scatters ($\sigma_{intr}\lesssim 0.3$~dex). Instead, the SSFR appears to be weakly anti-correlated with $L^{\prime}_{\rm CO(1-0)}$ but with a very high $\sigma_{intr}$, essentially confirming the absence of a strong link between $L^{\prime}_{\rm CO(1-0)}$ and SSFR in this sample of local star-forming galaxies. The relationships with SFR and $M_*$ are both super-linear (with a slope, $\beta>1$), and the relation with $M_*$ is closer to linear than the one with SFR. 

A close-to-linear relationship between $L^{\prime}_{\rm CO(1-0)}$ and the K-band luminosity $L_K$ (which traces the total $M_*$ of a galaxy) was already evidenced for isolated massive spirals (e.g. \citealt{Lisenfeld+11}). A strong correlation appears to hold for smaller galaxies and dwarfs \citep{Leroy+05,Amorin+16}, but it breaks down in early types \citep{Lisenfeld+11, Young+11}. The $L^{\prime}_{\rm CO(1-0)}$ vs SFR relation instead is very well known, and it has been extensively studied in the literature. Consistently with previous findings (e.g. \citealt{Verter88,Tacconi+Young87,Gao+Solomon04a,Leroy+05,Lisenfeld+11}), this is the strongest correlation among those investigated in this paper (if excluding the relations obtained separately for low-$M_*$ and high-$M_*$ galaxies, see Table~\ref{table:regression_summary}), with an intrinsic scatter of only $\sigma_{intr} = 0.21 \pm 0.04$. The $L^{\prime}_{\rm CO(1-0)}$ vs SFR relation is essentially the inverse of an integrated Schmidt-Kennicutt (S-K) law \citep{Schmidt59, Kennicutt98}, which is an empirical relation between the surface density of cold gas and that of SFR ($\Sigma_{SFR} \propto \Sigma_{mol}^N$). The S-K law has been identified for many years as a `fundamental' scaling relation between the amount of fuel available for star formation and the SFR itself, holding for several orders of magnitude in both $\Sigma_{SFR}$ and $\Sigma_{mol}$ (e.g. \cite{Wu+05}). 
A discussion on the S-K law is beyond the scope of this data release paper and it would require an additional step for converting the CO(1-0) luminosity into a molecular gas mass estimate, but it will be addressed in future publications by the team. 

The reasons for such a tight - and linear - correlation between $L^{\prime}_{\rm CO(1-0)}$ and $M_*$ (or $L_K$) observed consistently across different galaxy samples except in early type galaxies have been little explored in the literature. The interpretation favoured by \cite{Leroy+05} is that CO emission and stars are linked through the hydrostatic pressure in the disk, which depends mainly on the stellar surface density ($\Sigma_*$) and sets the rate at which H{\sc i} is converted into H$_2$. An alternative explanation that we propose here goes back to the nature of the optically thick $^{12}$CO emission and to the approximately linear relation between $^{12}$CO luminosity and virial mass found for GMCs (see e.g. \cite{Scoville+87,Solomon+87}, and \cite{Bolatto+13} for a more recent compilation). By extrapolating from the relation shown by \cite{Scoville+87}, our observed $L^{\prime}_{\rm CO(1-0)}-M_*$ correlation may be so tight and close to linear simply because the global $^{12}$CO luminosity is a very good tracer of the dynamical mass in star-forming galaxies, assuming that in this class of objects the bulk of the CO emission traces molecular gas clouds in virial motions (for example in a rotating disk). This explanation of course assumes that in most star-forming galaxies the stellar mass is also a good tracer of the dynamical mass. Following on from our hypothesis, a possible explanation for the break down of the $L^{\prime}_{\rm CO(1-0)}-M_*$ relation in early types is that in these sources, even when CO is detected, the motions of the CO-emitting clouds are poor tracers of the total dynamical mass of the system. Interferometric CO observations of large samples of early type galaxies have indeed shown that the CO emission in these objects tends to be rather compact (on average extending over $\sim1$~kpc) compared to the optical extent of the galaxy \citep{Alatalo+13,Davis+13}.

We note that our sample selection (which, as previously discussed, indirectly leads to a bias against sources significantly below the MS), combined with the well-known $L^{\prime}_{\rm CO(1-0)}$-SFR correlation (central panel of Fig.~\ref{fig:lco_plots_1}), may affect the slope of the observed $L^{\prime}_{\rm CO(1-0)}$ vs $M_*$ correlation. More specifically, the galaxies below the MS that are not properly sampled by our analysis are also the galaxies that, due to the $L^{\prime}_{\rm CO(1-0)}$-SFR correlation, are expected statistically to lie below the observed $L^{\prime}_{\rm CO(1-0)}$ vs $M_*$ relation. This selection effect may result in a dataset that is not fully representative of the underlying scaling relation by probing mostly its upper envelope. Inspired by the discussion in \cite{Andreon+13}, we tested for this selection effect by calculating the mode of the $L^{\prime}_{\rm CO(1-0)}$/$M_*$ distribution separately for the two samples of low-$M_*$ and high-$M_*$ galaxies. The underlying assumption is that the statistical mode is not affected by selection effects. We verified that the modes calculated for the two samples, when plotted in correspondence of the median $M_*$ of each sample, sit on the best-fit relation to the total sample (shown in Fig.~\ref{fig:lco_plots_1}). This simple test confirmed that our poor sampling of the low-SSFR tail of the star-forming galaxy distribution does not affect significantly the observed correlation.

The absence of a link between $L^{\prime}_{\rm CO(1-0)}$ and SSFR is somehow counterintuitive. The SSFR is often used to define whether a galaxy is passive (low SSFR) or strongly star-forming (high SSFR) and has been identified as one of the key parameters for galaxy evolution \citep{Lilly+13}. The right panel of Fig.~\ref{fig:lco_plots_1} shows that the SSFR and the global CO luminosity of a galaxy are apparently unrelated. However, as shown in Fig~\ref{fig:MS}, our sample does not include many strong outliers of the MS, and so probes a relatively narrow range in SSFR (as already noted in $\S$~\ref{sec:histo_1}), which could explain why the CO content is apparently independent of the SSFR.

\subsection{$L^{\prime}_{\rm CO}$ vs visual extinction $A_V$}\label{sec:lco_av}

\begin{figure*}[tbp]
\centering
    \includegraphics[clip=true,trim=9cm 4cm 2.cm 3cm,width=0.32\textwidth,angle=180]{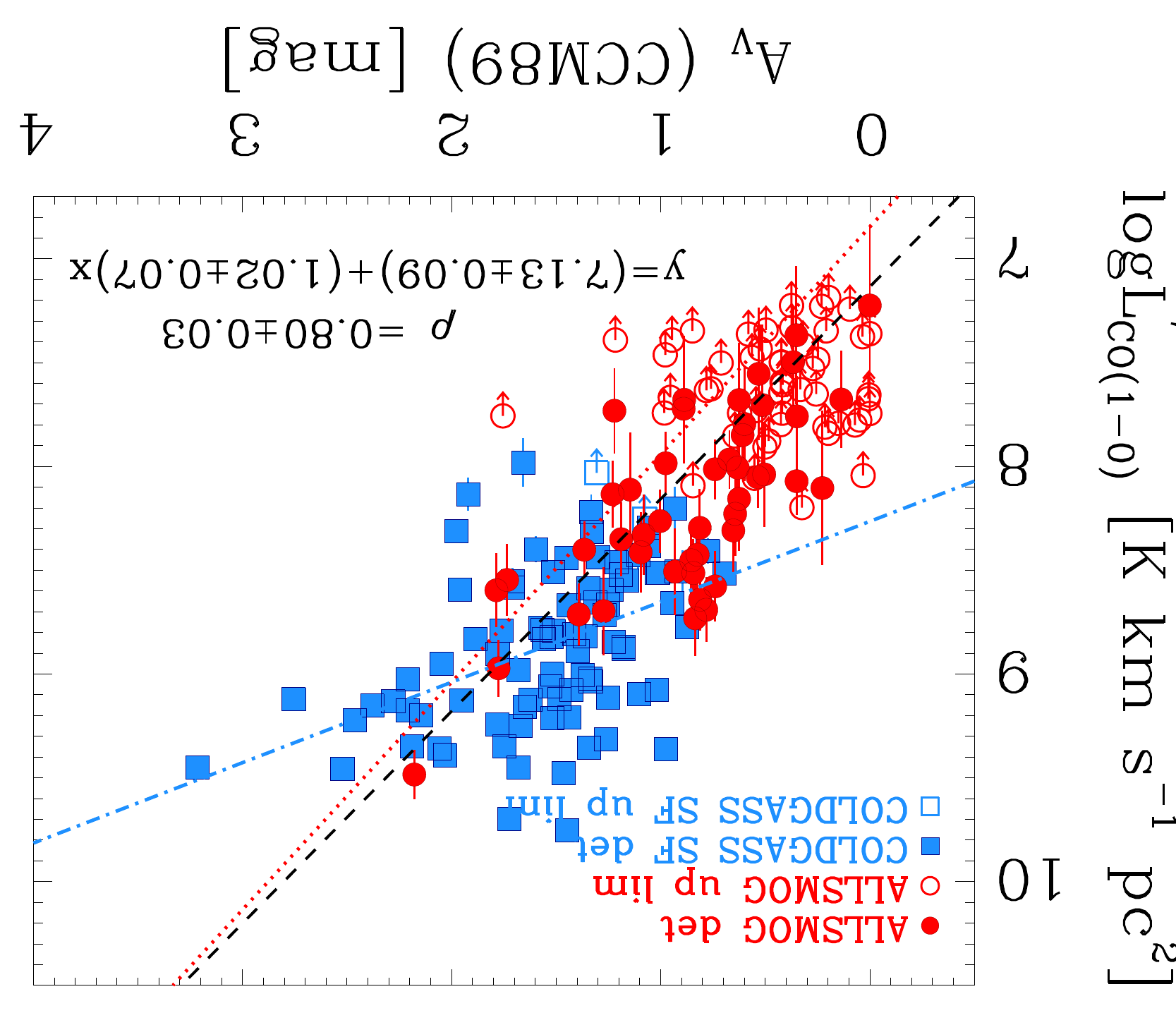}\quad    
    \includegraphics[clip=true,trim=9cm 4cm 2.cm 3cm,width=0.32\textwidth,angle=180]{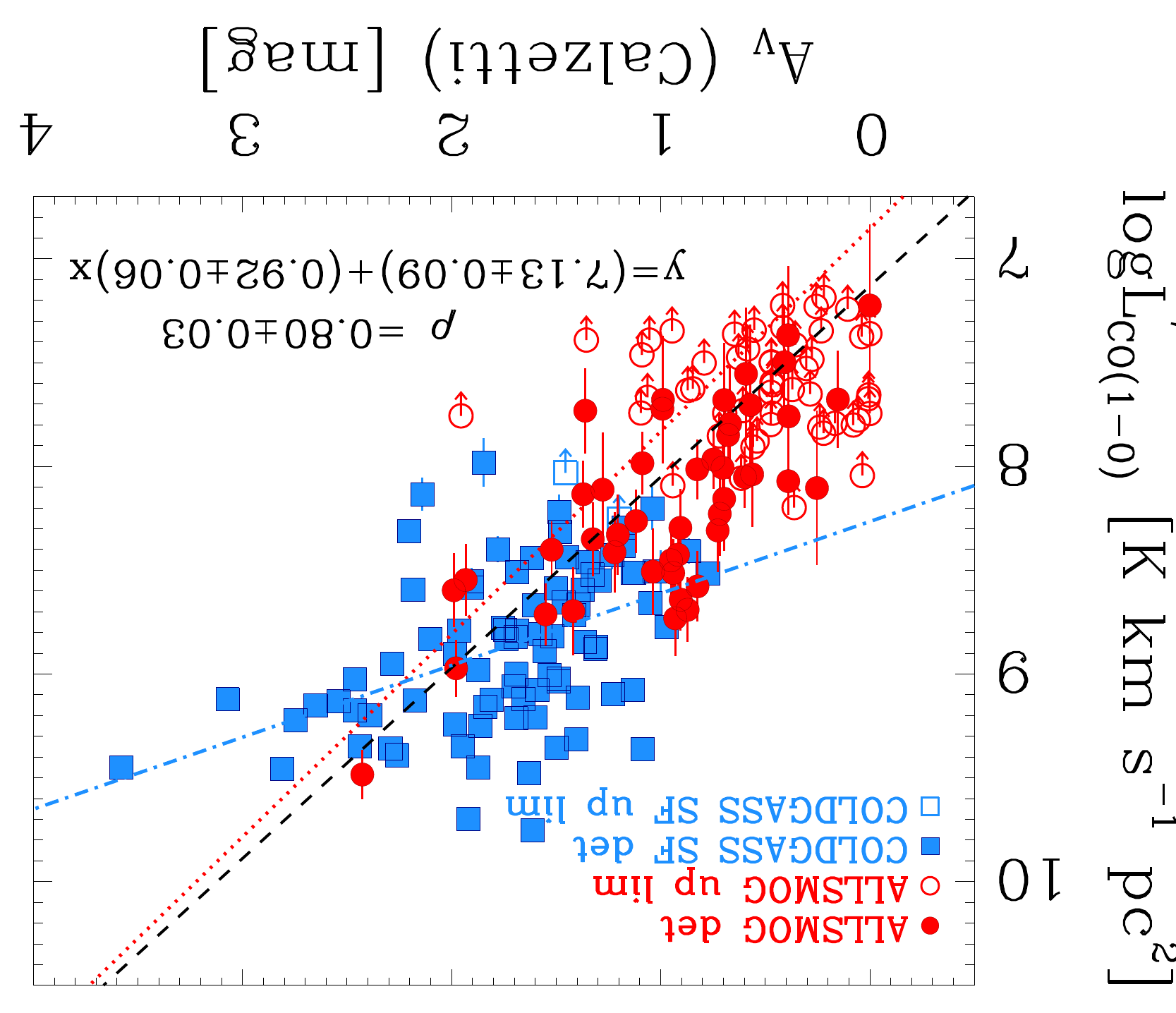}\\   
      \vspace{0.2cm} 
     \caption{CO(1-0) line luminosity as a function of the nebular visual extinction for the sample of local star-forming galaxies defined by the ALLSMOG survey and by the sub-sample of COLD GASS selected in $\S$~\ref{sec:coldgass}. The {\it left} and {\it right} panels shows the $A_V$ values obtained by using the attenuation curves proposed by \cite{CCM89} and \cite{Calzetti+00}, respectively. Further explanation relevant to the quantity plotted on the y-axis, the symbols and the regression analysis can be found in the caption of Fig.~\ref{fig:lco_plots_1} and in the text ($\S$~\ref{sec:lco_vs_prop}). Since the typical uncertainty on $A_V$ is very large, of the order of $\sim0.5~mag$, the fit was performed without accounting for the measurement error on $x$, in order to not bias the results (see discussion in \cite{Kelly07}). For the same reason the plots do not show error-bars in the $x$ direction.
       }
   \label{fig:lco_plots_3}
\end{figure*}

Figure~\ref{fig:lco_plots_3} shows $L^{\prime}_{\rm CO(1-0)}$  as a function of the nebular visual extinction calculated using the \cite{CCM89} and the \cite{Calzetti+00} attenuation curve models as explained in $\S$~\ref{sec:av_deriv}. To our knowledge, the dependency of $L^{\prime}_{\rm CO(1-0)}$ on the dust extinction has not been examined directly by any previous observational study of CO based on large galaxy samples. Within our sample we observe a positive correlation, with $\rho = 0.80 \pm 0.03$ and a slope roughly consistent with $\sim1$, although with a large intrinsic dispersion, $\sigma_{intr} = 0.49 \pm 0.03$ (Table~\ref{table:regression_summary}). Such an high $\sigma_{intr}$ output value could be partly due to the fact that we performed the Bayesian regression analysis without taking into account the error-bars on the $x$ variables, which are very large ($\Delta A_V\sim0.5~mag$) because of the large uncertainty in estimating the Balmer decrement using low S/N SDSS spectra. 
Large error-bars in the $x$ direction are known to bias the Bayesian regression analysis results \citep{Kelly07}. As a comparison, by including the errors on $A_V$ in the regression analysis, we obtain a significantly lower intrinsic scatter of $\sigma_{intr}\sim0.15$ for both the CCM89 and Calzetti models, and steeper relations with slopes of $\beta=1.70\pm0.14$ and $\beta=1.54\pm0.13$ respectively for the CCM89 and Calzetti models. In the analysis accounting for the measurement errors on $x$, the output coefficient parameter is significantly higher, $\rho \sim 0.98$. In summary, the large uncertainties on our $A_V$ estimates do not allow us to place strict constraints on the slope of the observed $L^{\prime}_{\rm CO(1-0)}$-$A_V$ relation.

Furthermore, we emphasise that the interpretation of Fig~\ref{fig:lco_plots_3} is hard not only because of the large error-bars on $A_V$, but also because of possible biases due to the sample selection and to the CO survey sensitivity limit \citep{Andreon+13}. We tested the $L^{\prime}_{\rm CO(1-0)}$ vs $A_V$ relation for selection effects by following the same strategy used for the $L^{\prime}_{\rm CO(1-0)}$ vs $M_*$ relation ($\S$~\ref{sec:lco_sfr}). By considering only the CO detections, we calculated the modes of the $L^{\prime}_{\rm CO(1-0)}$/$A_V$ distributions obtained for the low-$A_V$ (i.e. $A_V<1~mag$) and the high-$A_V$ (i.e. $A_V>1~mag$) galaxies in our sample, and we used these values to calculate the $L^{\prime}_{\rm CO(1-0)}$ expected in correspondence of the median $A_V$ of each sample. We thus found that the modes lie below the best-fit relation, by $\sim1$~dex (CCM89) - $1.3$~dex (Calzetti) for the low-$A_V$ sample and by $0.2$~dex (CCM89) - $0.5$~dex (Calzetti) for the high-$A_V$ sample. Besides being offset to lower $L^{\prime}_{\rm CO(1-0)}$ values, the modes indicate a slope that is steeper than one and closer to $\beta\sim2.5$.
This simple check highlights the strong role of selection effects in shaping the observed relation between $L^{\prime}_{\rm CO(1-0)}$ and $A_V$ (Figure~\ref{fig:lco_plots_3}). From the results of this test, we infer that the CO detections with $A_V<1~mag$ probe only the upper envelope of a broader distribution of $L^{\prime}_{\rm CO(1-0)}$ values. Instead, for higher $A_V$ values, the CO detections are much closer to probing the ridge of the `real' underlying $L^{\prime}_{\rm CO(1-0)}$ vs $A_V$ relation.

We think that the main factor responsible for this bias is the sensitivity limit of the CO observations, combined with an underlying steep dependency of $L^{\prime}_{\rm CO(1-0)}$ on $A_V$ for $A_V<1~mag$ values, which does not allow us to probe the CO luminosities that are best representative of the low-$A_V$ galaxy population. 
The presence of a steep $L^{\prime}_{\rm CO(1-0)}$-$A_V$ relation at low-$A_V$ is qualitatively consistent with the predictions of theoretical models investigating the relationship between molecular gas mass and CO emission \citep{Wolfire+10,Glover+MacLow11}. According to these models, below a certain threshold of mean visual extinction, the CO-to-H$_2$ conversion factor is strongly anti-correlated with $\langle A_V\rangle$ and rises steeply towards lower $\langle A_V\rangle$ values. \cite{Glover+MacLow11} determined such threshold to be around $\langle A_V\rangle\sim 3.5~mag$, but unfortunately it is not easy to translate this value into an observed nebular $A_V$ measured along one line-of-sight, which is the quantity that we measure (see also discussion in $\S$~\ref{sec:histo_av}).

Another element that possibly contributes to biasing the observed $L^{\prime}_{\rm CO(1-0)}$ vs $A_V$ relation in the low-$A_V$ regime is the metallicity cut applied to our sample selection ($\S$~\ref{sec:sample}). Indeed, 
because of the shape of the relation between gas-phase metallicity and nebular extinction \citep{Stasinska+04,Garn+Best10}, a side effect of the cut in metallicity is that very low-$A_V$ (i.e. $A_V\lesssim0.5~mag$) galaxies are underrepresented in our sample. In the hypothesis of a very steep $L^{\prime}_{\rm CO(1-0)}$-$A_V$ relation in the $0<A_V [mag] <1$ part of the diagram in Fig.~\ref{fig:lco_plots_3}, such undersampling at low-$A_V$ would `remove weight' to the low-$A_V$ end of the observed total relation, hence introducing an additional statistical bias in the direction of that observed. 

We therefore expect the real underlying relation between $L^{\prime}_{\rm CO(1-0)}$ and $A_V$ to be offset by $\sim0.5-1$~dex in $L^{\prime}_{\rm CO(1-0)}$ with respect to the best-fit relation to the total sample obtained through the Bayesian regression analysis, and to have a slope $\beta\gtrsim2$ consistent with that obtained from the Bayesian analysis conducted on the low-$M_*$ sample. This hypothesis can only be confirmed by much deeper CO observations of galaxies characterised with $A_V<1~mag$.

We note that a strong correlation exists between $M_*$ and $A_V$ in our sample ($\rho=0.82\pm0.02$, $\alpha=-6.3- -7.0$, $\beta=0.75-0.84$ and $\sigma_{intr}=0.37-0.41$~dex for the $A_V$ vs $\log M_*$ relation\footnote{The range in best-fit parameters indicated is due to the different attenuation curves adopted.}), consistent with previous findings based on local SDSS star-forming galaxies \citep{Stasinska+04,Garn+Best10}. In addition, \cite{Garn+Best10} demonstrated that $M_*$ is the primary galaxy parameter responsible for the variations of $A_V$ among star-forming galaxies. In the light of this result, it is reasonable to hypothesise that the strong correlation between $L^{\prime}_{\rm CO(1-0)}$ and $A_V$ (Fig.~\ref{fig:lco_plots_3}) is a result, at least in part, of the collinearity between $A_V$ and $M_*$, and so that the $L^{\prime}_{\rm CO(1-0)}$ vs $M_*$ relation enters in the dependency of 
$L^{\prime}_{\rm CO(1-0)}$ on $A_V$. To test this hypothesis, we have analysed the mass-independent quantity $\log L^{\prime}_{\rm CO(1-0)}-1.1\log M_*$ (i.e. the residuals of the power-law fit to the $L^{\prime}_{\rm CO(1-0)}$ vs $M_*$ relation shown in Fig~\ref{fig:lco_plots_1}) as a function of $A_V$, and we found no significant residual dependence on $A_V$. The relevant plots and tables are shown in Appendix~\ref{sec:appendix_ratios}. This result would suggest that the scaling with $M_*$ alone may account for all of the observed variation of $L^{\prime}_{\rm CO(1-0)}$ as a function of $A_V$. However, in the above discussion about selection effects we argued that the slope of the underlying $L^{\prime}_{\rm CO(1-0)}$ vs $A_V$ relation is likely steeper than the one observed, and so the existence of a secondary dependence of $L^{\prime}_{\rm CO(1-0)}$ on $A_V$ cannot be entirely ruled out based on our data.


\subsection{$L_{\rm CO}^{\prime}$ vs gas-phase metallicity}\label{sec:lco_met}

\begin{figure*}[tbp]
\centering
    \includegraphics[clip=true,trim=9cm 4cm 2.cm 3cm,width=0.32\textwidth,angle=180]{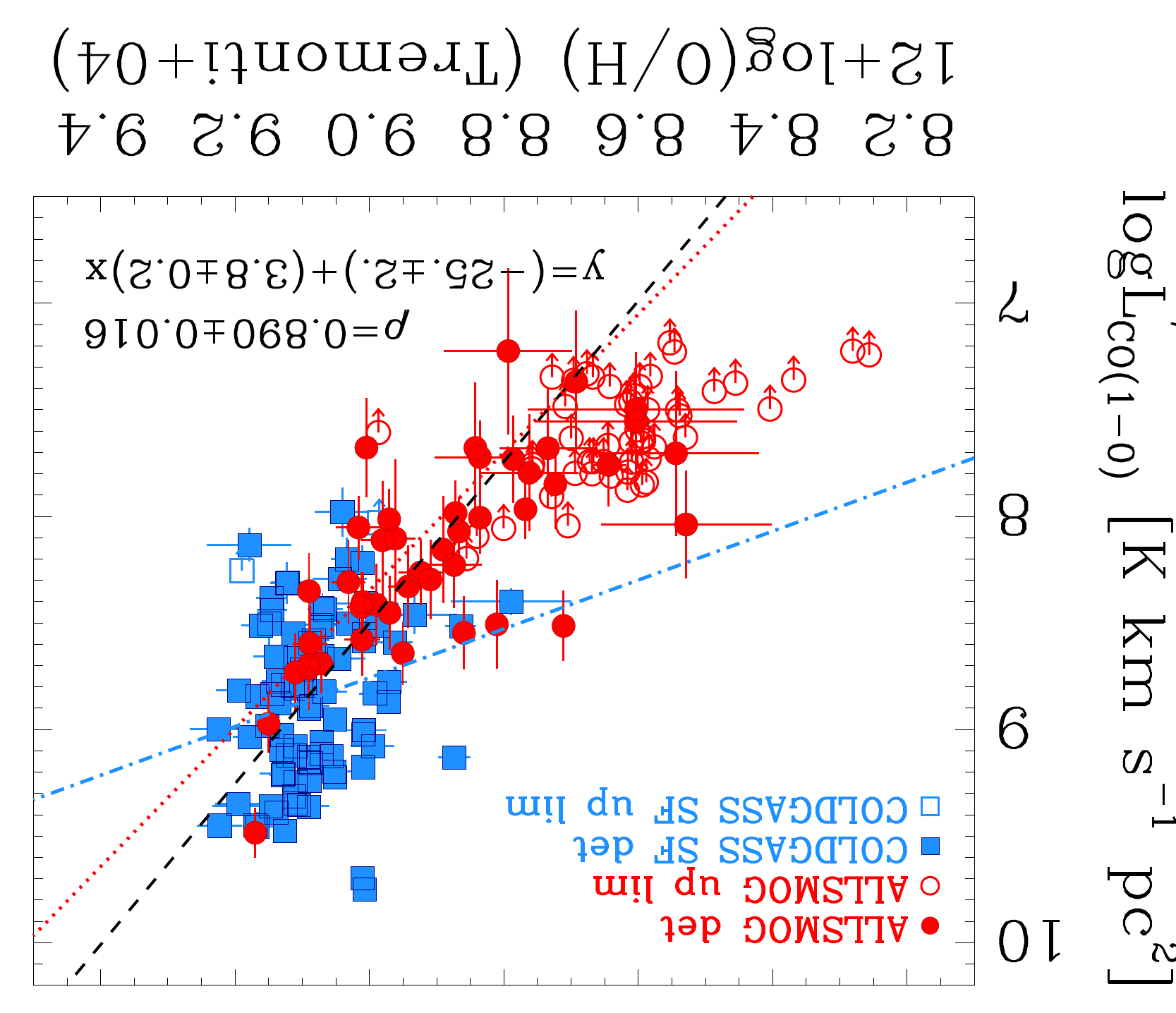}\quad 
      \includegraphics[clip=true,trim=9cm 4cm 2.cm 3cm,width=0.32\textwidth,angle=180]{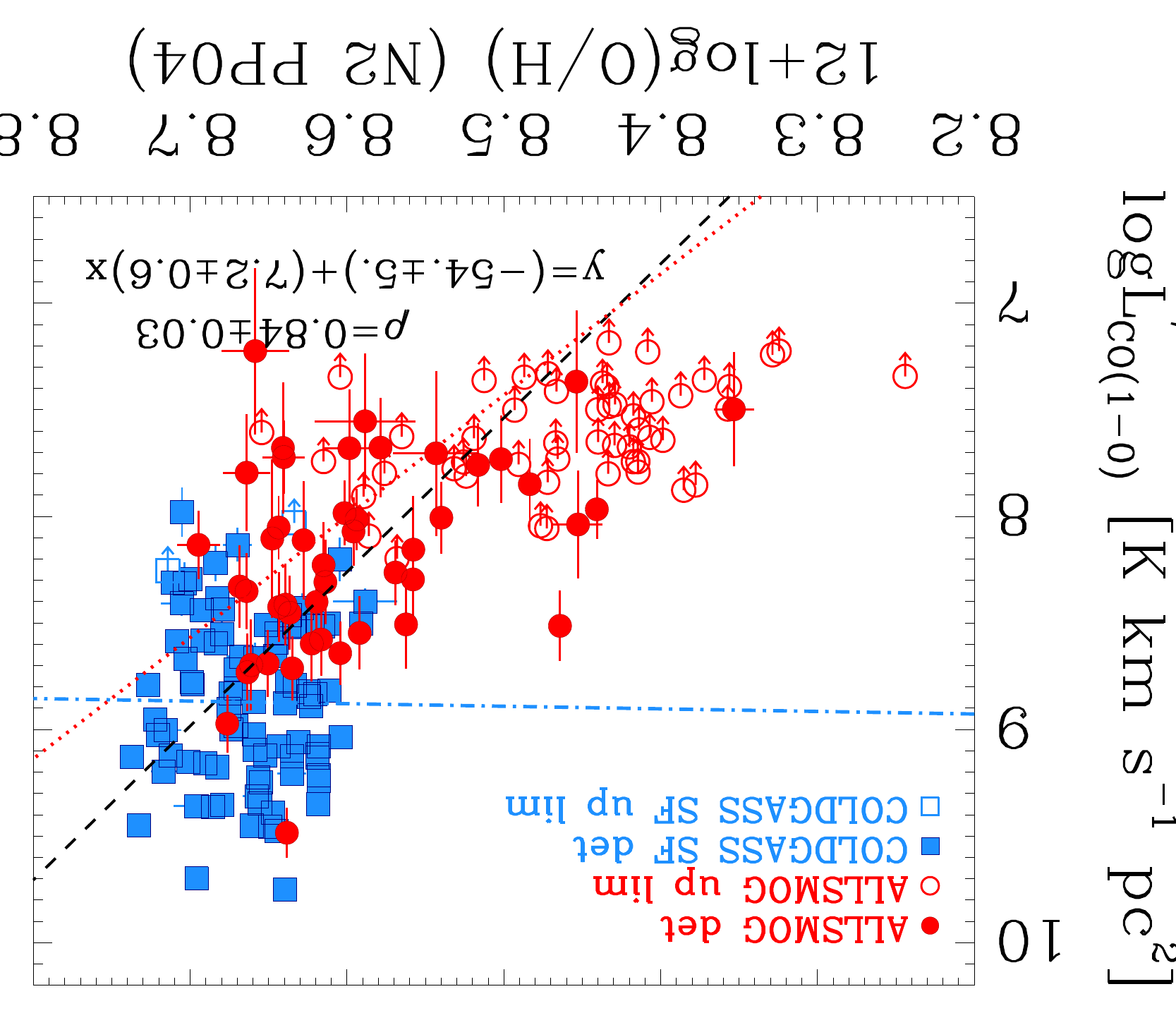}\quad 
       \includegraphics[clip=true,trim=9cm 4cm 2.cm 3cm,width=0.32\textwidth,angle=180]{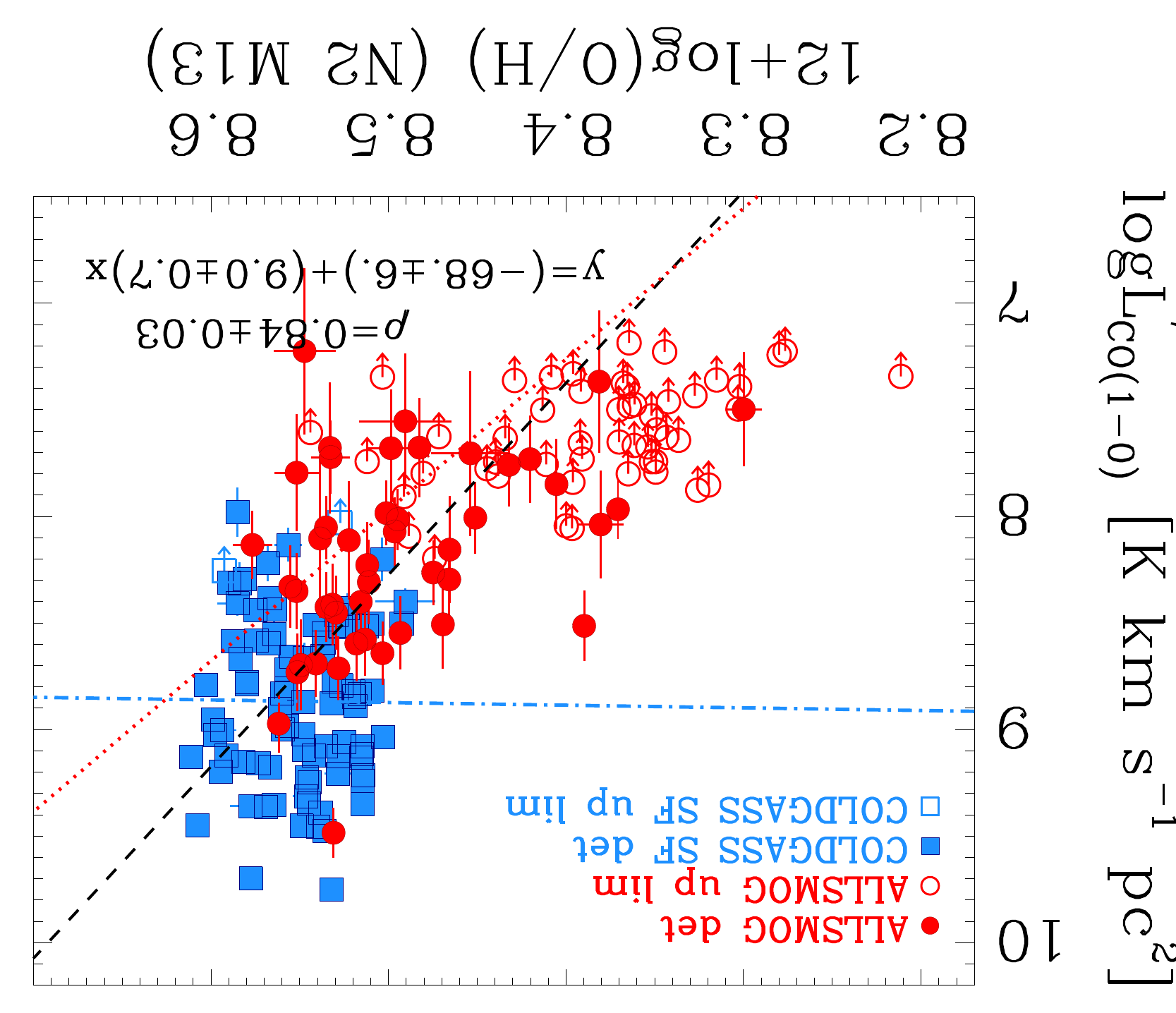}\\ 
        \vspace{0.2cm} 
          \includegraphics[clip=true,trim=9cm 4cm 2.cm 3cm,width=0.32\textwidth,angle=180]{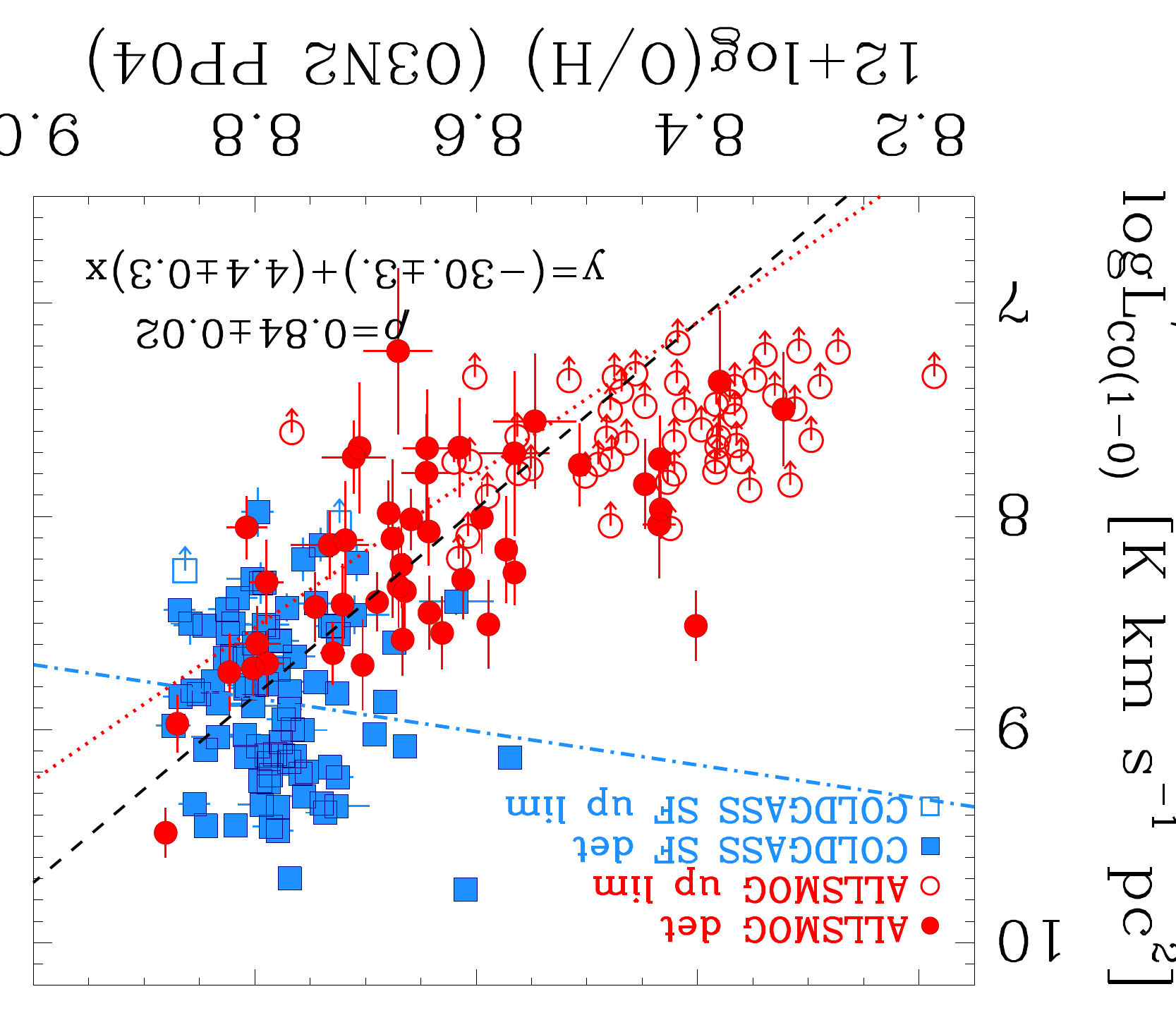}\quad 
       \includegraphics[clip=true,trim=9cm 4cm 2.cm 3cm,width=0.32\textwidth,angle=180]{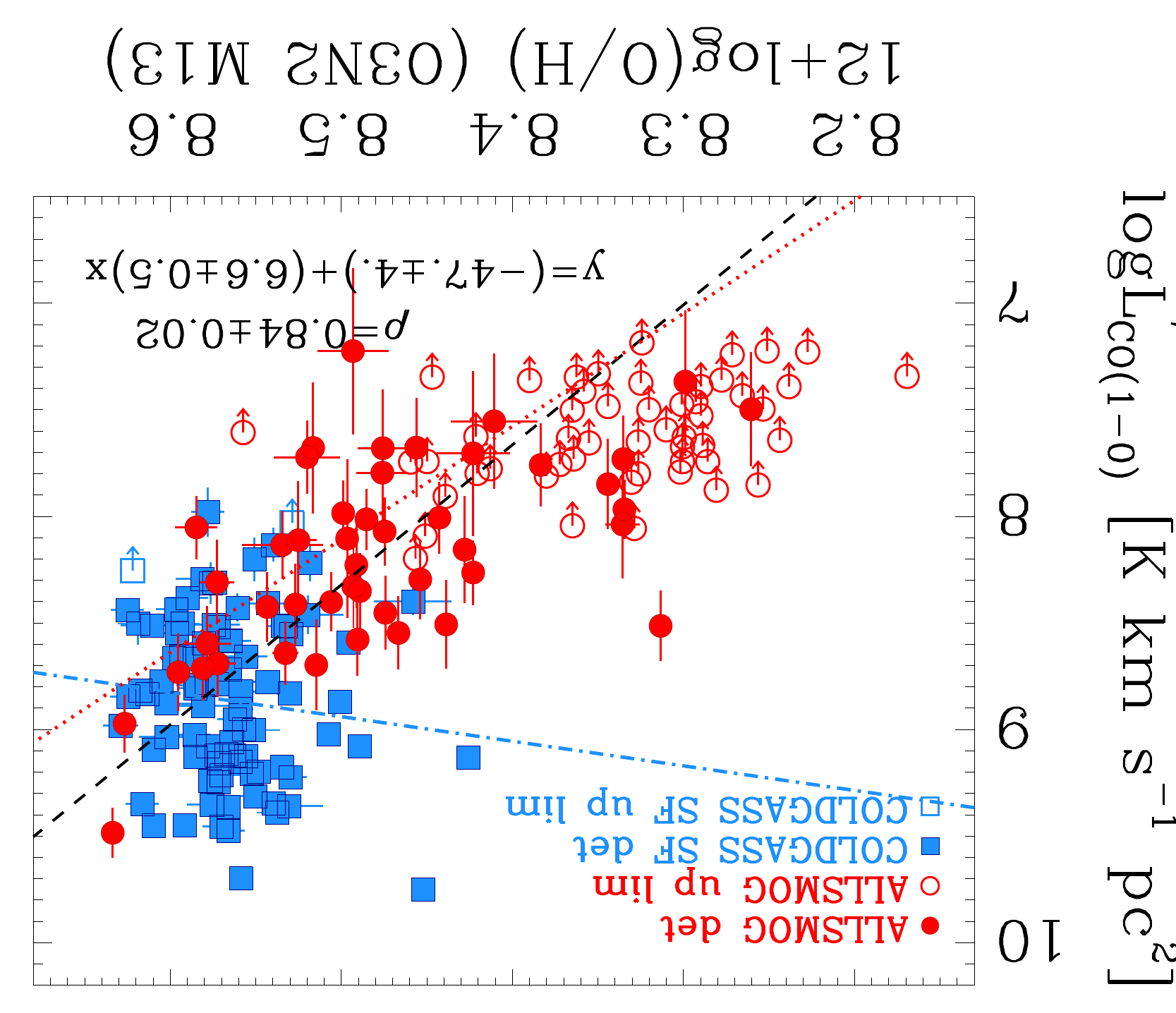}\\ 
     \vspace{0.2cm} 
     \caption{CO(1-0) line luminosity as a function of the Oxygen abundance in log-log scale for the sample of local star-forming galaxies defined by the ALLSMOG survey and by the sub-sample of COLD GASS selected in $\S$~\ref{sec:coldgass}. Each plot corresponds to a different strong-line metallicity calibration method (see $\S$~\ref{sec:metallicity}).
 Further explanation relevant to the quantity plotted on the y-axis, the symbols and the regression analysis can be found in the caption of Fig.~\ref{fig:lco_plots_1} and in the text ($\S$~\ref{sec:lco_vs_prop}).}
   \label{fig:lco_plots_2}
\end{figure*}

The relations observed between the CO(1-0) line luminosity and the five gas-phase metallicity estimates explored in this work ($\S$~\ref{sec:metallicity}) are shown in Figure~\ref{fig:lco_plots_2}. Figure~\ref{fig:lco_plots_2} clearly demonstrates that the range in metallicity spanned by ALLSMOG is significantly broader than that covered by COLD GASS, and this makes ALLSMOG a unique sample to investigate possible relations between the CO luminosity and the metallicity of the ISM. A high degree of correlation between $L^{\prime}_{\rm CO(1-0)}$ and Oxygen abundance is evident in all five plots shown in Fig.~\ref{fig:lco_plots_2}, with the correlation coefficient measured for the \cite{Tremonti+04} metallicity calibration ($\rho = 0.890 \pm 0.016$) being slightly higher than the one measured for the other four empirical calibrations ($\rho = 0.84$). The Bayesian regression analysis returns super-linear slopes for all relations, ranging between $\beta \in (3.8, 9.0)$, and large intrinsic scatters of $\sigma_{intr} \in (0.39, 0.50)$ (Table~\ref{table:regression_summary}), where the lowest value of $\sigma_{intr}\sim 0.39$ refers to the relation obtained using the \cite{Tremonti+04} calibration. Both the high correlation coefficient and the high slopes obtained for the $L^{\prime}_{\rm CO(1-0)}$ vs metallicity relation are consistent with previous findings \citep{Schruba+12,Amorin+16,Kepley+16}

We note that, from Fig~\ref{fig:lco_plots_2}, it may appear that the relation flattens towards lower metallicities. Such effect is more pronounced in the plots obtained using the empirical metallicity calibrations by \cite{PP04} and \cite{Marino+13}, where at abundances of $12+\log(O/H)\lesssim 8.5$ (corresponding to $\mathcal{Z}<0.6~\mathcal{Z}_{\odot}$), the detections exhibit a large scatter and show apparently no correlation between $L^{\prime}_{\rm CO(1-0)}$ and $12+\log(O/H)$. However, we stress that at these low metallicities, more than half of the data points are 3$\sigma$ upper limits, and so the seeming flattening is likely just a result of an underlying steep $L^{\prime}_{\rm CO(1-0)}$ - metallicity relation (as determined by the Bayesian regression analysis, which takes into account the upper limits) combined with the use of a flux-limited CO dataset. 

Similar to the other relations investigated in the previous sections, we checked for possible selection effects affecting our analysis by using the mode of the $L^{\prime}_{\rm CO(1-0)}/12+\log(O/H)$ distribution, calculated separately for the low-metallicity and the high-metallicity halves of the sample. We find that the mode of the distribution is consistent with the best-fit relation shown in Fig.~\ref{fig:lco_plots_2} (for all five metallicity calibration methods), hence suggesting that the Bayesian regression analysis provides a reliable description of the underlying relation - although obviously limited to the metallicity range considered in this work ($\S$~\ref{sec:sample}). This test also corroborates the hypothesis that the flattening at low metallicities is not real. Deeper CO observations will certainly provide more stringent constraints on the low-metallicity end of this relation.

We tested to what extent the strong and steep correlations observed in Figure~\ref{fig:lco_plots_2} may be due to the well known mass-metallicity relation \citep{Tremonti+04}, which of course holds true for our sample of local star-forming galaxies ($\rho=0.86-0.90$, $\alpha=6.5-7.5$, $\beta=0.11-0.27$, and $\sigma_{intr}\sim0.04-0.09$, for the 12+$\log(O/H)$ vs $\log M_*$ relation\footnote{The range in best-fit parameters indicated is due to the different metallicity calibrations adopted.}). This was done by examining any residual dependence of the quantity $\log L^{\prime}_{\rm CO(1-0)} - 1.1\log M_*$ (which is independent of $M_*$, see results of the best fit to the $L^{\prime}_{\rm CO(1-0)}$ vs $M_*$ relation shown in Fig.~\ref{fig:lco_plots_1}) on the Oxygen abundance, similar to what we did in $\S$~\ref{sec:lco_av} for the relation with $A_V$. The relevant plots and fit results are reported in Appendix~\ref{sec:appendix_ratios}. We found that removing the scaling with $M_*$ has the effect of significantly weakening the dependence on metallicity, although a residual positive trend can still be observed as a function of O/H, independently of the metallicity calibration adopted. 
This residual dependence of CO emission on O/H is probably due an increased rate of CO photodissociation in low metallicity environments \citep{Wolfire+10,Glover+MacLow11}.

\subsection{$L_{\rm CO}^{\prime}$ vs H{\sc i} gas mass}\label{sec:lco_hi}

\begin{figure}[tbp]
\centering
   \includegraphics[clip=true,trim=9cm 4cm 2.cm 3cm,width=0.32\textwidth,angle=180]{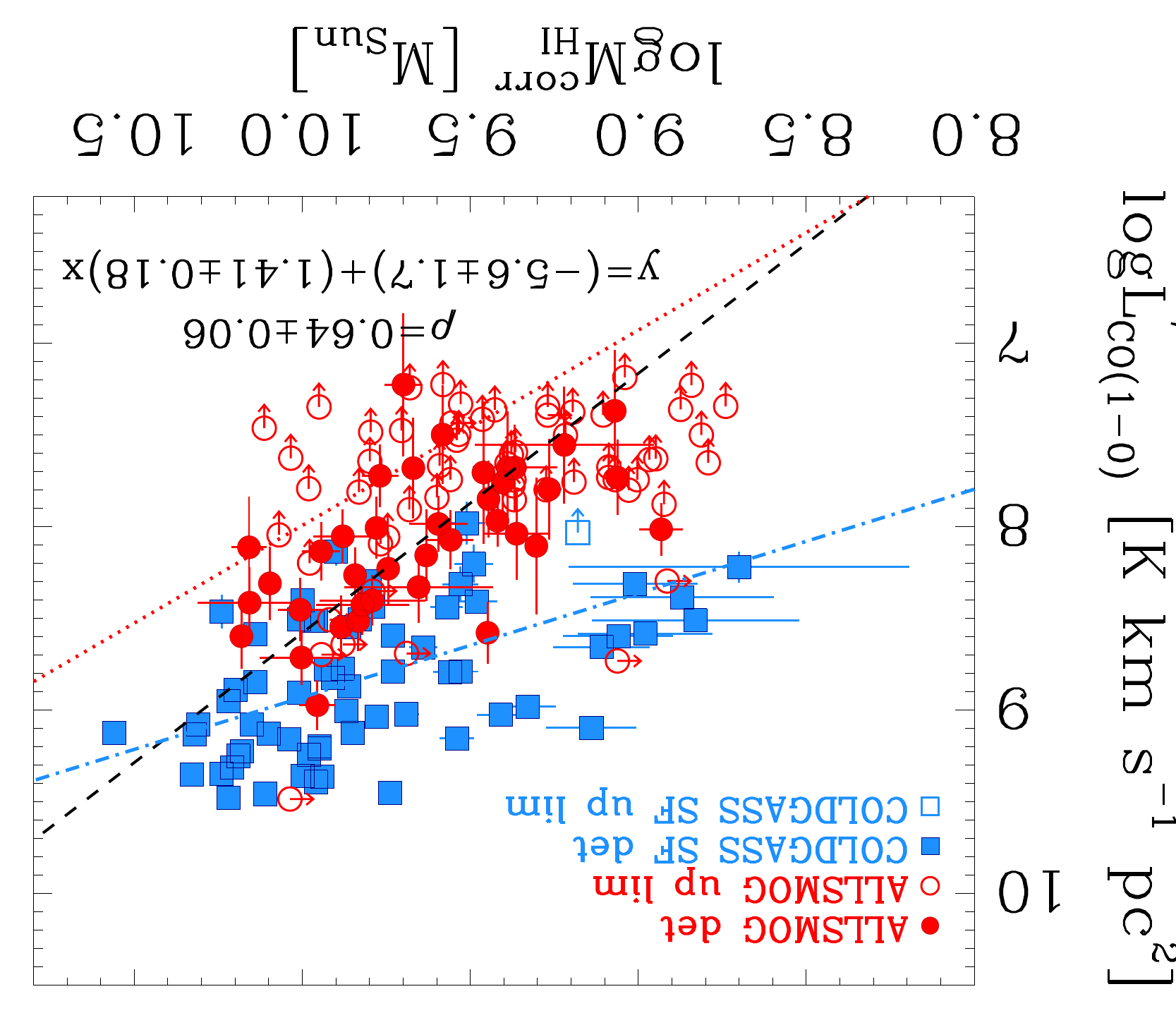}\quad   
      \vspace{0.2cm} 
      \centering
       \includegraphics[clip=true,trim=5cm 4cm 0cm 3cm,width=0.4\textwidth,angle=180]{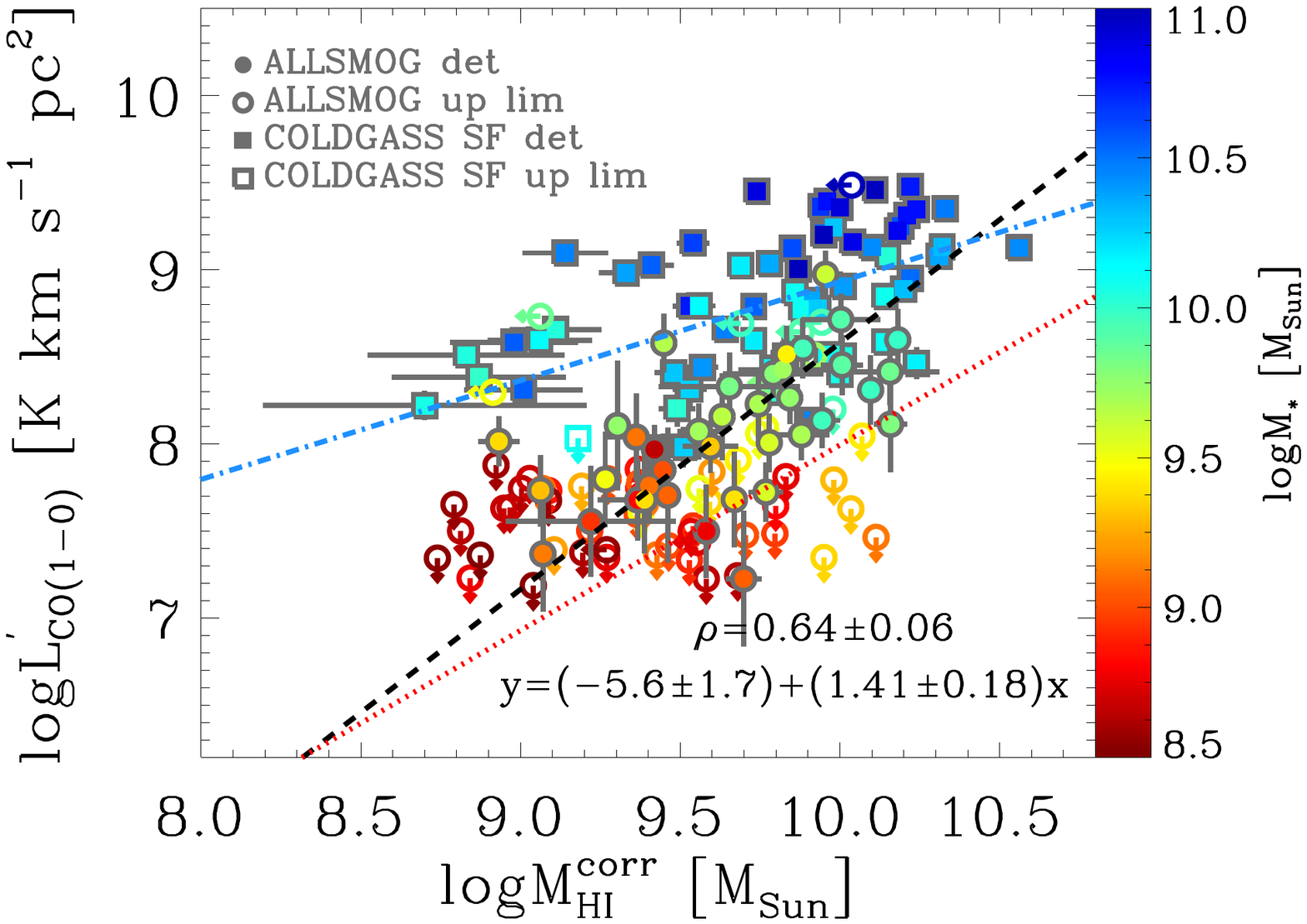}\\    
     \caption{{\it Top panel:} CO(1-0) line luminosity as a function of H{\sc i} gas mass in log-log scale for the sample of local star-forming galaxies defined by the ALLSMOG survey and by the sub-sample of COLD GASS selected in $\S$~\ref{sec:coldgass}. For ALLSMOG galaxies, H{\sc i} gas masses or 3$\sigma$ upper limits (indicated by leftwards arrows) were estimated using publicly available H{\sc i}~21cm observations as explained in $\S$~\ref{sec:HIdata}. For the COLD GASS sample, we used the $M_{\rm HI}$ values listed in the COLD GASS DR3 catalogue, which however does not provide upper limits for non-detections in H{\sc i} (23 out of the 88 galaxies considered in this work), so the COLD GASS sources not detected in H{\sc i} are not shown in the plot. Further explanation relevant to the quantity plotted on the y-axis, the symbols and the regression analysis can be found in the caption of Fig.~\ref{fig:lco_plots_1} and in the text ($\S$~\ref{sec:lco_vs_prop}). The Bayesian analysis was performed by including only the sources with an H{\sc i} line detection. {\it Bottom panel:} Same plot shown in the top panel, but with symbols colour-coded according to their stellar mass. }
   \label{fig:lco_plots_4}
\end{figure}

The last relation that we examine in this paper is the one between $L^{\prime}_{\rm CO(1-0)}$ and the total H{\sc i} gas mass, plotted in Fig~\ref{fig:lco_plots_4}. As we pointed out in $\S$~\ref{sec:HIdata}, due to the non-homogeneous sensitivity of the H{\sc i} data available for ALLSMOG galaxies, the upper limits on $M_{\rm HI}$ that we have estimated for the ten sources without a detection in H{\sc i} are not very informative. Furthermore, although the H{\sc i} coverage of the COLD GASS sample is much more uniform in terms of sensitivity than the ALLSMOG one, unfortunately no upper limits
on $M_{\rm HI}$ (or on the H{\sc i}~21cm flux) are provided for the galaxies not detected in H{\sc i} in the COLD GASS DR3 catalogue. For these reasons we excluded the sources without an H{\sc i} detection from the Bayesian regression analysis conducted on the $L^{\prime}_{\rm CO(1-0)}$-$M_{\rm HI}$ relation. 

With these premises in mind, our analysis evidences a positive correlation between $L^{\prime}_{\rm CO(1-0)}$ and $M_{\rm HI}$, with a correlation coefficient of $\rho=0.64\pm0.06$ (Table~\ref{table:regression_summary}). The intrinsic scatter is quite large, $\sigma_{intr} = 0.67\pm0.05$, that is the largest among all relationships explored in this paper (excluding the one between $L^{\prime}_{\rm CO(1-0)}$ and SSFR). In fact, we note that there is a significant number of outliers of the best-fit relation. By looking at the bottom panel of Fig~\ref{fig:lco_plots_4}, where the symbols are colour-coded by $M_*$, we infer that the galaxies lying significantly above the best-fit are mostly massive objects ($M_*\gtrsim10^{9.5}~M_{\odot}$) that are deficient in H{\sc i} compared to galaxies with a similar $M_*$. 
The large scatter of the CO-H{\sc i} relation was already noted by several previous studies \citep{Verter88,Young+Scoville91,Leroy+05,Lisenfeld+11,Saintonge+11a}. In particular, \cite{Lisenfeld+11} identified the sample selection method as a major factor determining the observed distribution in $L^{\prime}_{\rm CO(1-0)}/M_{\rm HI}$ values.

Similar to the previous relations, we tested our analysis for selection effects by calculating the mode of the $L^{\prime}_{\rm CO(1-0)}/M_{\rm HI}$ distribution separately for the low-$M_{\rm HI}$ and the high-$M_{\rm HI}$ samples, by including only detections in both CO and H{\sc i}. The test returned values that are in very good agreement with the best-fit relation obtained from the Bayesian regression analysis conducted on the total sample, suggesting that the latter provides a reliable description of the underlying relation.
On the contrary, the analyses conducted separately on low-$M_*$ and high-$M_*$ galaxies (see Table~\ref{table:regression_summary}), return relations that are significantly offset from that indicated by the mode of the $L^{\prime}_{\rm CO(1-0)}/M_{\rm HI}$ distribution, confirming that sample selection effects are especially important for the $L^{\prime}_{\rm CO(1-0)}$ vs $M_{\rm HI}$ relation \citep{Leroy+05,Lisenfeld+11}.
In summary, our analysis shows that, although there is a clear correlation between $L^{\prime}_{\rm CO(1-0)}$ and $M_{\rm HI}$, it has a much larger scatter than other scaling relations examined in this paper. Hence we infer that the H{\sc i} gas mass is not a very good predictor of the CO luminosity in typical local star-forming galaxies. 

We note that the H{\sc i} gas mass exhibits a positive correlation with stellar mass in star-forming galaxies (e.g. \citealt{Huang+12, Catinella+10}), although weaker and more scattered than the ones between $A_V$ or gas-phase metallicity and $M_*$ (discussed in $\S$~\ref{sec:lco_av} and \ref{sec:lco_met}). We confirmed that such correlation is present also within our sample, with $\rho=0.60\pm0.05$, $\alpha=6.4\pm0.4$, $\beta=0.34\pm0.04$, and $\sigma_{intr}=0.32\pm 0.02$, where these values refer to the $\log(M_{\rm HI})$ vs $\log(M_*)$ relation. To check whether the observed dependence of $L^{\prime}_{\rm CO(1-0)}$ on the H{\sc i} gas mass is the result of the scaling of $M_{\rm HI}$ with $M_*$, we studied the $M_*$-independent quantity $\log L^{\prime}_{\rm CO(1-0)} - 1.1\log M_*$ as a function of $M_{\rm HI}$ (Appendix~\ref{sec:appendix_ratios}). We found a positive residual correlation with $\rho=0.8\pm0.2$, and a shallow slope of $\beta = 0.27 \pm 0.12$. We infer that most of the scatter observed in Fig.~\ref{fig:lco_plots_4} is ascribable to variations in stellar mass, but stellar mass alone cannot account for all of the observed correlation between $L^{\prime}_{\rm CO(1-0)}$ and $M_{\rm HI}$. The residual dependence of the quantity $\log L^{\prime}_{\rm CO(1-0)} - 1.1\log M_*$ on $M_{\rm HI}$ may reflect an increased molecular gas fraction for the galaxies with a larger H{\sc i} gas reservoir in our sample.

\begin{table*}
\caption{Analysis of the dependencies of the CO(1-0) line luminosity on galaxy properties} 
\label{table:regression_summary}
\centering
\begin{tabular}{lcccc}
\hline\hline
\multicolumn{5}{c}{Model: $y = \alpha + \beta x$, with $y = \log L^{\prime}_{\rm CO(1-0)} [{\rm K~km~s^{-1}~pc^2}]$} \\
\hline
\multicolumn{5}{c}{Total sample:} \\
$x$ & $\alpha$ & $\beta$ & $\sigma_{intr}$ & $\rho$  \\
\hline
$\log M_{*}~[M_{\odot}]$ & $-2.5 \pm 0.5$ & $1.10 \pm 0.05$ & $0.31 \pm 0.02$ & $0.926 \pm 0.012$ \\
$\log SFR~[M_{\odot}~yr^{-1}]$ & $8.16 \pm 0.04$  & $1.34 \pm 0.07$ & $0.21 \pm 0.04$ & $0.964 \pm 0.013$ \\
$\log SSFR~[yr^{-1}]$ & $1.0 \pm 3.0$ & $-0.7 \pm 0.3$ & $0.83 \pm 0.06$ & $-0.25 \pm 0.09$ \\
$A_V~[mag]$ $^{\dag}$ (CCM89)  & $7.13 \pm 0.09$ & $1.02 \pm 0.07$ & $0.49 \pm 0.03$ & $0.80 \pm 0.03$ \\
$A_V~[mag]$ $^{\dag}$ (Calzetti)  & $7.13 \pm 0.09$ & $0.92 \pm 0.06$ & $0.49 \pm 0.03$ & $0.80 \pm 0.03$ \\
$12 + \log (O/H)$ (Tremonti+04) & $-25 \pm 2$ & $3.8 \pm 0.2$ & $0.39 \pm 0.03$ & $0.890 \pm 0.016$ \\
$12 + \log (O/H)$ N2 PP04 & $-54 \pm 5$ & $7.2 \pm 0.6$ & $0.50 \pm 0.04$ & $0.84 \pm 0.03$ \\
$12 + \log (O/H)$ N2 M13 & $-68 \pm 6$ & $9.0 \pm 0.7$ & $0.51 \pm 0.03$ & $0.84 \pm 0.03$ \\
$12 + \log (O/H)$ O3N2 PP04 & $-30 \pm 3$ & $4.4 \pm 0.3$ & $0.48 \pm 0.03$ & $0.84 \pm 0.02$ \\
$12 + \log (O/H)$ O3N2 M13 & $-47 \pm 4$ & $6.6 \pm 0.5$ & $0.48 \pm 0.03$ & $0.84 \pm 0.02$ \\
$\log M_{\rm HI}^{corr}$ $^{\ddag}$ & $ -5.6 \pm 1.7 $ & $1.41 \pm 0.18 $ & $ 0.67\pm 0.05$ & $0.64 \pm 0.06$ \\
\hline
\multicolumn{5}{c}{Low-$M_*$ sample ($\log M_* [M_{\odot}] < 10.0$):} \\
$x$ & $\alpha$ & $\beta$ & $\sigma_{intr}$ & $\rho$  \\
\hline
$\log M_{*}~[M_{\odot}]$ & $-4.5 \pm 1.2$ & $1.31 \pm 0.13$ & $0.29 \pm 0.05$ & $0.90 \pm 0.03$ \\
$\log SFR~[M_{\odot}~yr^{-1}]$ & $8.20 \pm 0.07$  & $1.56 \pm 0.17$ & $0.18 \pm 0.08$ & $0.96 \pm 0.04$ \\
$\log SSFR~[yr^{-1}]$ & $-6 \pm 14$ & $-1.4 \pm 1.5$ & $0.71 \pm 0.11$ & $-0.3 \pm 0.3$ \\
 $A_V~[mag]$ $^{\dag}$ (CCM89)  &  $6.85 \pm 0.16$ &  $1.10 \pm 0.17$ & $0.54 \pm 0.07$ & $0.66 \pm 0.08$ \\
$A_V~[mag]$ $^{\dag}$ (Calzetti)  & $6.86 \pm 0.16$ & $0.98 \pm 0.15$ & $0.54 \pm 0.07$ & $0.66 \pm 0.07$ \\
$12 + \log (O/H)$ (Tremonti+04) & $-21 \pm 3$ & $3.2 \pm 0.3$ & $0.30 \pm 0.05$ & $0.88 \pm 0.03$ \\
$12 + \log (O/H)$ N2 PP04 & $-41 \pm 7$ & $5.7 \pm 0.8$ & $0.44 \pm 0.06$ & $0.80 \pm 0.05$ \\
$12 + \log (O/H)$ N2 M13 & $-52 \pm 8$ & $7.1 \pm 1.0$ & $0.45 \pm 0.07$ & $0.80 \pm 0.05$ \\
$12 + \log (O/H)$ O3N2 PP04 & $-23 \pm 4$ & $3.6 \pm 0.4$ & $0.41 \pm 0.06$ & $0.81 \pm 0.04$ \\
$12 + \log (O/H)$ O3N2 M13 & $-37 \pm 5$ & $5.3 \pm 0.6$ & $0.40 \pm 0.06$ & $0.81 \pm 0.05$ \\
$\log M_{\rm HI}^{corr}$ $^{\ddag}$ & $ -3 \pm 3 $ & $1.1 \pm 0.3 $ & $ 0.59\pm 0.09$ & $0.56 \pm 0.10$ \\
\hline
\multicolumn{5}{c}{High-$M_*$ sample ($\log M_* [M_{\odot}] \geq 10.0$):} \\
$x$ & $\alpha$ & $\beta$ & $\sigma_{intr}$ & $\rho$  \\
\hline
$\log M_{*}~[M_{\odot}]$ & $-0.5 \pm 1.5$ & $0.91 \pm 0.15$ & $0.32 \pm 0.03$ & $0.60 \pm 0.08$ \\
$\log SFR~[M_{\odot}~yr^{-1}]$ & $8.30 \pm 0.09$  & $1.03 \pm 0.15$ & $0.24 \pm 0.03$ & $0.80 \pm 0.07$ \\
$\log SSFR~[yr^{-1}]$ & $12.0 \pm 1.5$ & $0.32 \pm 0.15$ & $0.39 \pm 0.03$ & $0.28 \pm 0.13$ \\
$A_V~[mag]$ $^{\dag}$ (CCM89)  & $8.26 \pm 0.14$ & $0.39 \pm 0.09$ & $0.36 \pm 0.03$ & $0.46 \pm 0.09$ \\
$A_V~[mag]$ $^{\dag}$ (Calzetti)  & $8.27 \pm 0.14$ & $0.35 \pm 0.08$ & $0.36 \pm 0.03$ & $0.45 \pm 0.09$ \\
$12 + \log (O/H)$ (Tremonti+04) & $-2 \pm 6$ & $1.1 \pm 0.7$ & $0.39 \pm 0.03$ & $0.20 \pm 0.12$ \\
$12 + \log (O/H)$ N2 PP04 & $10 \pm 11$ & $-0.1 \pm 1.3$ & $0.40 \pm 0.03$ & $-0.01 \pm 0.11$ \\
$12 + \log (O/H)$ N2 M13 & $10 \pm 13 $ & $-0.1 \pm 1.6$ & $0.40 \pm 0.03$ & $-0.01 \pm 0.11$ \\
$12 + \log (O/H)$ O3N2 PP04 & $16 \pm 7$ & $-0.8 \pm 0.8$ & $0.40 \pm 0.03$ & $-0.11 \pm 0.12$ \\
$12 + \log (O/H)$ O3N2 M13 & $19 \pm 11$ & $-1.2 \pm 1.3$ & $0.40 \pm 0.03$ & $-0.11 \pm 0.12$ \\
$\log M_{\rm HI}^{corr}$ $^{\ddag}$ & $ 3.3 \pm 1.2 $ & $0.57 \pm 0.13 $ & $ 0.34\pm 0.03$ & $0.53 \pm 0.10$ \\
\hline

\end{tabular}
\tablefoot{$(\alpha, \beta)$ are the best-fit linear regression coefficients, $\sigma_{intr}$ is the intrinsic scatter about the best-fit regression line, and $\rho$ is the correlation coefficient. The analysis has been executed using the \texttt{IDL} code \texttt{linmix\_err.pro} developed by \cite{Kelly07}, which employes a Bayesian approach to perform a linear regression on a dataset that includes upper limits in $y$ as well as measurement errors on both $x$ and $y$. For each parameter of interest, the value and the associated error reported in the table correspond respectively to the median and to a robust estimate of the standard deviation of the posterior distribution returned by the code.
$^{\dag}$ Because of the large measurement errors on $A_V$, of the order of $\sim0.5~mag$, the regression analysis for the $L^{\prime}_{\rm CO(1-0)}$ vs $A_V$ relationships was performed without taking into account the errors on $x$, in order not to bias the results.
$^{\ddag}$ In this case the regression analysis was performed by considering only the sources with an H{\sc i} line detection.}
\end{table*}


\section{Discussion}\label{sec:discussion}
\subsection{The importance of sample selection to study the effect of a low metal and dust content on CO emission}\label{sec:alphaco_note}

In the low-$M_*$ regime of normal, MS star-forming galaxies probed by ALLSMOG, the dust and metal content of the ISM plays an essential role in the interpretation of CO observations. 
Theoretical work has provided evidence that a low metal ISM, with its consequently low dust extinction, can induce a significant suppression of CO luminosity in molecular clouds, because CO molecules are easily photo-dissociated in clouds that are not dust-shielded from the UV radiation field \citep{Wolfire+10, Glover+MacLow11}. 
Observationally, it is well known that CO emission is very faint in low-metallicity environments \citep{Elmegreen+80,Verter+Hodge95,Taylor+98}, and, statistically, low-$M_*$ galaxies are low in metallicity \citep{Tremonti+04}. However, it is still debated whether the faintness of CO emission in low-$M_*$ galaxies reflects a genuinely lower H$_2$ gas content compared to massive galaxies \citep{Leroy+05}, or if it is mainly a result of their low metal content and poor shielding by dust which makes CO easier to photo-dissociate in their ISM \citep{Taylor+98,Schruba+12}. The former scenario brings along a second question, that is: is the reduced molecular gas reservoir of smaller galaxies consistent with simple scaling effects with the galaxy mass or is it due to additional differences in the physical properties of low-$M_*$ and high-$M_*$ galaxies \citep{Leroy+05}? The latter scenario instead implies that CO is blind to a significant amount of molecular gas in these sources. The two effects can obviously coexist, making it harder to identify which one prevails (e.g. \citealt{Hunt+15}). 

In the next section ($\S$~\ref{sec:lowM_vs_highM}) we will argue that much of the discrepancies between the findings of previous observational studies investigating the effects of a low metal and dust content on the observed CO emission may be ascribed to selection effects. When studying scaling relations, the most crucial task is to designate the target population that one wishes to study, which in our case is the population of normal star-forming galaxies in the local Universe. Following this, one needs to construct a sample that is both representative of the underlying target population and well-characterised in terms of the physical properties that may affect the observable, that is the CO luminosity. We suggest that in order to isolate the effects of a metallicity-dependent CO-to-H$_2$ conversion factor (i.e. $\alpha_{\rm CO}\equiv M_{mol}/L^{\prime}_{\rm CO(1-0)}$) on the observed CO luminosity, one would need to select a sample of galaxies that is as uniform as possible in star formation efficiency (${\rm SFE \equiv SFR}/M_{mol}$, i.e. the SFR per unit molecular gas mass), such as a sample of galaxies lying on the MS. In fact, galaxies well above the MS (characterised by high SSFRs) are known to have an enhanced SFE due to higher concentrations of dense molecular gas resulting from the compression of gas due to mergers and interactions \citep{Young+Scoville91,Lisenfeld+11}. At a given SFR, a higher SFE would imply a lower molecular mass, hence mimicking the effect of a higher $\alpha_{CO}$ \citep{McQuinn+12, Hunt+15, Amorin+16}. On the other hand, further complicating the interpretation of CO observations in galaxies above the MS, turbulent non-virial motions associated with mergers and galaxy interactions result in large gas velocity dispersions that can significantly lower the $\alpha_{\rm CO}$ \citep{Downes+Solomon98,Yao+03}. All the different factors mentioned above may complicate the interpretation of CO observations in blue compact dwarfs (BCDs) \citep{Kepley+16, Hunt+15, Amorin+16}. In these cases, using alternative H$_2$ tracers such as [CI] \citep{Papadopoulos+04a, Papadopoulos+04b, Gullberg+16, Krips+16,Bothwell+17} can certainly help solve the degeneracy between $\alpha_{CO}$ and SFE.

On the one hand, ALLSMOG galaxies are characterised by a normal, MS-like star formation activity (Figure~\ref{fig:MS}), spanning a rather narrow range of SSFRs (Figure~\ref{fig:histo}), hence they are unlikely to show the strong variations in SFE exhibited by starburst dwarfs that can complicate the interpretation of CO observations. On the other hand, ALLSMOG significantly extends the dynamic range of $M_*$, SFR, gas-phase metallicity, and $A_V$ probed by previous CO surveys targeting star-forming galaxies on the MS, as demonstrated by Figs~\ref{fig:lco_plots_1}, \ref{fig:lco_plots_3} and \ref{fig:lco_plots_2}. 
In particular, as shown in $\S$~\ref{sec:lco_vs_prop}, ALLSMOG can be used in combination with COLD GASS to construct a sample that is highly representative of typical star-forming galaxies, and little biased (within the range probed) in terms of $M_*$, SFR, metallicity, $M_{\rm HI}$ and CO luminosity, although obviously not completely free from selection effects (see discussion in $\S$~\ref{sec:lco_vs_prop}). In conclusion, ALLSMOG represents the ideal starting point to investigate - in a statistically-sound way for typical star-forming galaxies - scaling relations between gas content and other galaxy properties, including the effect of a moderately low metal and dust content on the observed CO luminosity. 

\subsection{Are low-$M_*$ and high-$M_*$ galaxies different in their gas properties?}\label{sec:lowM_vs_highM}

The question of whether low-$M_*$ galaxies trace a different population in terms of gas properties from massive star-forming galaxies has been explored by various groups (e.g. \citealt{Leroy+05,Schruba+12}). Earlier studies based on CO observations of local dwarfs suggested that they are scaled-down versions of large spirals, with the only major difference being their apparently larger H{\sc i} gas reservoir \citep{Leroy+05}. According to this scenario, the faintness of CO emission in small galaxies and their consequently low CO detection rate compared to more massive objects are fully consistent with a mass-scaling of the CO luminosity, hence ruling out both strong variations in the $\alpha_{\rm CO}$ factor and a significant dependence of the molecular gas content on galaxy properties other than $M_*$. More specifically, \cite{Leroy+05} noted that in their sample of star-forming galaxies the CO line luminosity\footnote{These authors adopt a constant $\alpha_{\rm CO}$ for all sources independently of their $M_*$ and so they are effectively studying the CO luminosity rather than the molecular gas content.} correlates strongly with the luminosity of the stellar component, and that once this dependance is accounted for (for example by normalising $L^{\prime}_{\rm CO}$ by the K-band luminosity, i.e. $L^{\prime}_{\rm CO}/L_{K}$), most of the other dependencies between the CO content and galaxy properties are removed. 

The more recent stacking analysis by \cite{Schruba+12} has challenged the scenario depicted above by implying that mass scaling alone cannot explain the significantly lower CO content of dwarfs compared to massive spirals. \cite{Schruba+12} find that dwarfs lie significantly below the trends between CO luminosity and galaxy properties defined by massive sources, and this result persists when $L_{\rm CO}^{\prime}$ is normalised by the stellar light or the SFR. By attributing all of the observed variation in the CO brightness per unit SFR to metallicity-dependent variations in $\alpha_{\rm CO}$, \cite{Schruba+12} estimate the $\alpha_{\rm CO}$ in these objects to be up to more than one order of magnitude higher than the Galactic value.

In our view, much of the discrepancies between the findings of \cite{Leroy+05} and \cite{Schruba+12} are due to sample selection effects, as anticipated in $\S$~\ref{sec:alphaco_note}. The dwarf galaxy sample of \cite{Leroy+05} includes mostly highly star-forming objects, selected among those detected by the Infrared astronomical satellite ({\it IRAS}) in order to ``maximise the chances of detecting CO emission'' \citep{Leroy+05}. Furthermore, their analysis of scaling relations accounts only for the CO detections, neglecting the upper limits given by the non-detections. Hence it is reasonable to suppose that the study by \cite{Leroy+05} probed primarily the CO-rich, highly star-forming tail of the local low-$M_*$ galaxy population. In contrast, the sample of \cite{Schruba+12} includes many truly metal-poor dwarfs, with Oxygen abundances down to $12+\log(O/H) \simeq 7.5$, in the majority of which CO remains undetected even after stacking. It is therefore plausible that \cite{Schruba+12} reached their conclusions because they were focussing on the very low-metallicity tail of the distribution of local star-forming galaxies. In summary, \cite{Leroy+05} and \cite{Schruba+12} probed the two `extremes' of the local low-$M_*$ galaxy population, and so reached apparently discrepant conclusions.

ALLSMOG, sampling the low-$M_*$ end of the local MS, bridges the gap between IR-luminous CO-rich objects and extremely metal-poor dwarfs. Therefore, as we have already suggested in $\S$~\ref{sec:alphaco_note}, the ALLSMOG sample is particularly appropriate to investigate possible differences in the gas properties of typical low-$M_*$ and high-$M_*$ local star-forming galaxies. In $\S$~\ref{sec:lco_vs_prop} we have presented the results of a Bayesian linear regression analysis performed on the relationships between $L_{\rm CO}^{\prime}$ and various galaxy-integrated physical properties. The analysis has been carried out both on the total sample (defined by ALLSMOG and by the star-forming subsample of COLD GASS) and on the two sub-samples of low-$M_*$ ($M_{*}<10^{10}~M_{\odot}$) and high-$M_*$ ($M_{*} \geq 10^{10}~M_{\odot}$) star-forming galaxies separately. 

The results of the regression analysis conducted on the different samples, summarised in Table~\ref{table:regression_summary}, show that in most cases where a strong and tight correlation is identified by the Bayesian analysis, a unique relation provides a good fit to the entire sample. However, the correlations between $L_{\rm CO}^{\prime}$ and many galaxy properties (namely: $M_*$, SFR, $A_V$, $12+\log(O/H)$) are systematically stronger (higher $\rho$), tighter (smaller $\sigma_{intr}$) and steeper (higher $\beta$) for the low-$M_*$ than for the high-$M_*$ sample. In particular, the correlation with the metallicity, which is strong and very steep for the low-$M_*$ sample, disappears for massive galaxies ($\S$~\ref{sec:lco_met}). By looking at Fig~\ref{fig:lco_plots_2}, it is evident that the very narrow range of metallicities probed by massive galaxies is the reason why the Bayesian analysis cannot retrieve a correlation between $L_{\rm CO}^{\prime}$ and Oxygen abundance based on the COLD GASS sample alone. A steep relation, consistent with that found by \cite{Schruba+12} (see their Fig.~4), can be retrieved only by including the low-$M_*$ (and low-metallicity) ALLSMOG galaxies and by taking into account the upper limits. Further supporting this hypothesis, we note that, for metallicities where a few low-$M_*$ ALLSMOG sources overlap with more massive COLD GASS galaxies (Figure~\ref{fig:lco_plots_2}), the two sub-samples are virtually indistinguishable in terms of their CO luminosities. 

In summary, within our sample, similar scaling relations between CO luminosity and several galaxy-averaged properties (e.g. $M_*$, SFR, $A_V$, $12+\log(O/H)$, $M_{\rm HI}$) are followed by galaxies spanning over two orders of magnitude in stellar mass.
When the results of the Bayesian regression analysis performed on the massive galaxy sample are discrepant from those given by the low-$M_*$ sample (such as for $L_{\rm CO}^{\prime}$-metallicity relations in Fig.~\ref{fig:lco_plots_2}, or the $L_{\rm CO}^{\prime}$-$M_{\rm HI}$ relation in Fig.~\ref{fig:lco_plots_4}), the discrepancy can be entirely explained by selection effects.  

 Among all CO scaling relations examined in this work, the ones with SFR and $M_*$ exhibit the smallest scatters ($\sigma_{intr}\sim0.3$~dex, Table~\ref{table:regression_summary}). For this reason, and because $M_*$ is known to be positively correlated with $A_V$, gas-phase metallicity and $M_{HI}$ in star-forming galaxies \citep{Garn+Best10, Tremonti+04, Huang+12, Catinella+10} and hence also within our sample, it is reasonable to expect that the underlying scaling of these parameters with $M_*$ is responsible for their correlation with CO luminosity (Figures~\ref{fig:lco_plots_3}-\ref{fig:lco_plots_4}). However, a simple analysis of the quantity $\log L_{CO(1-0)}^{\prime} - 1.1\log M_*$ (where the power-law scaling of $L_{CO(1-0)}^{\prime}$ with $M_*$ has been removed) as a function of O/H and $M_{\rm HI}$ shows a residual positive correlation in both cases (Appendix~\ref{sec:appendix_ratios}), hence ruling out that the scaling with stellar mass alone can account for all of the observed variations of $L_{CO(1-0)}^{\prime}$ as a function of these parameters. We stress, however, that an identification of the most fundamental parameters driving the observed scaling relations is beyond the scope of this data release paper, and we refer to previous \citep{Bothwell+16b,Bothwell+16a} and future works for a more rigorous analysis of the problem.


\section{Summary and conclusions}

We have presented the final data release of the ALLSMOG survey, comprising APEX CO(2-1) emission line observations 
of 88 local, low-$M_*$ ($10^{8.5} < M_{*} [M_{\odot}] <10^{10}$) star-forming galaxies (including the 42 sources presented in \cite{Bothwell+14}), and an additional sample of nine low-$M_*$ ($M_* < 10^9~M_{\odot}$) star-forming galaxies observed in CO(1-0) and CO(2-1) line emission with the IRAM~30m telescope, for a total of 97 sources.
We have described in detail the sample selection, the observations, the data reduction and analysis methods, as well as the ancillary optical and H{\sc i} observations available for the full sample.  

At the sensitivity limit of our survey, we have registered a total CO detection rate of 47\% (46/97 detections). Galaxies with higher $M_*$, SFR, $A_V$, $12+\log (O/H)$ and $M_{\rm HI}$ have systematically higher CO detection rates. In particular, a two-sample K-S test has indicated that the galaxy parameter that shows the strongest statistical differences between CO detections and non detections is the gas-phase metallicity, for any of the five metallicity calibrations examined in this work. On the contrary, we have not found any strong bias of CO detections as a function of SSFR or redshift, most likely because ALLSMOG galaxies by construction span only a narrow range in these parameters.

We have then investigated scaling relations between the CO(1-0) line luminosity and galaxy-averaged properties using a sample of 185 local star-forming galaxies comprising the ALLSMOG survey and a sub-sample drawn from the COLD GASS survey selected to complement ALLSMOG at $M_*>10^{10}~M_{\odot}$. Our aim was to construct a statistically-sound sample that is highly representative of the local actively star-forming galaxy population, probing the local MS and its $\pm 0.3$~dex intrinsic scatter in the stellar mass range, $10^{8.5}<M_*[M_{\odot}]\lesssim10^{11}$. We have used a Bayesian regression analysis method that takes into account the non-detections to (i) identify the global galaxy parameters that are best predictors of the CO luminosity within our sample, and to (ii) explore possible differences between the scaling relations defined by low-$M_*$ and high-$M_*$ galaxies. 

We have found a strong correlation between $L^{\prime}_{\rm CO(1-0)}$ and the following galaxy-averaged properties: $M_*$, SFR, $A_V$ and gas-phase metallicity (for all five metallicity calibrations explored), and a weaker correlation with $M_{\rm HI}$. Not surprisingly in light of previous studies, the strongest correlation is the one with SFR, but we have found that the $L^{\prime}_{\rm CO(1-0)}$-$M_*$ relation is almost comparably tight and significantly closer to linear. While the $L^{\prime}_{\rm CO(1-0)}$-SFR correlation has been identified for many years as a fundamental relation between the fuel available for star formation and the SFR itself (e.g. the S-K law), the tight and almost linear relation between $L^{\prime}_{\rm CO(1-0)}$ and $M_*$ is more puzzling. We have ruled out that the $\beta\sim1$ slope is a result of selection effects, and we have proposed that this relation may be so tight and linear because the luminosity of optically thick low-$J$ CO transitions is an excellent tracer of the dynamical mass in star-forming galaxies, assuming that in this class of objects the bulk of CO probes molecular clouds in virial motions (see also the linear relation between $^{12}$CO luminosity and virial mass found for GMCs by \cite{Scoville+87}). This explanation assumes that the stellar mass is a good tracer of the dynamical mass in our sample. The steepest relation is the one between $L^{\prime}_{\rm CO(1-0)}$ and metallicity, but we have noted that the relation with $A_V$ may be also very steep once selection effects are accounted for.

Our results suggest that local star-forming galaxies spanning over two orders of magnitude obey to similar scaling relations between CO luminosity and several galaxy-averaged parameters (e.g. $M_*$, SFR, $A_V$, O/H and $M_{\rm HI}$). 
In those cases where the Bayesian regression analysis has returned best-fit relations for the low-$M_*$ and the high-$M_*$ samples that are inconsistent with each other, we have identified sample selection effects to be responsible for the apparent discrepancy. The small intrinsic scatters ($\sigma_{intr}\sim0.3$~dex) of the $L_{CO(1-0)}^{\prime}$ vs SFR and $M_*$ relations suggest that SFR and stellar mass are fundamental parameters in setting the observed variations of CO luminosity in local star-forming galaxies. However, our preliminary analysis of secondary dependencies suggests that the scaling with $M_*$ alone cannot account for all of the observed variations of $L_{CO(1-0)}^{\prime}$ as a function of metallicity and H{\sc i} gas mass. 

The fully-reduced APEX data products that are part of the ALLSMOG survey are released to the public via the ESO Phase 3 platform, and we strongly encourage the astronomical community to exploit them for their own investigation. Furthermore, all data products are made publicly available through the ALLSMOG website\footnote{\texttt{http://www.mrao.cam.ac.uk/ALLSMOG/}}.

\begin{acknowledgements}

The ALLSMOG team thanks immensely the APEX and ESO staff members for their help and continuous support 
during the three years of APEX observations. We also thank the IRAM staff for help provided during the observations. We are grateful to Stefano Andreon for fruitful discussions about Bayesian analysis and selection effects, to Ivy Wong for the help with the HIPASS catalogue, and to Myha Vuong for helping us with the ESO Phase 3 preparation. We thank the anonymous referee for providing useful comments that helped improve the paper.
CC acknowledges funding from the European Union's Horizon 2020 research and innovation programme under the Marie Sklodowska-Curie grant agreement No 664931. 
KS and CC acknowledge support from Swiss National Science Foundation Grants PP00P2\_138979 and PP00P2\_166159.
RM acknowledges support from the ERC Advanced Grant 695671
``QUENCH'' and from the Science and Technology Facilities Council (STFC).
Z-Y.Z. acknowledges support from ERC in the form of the Advanced
Investigator Programme, 321302, COSMICISM.
DR acknowledges support from the National Science
Foundation under grant number AST-1614213 to Cornell University.
MA acknowledges partial support from FONDECYT through grant 1140099.
YP acknowledges support from the National Key Program for Science and
Technology Research and Development under grant number 2016YFA0400702, and
from the Thousand Youth Talents Program of China.
FB acknowledges support from the Science and Facilities Research Council (STFC).
This publication is based on data acquired with the Atacama Pathfinder EXperiment
(APEX) under programme ID 192.A-0359. APEX is a collaboration between the
Max-Planck-Institut fur Radioastronomie, the European Southern Observatory,
and the Onsala Space Observatory. This work is also based on observations carried out under project number 188-14 with the IRAM 30m Telescope. IRAM is supported by INSU/CNRS (France), MPG (Germany) and IGN (Spain). We acknowledge the usage of the following public databases and catalogues: HyperLeda (http://leda.univ-lyon1.fr); the MPA-JHU release 
of spectral measurements for the SDSS DR7 (http://wwwmpa.mpa-garching.mpg.de/SDSS/DR7/); 
the HIPASS catalogue (http://www.atnf.csiro.au/research/multibeam/release/); the
COLD GASS catalogue (http://wwwmpa.mpa-garching.mpg.de/COLD\_GASS/).
This research has made use of the NASA/IPAC Extragalactic Database (NED) which is operated by the Jet Propulsion Laboratory, 
California Institute of Technology, under contract with the National Aeronautics and Space Administration.
{\it Software:} GILDAS/CLASS, IDL, Python.
\end{acknowledgements}

\bibliography{allsmog_bib}
\bibliographystyle{aa}


\begin{appendix}
\section{The role of stellar mass-scaling in setting the observed correlations}\label{sec:appendix_ratios}
In $\S$~\ref{sec:lco_vs_prop} we have demonstrated that in our sample of local star-forming galaxies the CO luminosity is not only strongly correlated with $M_*$ and SFR (Figure~\ref{fig:lco_plots_1}), but also with $A_V$, O/H and $M_{\rm HI}$ (Figures~\ref{fig:lco_plots_2}-\ref{fig:lco_plots_4}). However, because all of the above parameters are also known to correlate with stellar mass in local star-forming galaxies (see e.g. \citealt{Garn+Best10} for the $A_V$-$M_*$ relation,  \citealt{Tremonti+04} for the O/H-$M_*$ relation, and \citealt{Huang+12, Catinella+10} for the $M_{HI}$-$M_*$ relation), it is reasonable to inquire which among the observed dependences of $L^{\prime}_{\rm CO(1-0)}$ on such parameters will survive once the scaling with $M_*$ is removed. Answering to this question constitutes a (trivial) first step towards an analysis of the mutual correlations between galaxy parameters. However, we stress that the identification of the most fundamental parameters driving galaxy scaling relations goes beyond the scope of this data release paper, and we refer to previous \citep{Bothwell+16b, Bothwell+16a} and future works by our team for a more rigorous analysis of the problem.

The best-fit to the $L^{\prime}_{\rm CO(1-0)}$ vs $M_*$ relation is a power law with a slightly non-linear slope of $1.10\pm0.05$ (Figure~\ref{fig:lco_plots_1}). Hence, the quantity $\log L^{\prime}_{\rm CO(1-0)} - 1.1\log M_*$ is independent of $M_*$. We can then test the presence of a residual dependence of this quantity on $A_V$, O/H and $M_{\rm HI}$. We note that, by doing so, we are assuming a priori that $M_*$ is the primary factor in determining the variations of $L^{\prime}_{\rm CO(1-0)}$, and that the other dependencies are secondary effects. A similar approach was followed by \cite{Leroy+05} to investigate residual variations in the normalised CO content of galaxies on other properties than galaxy mass. 

\begin{figure}[tbp]
\centering
    \includegraphics[clip=true,trim=9cm 4cm 2.cm 3cm,width=0.3\textwidth,angle=180]{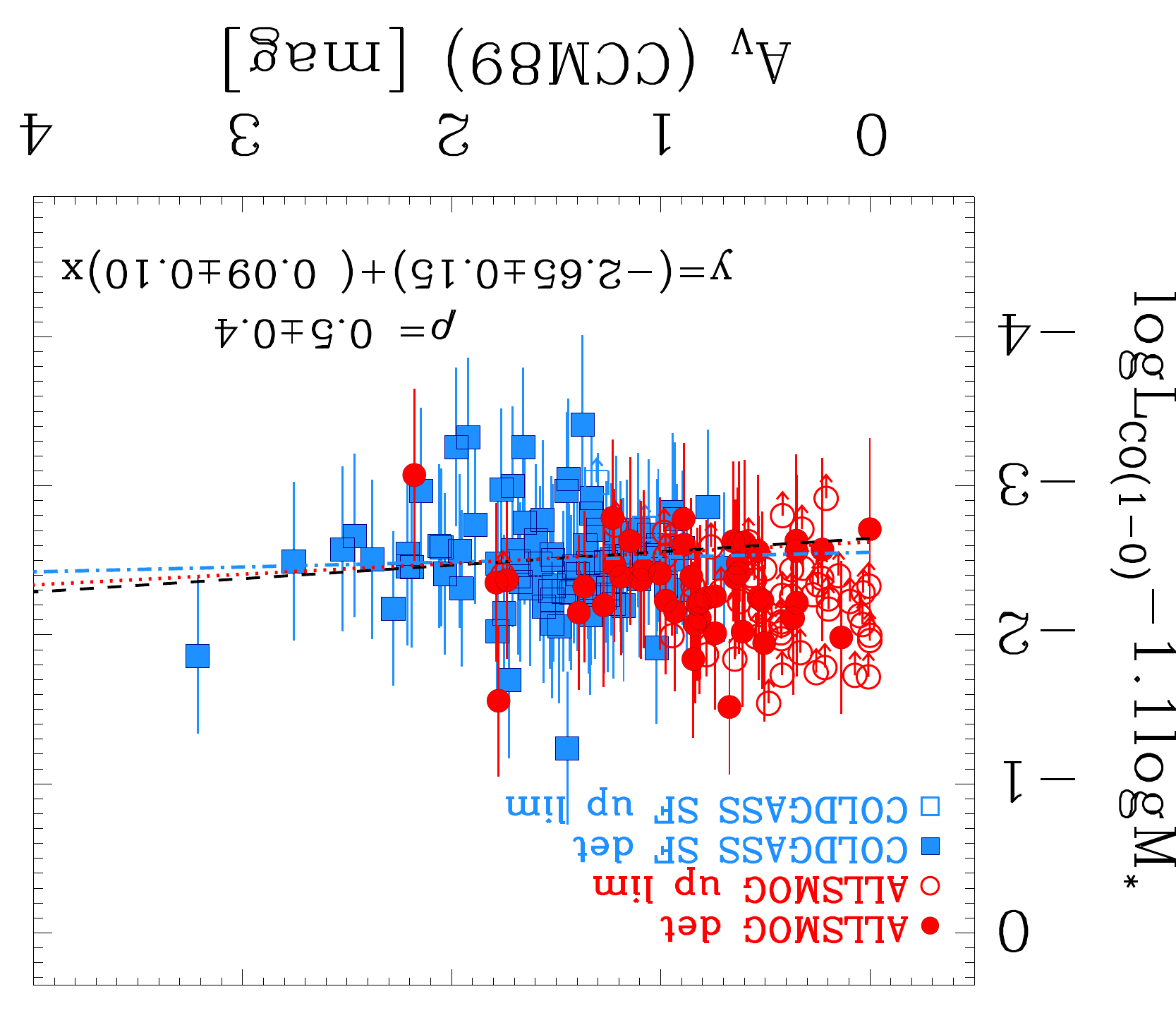}\\
     \includegraphics[clip=true,trim=9cm 4cm 2.cm 3cm,width=0.3\textwidth,angle=180]{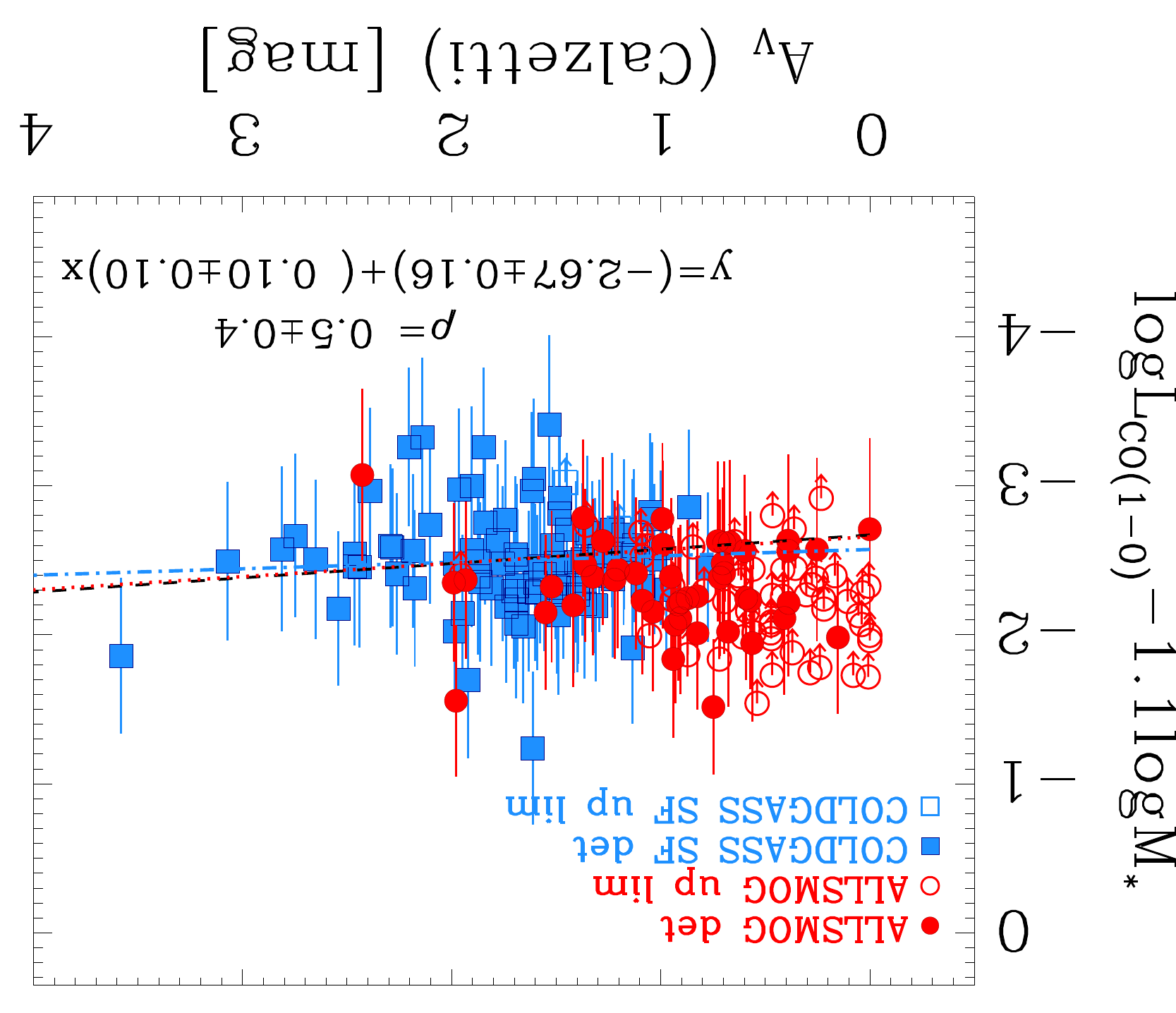}\\
      \vspace{0.2cm} 
     \caption{The $M_*$-independent quantity $\log L^{\prime}_{\rm CO(1-0)} - 1.1\log M_*$ as a function of nebular visual extinction for the sample of local star-forming galaxies defined by the ALLSMOG survey and by the sub-sample of COLD GASS selected in $\S$~\ref{sec:coldgass}. The {\it top} and {\it bottom} plots were produced by using the $A_V$ values obtained by employing the  \cite{CCM89}'s and \cite{Calzetti+00}'s attenuation curves, respectively. We refer to the caption of Fig.~\ref{fig:lco_plots_1} for an explanation of the symbols and lines shown in the plot. The results of the regression analysis are reported in Table~\ref{table:regression_summary_appendix}. As for the relations in Fig.~\ref{fig:lco_plots_3}, the fit was performed without accounting for the measurement error on $A_V$ (of the order of $\sim0.5~mag$).
       }
   \label{fig:plot_av_resfitlcomstar}
\end{figure}

\begin{figure*}[tbp]
\centering
    \includegraphics[clip=true,trim=9cm 4cm 2.cm 3cm,width=0.3\textwidth,angle=180]{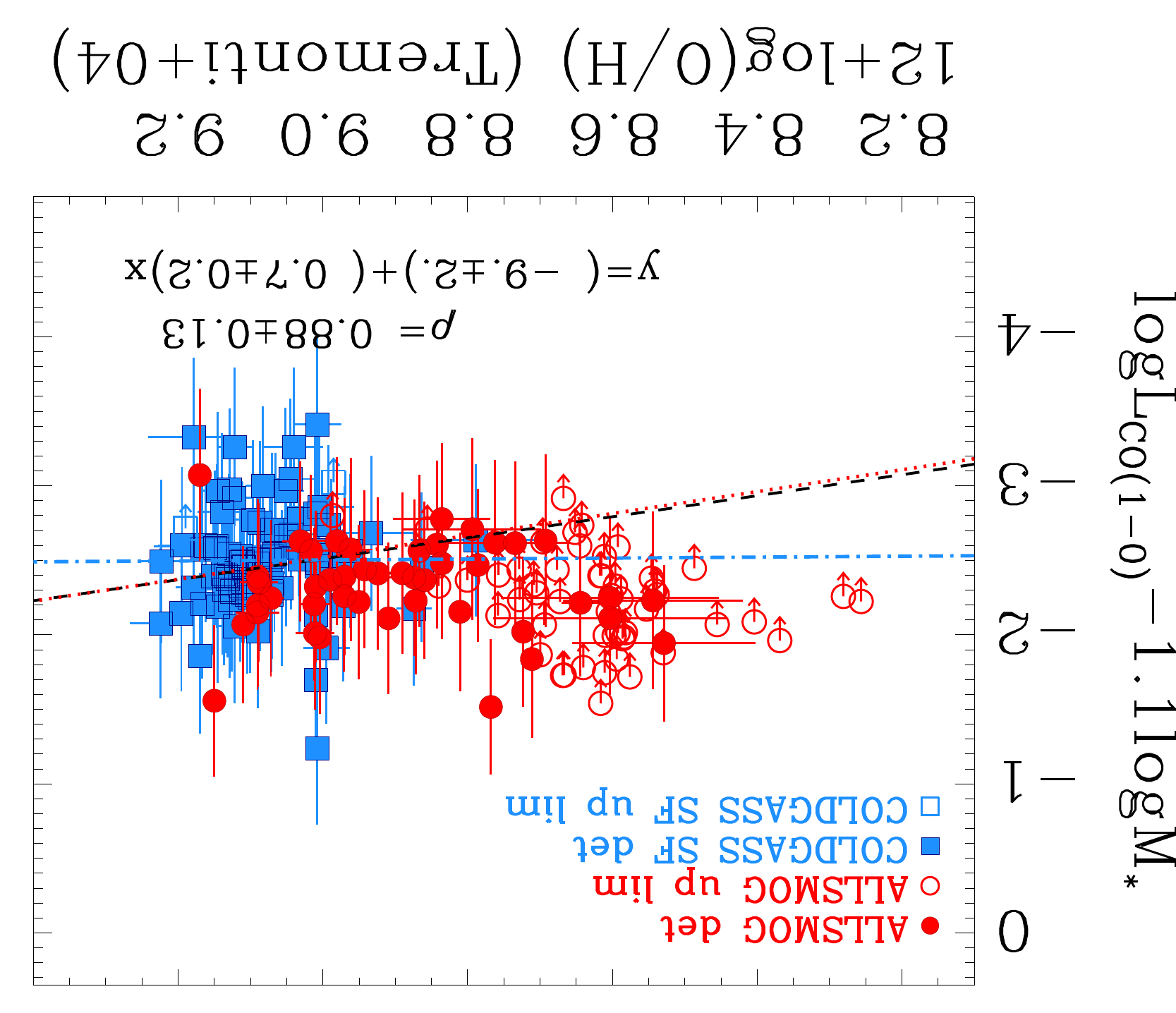}\quad
    \includegraphics[clip=true,trim=9cm 4cm 2.cm 3cm,width=0.3\textwidth,angle=180]{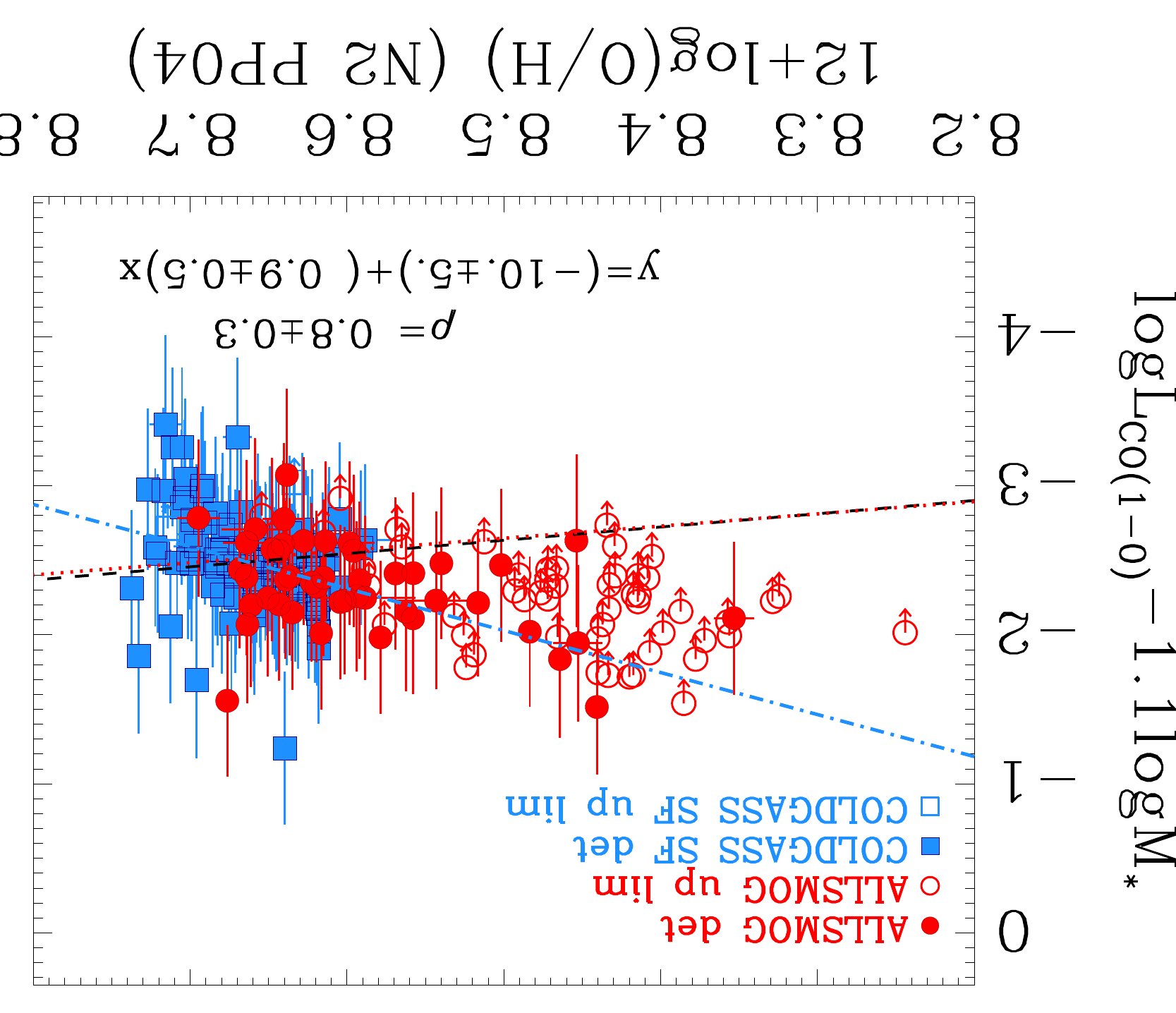}\quad
     \includegraphics[clip=true,trim=9cm 4cm 2.cm 3cm,width=0.3\textwidth,angle=180]{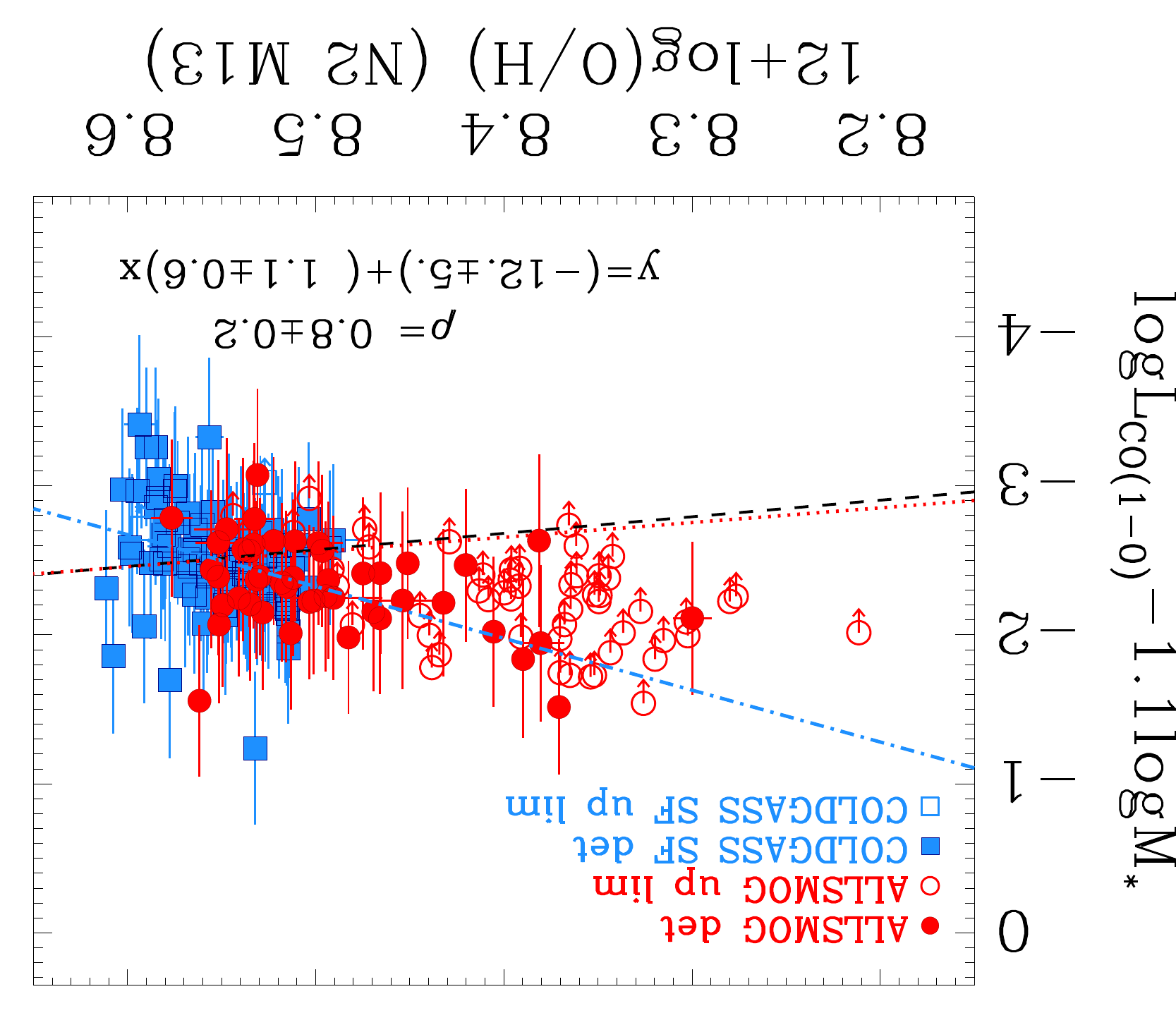}\\
        \includegraphics[clip=true,trim=9cm 4cm 2.cm 3cm,width=0.3\textwidth,angle=180]{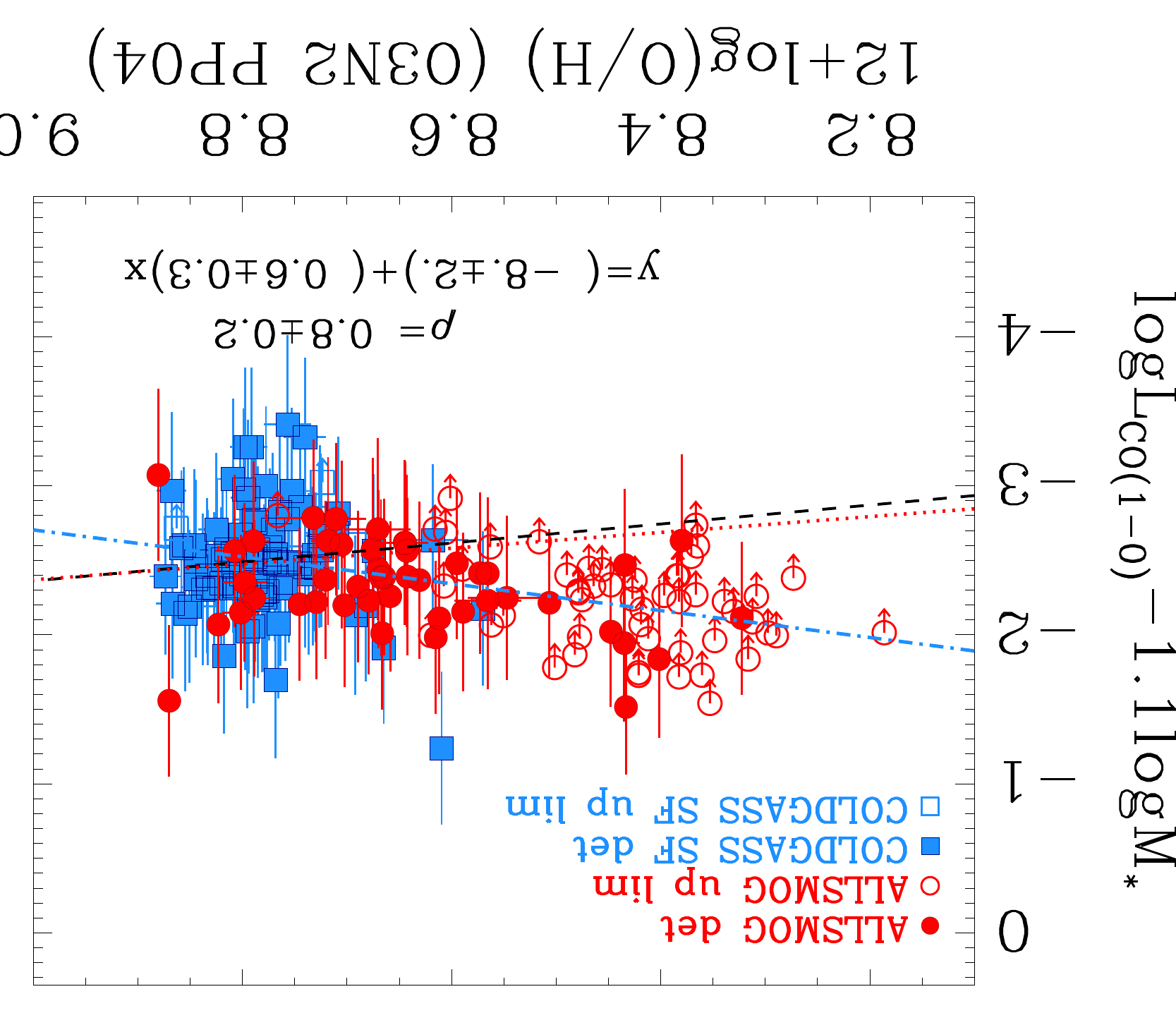}\quad
         \includegraphics[clip=true,trim=9cm 4cm 2.cm 3cm,width=0.3\textwidth,angle=180]{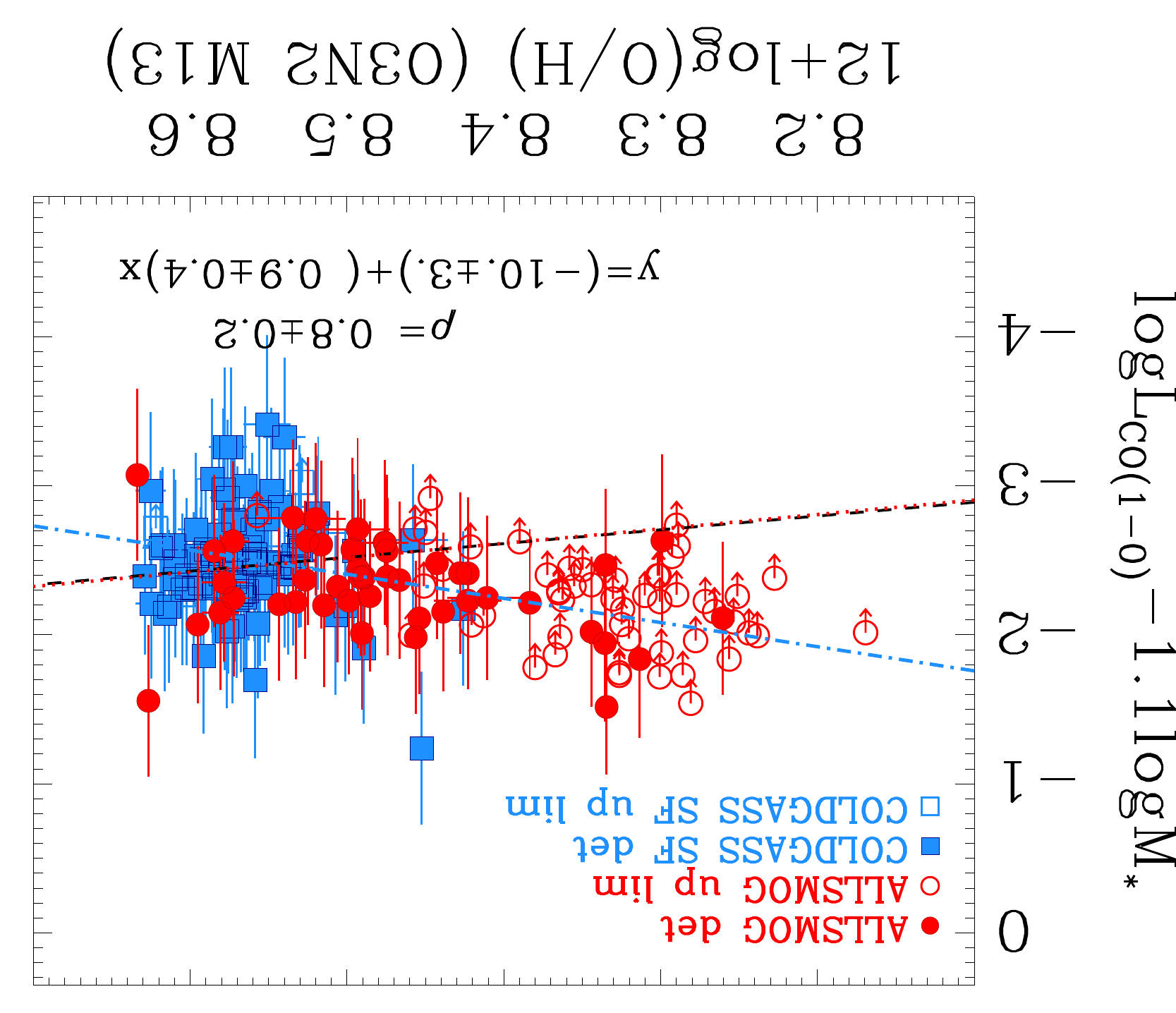}\\
      \vspace{0.2cm} 
     \caption{The M$_*$-independent quantity $\log L^{\prime}_{\rm CO(1-0)} - 1.1\log M_*$ as a function of Oxygen abundance for the sample of local star-forming galaxies defined by the ALLSMOG survey and by the sub-sample of COLD GASS selected in $\S$~\ref{sec:coldgass}. Each panel corresponds to a different strong-line metallicity calibration method (see $\S$~\ref{sec:metallicity}). We refer to the caption of Fig.~\ref{fig:lco_plots_1} for an explanation of the symbols and lines shown in the plot. The results of the regression analysis are reported in Table~\ref{table:regression_summary_appendix}.
       }
   \label{fig:plot_met_resfitlcomstar}
\end{figure*}

\begin{figure}[tbp]
\centering
    \includegraphics[clip=true,trim=9cm 4cm 2.cm 3cm,width=0.3\textwidth,angle=180]{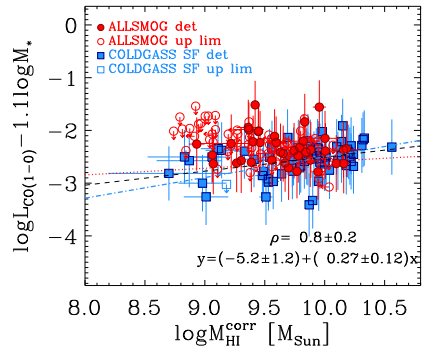}\\
      \vspace{0.2cm} 
     \caption{The M$_*$-independent quantity $\log L^{\prime}_{\rm CO(1-0)} - 1.1\log M_*$ as a function of H{\sc i} gas mass for the sample of galaxies shown in Fig.~\ref{fig:lco_plots_4}. As explained in the caption of Fig.~\ref{fig:lco_plots_4}, the COLD GASS galaxies with a non detection in H{\sc i} are not shown in this plot, because upper limits on $M_{\rm HI}$ are not available for these sources. For an explanation of the symbols and lines drawn on the plot, we refer to the caption of Fig.~\ref{fig:lco_plots_1}. The results of the regression analysis are reported in Table~\ref{table:regression_summary_appendix}. As for the relation shown in Fig.~\ref{fig:lco_plots_4}, the analysis was performed by including only the detections in H{\sc i}.
       }
   \label{fig:plot_mHI_resfitlcomstar}
\end{figure}

Figures~\ref{fig:plot_av_resfitlcomstar}, \ref{fig:plot_met_resfitlcomstar} and \ref{fig:plot_mHI_resfitlcomstar} show the quantity $\log L^{\prime}_{\rm CO(1-0)} - 1.1\log M_*$ as a function of $A_V$, gas-phase metallicity and $M_{\rm HI}$, respectively. The results of the Bayesian regression analysis performed on these relations, reported in Table~\ref{table:regression_summary_appendix}, are briefly discussed also in the main text ($\S$~\ref{sec:lco_av}, \ref{sec:lco_met} and \ref{sec:lco_hi}). 
Figure~\ref{fig:plot_av_resfitlcomstar} does not show a significant residual dependence on $A_V$, suggesting that the strong correlation between $L^{\prime}_{\rm CO(1-0)}$ and $A_V$ observed in Fig.~\ref{fig:lco_plots_3} is completely driven by the collinearity between $A_V$ and $M_*$. However, as noted in Sec~\ref{sec:lco_av}, after considering selection effects we expect the real underlying relation between $L^{\prime}_{\rm CO(1-0)}$ and $A_V$ to be significantly steeper than the one observed. Therefore, based on our data, we cannot fully rule out the existence of a residual secondary dependence of $L^{\prime}_{\rm CO(1-0)}$ on $A_V$. In fact, extinction and metallicity are strictly related in star-forming galaxies \citep{Garn+Best10}, and a residual positive trend of the quantity $\log L^{\prime}_{\rm CO(1-0)} - 1.1\log M_*$ as a function of metallicity is clearly suggested by the results of the Bayesian regression analysis performed on the plots shown in Fig.~\ref{fig:plot_met_resfitlcomstar}, with $\rho\sim0.8$ and slopes, $\beta\sim0.6-1.1$.
We suggest that the positive correlation with Oxygen abundance traces the increased rate of CO photodissociation - resulting in a reduced CO luminosity per unit molecular gas mass - which is expected for low metallicity (and low extinction) environments \citep{Wolfire+10, Glover+MacLow11}.

Figure~\ref{fig:plot_mHI_resfitlcomstar} shows a residual positive trend also as a function of  $M_{\rm HI}$ ($\rho\sim0.8$), although with a shallow slope ($\beta=0.27\pm0.12$), implying that the scaling of $M_{\rm HI}$ with $M_*$ cannot entirely account for the variations of $L^{\prime}_{\rm CO(1-0)}$ observed as a function of $M_{\rm HI}$ (Figure~\ref{fig:lco_plots_4}). The positive trend in Fig.~\ref{fig:plot_mHI_resfitlcomstar} may be due to higher molecular gas  fractions in galaxies with larger H{\sc i} gas reservoirs.

\begin{table*}
\caption{Analysis of the dependencies of the quantity $\log L^{\prime}_{\rm CO(1-0)} - 1.1\log M_*$ on $A_V$, O/H and $M_{\rm HI}$} 
\label{table:regression_summary_appendix}
\centering
\begin{tabular}{lcccc}
\hline\hline
\multicolumn{5}{c}{Model: $y = \alpha + \beta x$, with $y = \log L^{\prime}_{\rm CO(1-0)} - 1.1\log M_*$} \\
\hline
\multicolumn{5}{c}{Total sample:} \\
$x$ & $\alpha$ & $\beta$ & $\sigma_{intr}$ & $\rho$  \\
\hline
$A_V~[mag]$ $^{\dag}$ (CCM89)  & $-2.65\pm0.15$ & $0.09\pm0.10$ & $0.10\pm0.10$ & $0.5\pm0.4$ \\  
$A_V~[mag]$ $^{\dag}$ (Calzetti)  & $-2.67\pm0.16$ & $0.10\pm0.10$ & $0.09\pm0.10$ & $0.5\pm0.4$ \\ 
$12 + \log (O/H)$ (Tremonti+04) & $-9\pm2$ & $0.7\pm0.2$ & $0.07\pm0.10$ & $0.88\pm0.13$ \\ 
$12 + \log (O/H)$ N2 PP04 & $-10\pm5$ & $0.9\pm0.5$ & $0.06\pm0.10$ & $0.8\pm0.3$ \\ 
$12 + \log (O/H)$ N2 M13 & $-12\pm5$ & $1.1\pm0.6$ & $0.08\pm0.09$ & $0.8\pm0.2$ \\ 
$12 + \log (O/H)$ O3N2 PP04 & $-8\pm2$ & $0.6\pm0.3$ & $0.08\pm0.09$ & $0.8\pm0.2$ \\ 
$12 + \log (O/H)$ O3N2 M13 & $-10\pm3$ & $0.9\pm0.4$ & $0.07\pm0.10$ & $0.8\pm0.2$ \\ 
$\log M_{\rm HI}^{corr}$ $^{\ddag}$ & $-5.2\pm1.2$ & $0.27\pm0.12$ & $0.09\pm0.07$ & $0.8\pm0.2$ \\  
\hline
\multicolumn{5}{c}{Low-$M_*$ sample ($\log M_* [M_{\odot}] < 10.0$):} \\
$x$ & $\alpha$ & $\beta$ & $\sigma_{intr}$ & $\rho$  \\
\hline
$A_V~[mag]$ $^{\dag}$ (CCM89)  & $-2.62\pm0.17$ & $0.07\pm0.11$ & $0.10\pm0.10$ & $0.4\pm0.5$ \\ 
$A_V~[mag]$ $^{\dag}$ (Calzetti)  & $-2.66\pm0.15$ & $0.09\pm0.09$ & $0.10\pm0.10$ & $0.5\pm0.4$ \\ 
$12 + \log (O/H)$ (Tremonti+04) & $-9\pm3$ & $0.7\pm0.3$ & $0.07\pm0.09$ & $0.90\pm0.11$ \\ 
$12 + \log (O/H)$ N2 PP04 & $-10\pm4$ & $0.8\pm0.4$ & $0.07\pm0.04$ & $0.8\pm0.2$ \\ 
$12 + \log (O/H)$ N2 M13 & $-11\pm5$ & $1.0\pm0.6$ & $0.06\pm0.05$ & $0.8\pm0.2$ \\ 
$12 + \log (O/H)$ O3N2 PP04 & $-7\pm3$ & $0.5\pm0.3$ & $0.06\pm0.04$ & $0.81\pm0.19$ \\ 
$12 + \log (O/H)$ O3N2 M13 & $-11\pm3$ & $1.0\pm0.4$ & $0.07\pm0.05$ & $0.83\pm0.19$ \\ 
$\log M_{\rm HI}^{corr}$ $^{\ddag}$ & $-4\pm2$ & $0.1\pm0.2$ & $0.16\pm0.10$ & $0.3\pm0.4$ \\ 
\hline
\multicolumn{5}{c}{High-$M_*$ sample ($\log M_* [M_{\odot}] \geq 10.0$):} \\
$x$ & $\alpha$ & $\beta$ & $\sigma_{intr}$ & $\rho$  \\
\hline
$A_V~[mag]$ $^{\dag}$ (CCM89)  & $-2.55\pm0.19$ & $0.03\pm0.12$ & $0.08\pm0.10$ & $0.2\pm0.5$ \\ 
$A_V~[mag]$ $^{\dag}$ (Calzetti)  & $-2.57\pm0.19$ & $0.04\pm0.11$ & $0.10\pm0.06$ & $0.2\pm0.5$ \\ 
$12 + \log (O/H)$ (Tremonti+04) & $-3\pm8$ & $0.0\pm0.9$ & $0.09\pm0.10$ & $0.0\pm0.5$ \\ 
$12 + \log (O/H)$ N2 PP04 & $22\pm17$ & $-3\pm2$ & $0.06\pm0.03$ & $-0.8\pm0.3$ \\  
$12 + \log (O/H)$ N2 M13 & $27\pm17$ & $-3\pm2$ & $0.09\pm0.05$ & $-0.7\pm0.2$ \\ 
$12 + \log (O/H)$ O3N2 PP04 & $5\pm9$ & $-0.9\pm1.0$ & $0.07\pm0.05$ & $-0.5\pm0.4$ \\ 
$12 + \log (O/H)$ O3N2 M13 & $11\pm14$ & $-1.6\pm1.7$ & $0.09\pm0.06$ & $-0.5\pm0.4$ \\ 
$\log M_{\rm HI}^{corr}$ $^{\ddag}$ & $-6.4\pm1.9$ & $0.4\pm0.2$ & $0.10\pm0.05$ & $0.8\pm0.2$ \\ 
\hline

\end{tabular}
\tablefoot{$(\alpha, \beta)$ are the best-fit linear regression coefficients, $\sigma_{intr}$ is the intrinsic scatter about the best-fit regression line, and $\rho$ is the correlation coefficient. We refer to $\S$~\ref{sec:lco_vs_prop} and to Table~\ref{table:regression_summary} for additional details on the regression analysis method. \\ 
$^{\dag}$ Because of the large measurement errors on $A_V$, of the order of $\sim0.5~mag$, the regression analysis for the $L^{\prime}_{\rm CO(1-0)}/M_*$ vs $A_V$ relationships was performed without taking into account the errors on $x$, in order not to bias the results. \\
$^{\ddag}$ In this case the regression analysis was performed by considering only the sources with an H{\sc i} line detection.}
\end{table*}

\section{CO and H{\sc i} spectra}\label{sec:appendix_spectra}

\begin{figure*}[tbp]
\centering
    \includegraphics[clip=true,trim=-0.4cm 0cm 0cm 0cm,width=0.18\textwidth,angle=90]{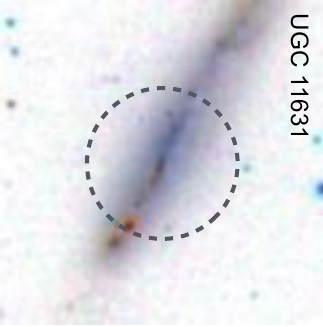}\quad
    \includegraphics[clip=true,trim=5.5cm 3.8cm 4cm 2cm,width=0.28\textwidth,angle=0]{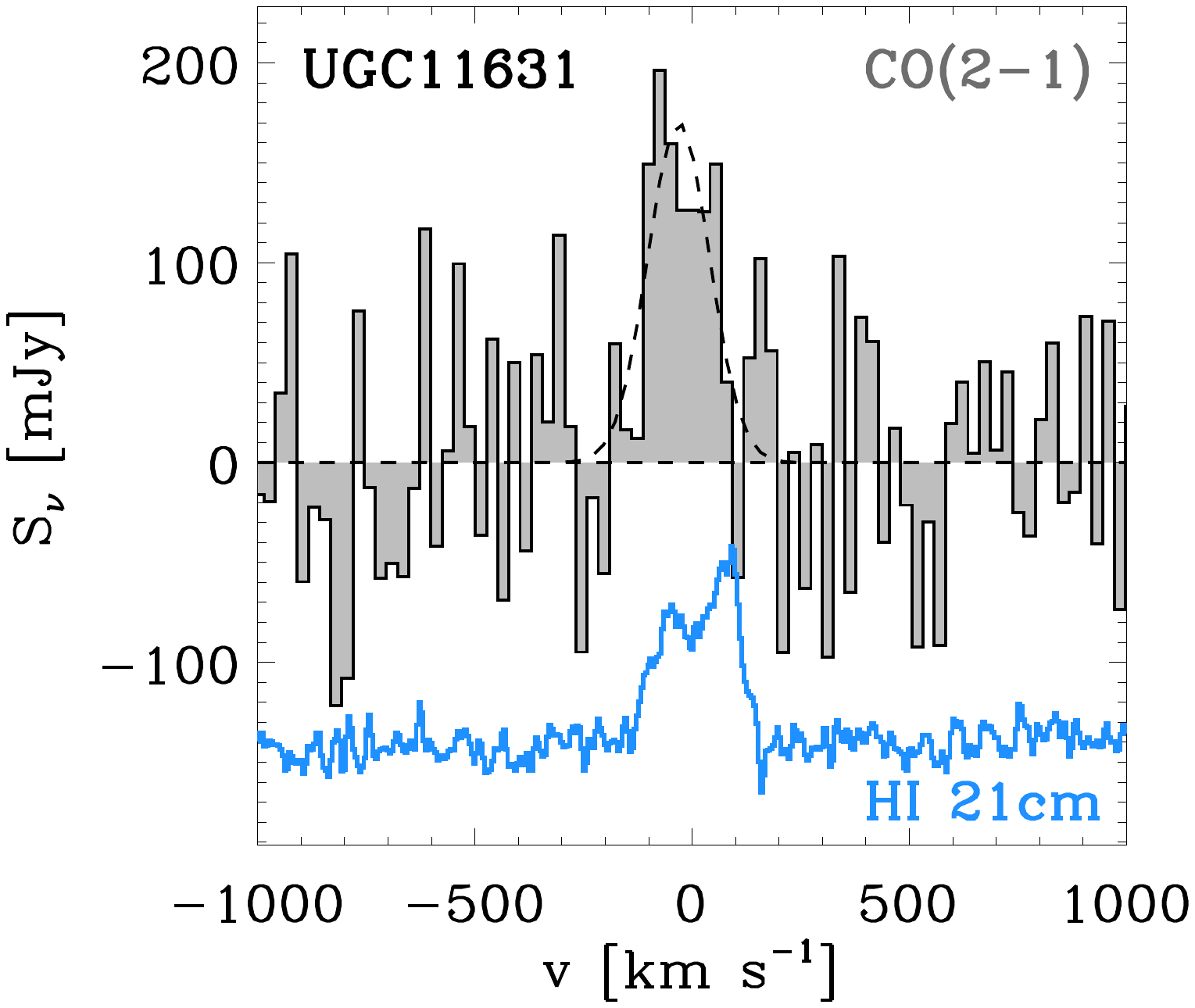}\quad 
    \includegraphics[clip=true,trim=-0.4cm 0cm 0cm 0cm,width=0.18\textwidth,angle=90]{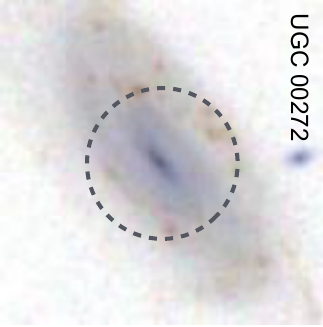}\quad
   \includegraphics[clip=true,trim=5.5cm 3.8cm 4cm 2cm,width=0.28\textwidth,angle=0]{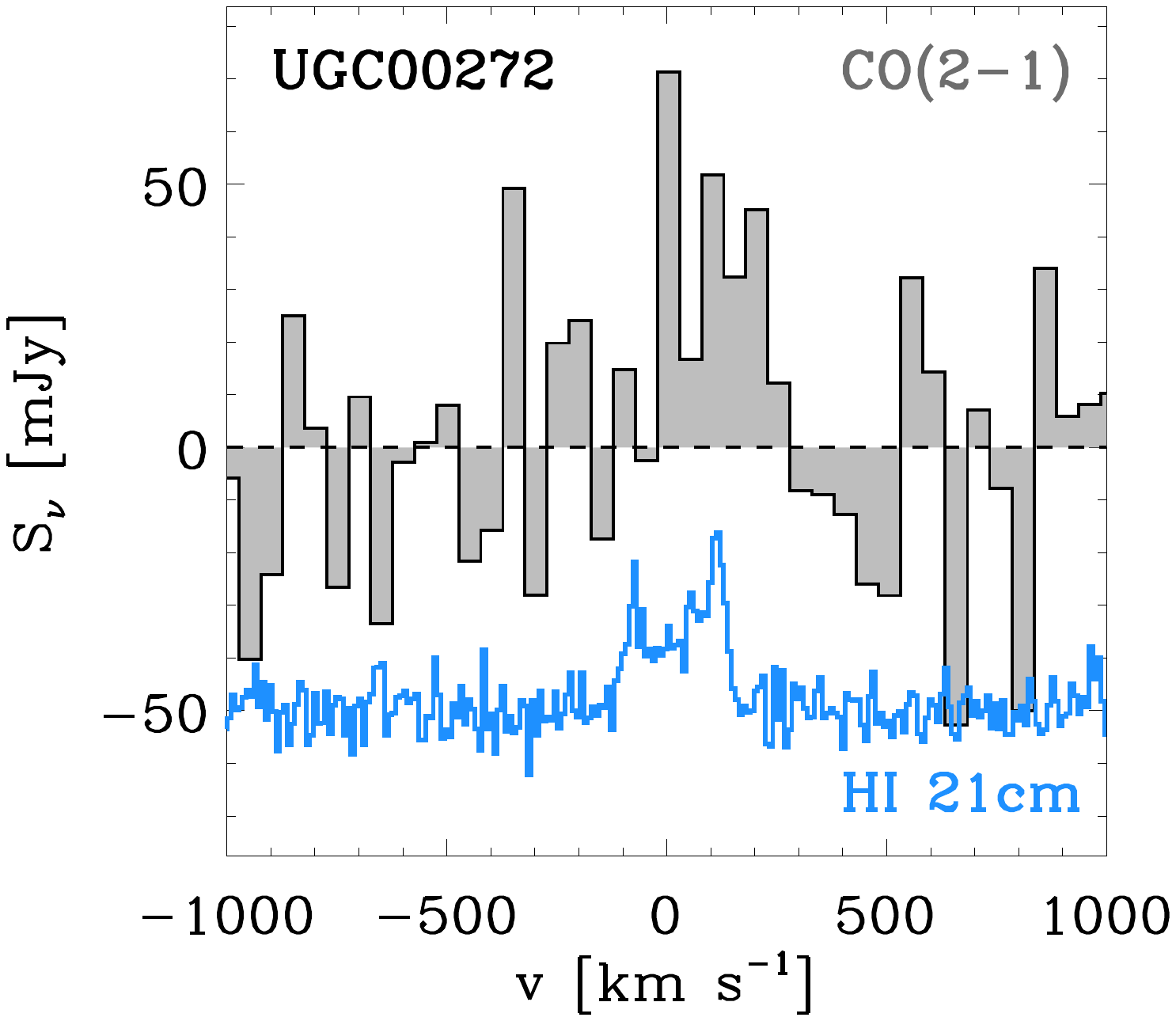}\\
    \includegraphics[clip=true,trim=-0.4cm 0cm 0cm 0cm,width=0.18\textwidth,angle=90]{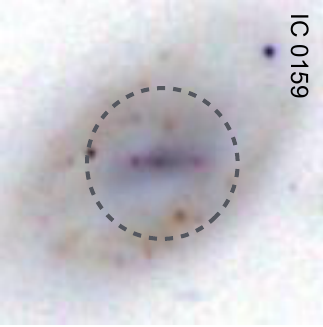}\quad
    \includegraphics[clip=true,trim=5.5cm 3.8cm 4cm 2cm,width=0.28\textwidth,angle=0]{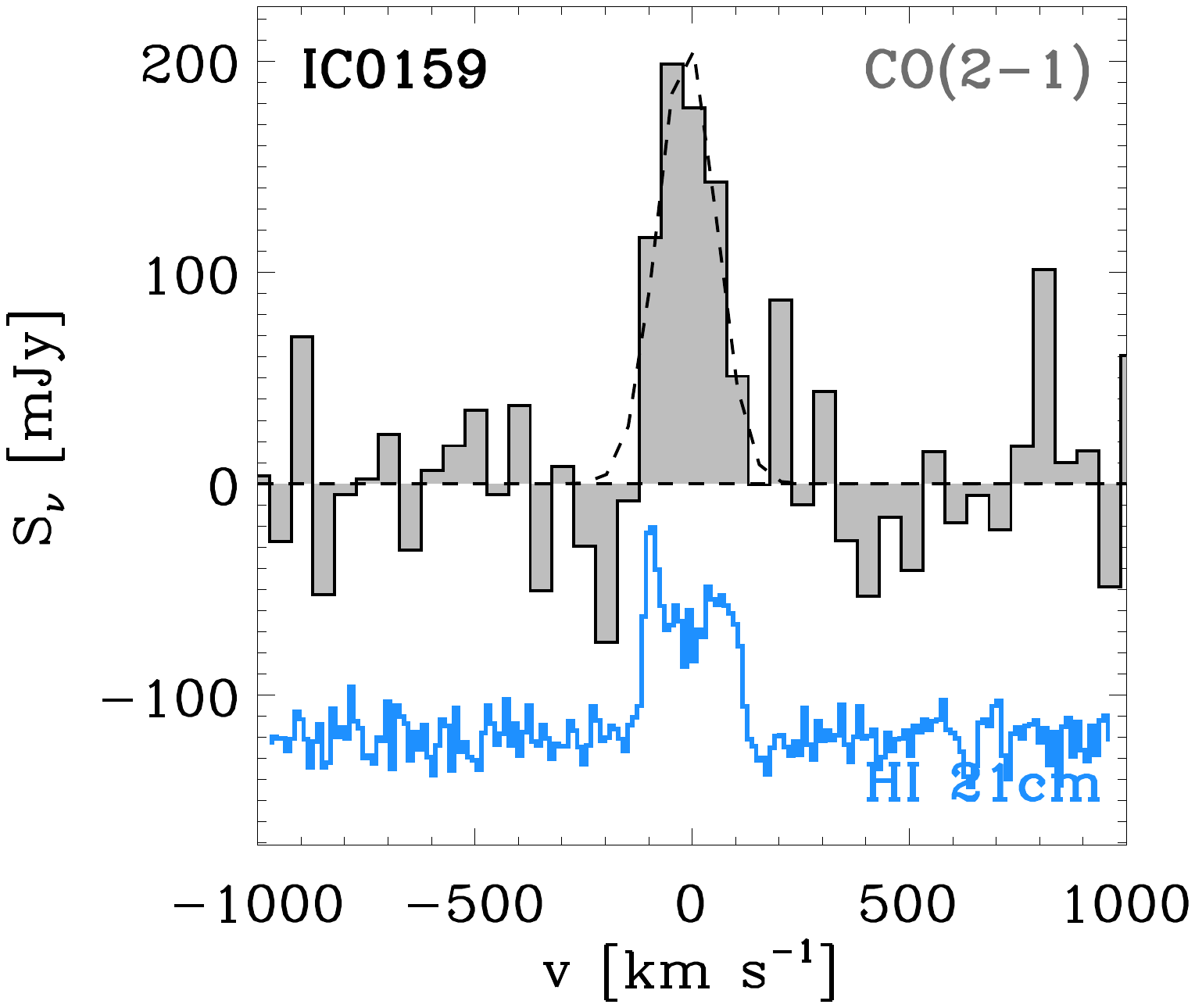}\quad
    \includegraphics[clip=true,trim=-0.4cm 0cm 0cm 0cm,width=0.18\textwidth,angle=90]{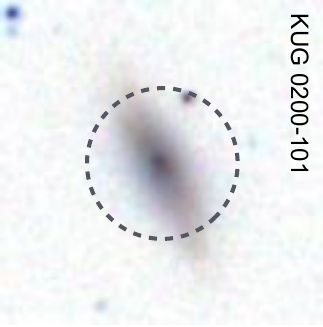}\quad
    \includegraphics[clip=true,trim=5.5cm 3.8cm 4cm 2cm,width=0.28\textwidth,angle=0]{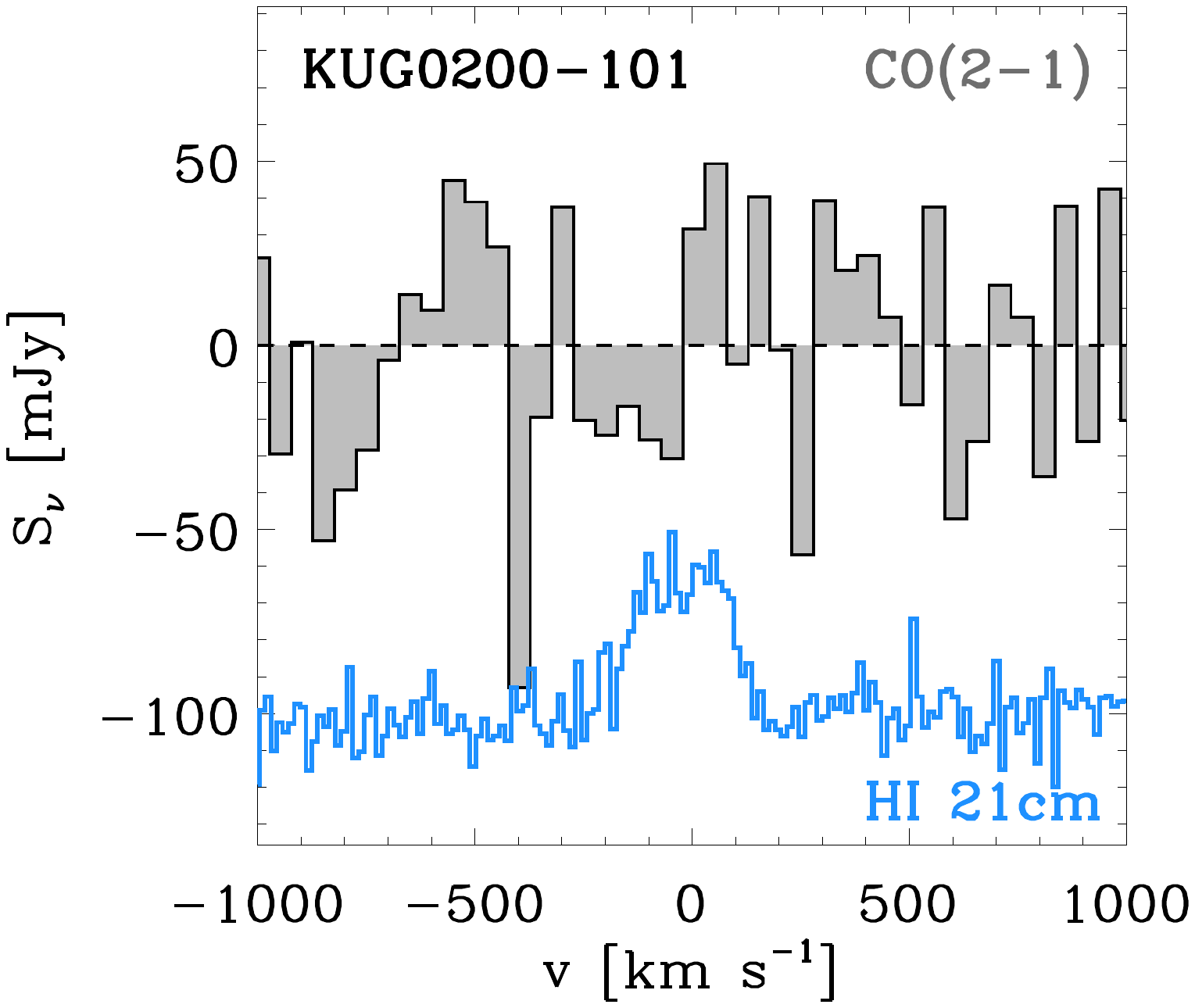}\\
      \includegraphics[clip=true,trim=-0.4cm 0cm 0cm 0cm,width=0.18\textwidth,angle=90]{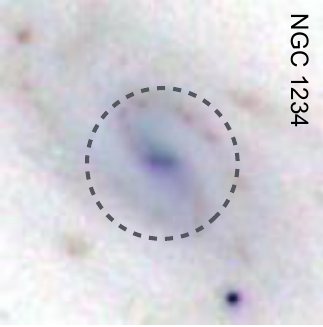}\quad
    \includegraphics[clip=true,trim=5.5cm 3.8cm 4cm 2cm,width=0.28\textwidth,angle=0]{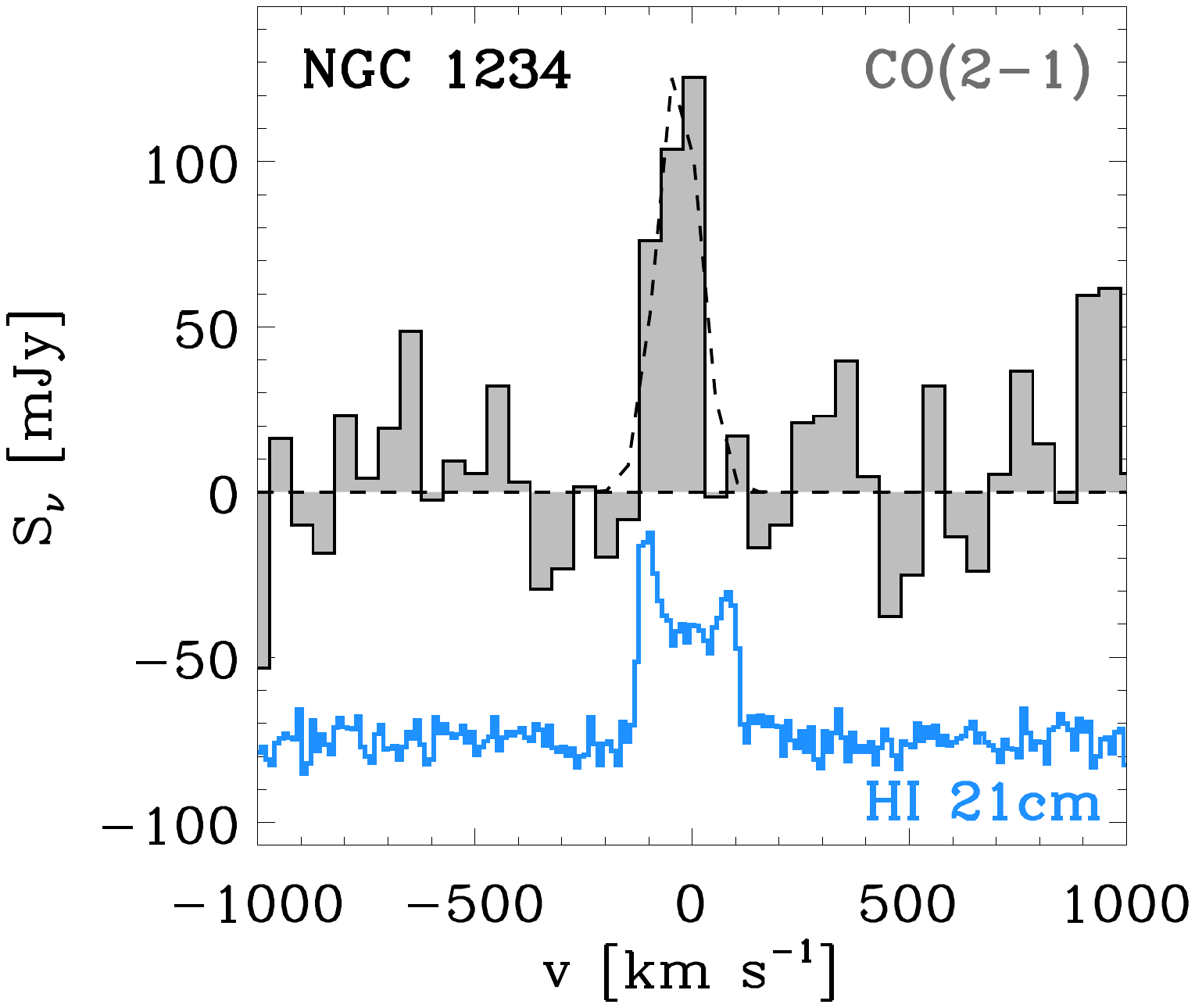}\quad
    \includegraphics[clip=true,trim=-0.4cm 0cm 0cm 0cm,width=0.18\textwidth,angle=90]{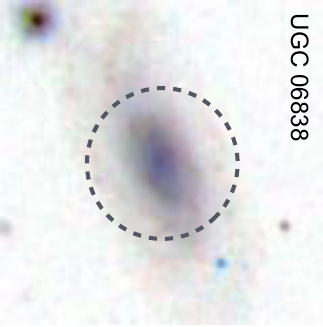}\quad
    \includegraphics[clip=true,trim=5.5cm 3.8cm 4cm 2cm,width=0.28\textwidth,angle=0]{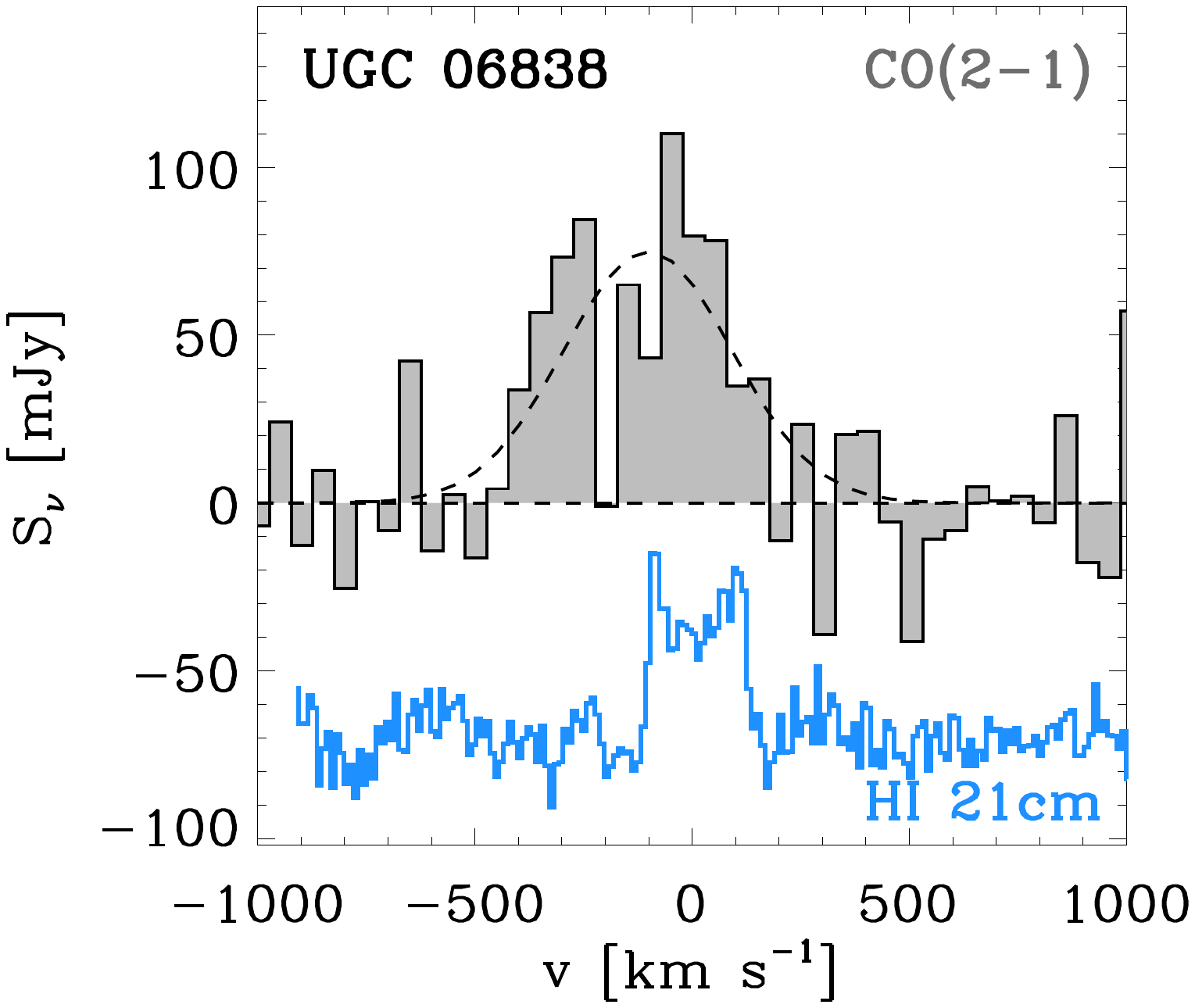}\\
      \includegraphics[clip=true,trim=-0.4cm 0cm 0cm 0cm,width=0.18\textwidth,angle=90]{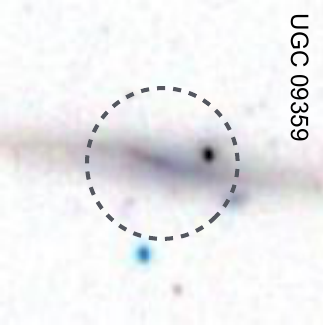}\quad
    \includegraphics[clip=true,trim=5.5cm 3.8cm 4cm 2cm,width=0.28\textwidth,angle=0]{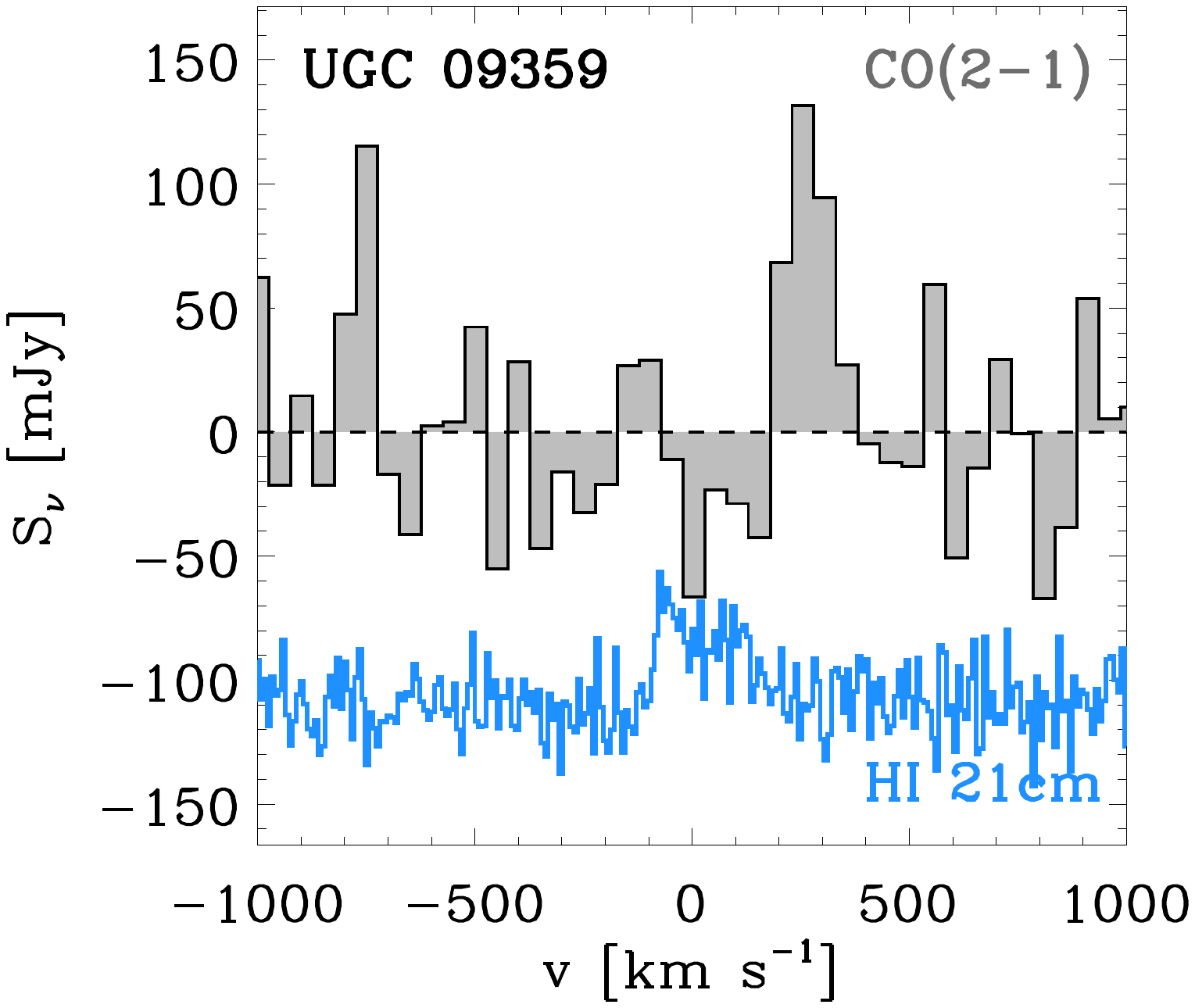}\quad
    \includegraphics[clip=true,trim=-0.4cm 0cm 0cm 0cm,width=0.18\textwidth,angle=90]{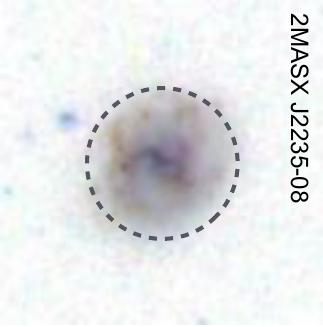}\quad
    \includegraphics[clip=true,trim=5.5cm 3.8cm 4cm 2cm,width=0.28\textwidth,angle=0]{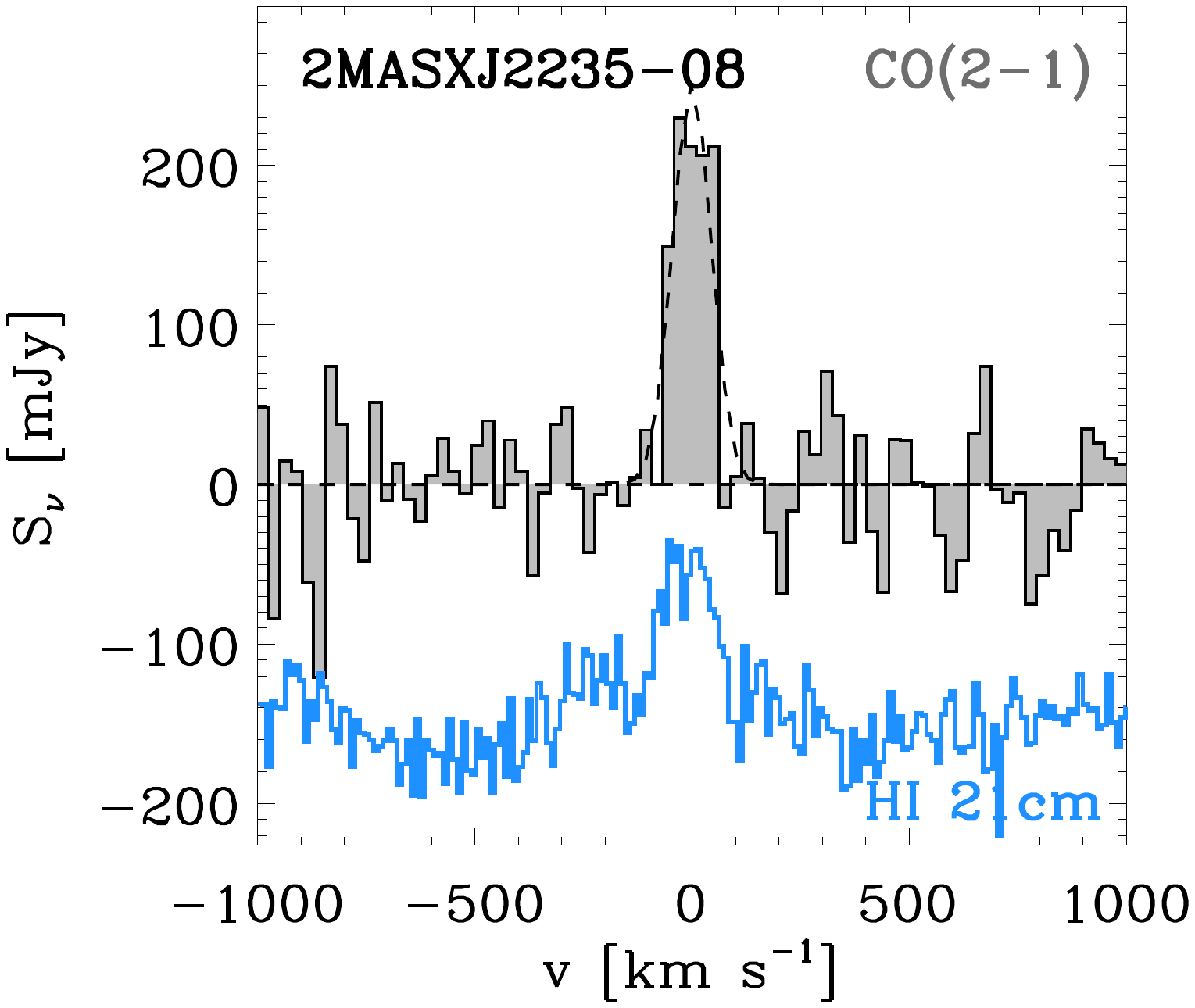}\\
      \includegraphics[clip=true,trim=-0.4cm 0cm 0cm 0cm,width=0.18\textwidth,angle=90]{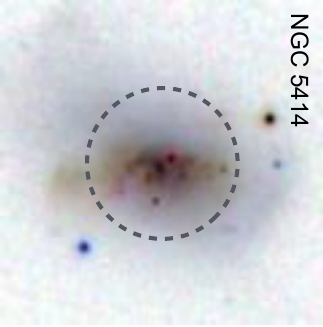}\quad
    \includegraphics[clip=true,trim=5.5cm 3.8cm 4cm 2cm,width=0.28\textwidth,angle=0]{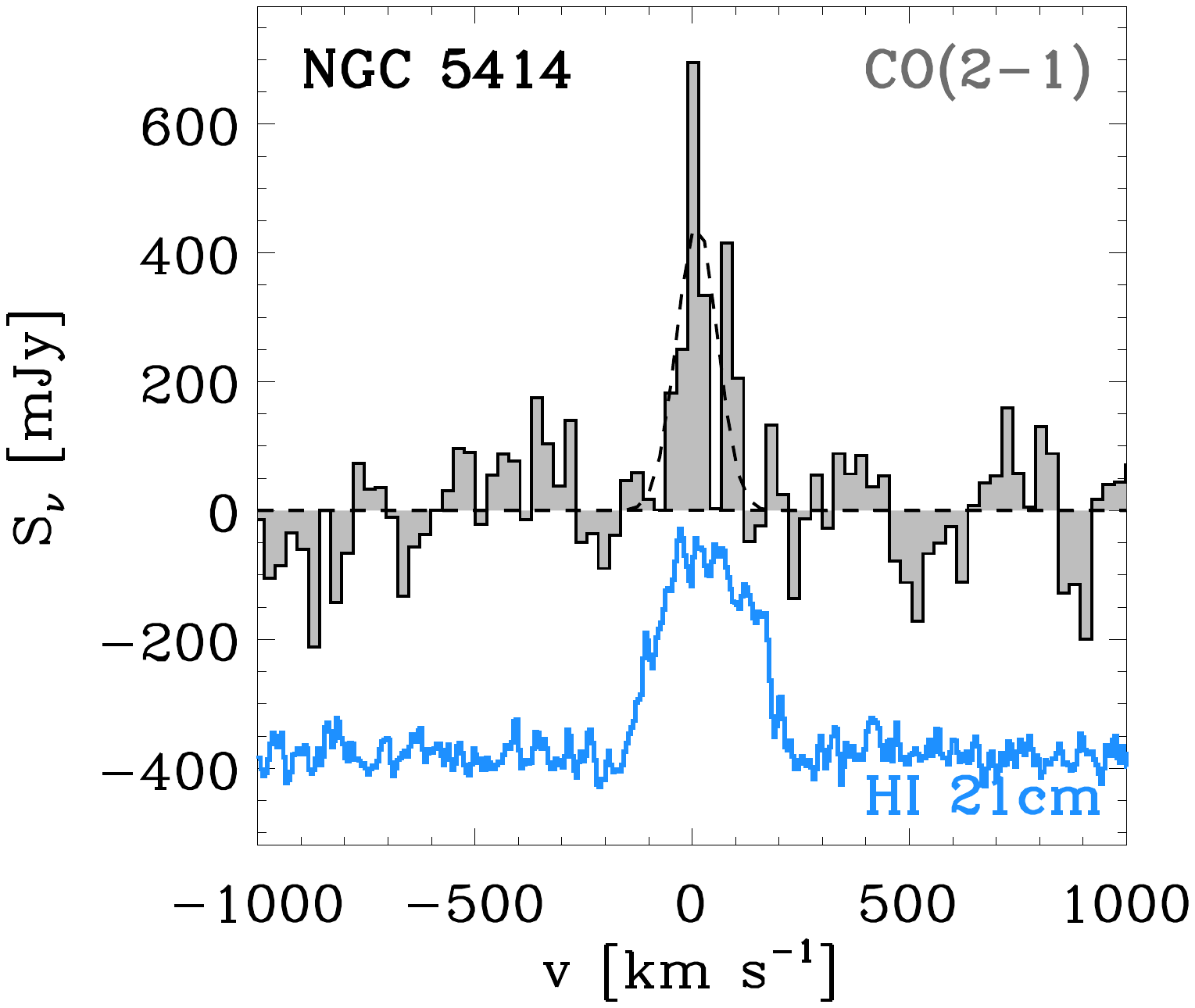}\quad
    \includegraphics[clip=true,trim=-0.4cm 0cm 0cm 0cm,width=0.18\textwidth,angle=90]{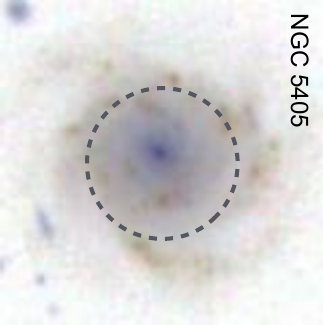}\quad
    \includegraphics[clip=true,trim=5.5cm 3.8cm 4cm 2cm,width=0.28\textwidth,angle=0]{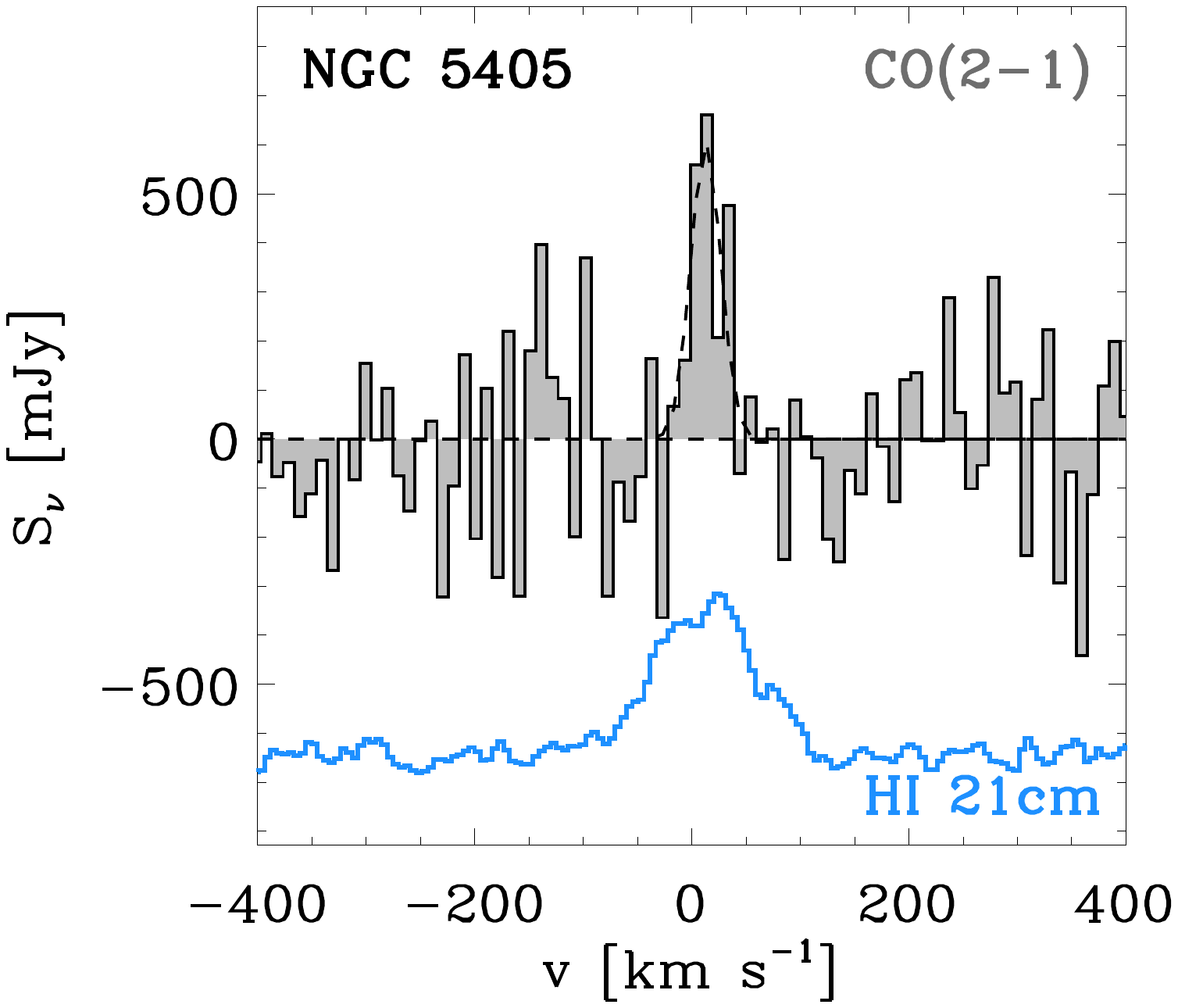}\\
    \caption{{\it Left panels:} SDSS cutout images ({\it g r i} composite, field of view = $60\arcsec\times60\arcsec$, scale = 0.5$\arcsec$/pixel, north is up and west is right) of ALLSMOG galaxies, showing the $27\arcsec$ APEX beam at 230 GHz. {\it Right panels:} APEX CO(2-1) baseline-subtracted spectra, rebinned in bins of $\delta \varv=50$~\kms (UGC00272, IC0159, KUG0200-101, NGC1234, UGC06838, UGC09359), 25~\kms (UGC11631, 2MASXJ2235-0845, NGC5414) or 10~\kms (NGC5405), depending on the width and S/N of the line. The corresponding H{\sc i}~21cm spectra are also shown for comparison, after having been renormalised for visualisation purposes (H{\sc i} references are given in Table~\ref{table:HI_parameters}). }
   \label{fig:spectra1}
\end{figure*}

\begin{figure*}[tbp]
\centering
    \includegraphics[clip=true,trim=-0.4cm 0cm 0cm 0cm,width=0.18\textwidth,angle=90]{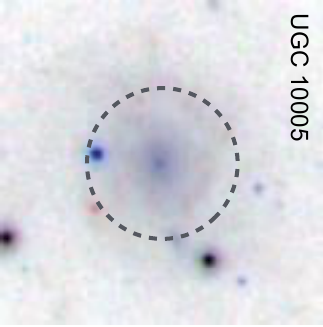}\quad
    \includegraphics[clip=true,trim=5.5cm 3.8cm 4cm 2cm,width=0.28\textwidth,angle=0]{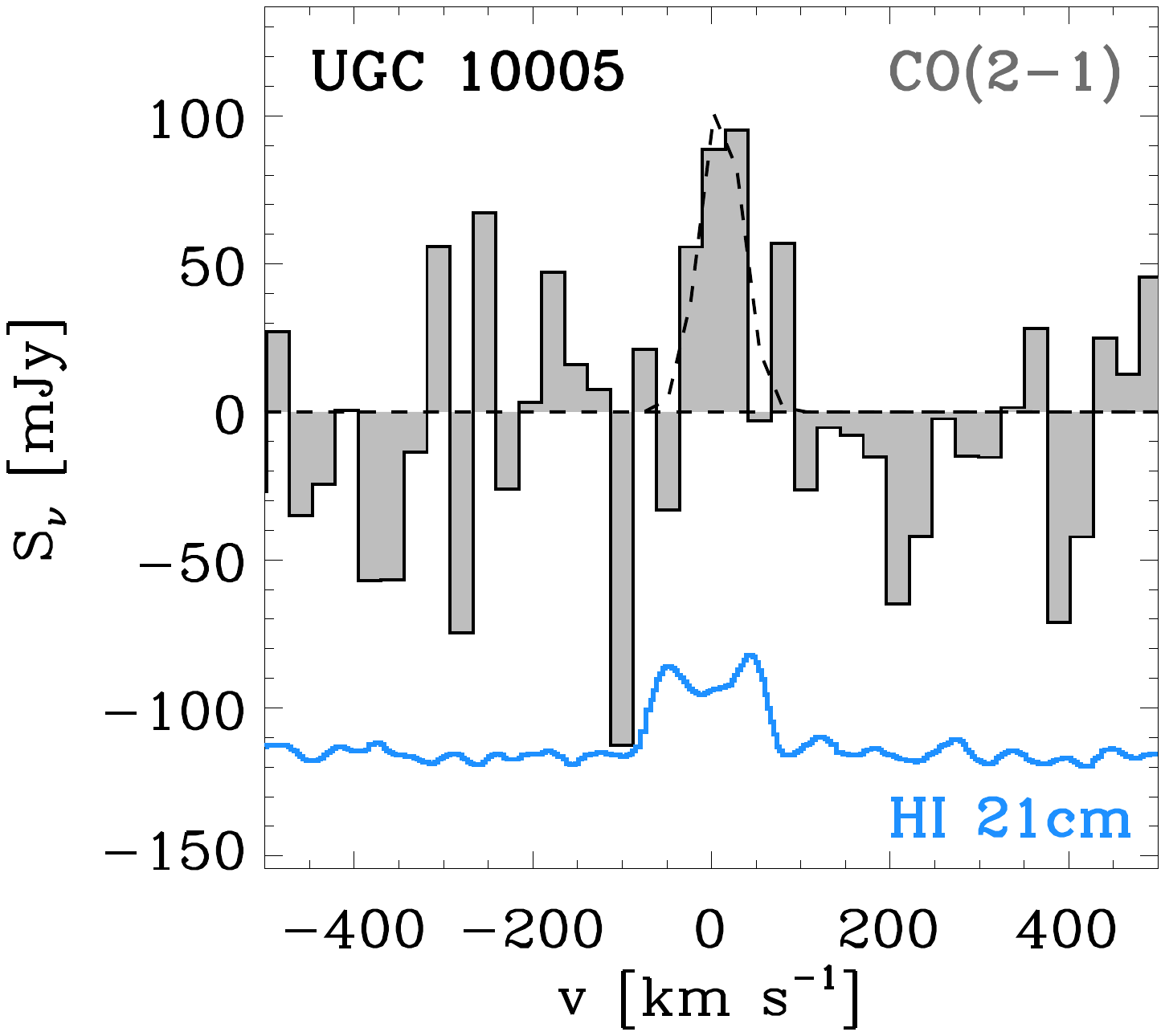}\quad
    \includegraphics[clip=true,trim=-0.4cm 0cm 0cm 0cm,width=0.18\textwidth,angle=90]{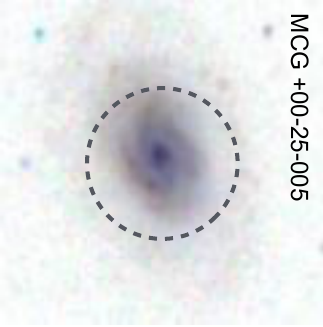}\quad
    \includegraphics[clip=true,trim=5.5cm 3.8cm 4cm 2cm,width=0.28\textwidth,angle=0]{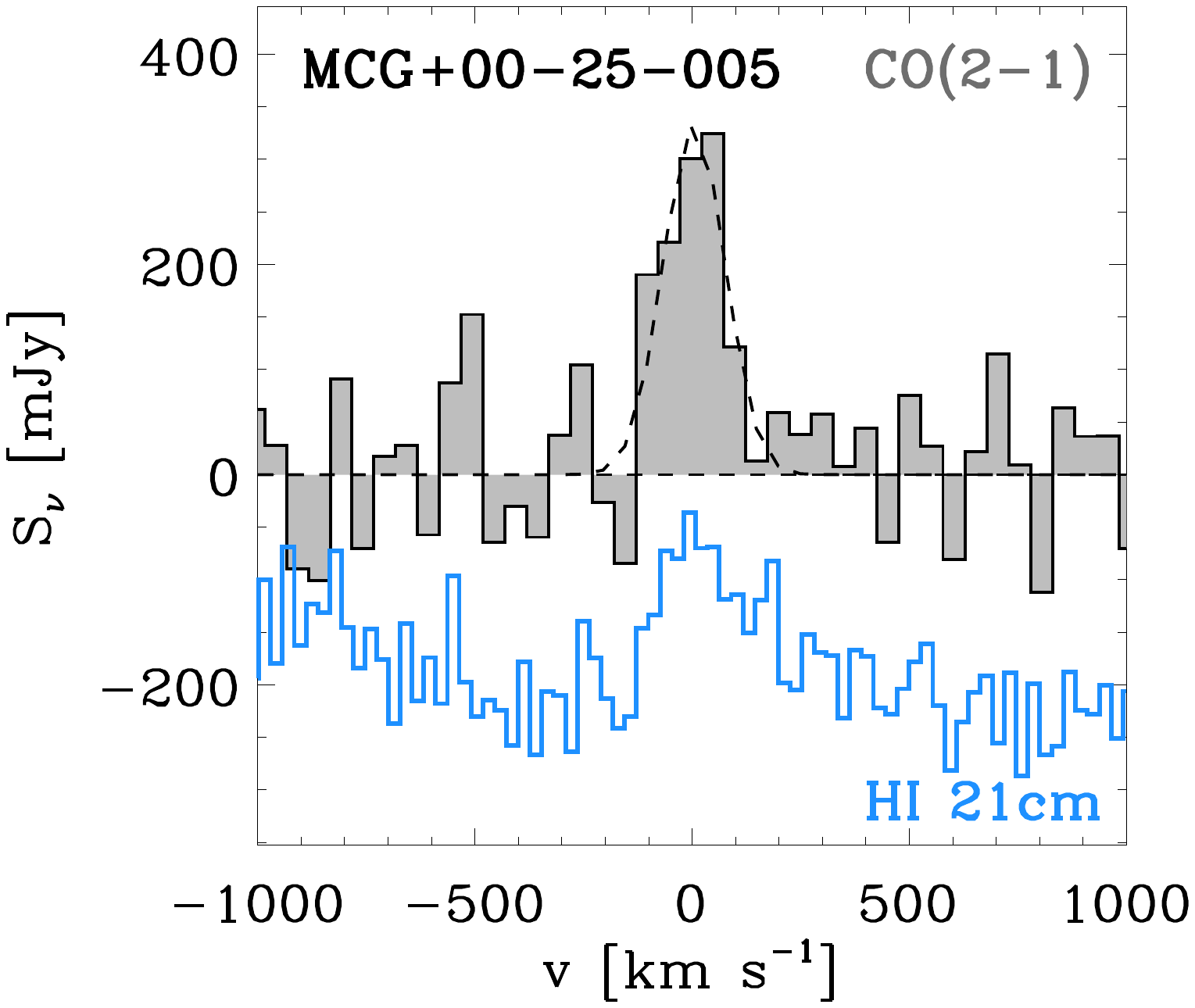}\\
    \includegraphics[clip=true,trim=-0.4cm 0cm 0cm 0cm,width=0.18\textwidth,angle=90]{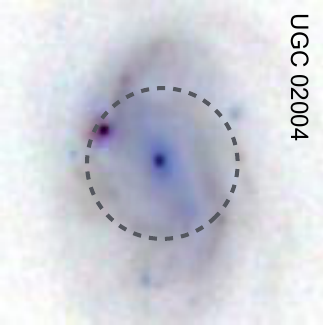}\quad
    \includegraphics[clip=true,trim=5.5cm 3.8cm 4cm 2cm,width=0.28\textwidth,angle=0]{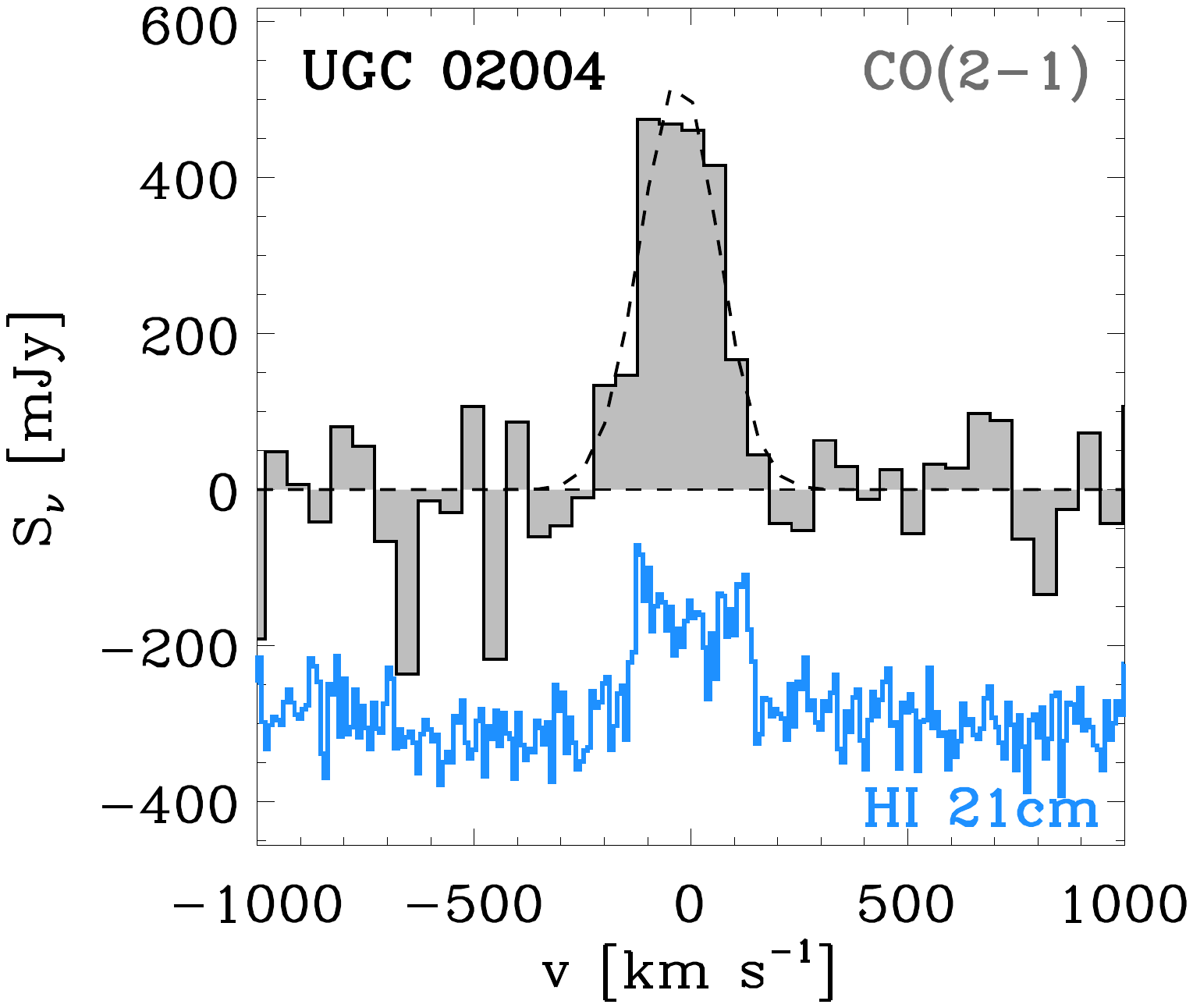}\quad
    \includegraphics[clip=true,trim=-0.4cm 0cm 0cm 0cm,width=0.18\textwidth,angle=90]{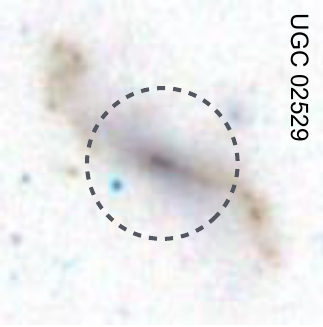}\quad
    \includegraphics[clip=true,trim=5.5cm 3.8cm 4cm 2cm,width=0.28\textwidth,angle=0]{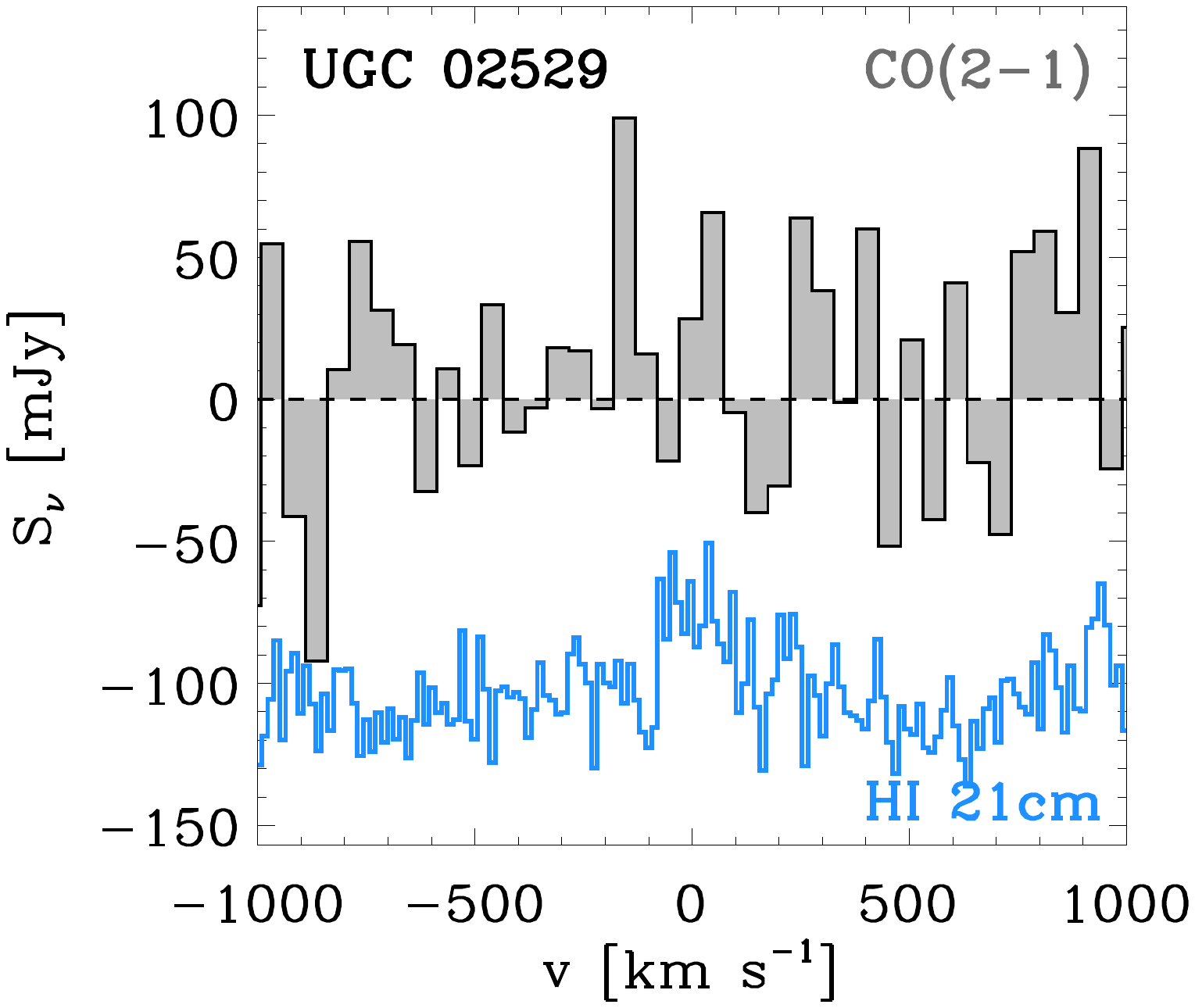}\\
      \includegraphics[clip=true,trim=-0.4cm 0cm 0cm 0cm,width=0.18\textwidth,angle=90]{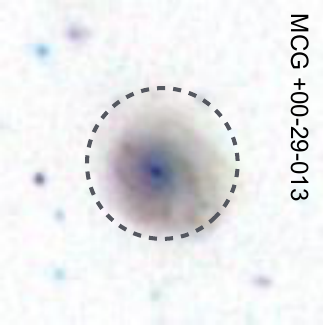}\quad
    \includegraphics[clip=true,trim=5.5cm 3.8cm 4cm 2cm,width=0.28\textwidth,angle=0]{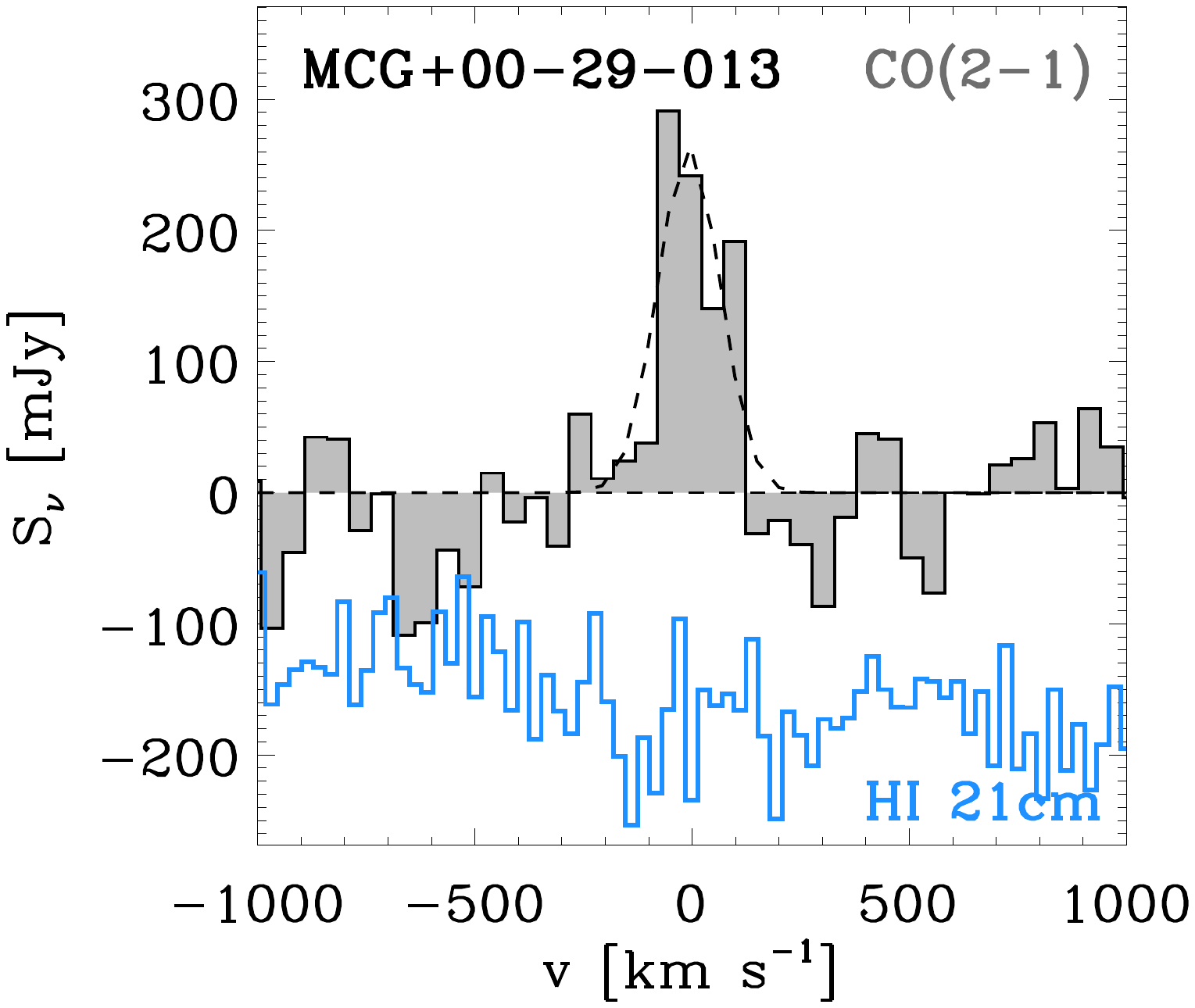}\quad
    \includegraphics[clip=true,trim=-0.4cm 0cm 0cm 0cm,width=0.18\textwidth,angle=90]{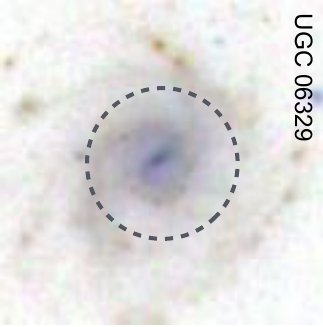}\quad
    \includegraphics[clip=true,trim=5.5cm 3.8cm 4cm 2cm,width=0.28\textwidth,angle=0]{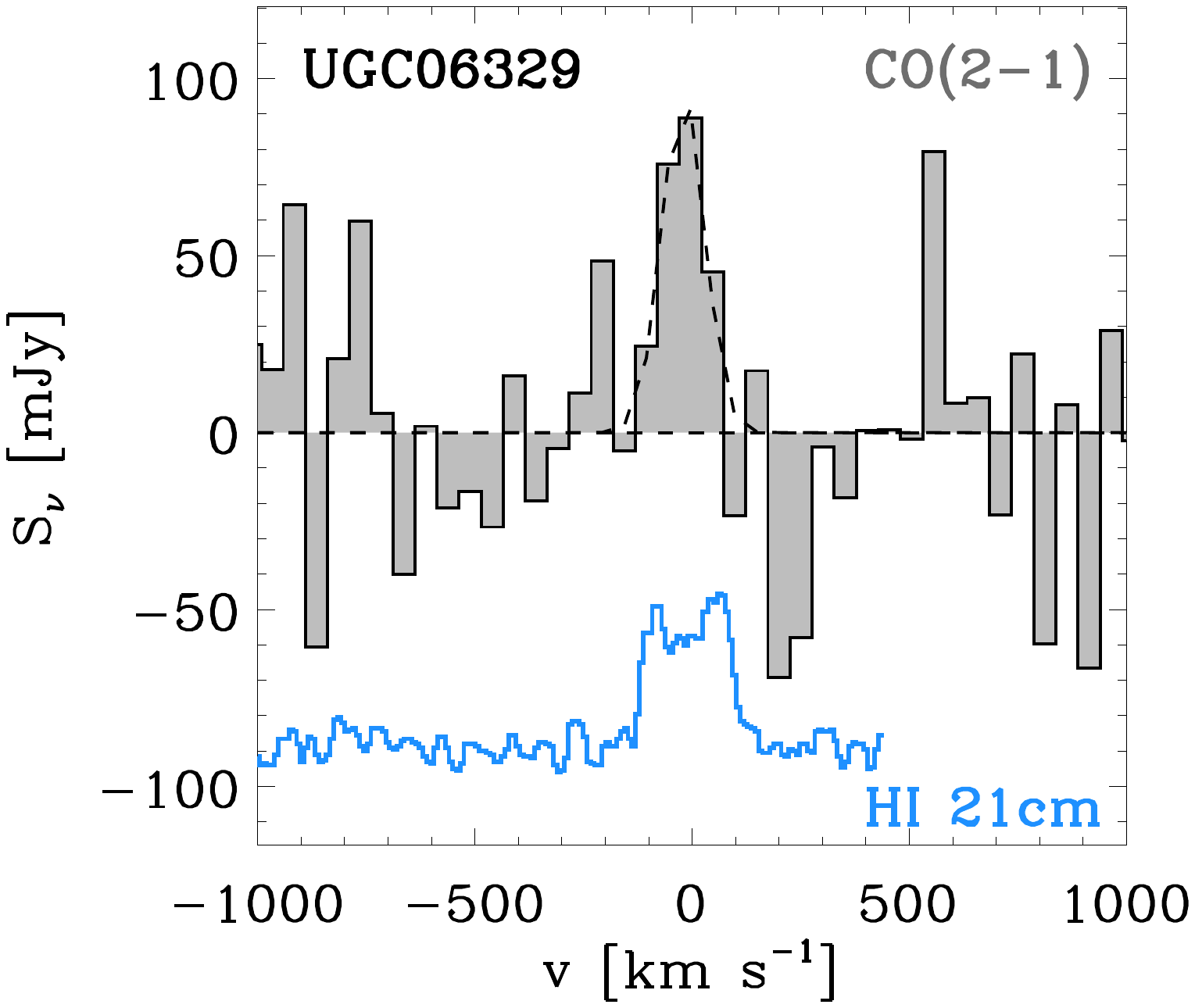}\\
      \includegraphics[clip=true,trim=-0.4cm 0cm 0cm 0cm,width=0.18\textwidth,angle=90]{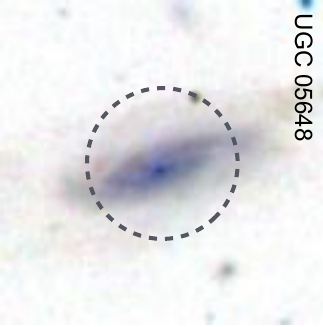}\quad
    \includegraphics[clip=true,trim=5.5cm 3.8cm 4cm 2cm,width=0.28\textwidth,angle=0]{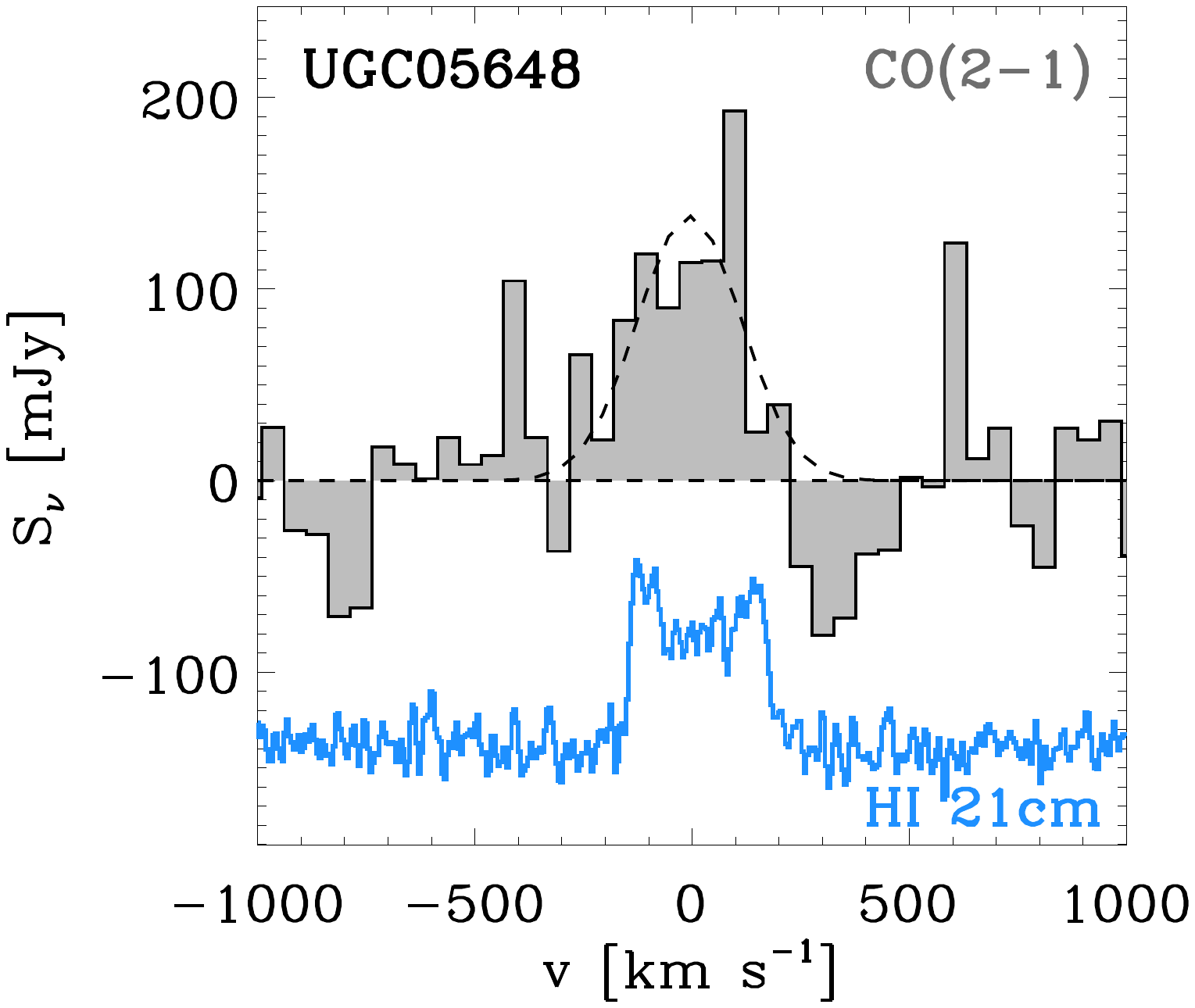}\quad
    \includegraphics[clip=true,trim=-0.4cm 0cm 0cm 0cm,width=0.18\textwidth,angle=90]{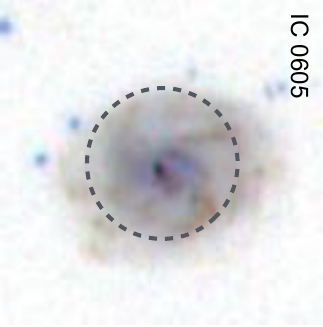}\quad
    \includegraphics[clip=true,trim=5.5cm 3.8cm 4cm 2cm,width=0.28\textwidth,angle=0]{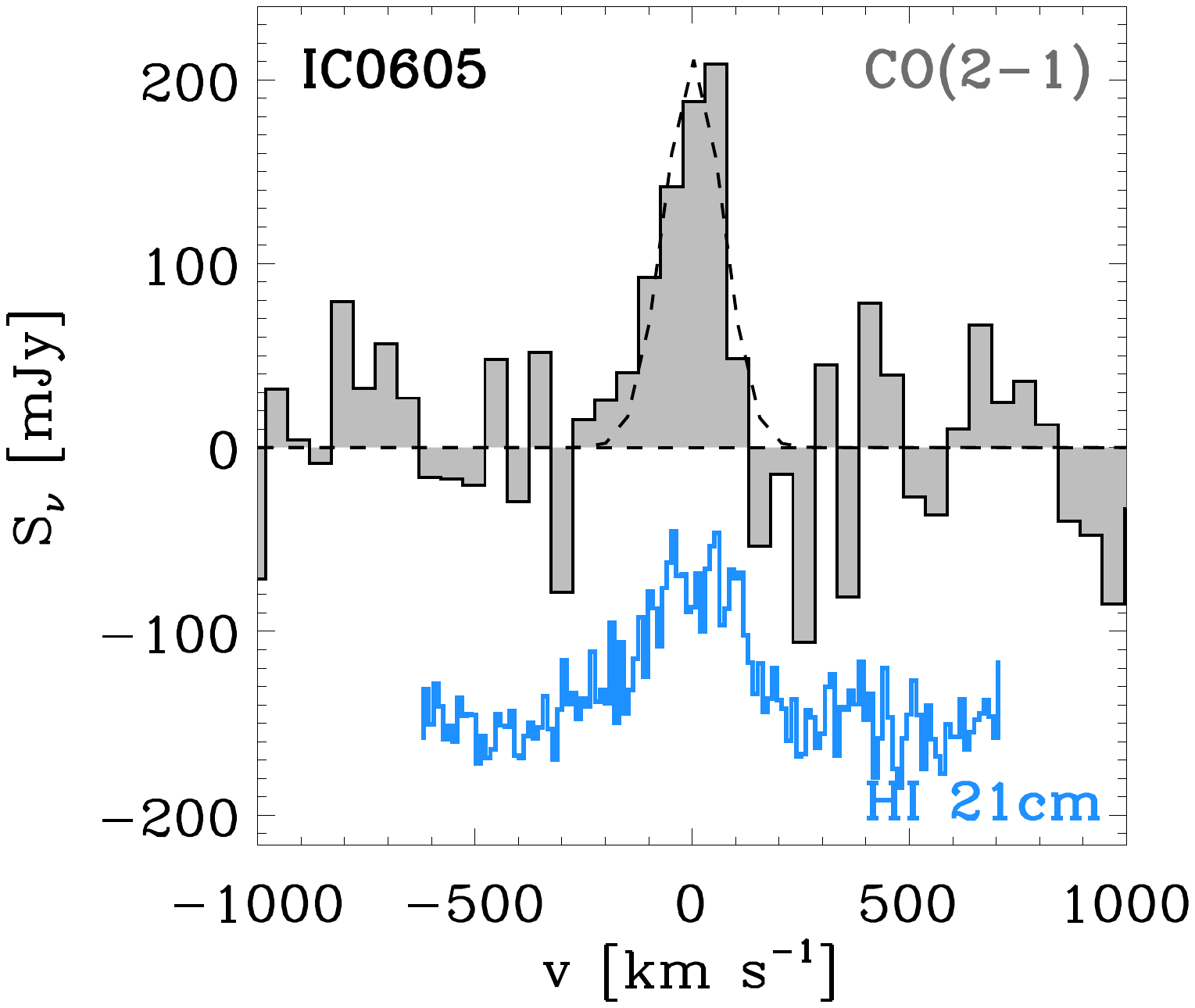}\\
      \includegraphics[clip=true,trim=-0.4cm 0cm 0cm 0cm,width=0.18\textwidth,angle=90]{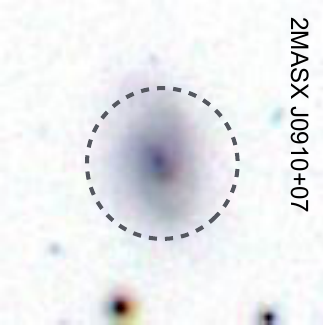}\quad
    \includegraphics[clip=true,trim=5.5cm 3.8cm 4cm 2cm,width=0.28\textwidth,angle=0]{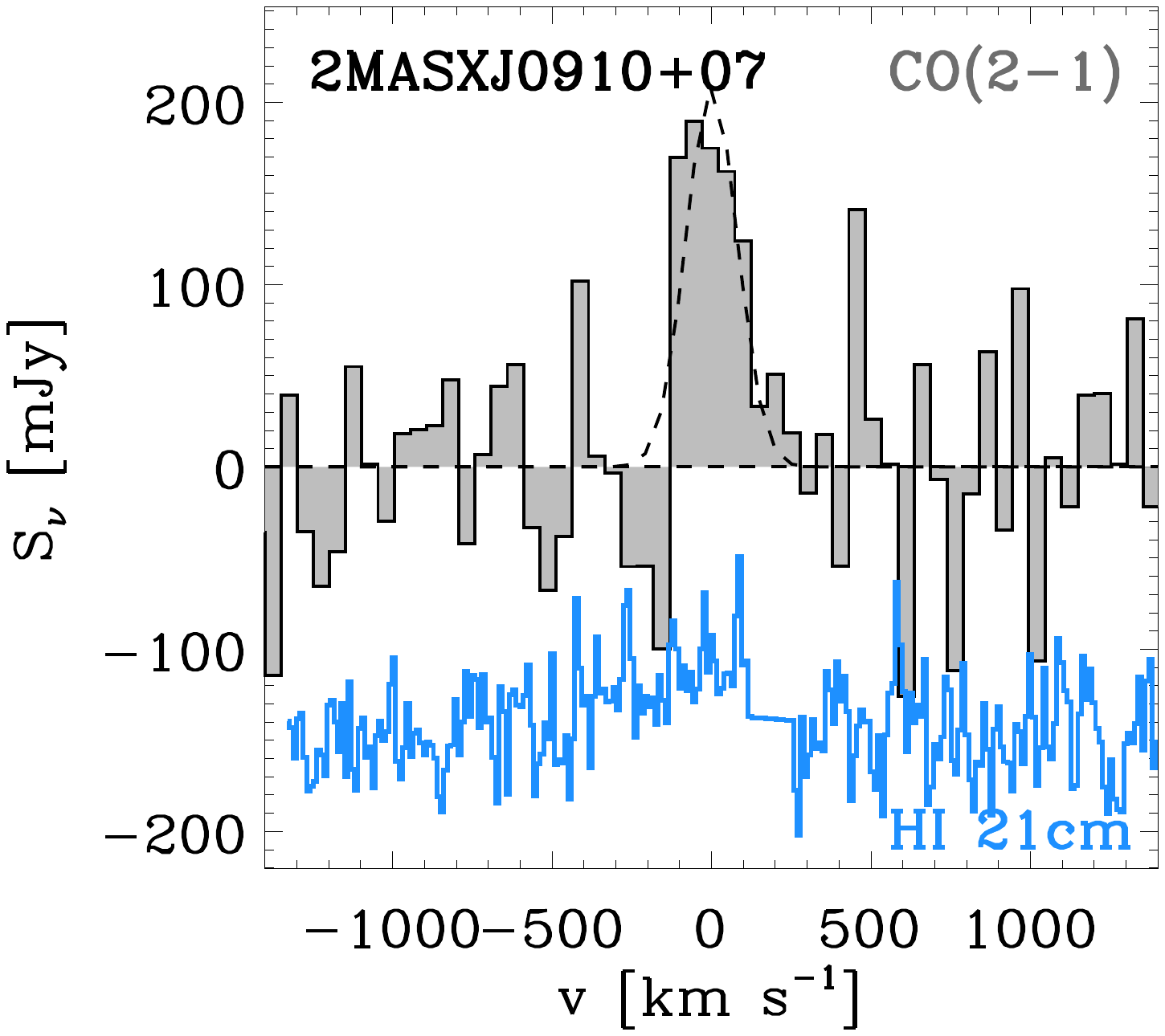}\quad
    \includegraphics[clip=true,trim=-0.4cm 0cm 0cm 0cm,width=0.18\textwidth,angle=90]{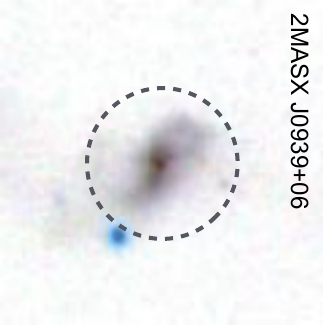}\quad
    \includegraphics[clip=true,trim=5.5cm 3.8cm 4cm 2cm,width=0.28\textwidth,angle=0]{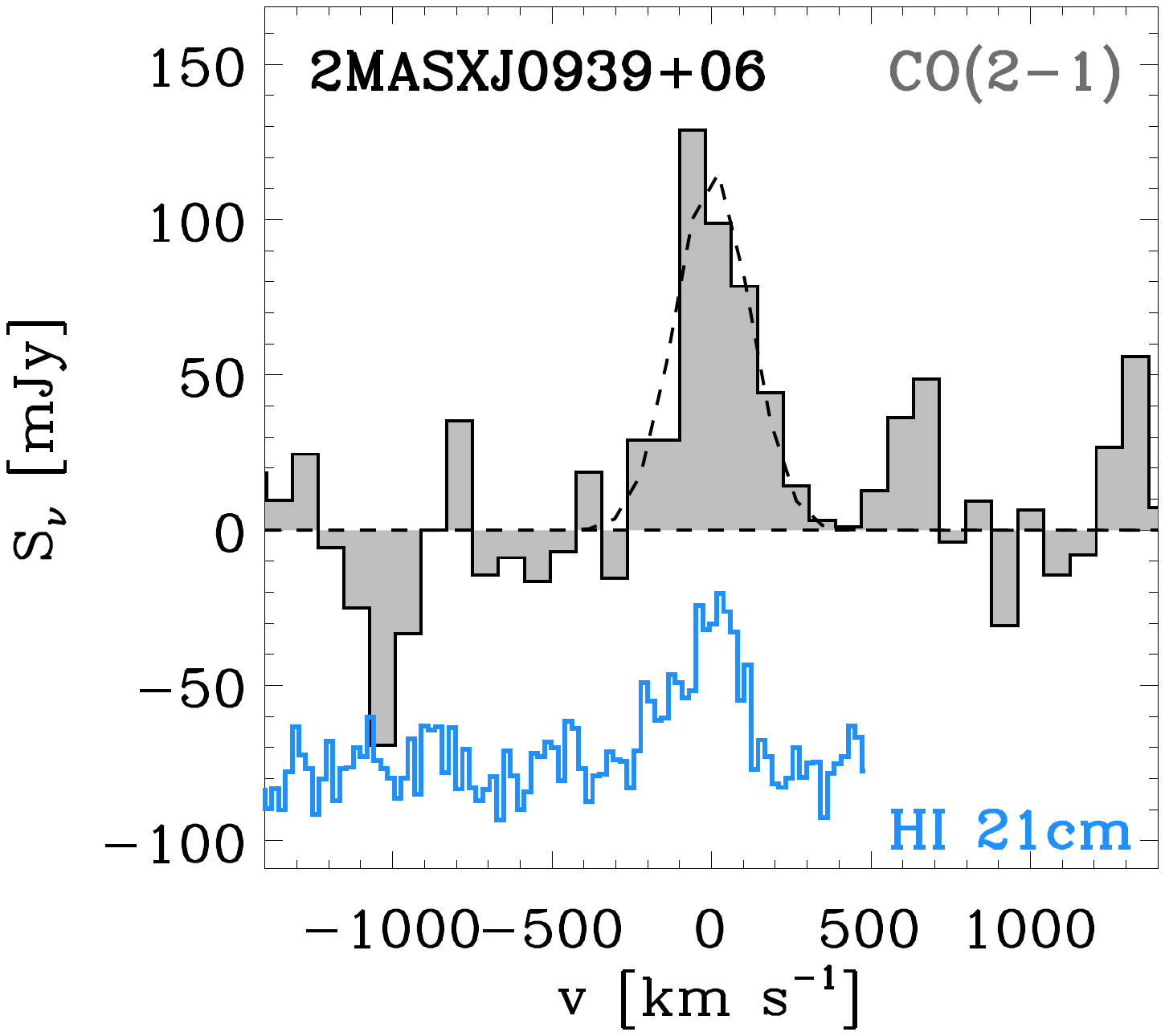}\\
     \caption{{\it Left panels:} SDSS cutout images ({\it g r i} composite, field of view = $60\arcsec\times60\arcsec$, scale = 0.5$\arcsec$/pixel, north is up and west is right) of ALLSMOG galaxies, showing the $27\arcsec$ APEX beam at 230 GHz. {\it Right panels:} APEX CO(2-1) baseline-subtracted spectra, rebinned in bins of $\delta \varv=80$~\kms (2MASXJ0939+0624),  50~\kms (MCG+00-25-005, UGC02004, UGC02529, MCG+00-29-013, UGC06329, UGC05648, IC0605, 2MASXJ0910+0752), or 25~\kms (UGC10005), depending on the width and S/N of the line. The corresponding H{\sc i}~21cm spectra are also shown for comparison, after having been renormalised for visualisation purposes (H{\sc i} references are given in Table~\ref{table:HI_parameters}).}
   \label{fig:spectra2}
\end{figure*}
\begin{figure*}[tbp]
\centering
    \includegraphics[clip=true,trim=-0.4cm 0cm 0cm 0cm,width=0.18\textwidth,angle=90]{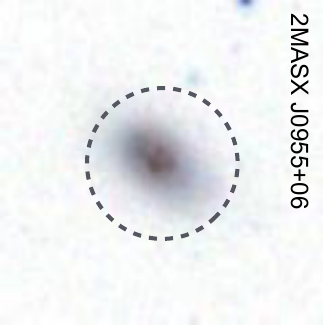}\quad
    \includegraphics[clip=true,trim=5.5cm 3.8cm 4cm 2cm,width=0.28\textwidth,angle=0]{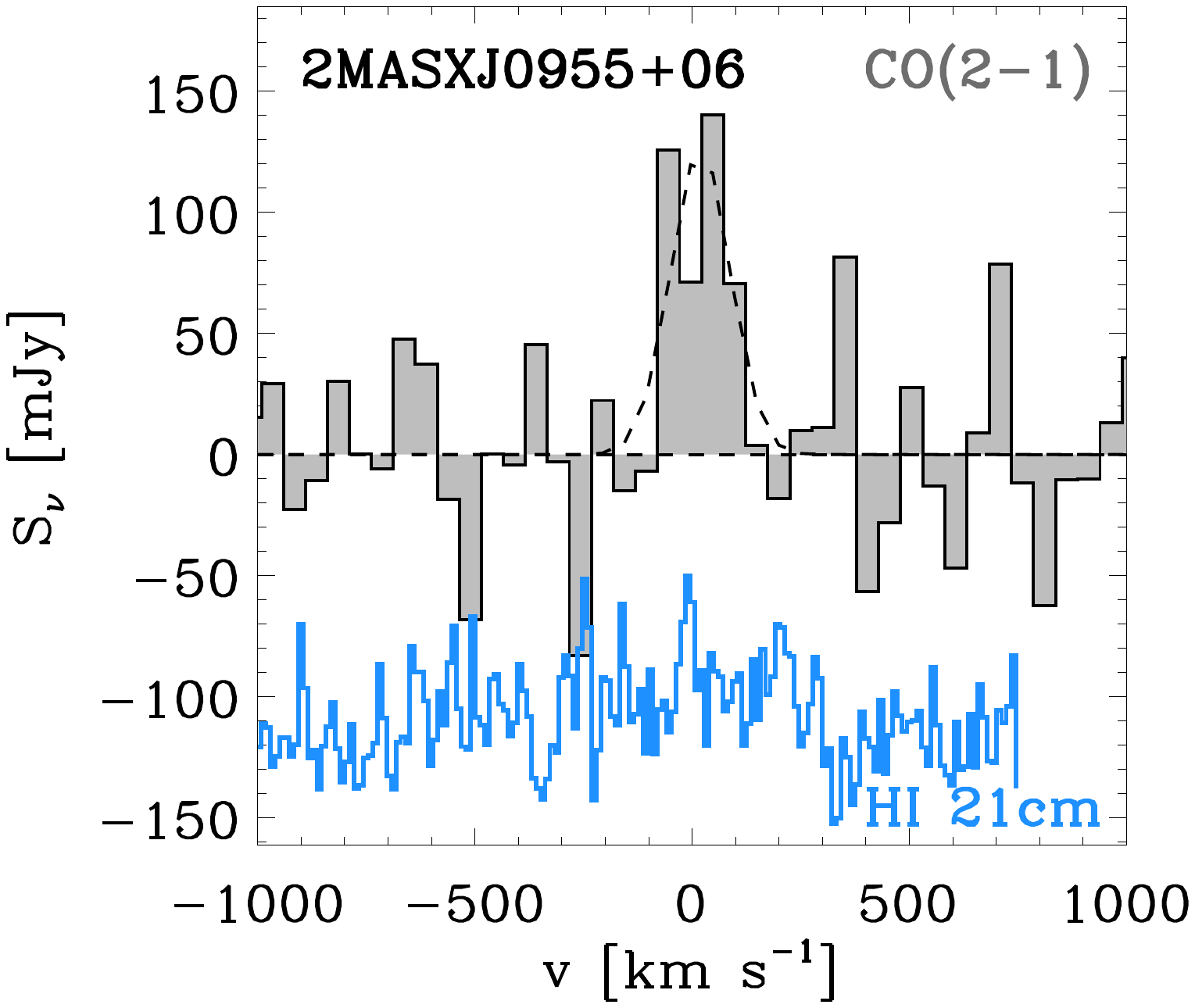}\quad
    \includegraphics[clip=true,trim=-0.4cm 0cm 0cm 0cm,width=0.18\textwidth,angle=90]{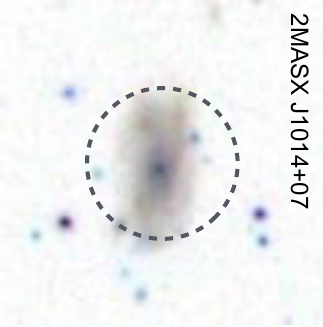}\quad
    \includegraphics[clip=true,trim=5.5cm 3.8cm 4cm 2cm,width=0.28\textwidth,angle=0]{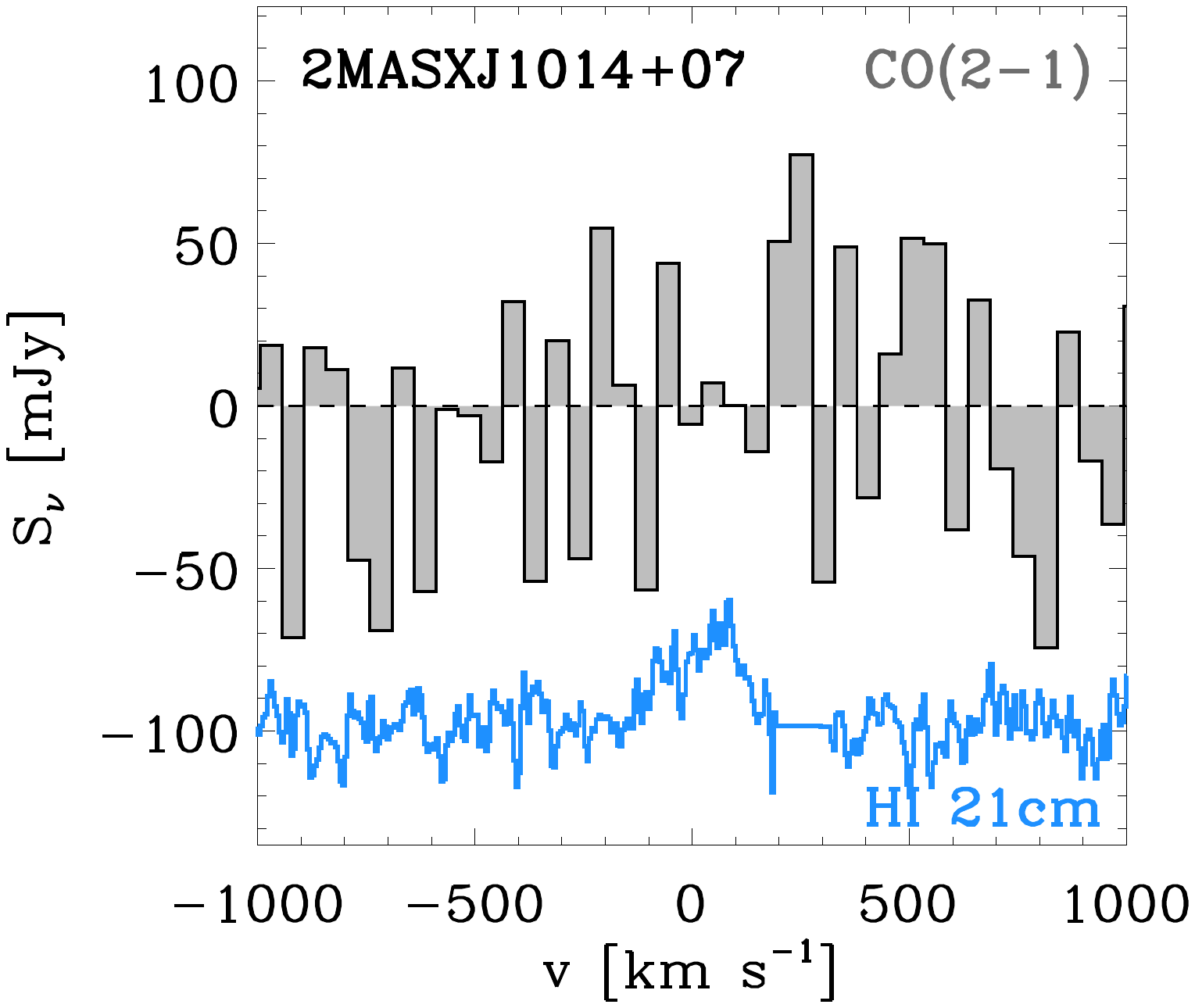}\\
    \includegraphics[clip=true,trim=-0.4cm 0cm 0cm 0cm,width=0.18\textwidth,angle=90]{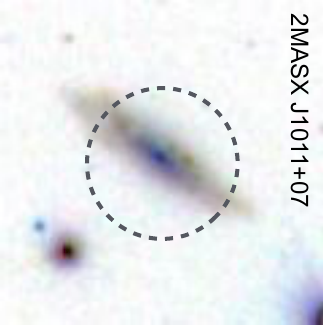}\quad
    \includegraphics[clip=true,trim=5.5cm 3.8cm 4cm 2cm,width=0.28\textwidth,angle=0]{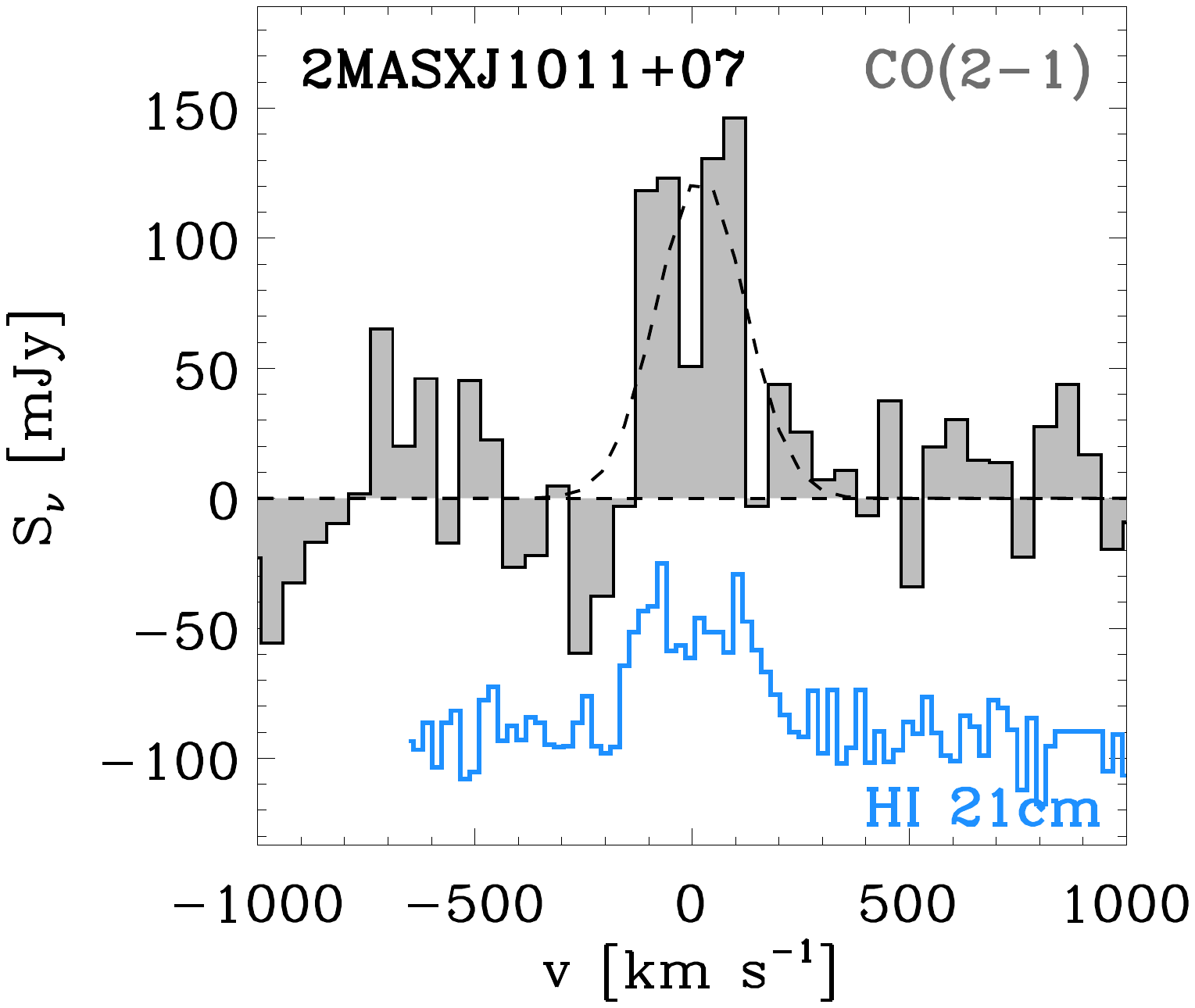}\quad
    \includegraphics[clip=true,trim=-0.4cm 0cm 0cm 0cm,width=0.18\textwidth,angle=90]{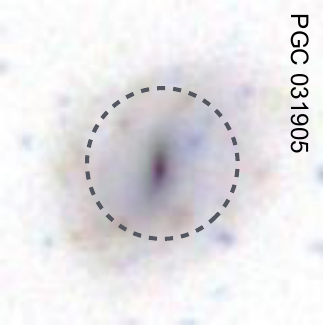}\quad
    \includegraphics[clip=true,trim=5.5cm 3.8cm 4cm 2cm,width=0.28\textwidth,angle=0]{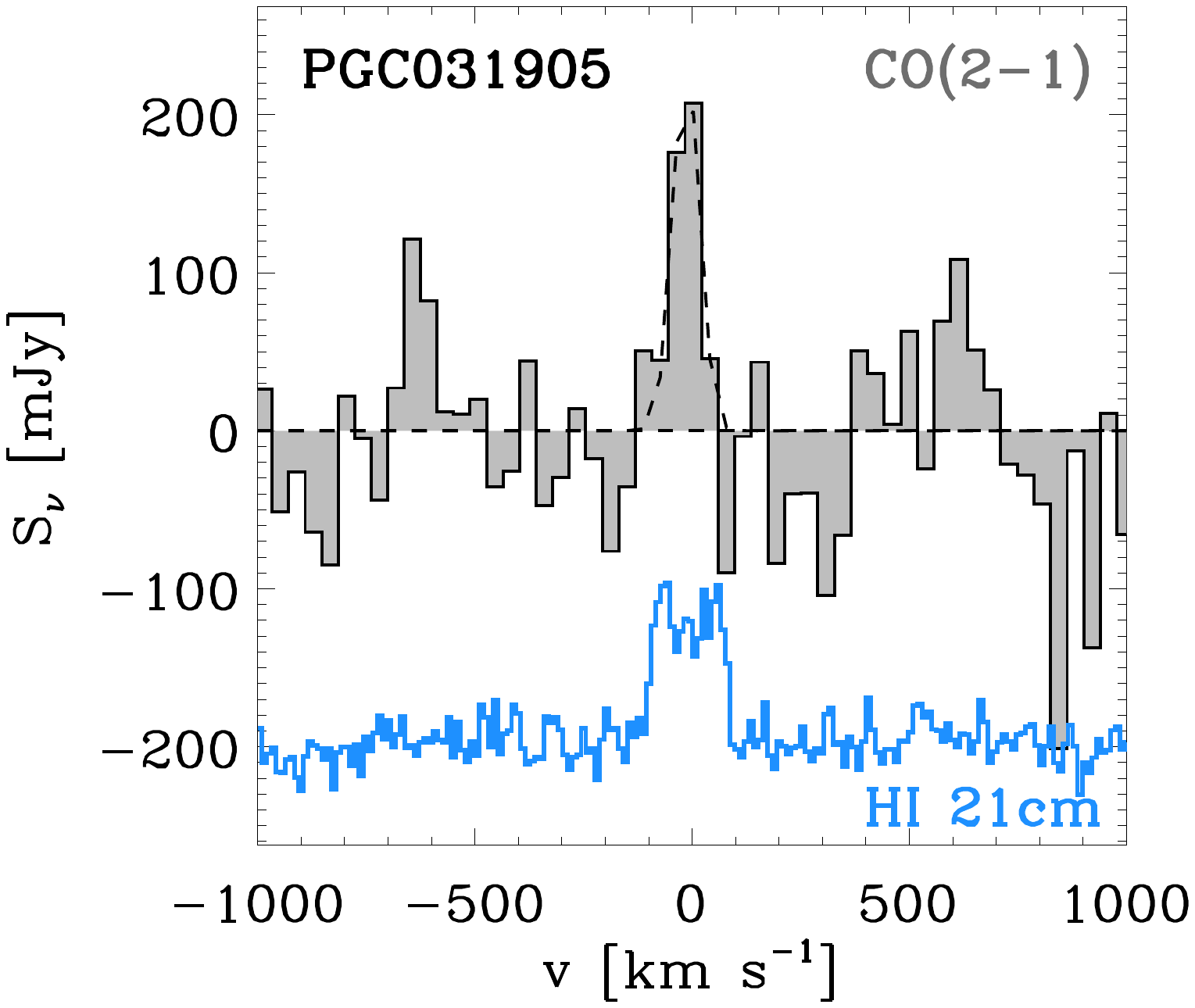}\\
      \includegraphics[clip=true,trim=-0.4cm 0cm 0cm 0cm,width=0.18\textwidth,angle=90]{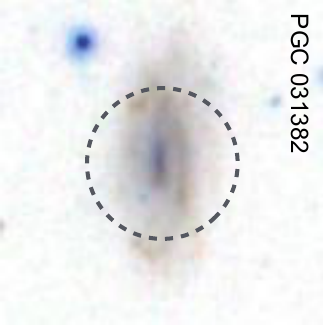}\quad
    \includegraphics[clip=true,trim=5.5cm 3.8cm 4cm 2cm,width=0.28\textwidth,angle=0]{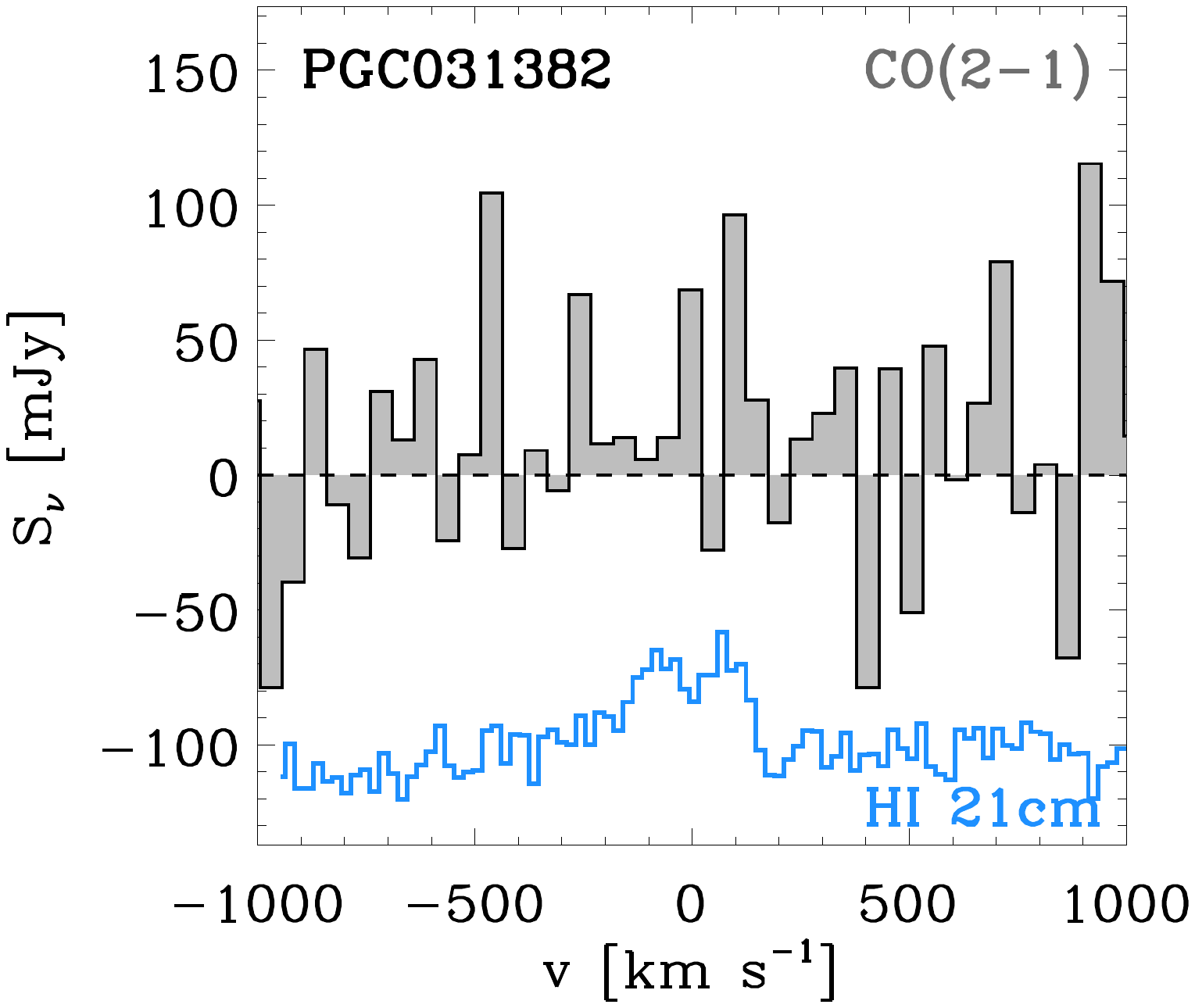}\quad
    \includegraphics[clip=true,trim=-0.4cm 0cm 0cm 0cm,width=0.18\textwidth,angle=90]{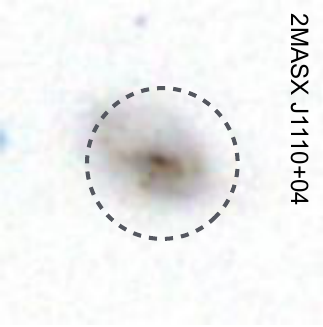}\quad
    \includegraphics[clip=true,trim=5.5cm 3.8cm 4cm 2cm,width=0.28\textwidth,angle=0]{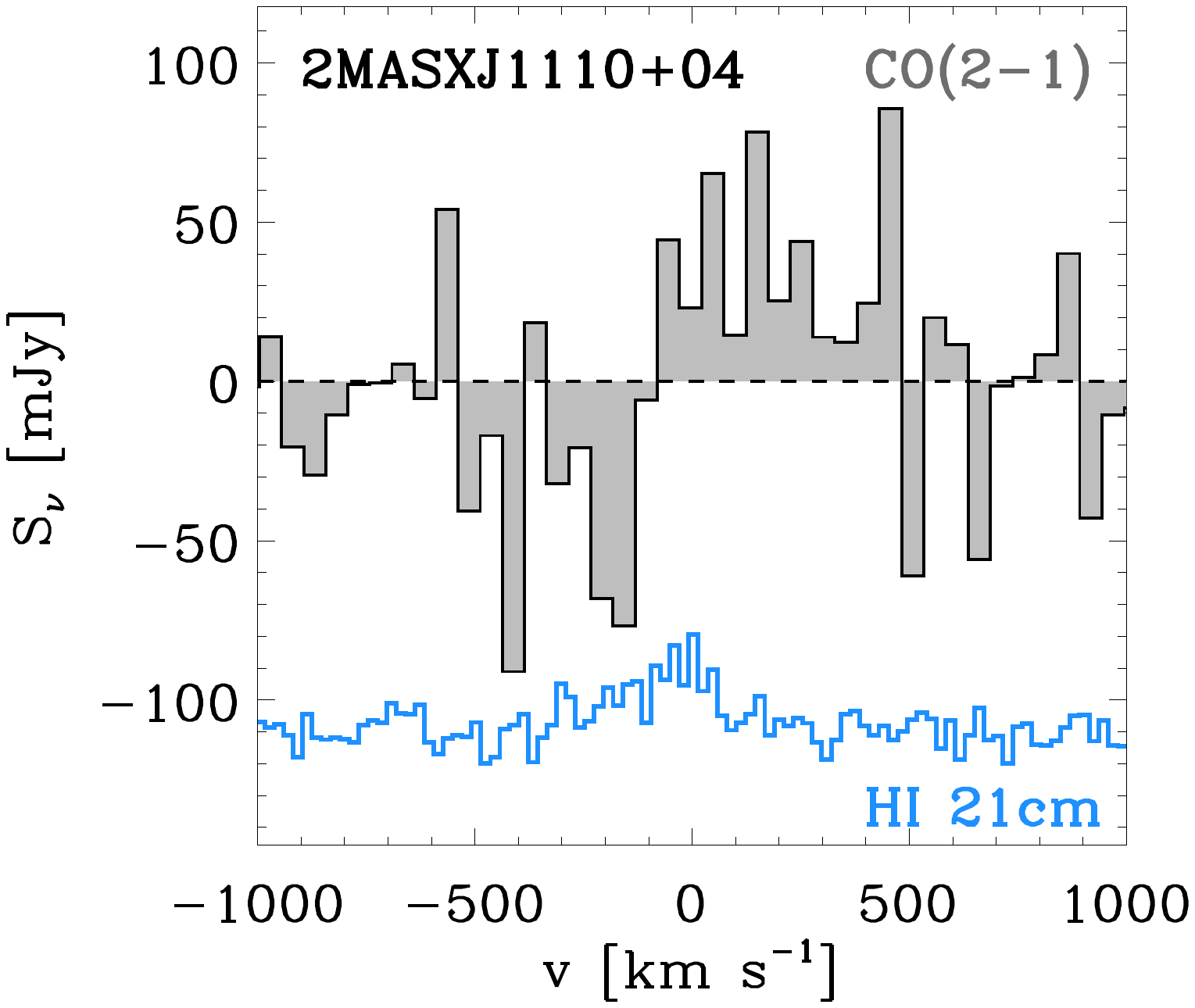}\\
     \includegraphics[clip=true,trim=-0.4cm 0cm 0cm 0cm,width=0.18\textwidth,angle=90]{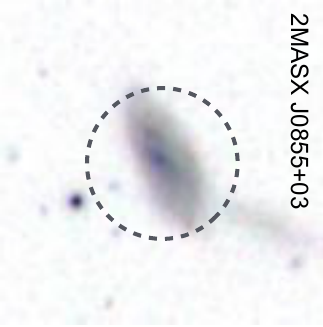}\quad
    \includegraphics[clip=true,trim=5.5cm 3.8cm 4cm 2cm,width=0.28\textwidth,angle=0]{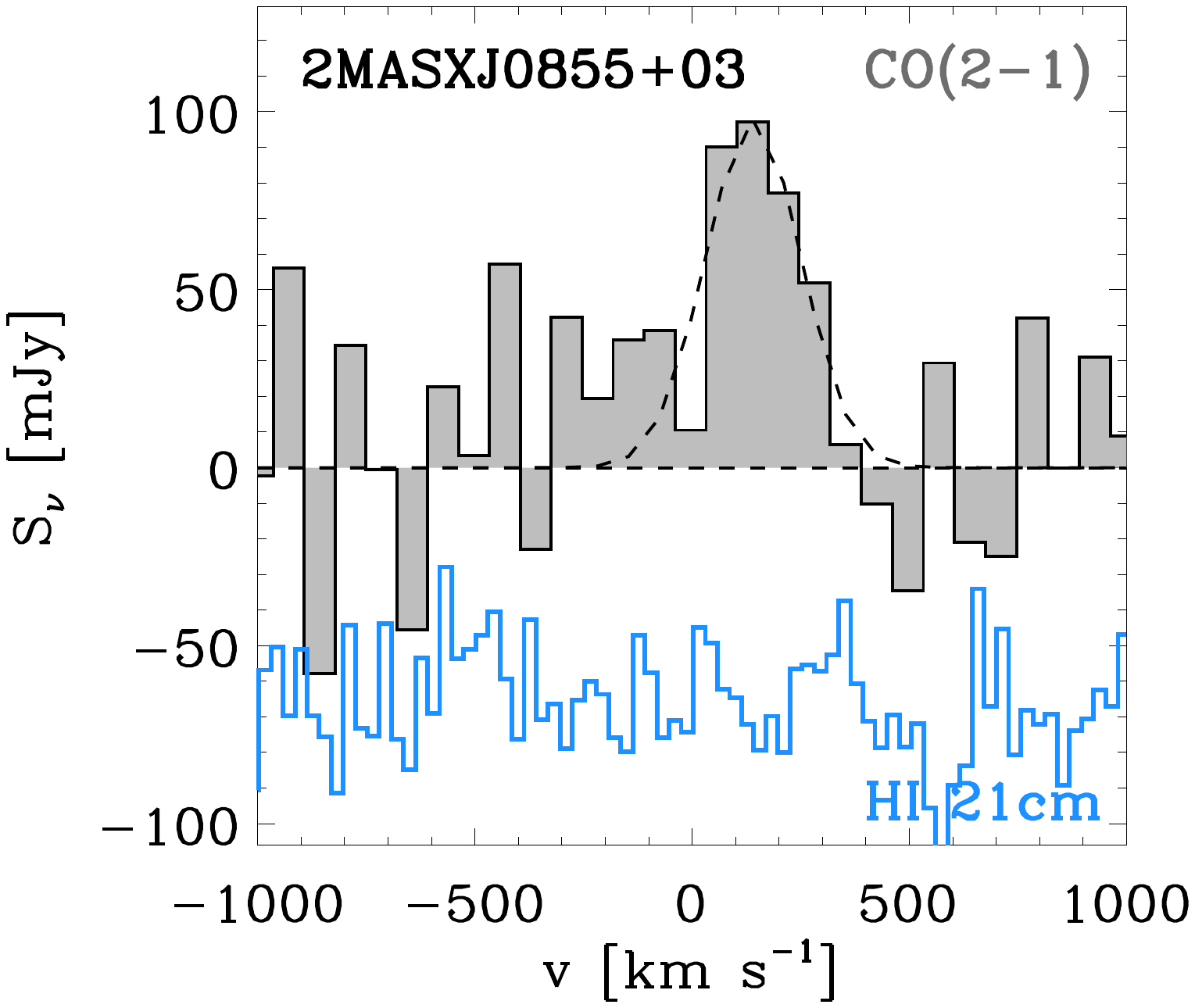}\quad
    \includegraphics[clip=true,trim=-0.4cm 0cm 0cm 0cm,width=0.18\textwidth,angle=90]{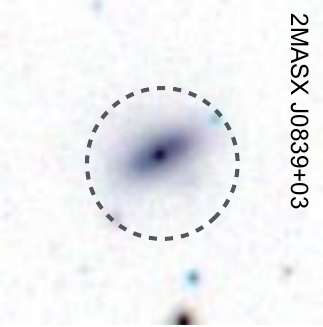}\quad
    \includegraphics[clip=true,trim=5.5cm 3.8cm 4cm 2cm,width=0.28\textwidth,angle=0]{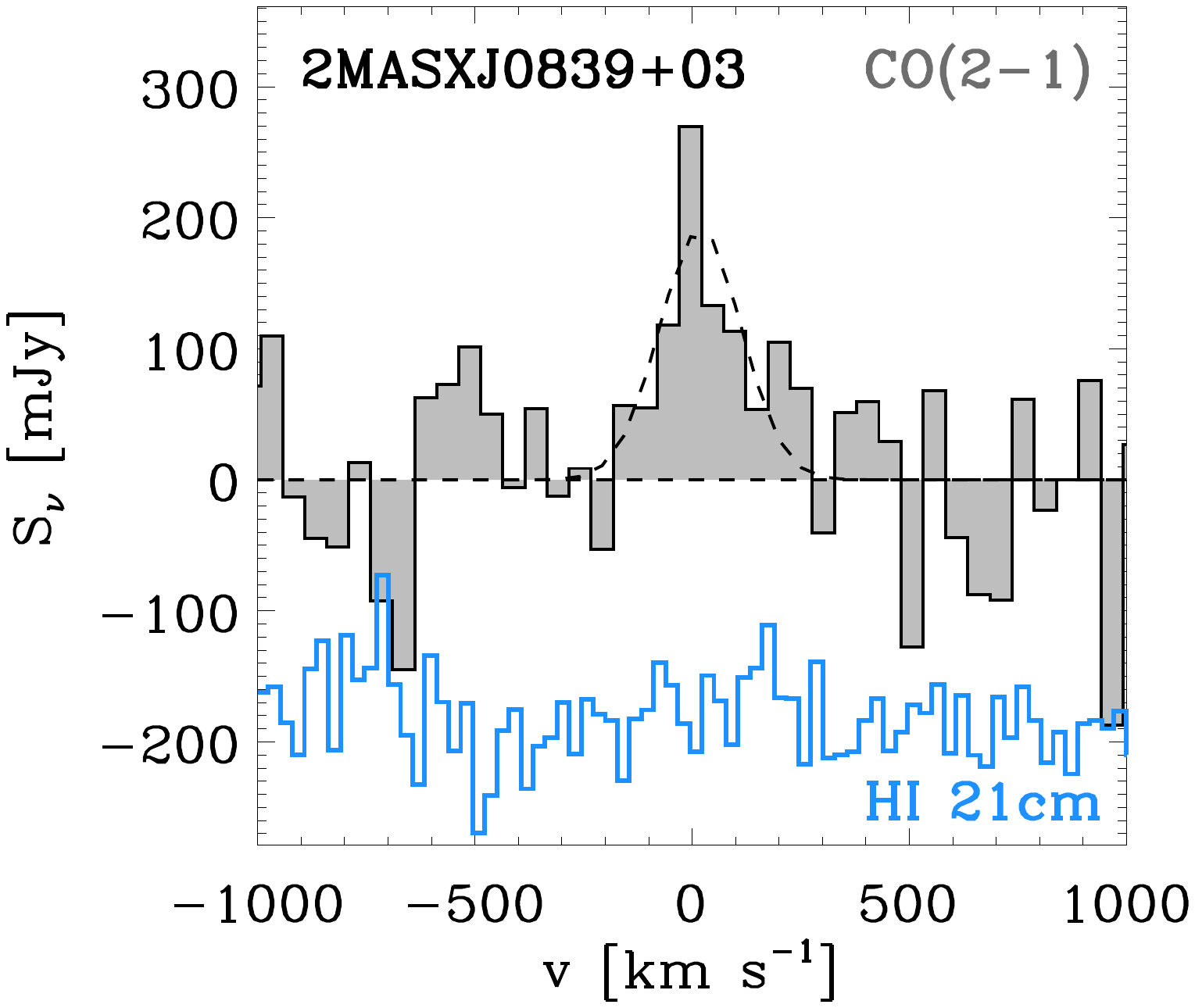}\\
      \includegraphics[clip=true,trim=-0.4cm 0cm 0cm 0cm,width=0.18\textwidth,angle=90]{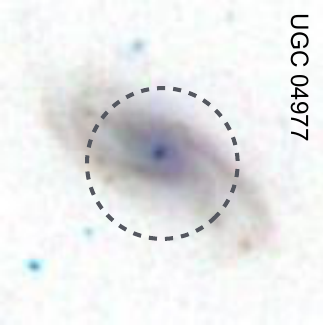}\quad
    \includegraphics[clip=true,trim=5.5cm 3.8cm 4cm 2cm,width=0.28\textwidth,angle=0]{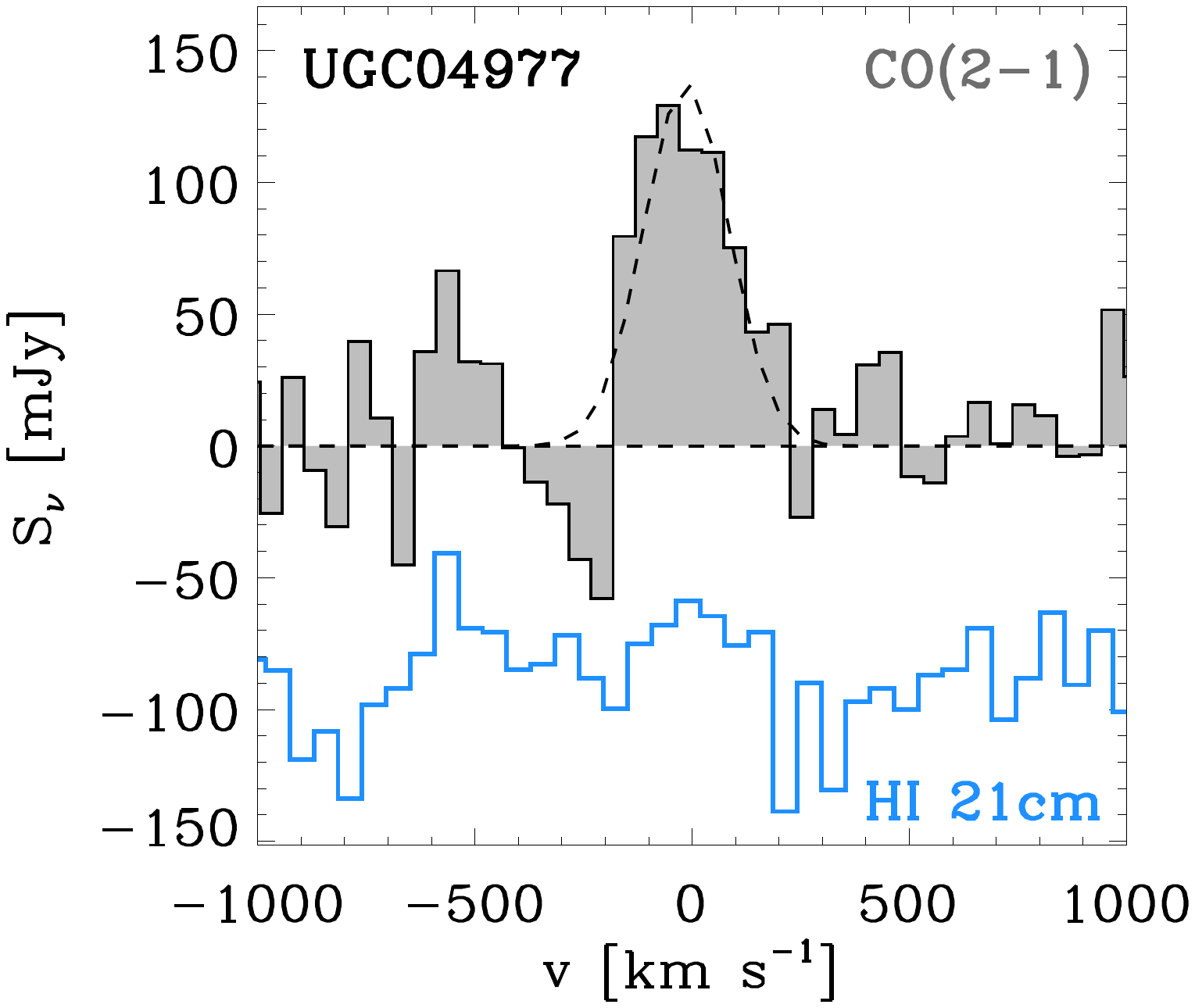}\quad
    \includegraphics[clip=true,trim=-0.4cm 0cm 0cm 0cm,width=0.18\textwidth,angle=90]{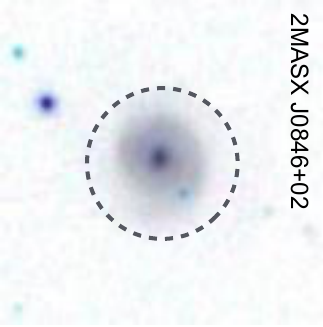}\quad
    \includegraphics[clip=true,trim=5.5cm 3.8cm 4cm 2cm,width=0.28\textwidth,angle=0]{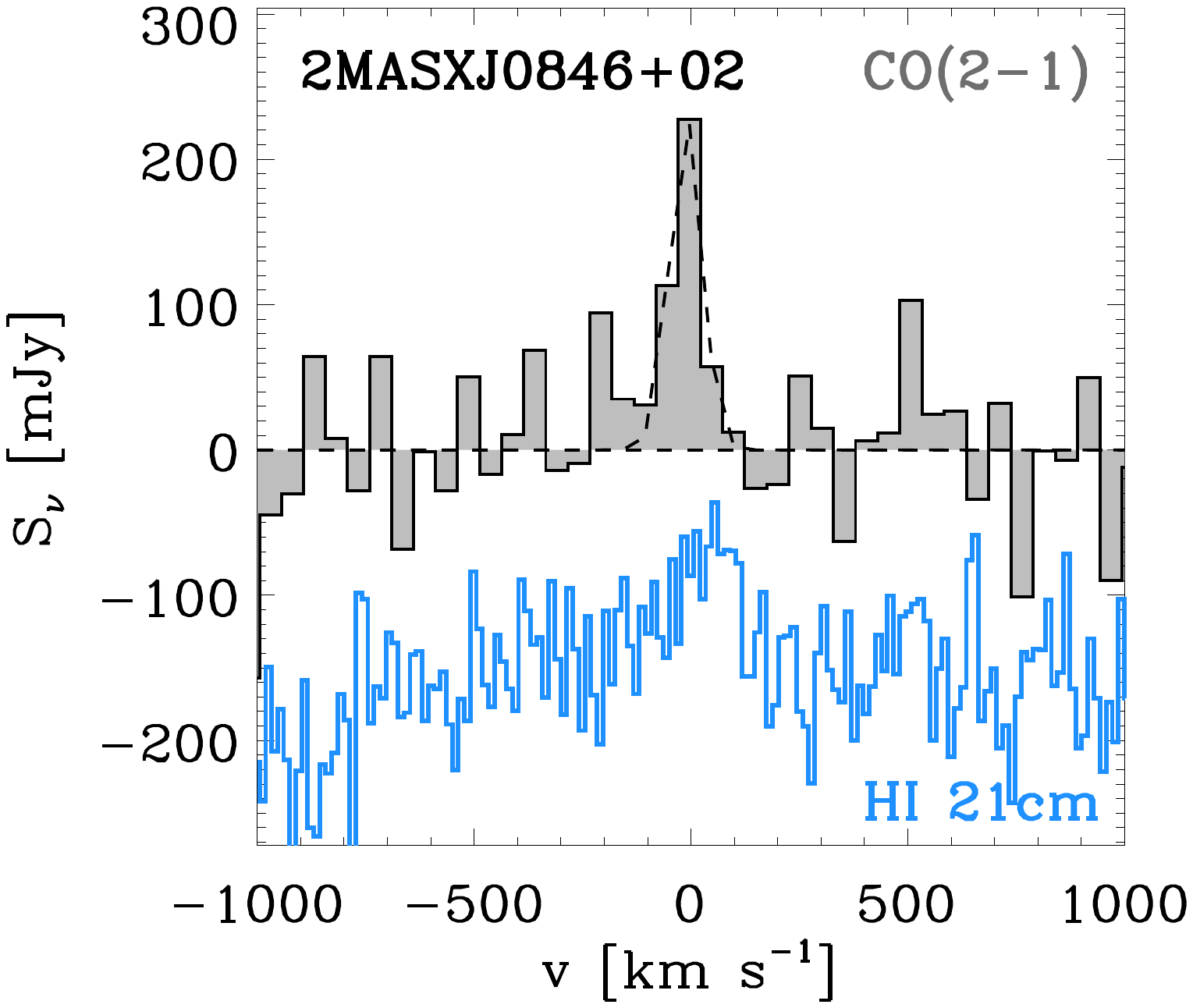}\\
     \caption{{\it Left panels:} SDSS cutout images ({\it g r i} composite, field of view = $60\arcsec\times60\arcsec$, scale = 0.5$\arcsec$/pixel, north is up and west is right) of ALLSMOG galaxies, showing the 27$''$ APEX beam at 230 GHz. {\it Right panels:} APEX CO(2-1) baseline-subtracted spectra, rebinned in bins of $\delta\varv=70$~\kms (2MASXJ0855+0345) or 50~\kms (2MASXJ0955+0632, 2MASXJ1014+0748, 2MASXJ1011+0746, PGC031905, PGC031382, 2MASXJ1110+0411, 2MASXJ0839+0349, UGC04977, 2MASXJ0846+0230), depending on the width and S/N of the line. The corresponding H{\sc i}~21cm spectra are also shown for comparison, after having been renormalised for visualisation purposes (H{\sc i} references are given in Table~\ref{table:HI_parameters}).}
   \label{fig:spectra3}
\end{figure*}
\begin{figure*}[tbp]
\centering
    \includegraphics[clip=true,trim=-0.4cm 0cm 0cm 0cm,width=0.18\textwidth,angle=90]{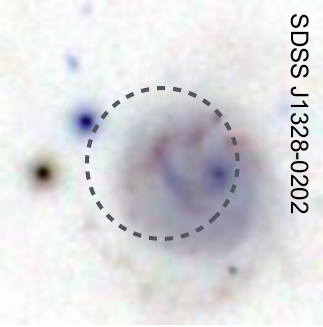}\quad
    \includegraphics[clip=true,trim=5.5cm 3.8cm 4cm 2cm,width=0.28\textwidth,angle=0]{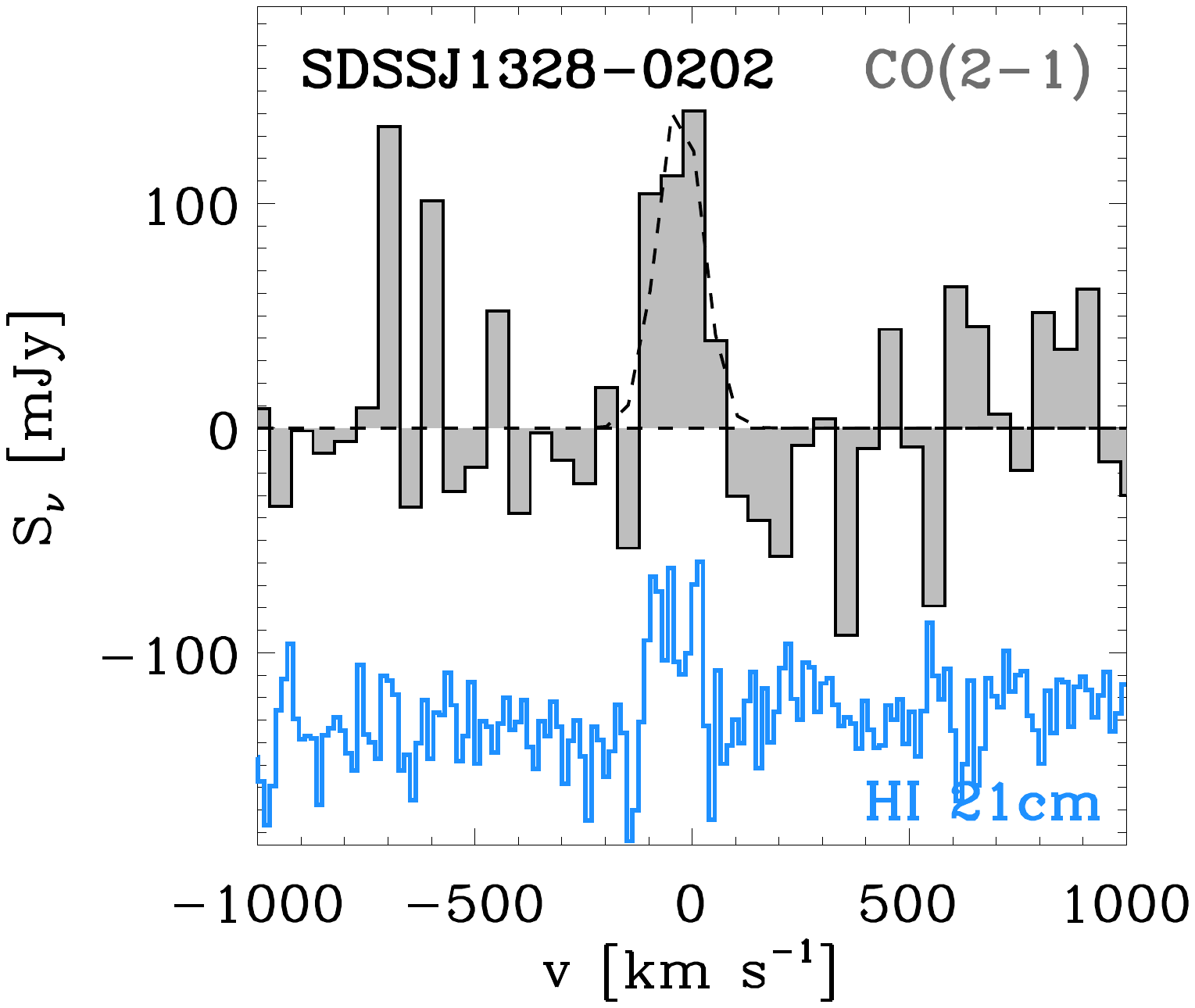}\quad
    \includegraphics[clip=true,trim=-0.4cm 0cm 0cm 0cm,width=0.18\textwidth,angle=90]{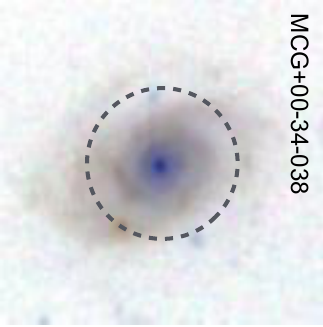}\quad
    \includegraphics[clip=true,trim=5.5cm 3.8cm 4cm 2cm,width=0.28\textwidth,angle=0]{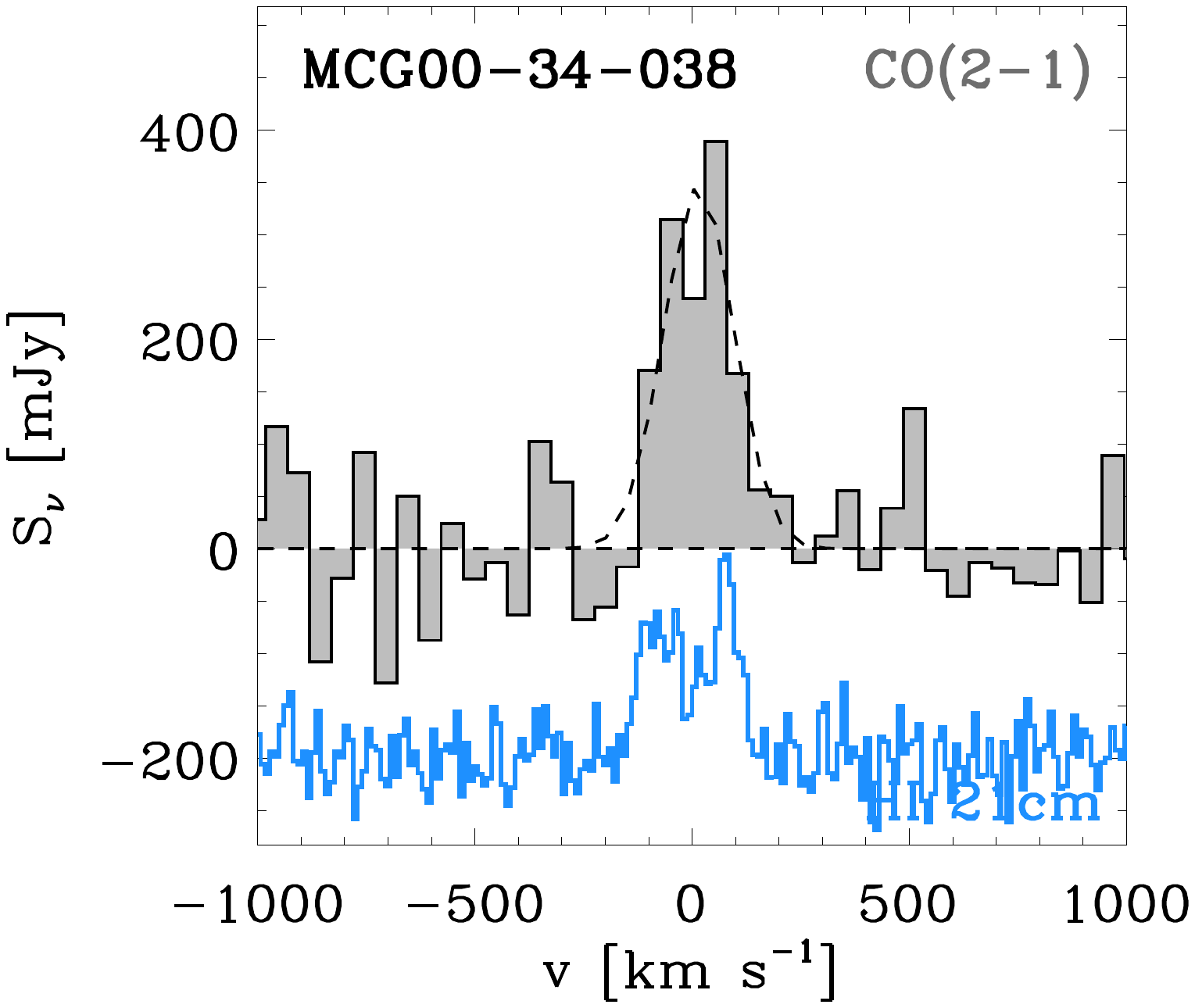}\\
    \includegraphics[clip=true,trim=-0.4cm 0cm 0cm 0cm,width=0.18\textwidth,angle=90]{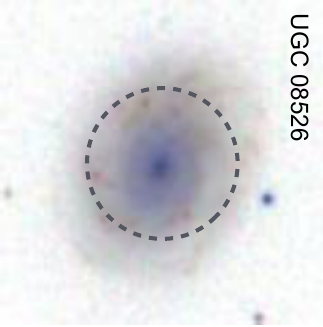}\quad
    \includegraphics[clip=true,trim=5.5cm 3.8cm 4cm 2cm,width=0.28\textwidth,angle=0]{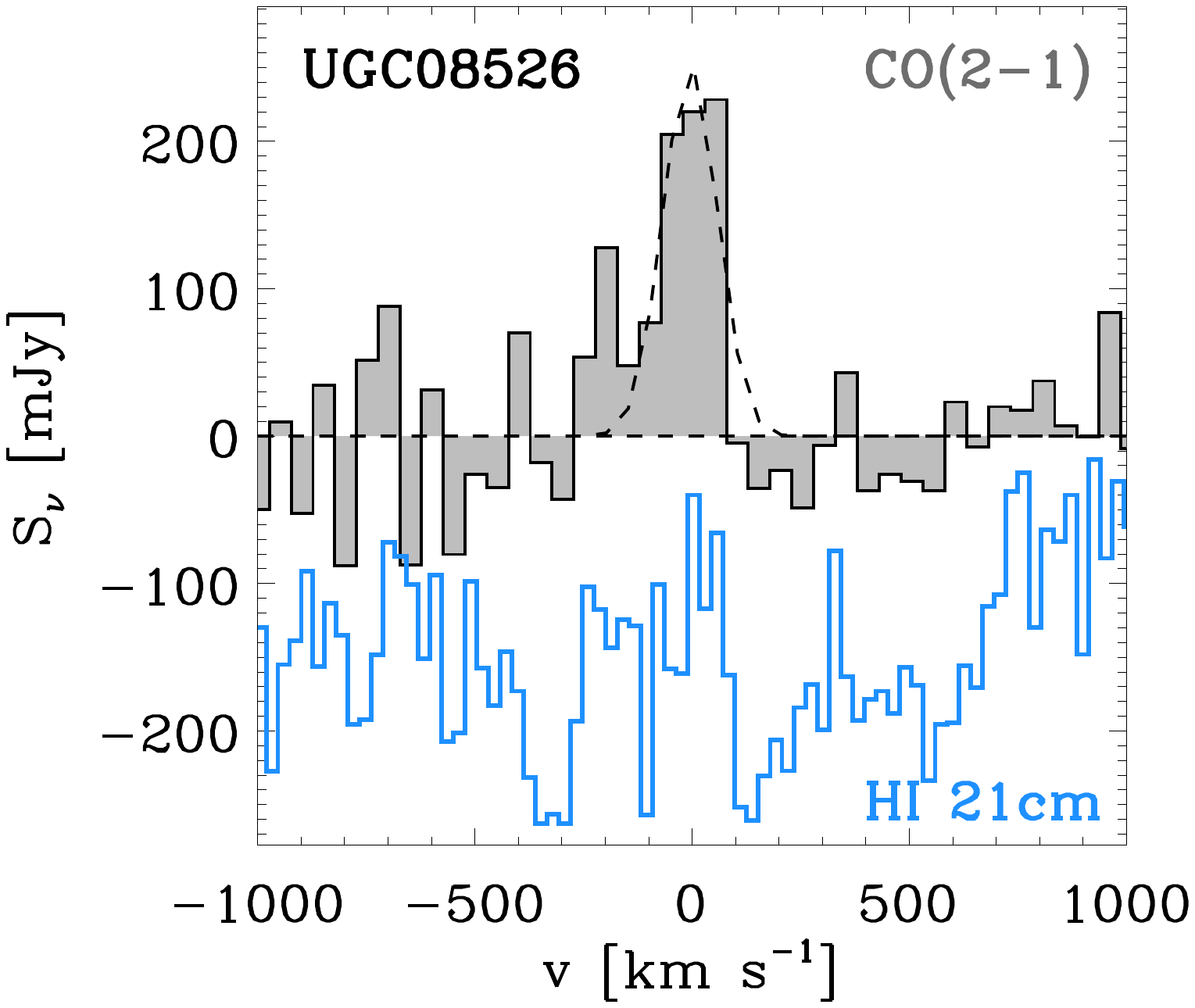}\quad
    \includegraphics[clip=true,trim=-0.4cm 0cm 0cm 0cm,width=0.18\textwidth,angle=90]{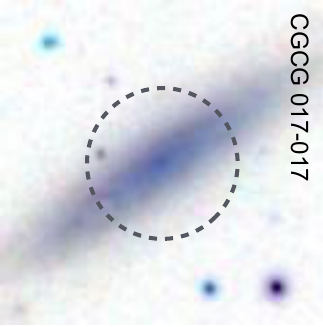}\quad
    \includegraphics[clip=true,trim=5.5cm 3.8cm 4cm 2cm,width=0.28\textwidth,angle=0]{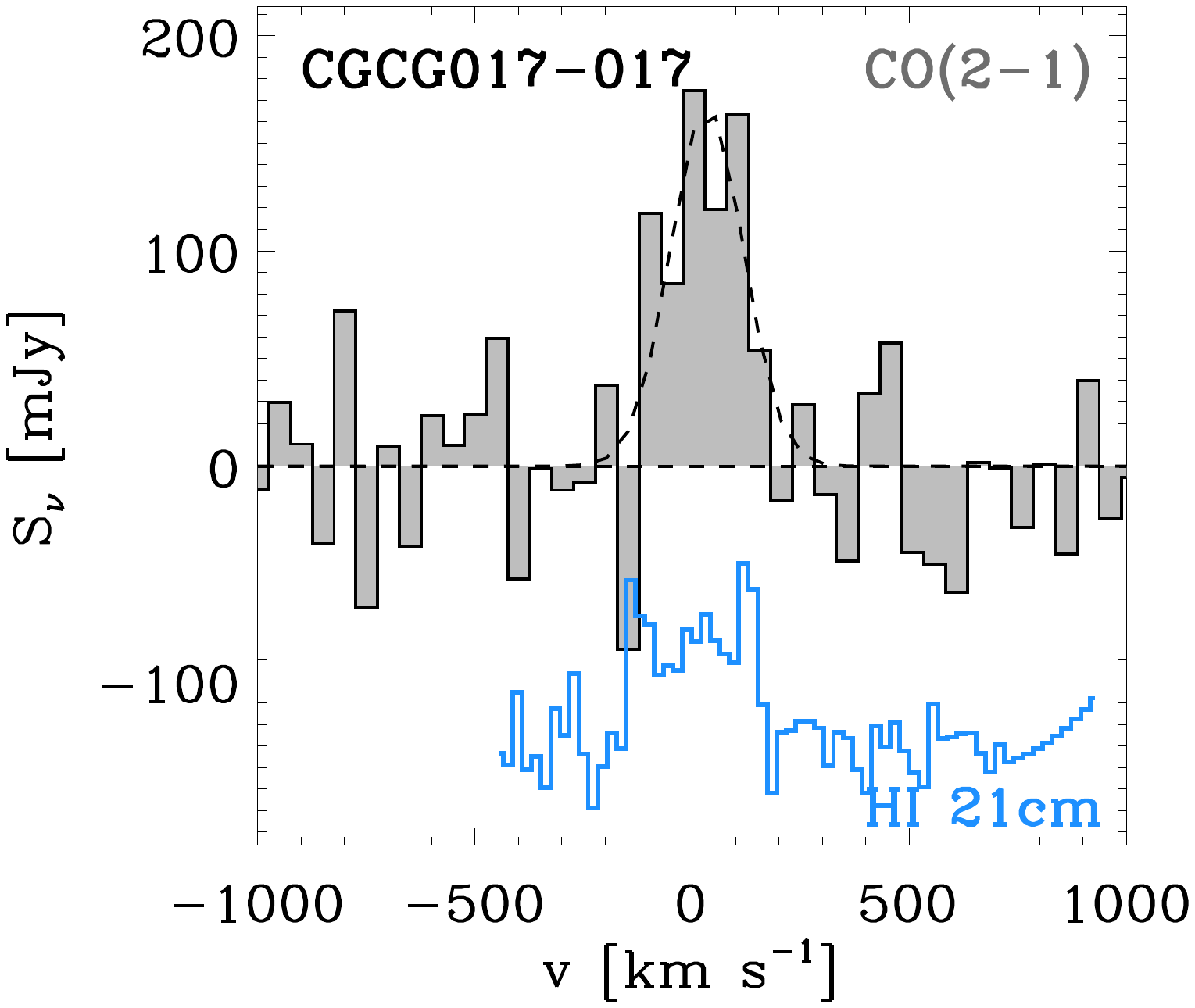}\\
      \includegraphics[clip=true,trim=-0.4cm 0cm 0cm 0cm,width=0.18\textwidth,angle=90]{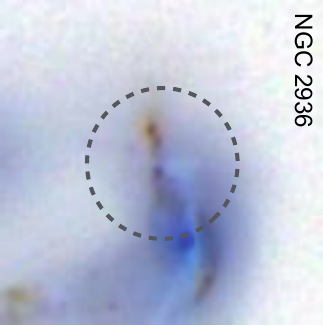}\quad
    \includegraphics[clip=true,trim=5.1cm 3.8cm 4cm 2cm,width=0.28\textwidth,angle=0]{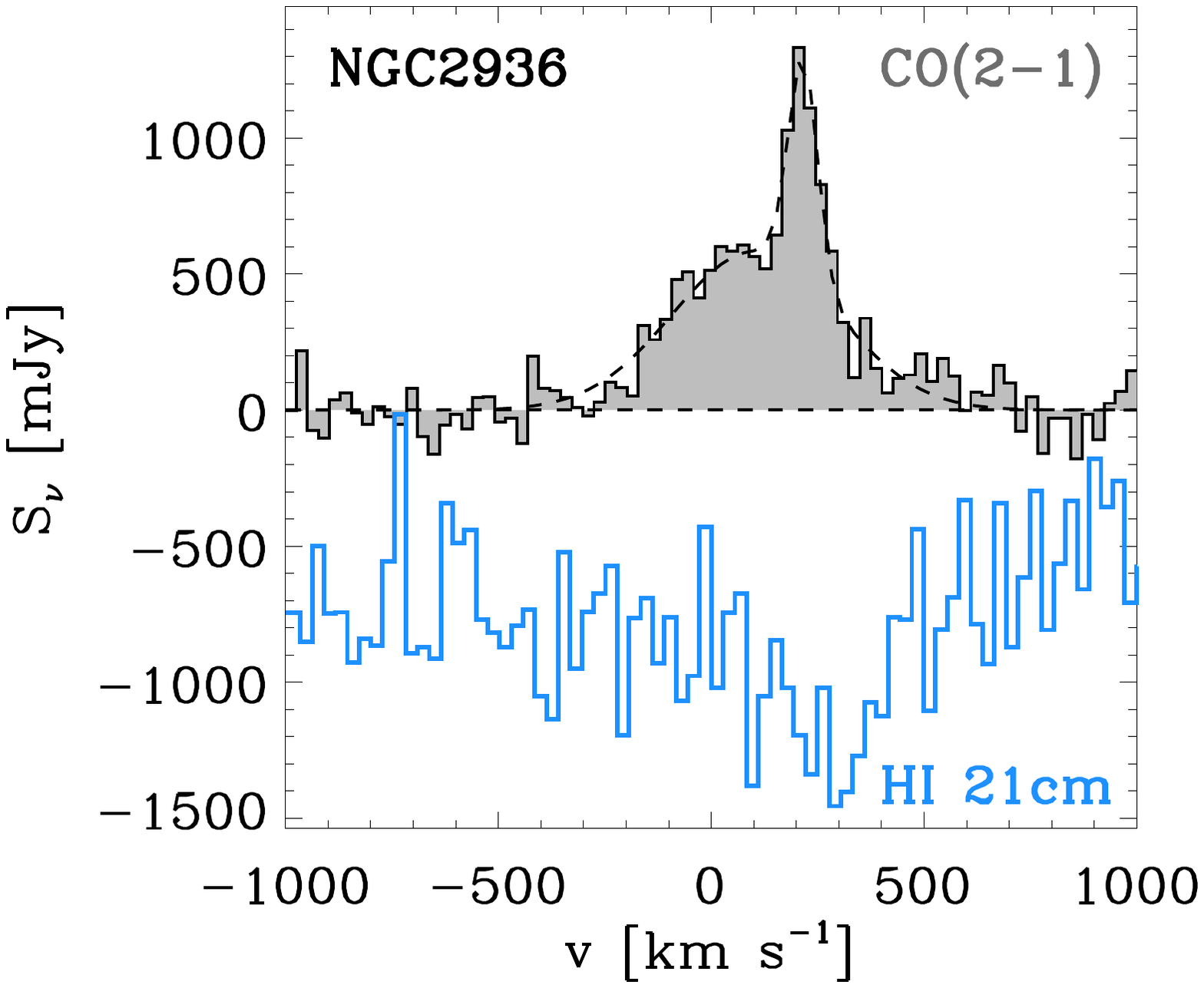}\quad
    \includegraphics[clip=true,trim=-0.4cm 0cm 0cm 0cm,width=0.18\textwidth,angle=90]{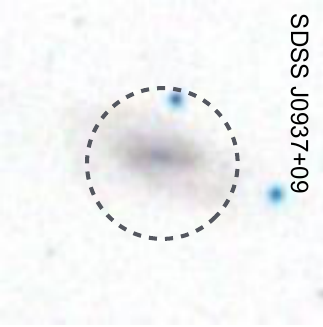}\quad
    \includegraphics[clip=true,trim=5.5cm 3.8cm 4cm 2cm,width=0.28\textwidth,angle=0]{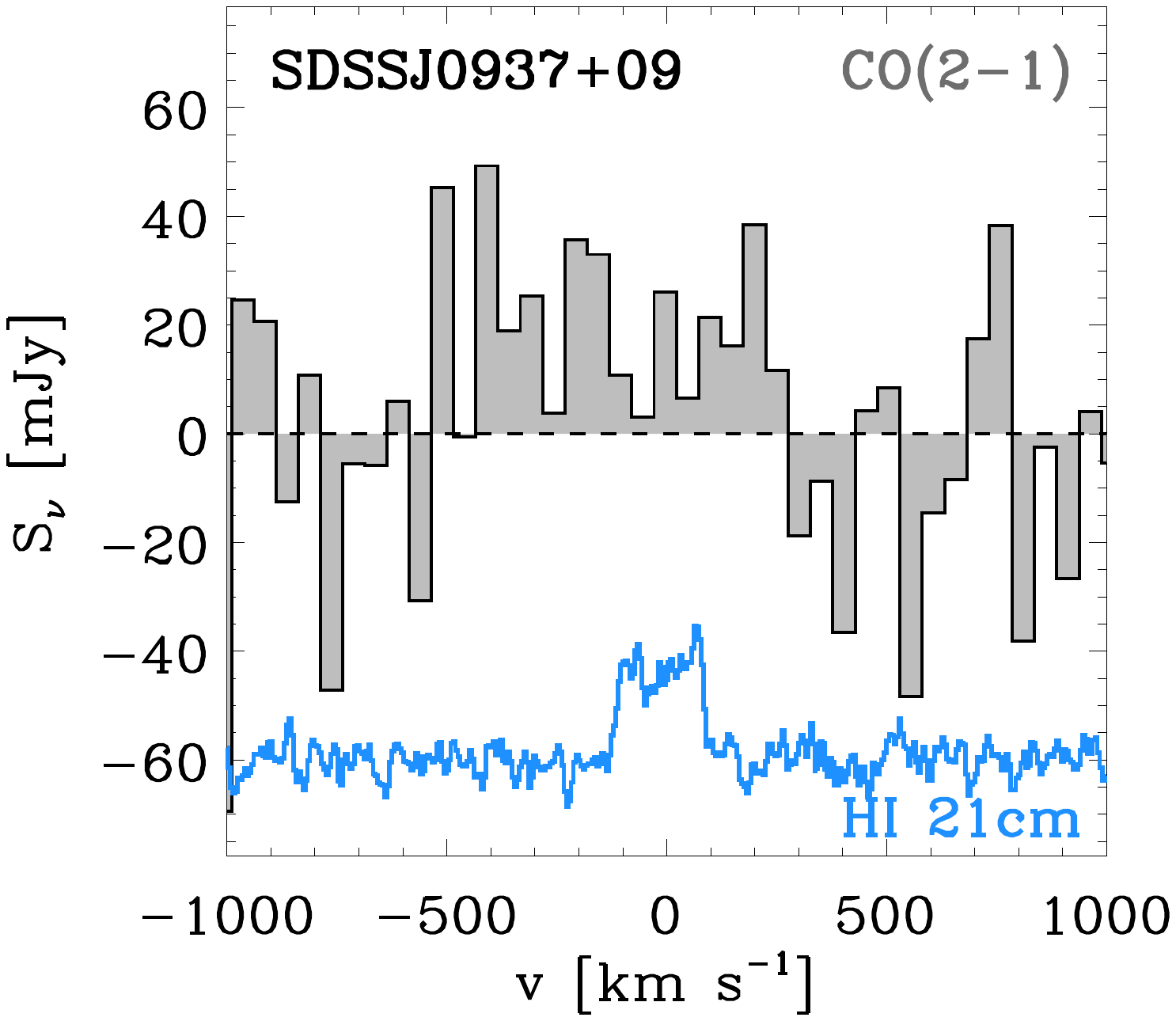}\\
      \includegraphics[clip=true,trim=-0.4cm 0cm 0cm 0cm,width=0.18\textwidth,angle=90]{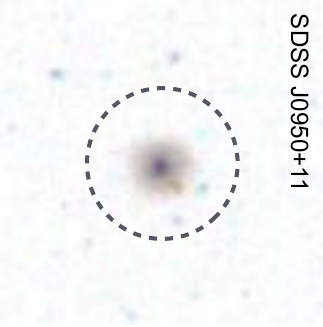}\quad
    \includegraphics[clip=true,trim=5.5cm 3.8cm 4cm 2cm,width=0.28\textwidth,angle=0]{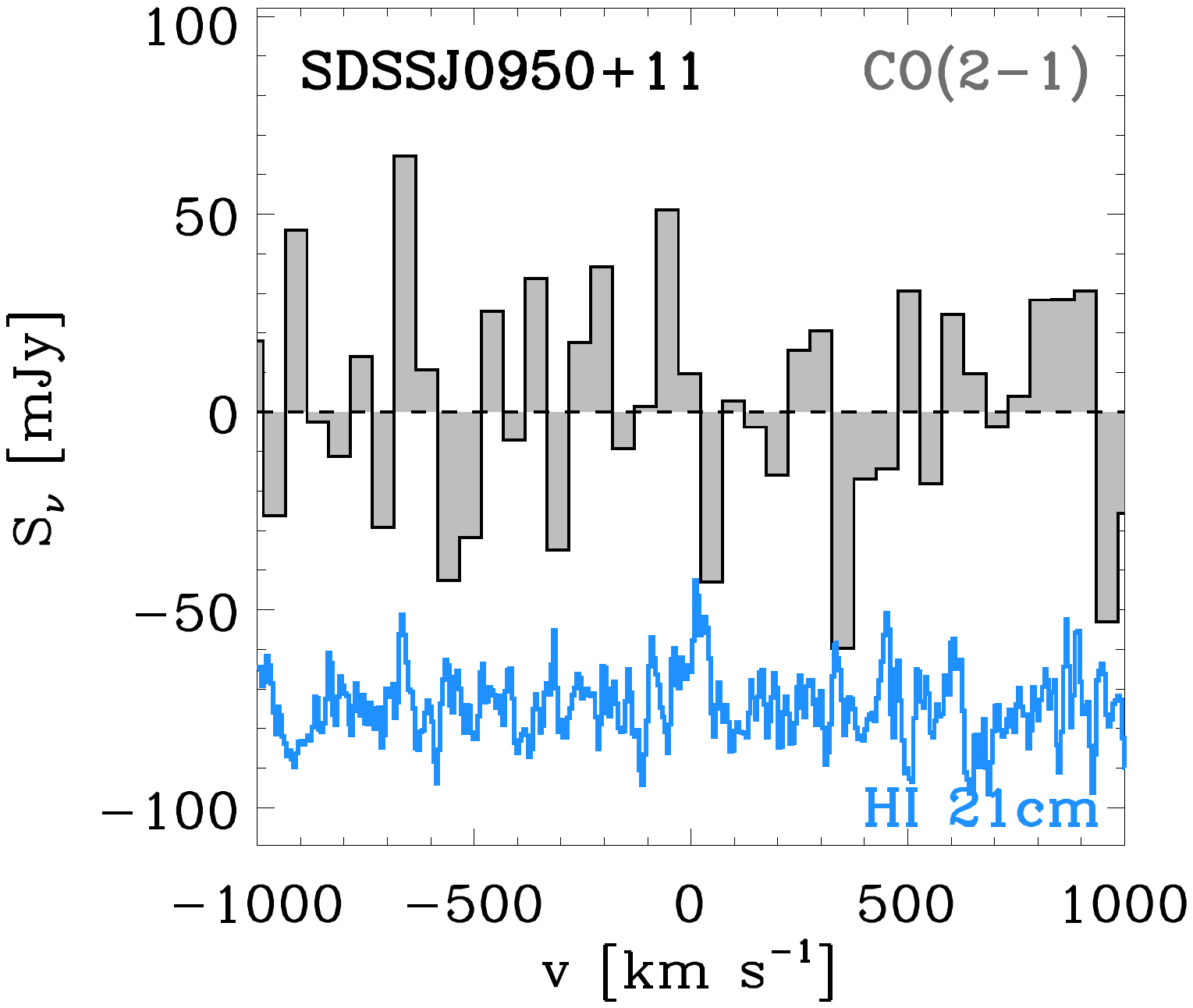}\quad
    \includegraphics[clip=true,trim=-0.4cm 0cm 0cm 0cm,width=0.18\textwidth,angle=90]{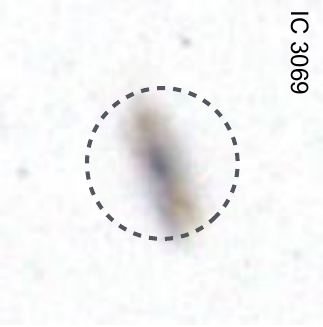}\quad
    \includegraphics[clip=true,trim=5.5cm 3.8cm 4cm 2cm,width=0.28\textwidth,angle=0]{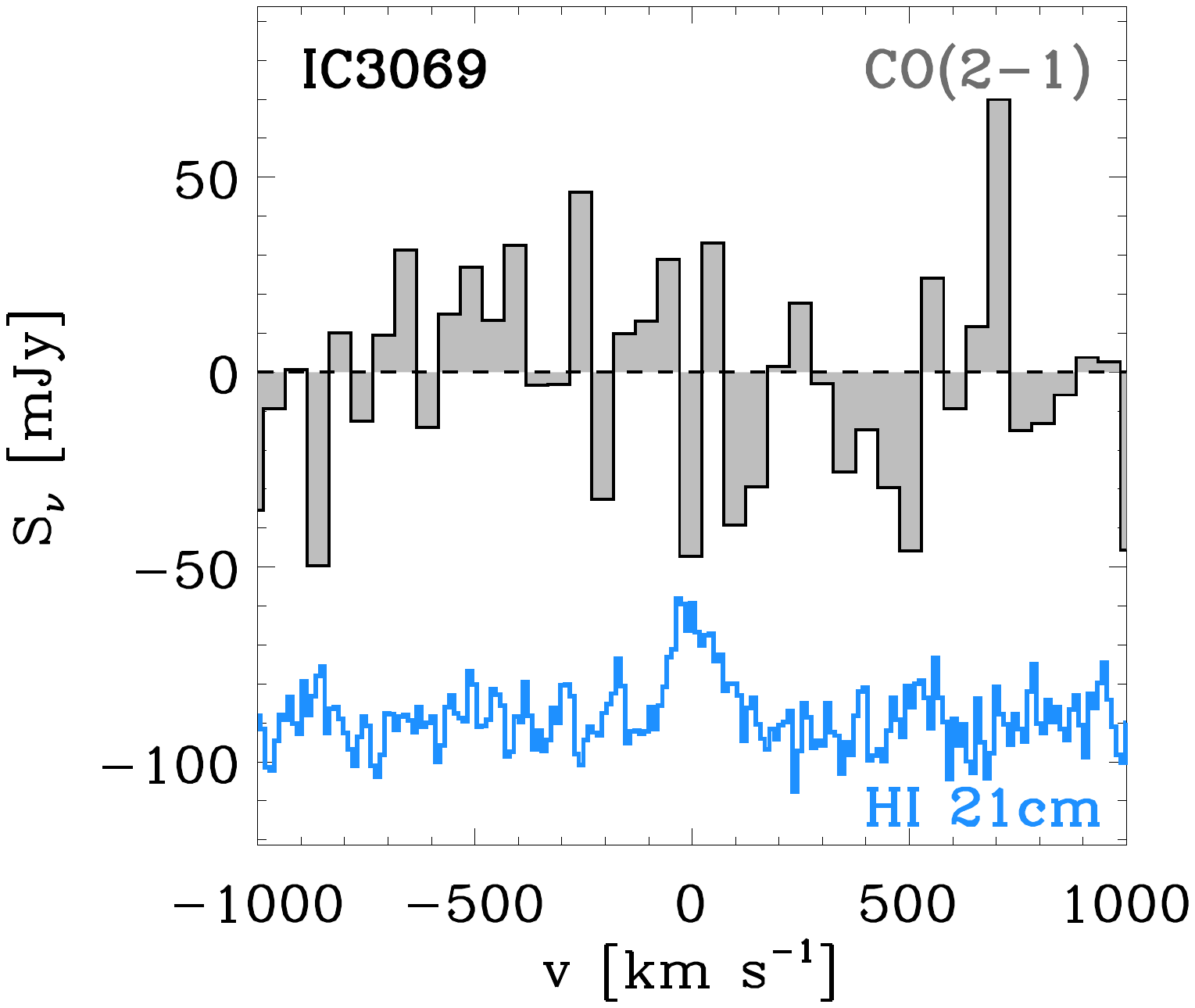}\\
      \includegraphics[clip=true,trim=-0.4cm 0cm 0cm 0cm,width=0.18\textwidth,angle=90]{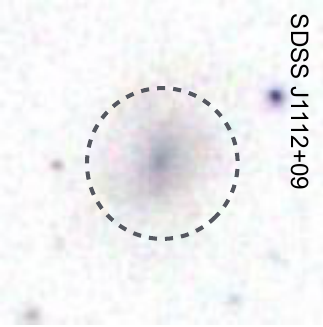}\quad
    \includegraphics[clip=true,trim=5.5cm 3.8cm 4cm 2cm,width=0.28\textwidth,angle=0]{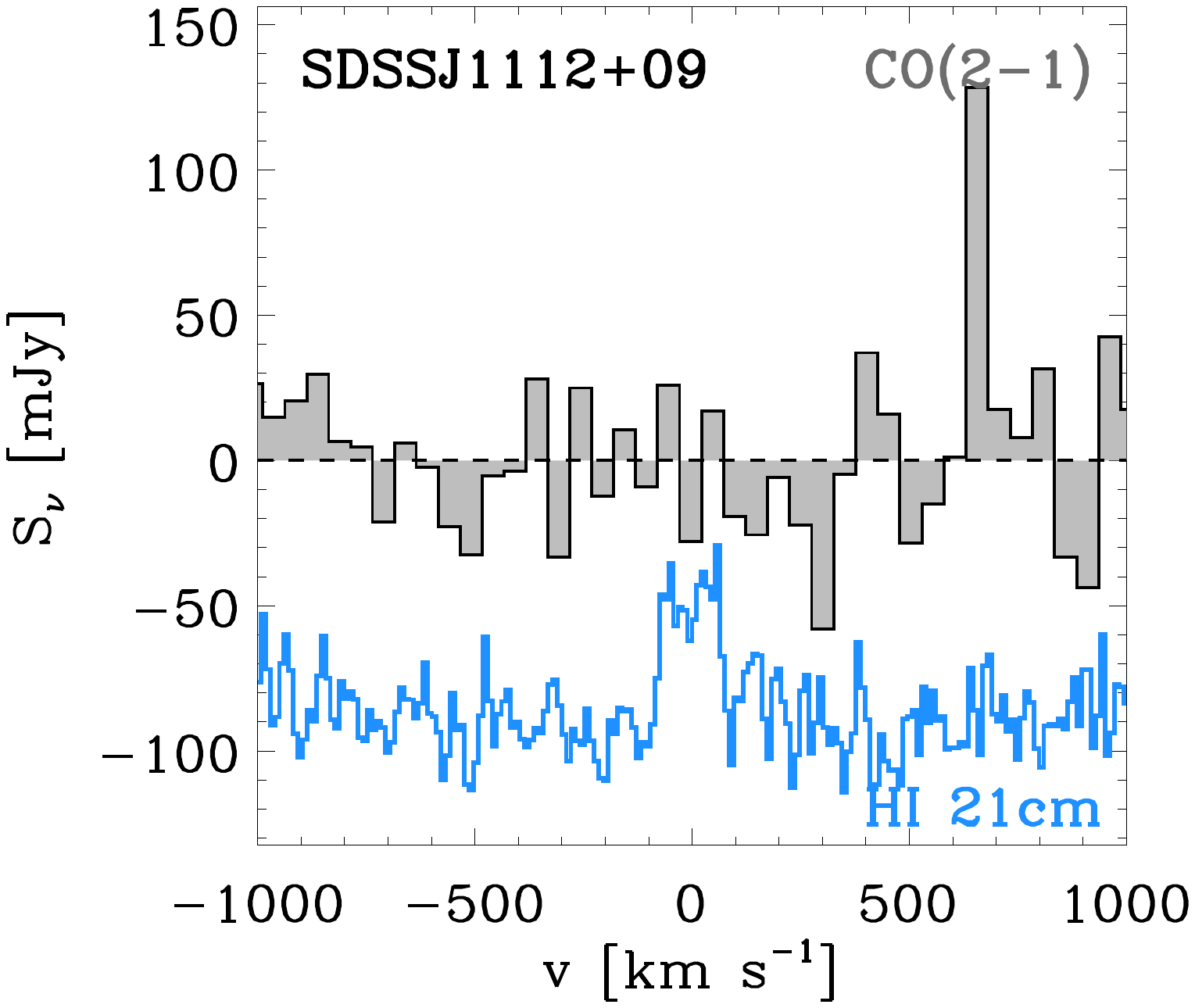}\quad
    \includegraphics[clip=true,trim=-0.4cm 0cm 0cm 0cm,width=0.18\textwidth,angle=90]{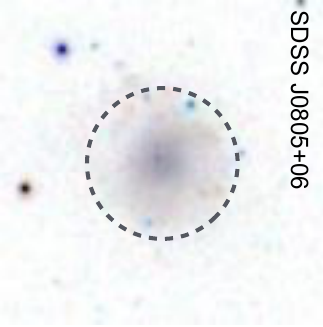}\quad
    \includegraphics[clip=true,trim=5.5cm 3.8cm 4cm 2cm,width=0.28\textwidth,angle=0]{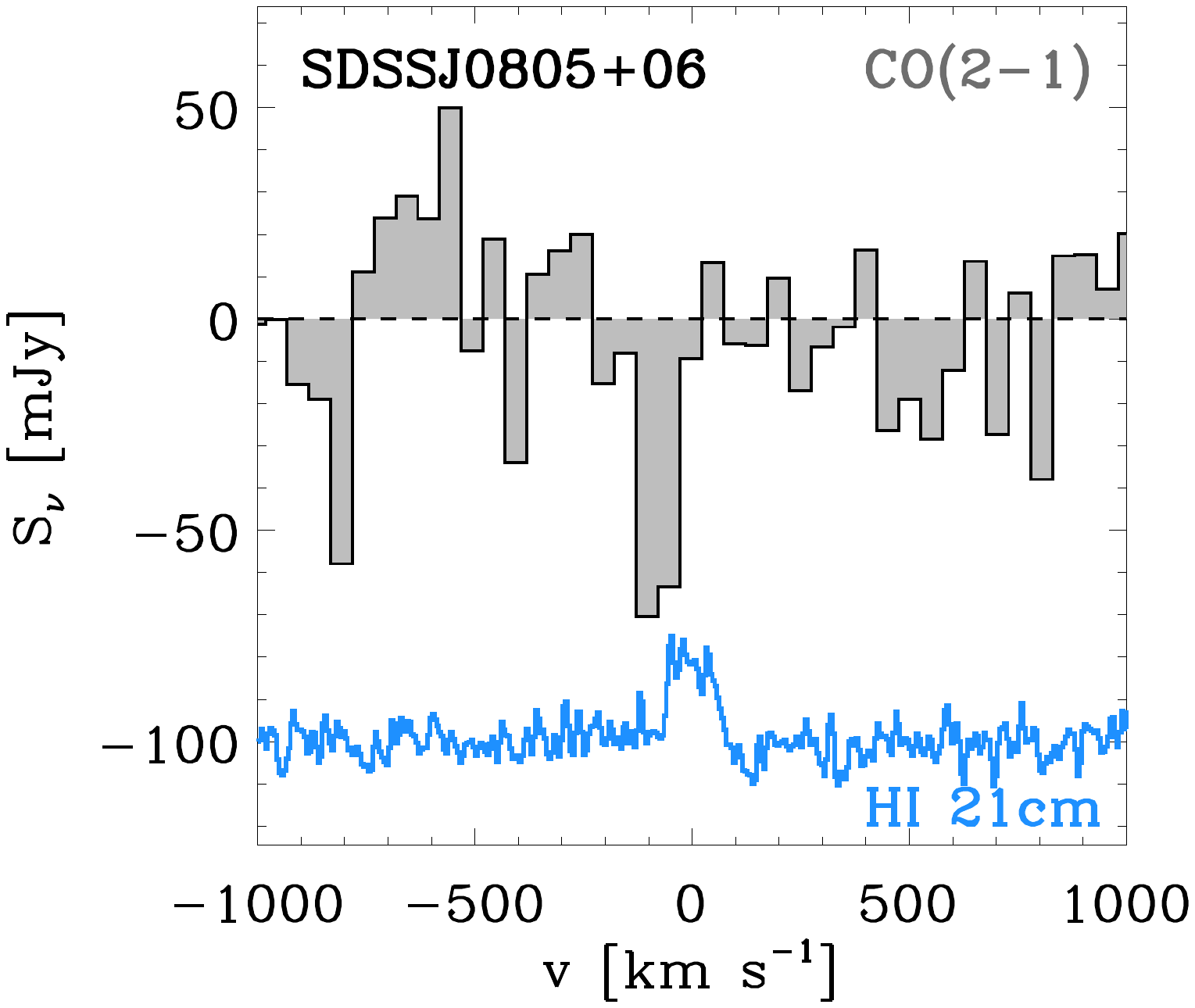}\\
     \caption{{\it Left panels:} SDSS cutout images ({\it g r i} composite, field of view = $60\arcsec\times60\arcsec$, scale = 0.5$\arcsec$/pixel, north is up and west is right) of ALLSMOG galaxies, showing the 27$''$ APEX beam at 230 GHz. {\it Right panels:} APEX CO(2-1) baseline-subtracted spectra, rebinned in bins of $\delta \varv=50$~\kms (SDSSJ1328-0202, MCG00-34-038, UGC08526, CGCG017-017, SDSSJ0937+0927, SDSSJ0950+1118, IC3069, SDSSJ1112+0931, SDSSJ0805+0659), or 25~\kms (NGC2936), depending on the width and S/N of the line. The corresponding H{\sc i}~21cm spectra are also shown for comparison, after having been renormalised for visualisation purposes (H{\sc i} references are given in Table~\ref{table:HI_parameters}).}
   \label{fig:spectra4}
\end{figure*}
\begin{figure*}[tbp]
\centering
    \includegraphics[clip=true,trim=-0.4cm 0cm 0cm 0cm,width=0.18\textwidth,angle=90]{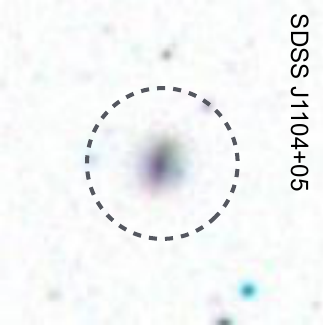}\quad
    \includegraphics[clip=true,trim=5.5cm 3.8cm 4cm 2cm,width=0.28\textwidth,angle=0]{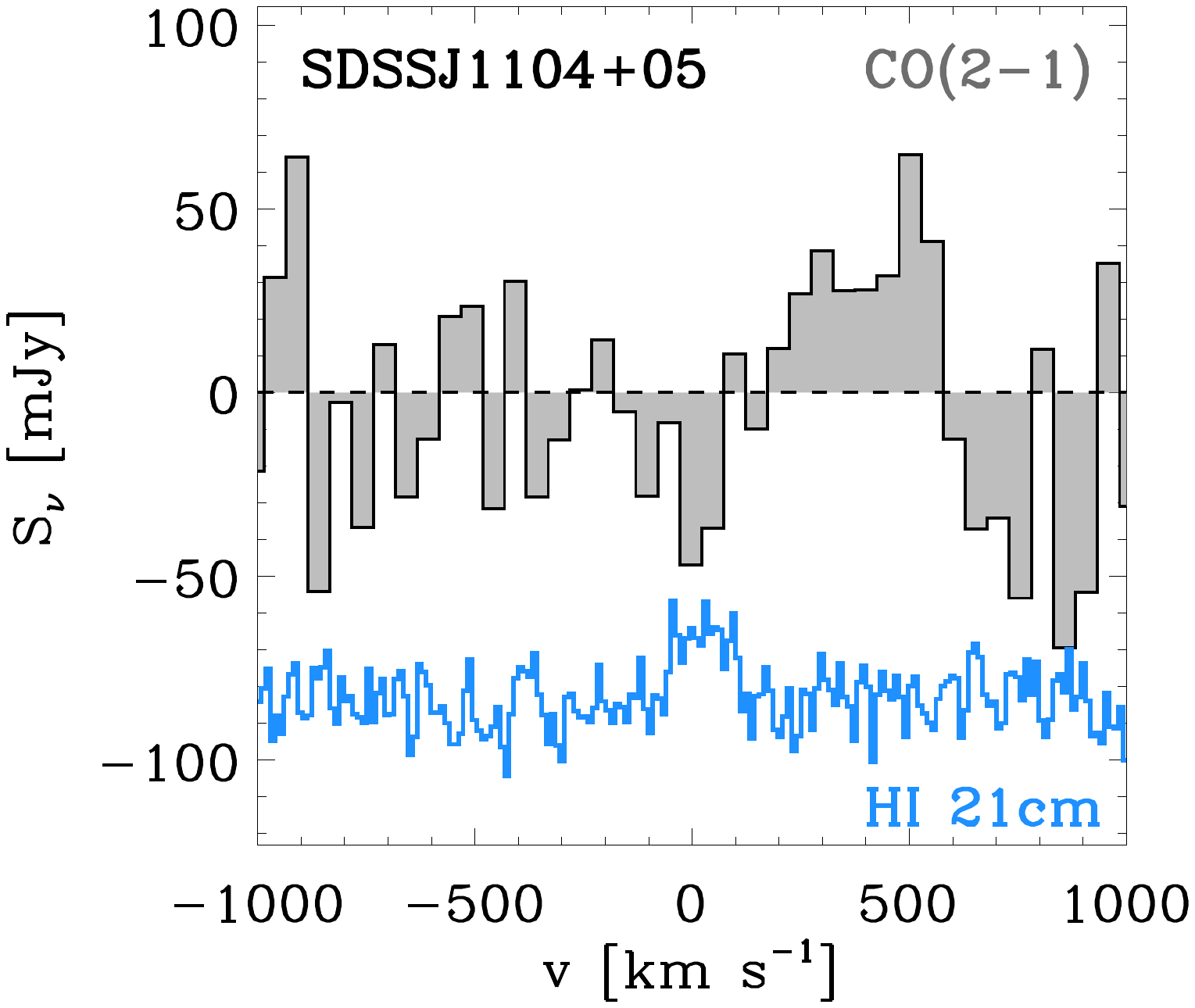}\quad
    \includegraphics[clip=true,trim=-0.4cm 0cm 0cm 0cm,width=0.18\textwidth,angle=90]{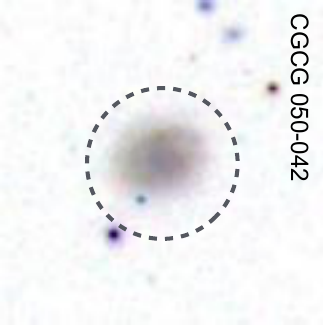}\quad
    \includegraphics[clip=true,trim=5.5cm 3.8cm 4cm 2cm,width=0.28\textwidth,angle=0]{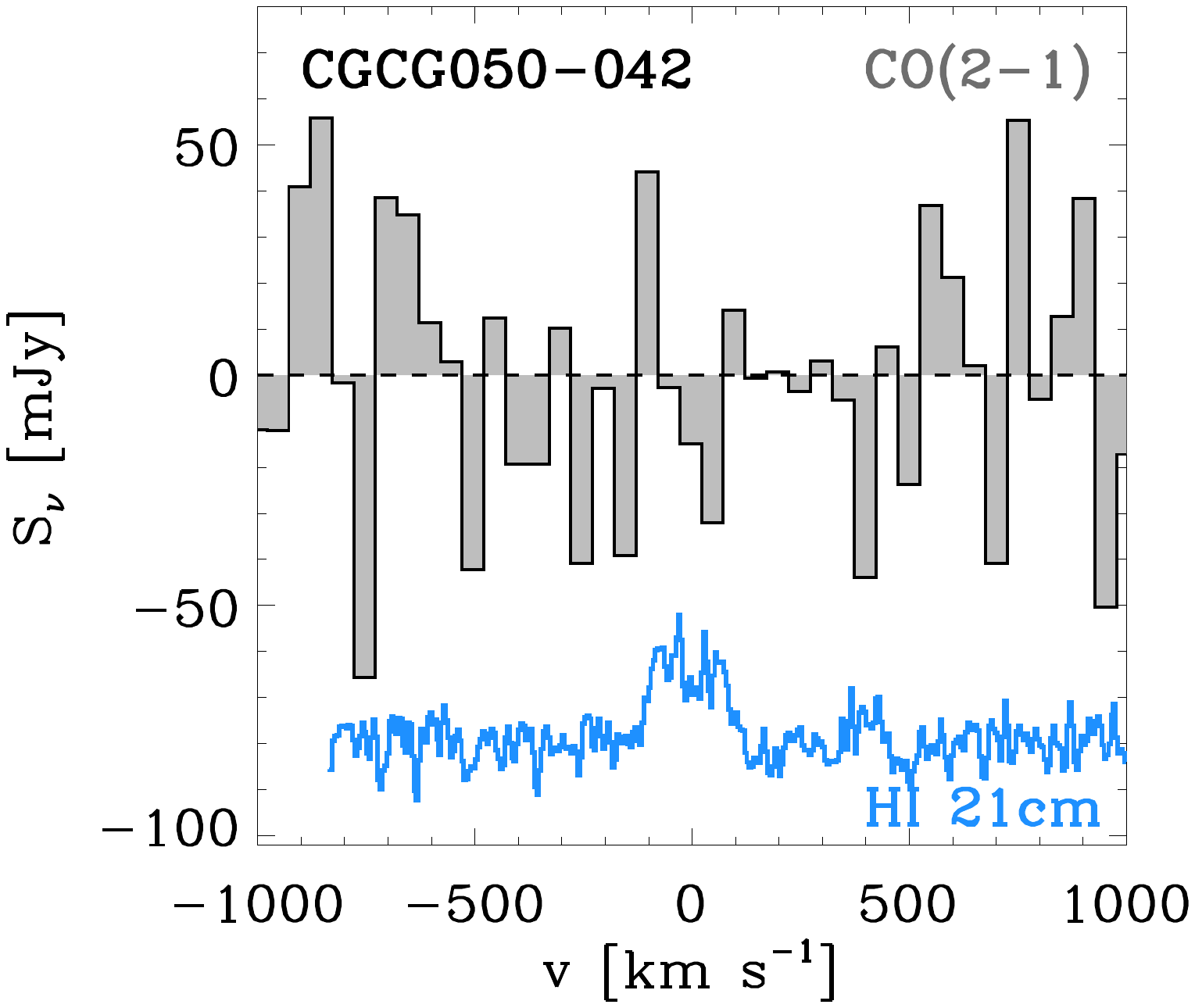}\\
    \includegraphics[clip=true,trim=-0.4cm 0cm 0cm 0cm,width=0.18\textwidth,angle=90]{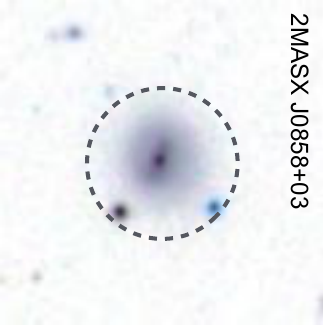}\quad
    \includegraphics[clip=true,trim=5.5cm 3.8cm 4cm 2cm,width=0.28\textwidth,angle=0]{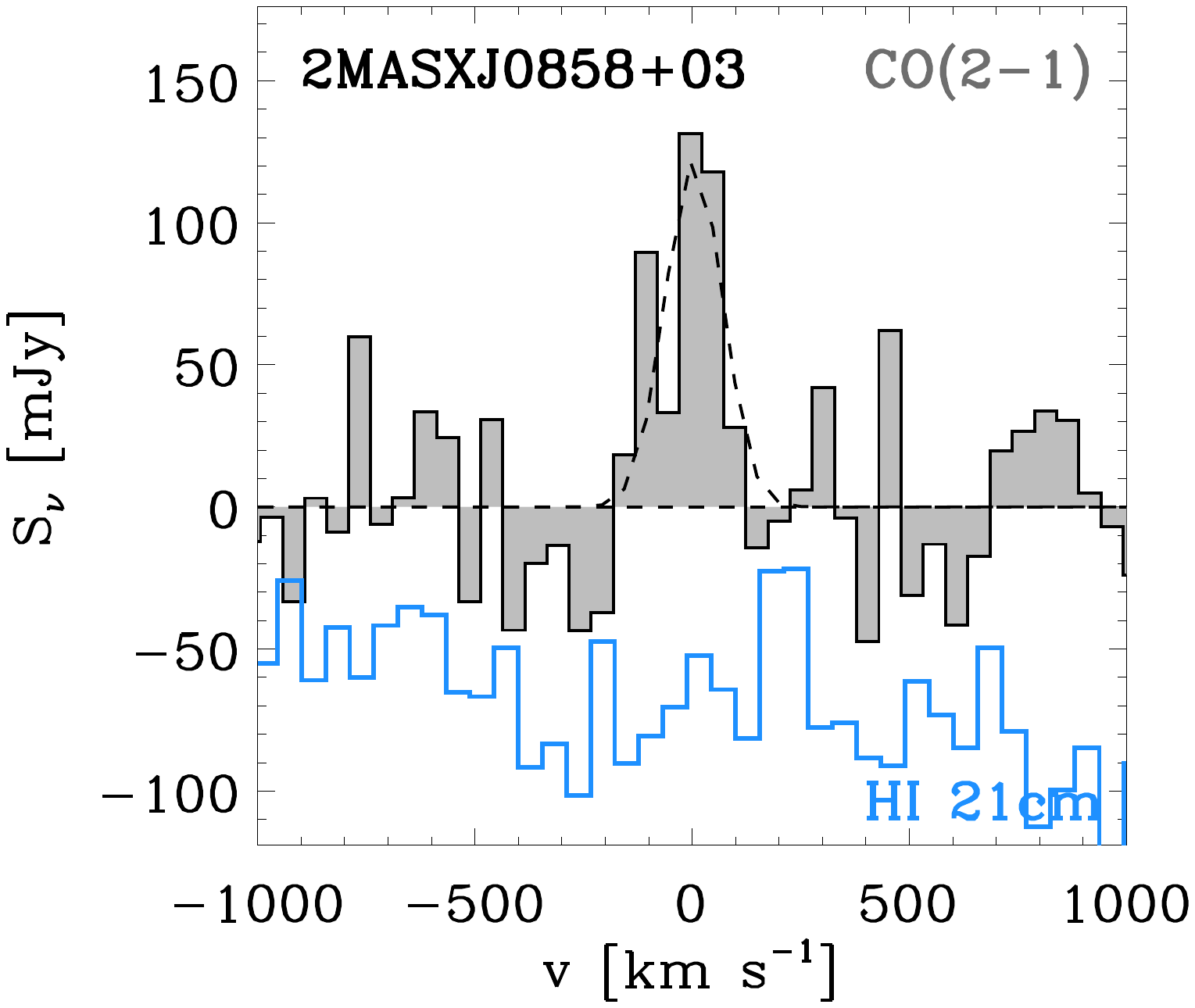}\quad
    \includegraphics[clip=true,trim=-0.4cm 0cm 0cm 0cm,width=0.18\textwidth,angle=90]{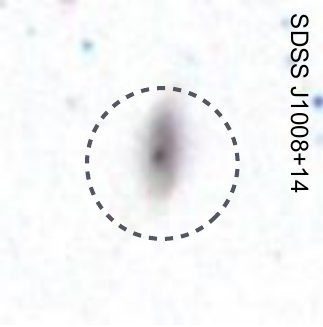}\quad
    \includegraphics[clip=true,trim=5.5cm 3.8cm 4cm 2cm,width=0.28\textwidth,angle=0]{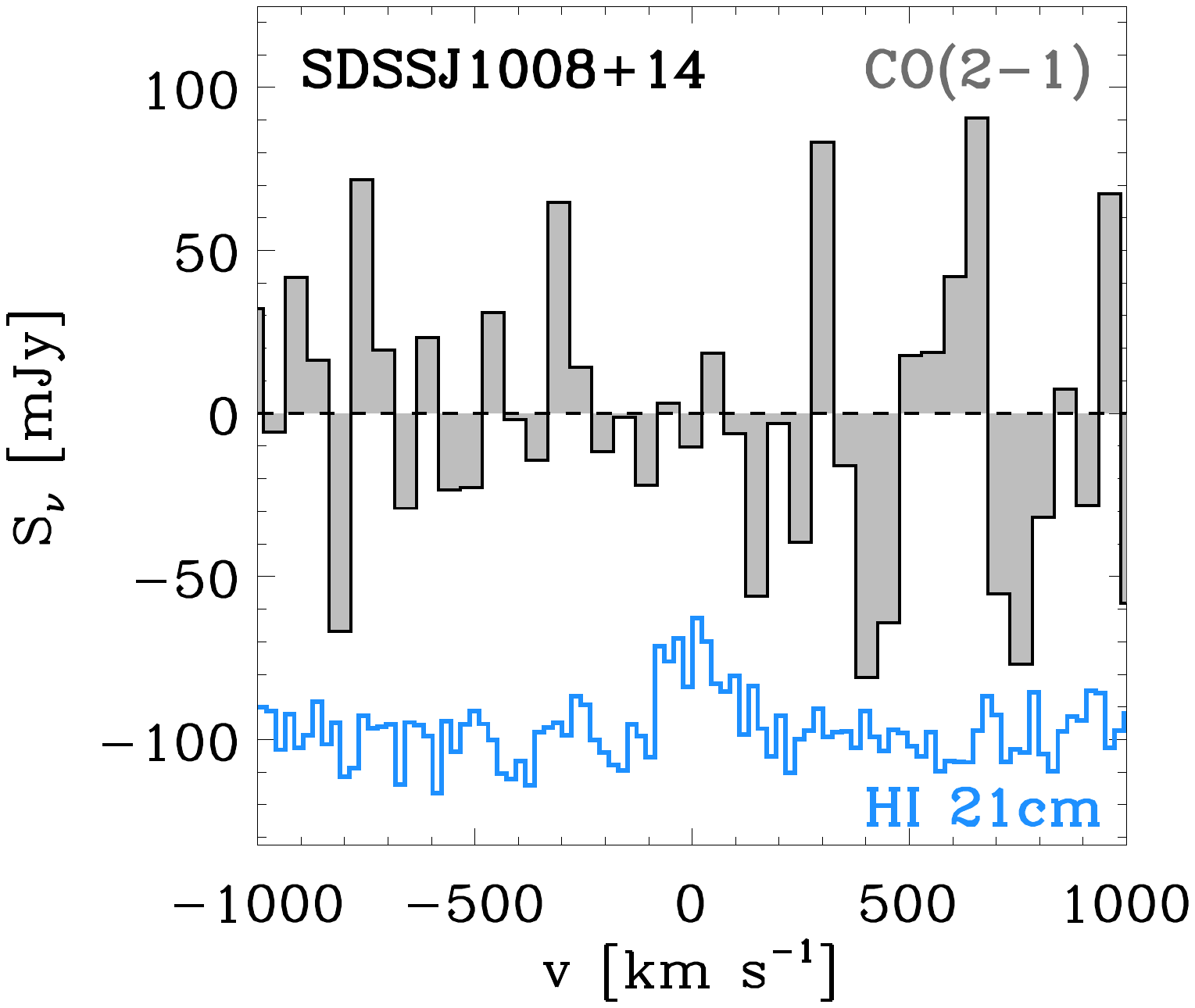}\\
      \includegraphics[clip=true,trim=-0.4cm 0cm 0cm 0cm,width=0.18\textwidth,angle=90]{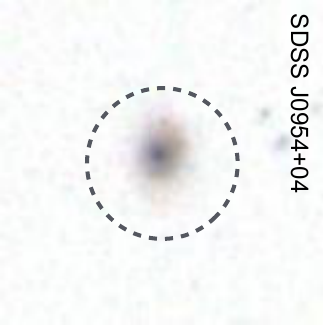}\quad
    \includegraphics[clip=true,trim=5.5cm 3.8cm 4cm 2cm,width=0.28\textwidth,angle=0]{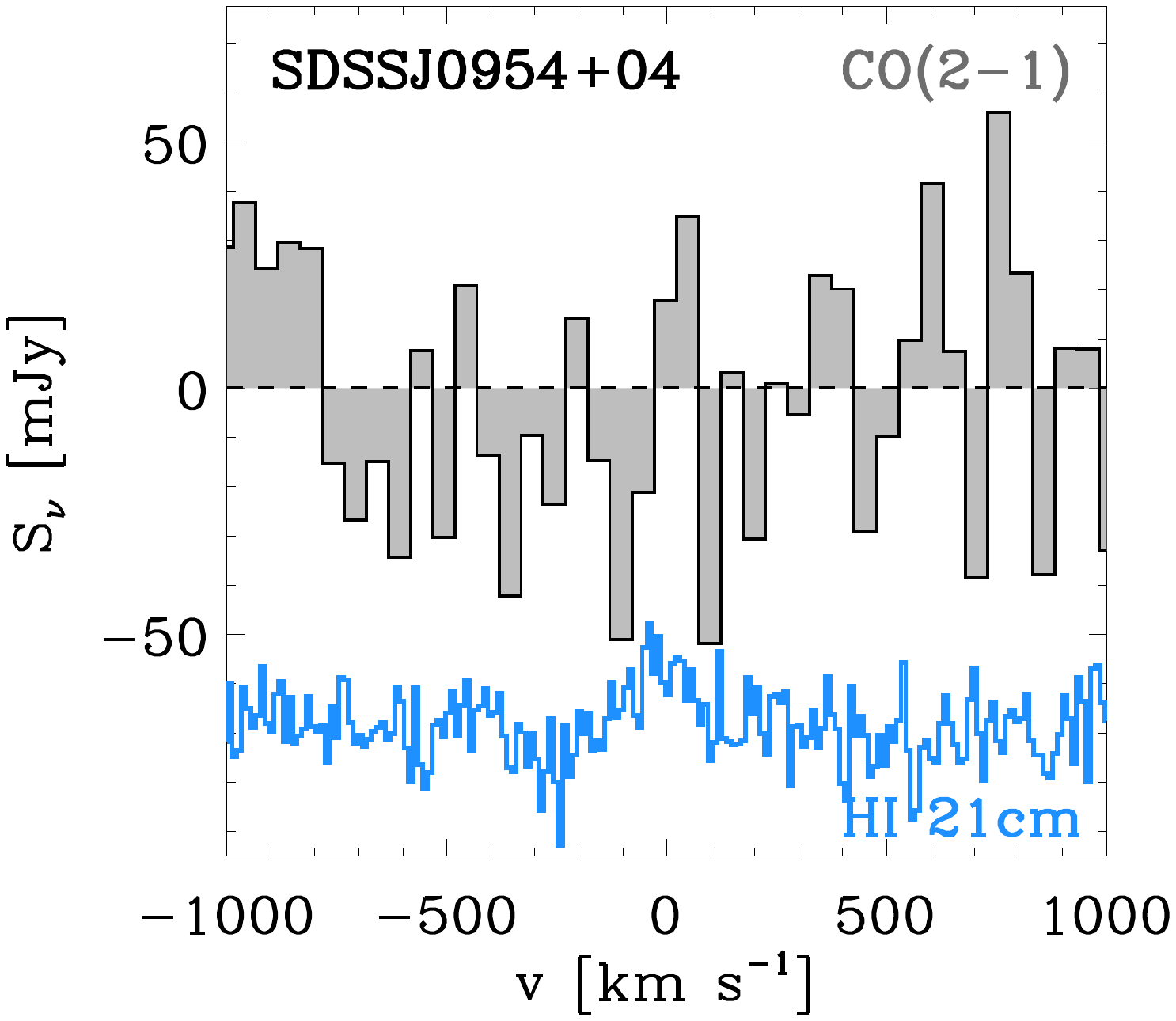}\quad
    \includegraphics[clip=true,trim=-0.4cm 0cm 0cm 0cm,width=0.18\textwidth,angle=90]{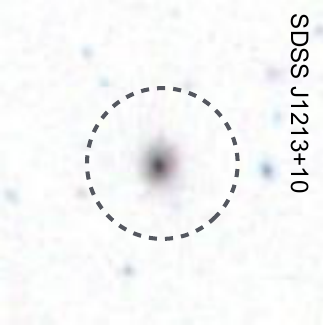}\quad
    \includegraphics[clip=true,trim=5.5cm 3.8cm 4cm 2cm,width=0.28\textwidth,angle=0]{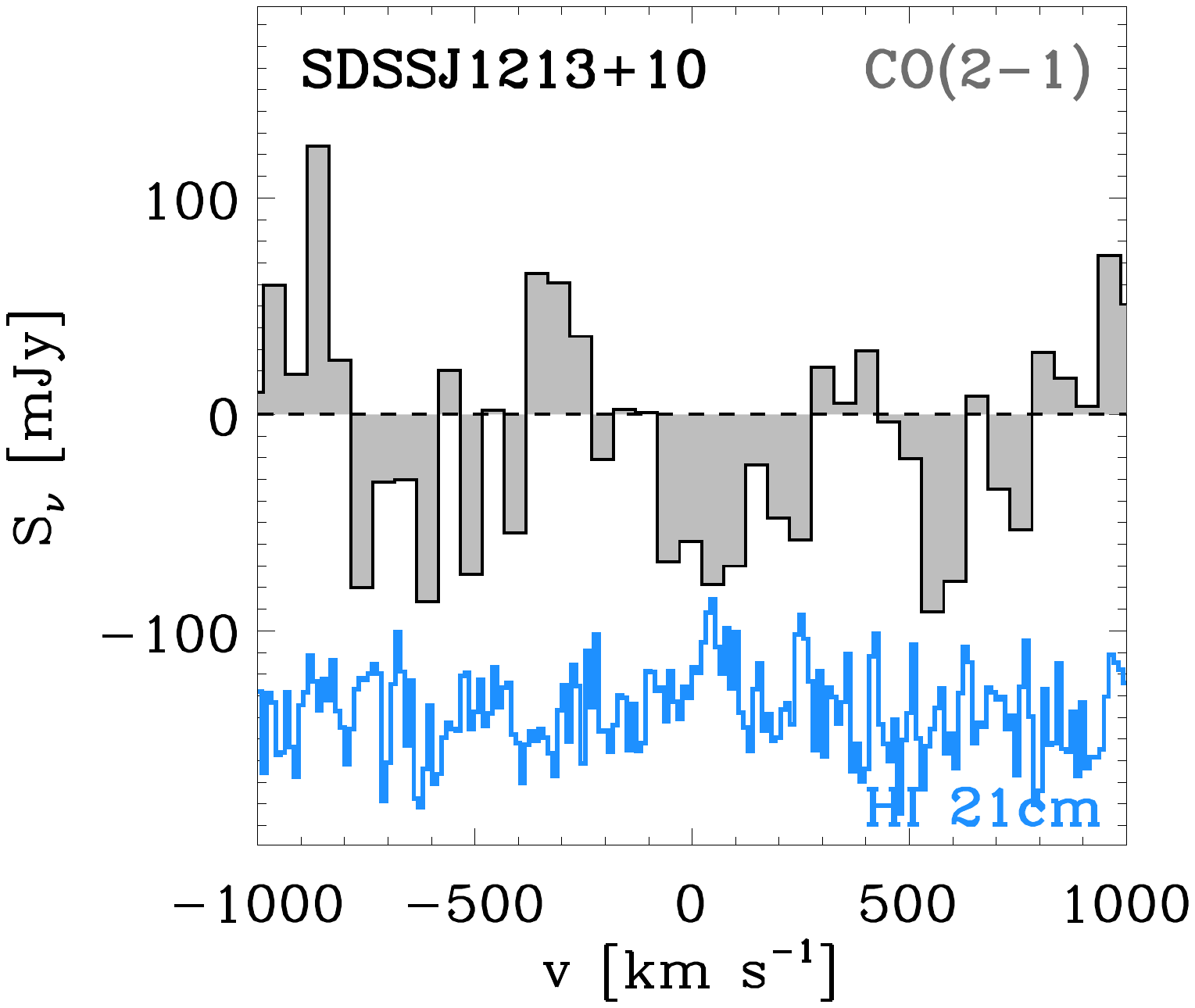}\\
      \includegraphics[clip=true,trim=-0.4cm 0cm 0cm 0cm,width=0.18\textwidth,angle=90]{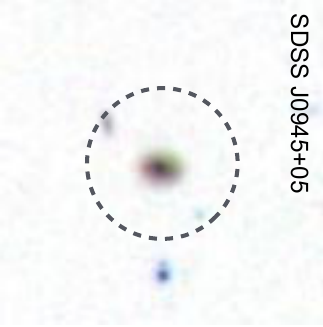}\quad
    \includegraphics[clip=true,trim=5.5cm 3.8cm 4cm 2cm,width=0.28\textwidth,angle=0]{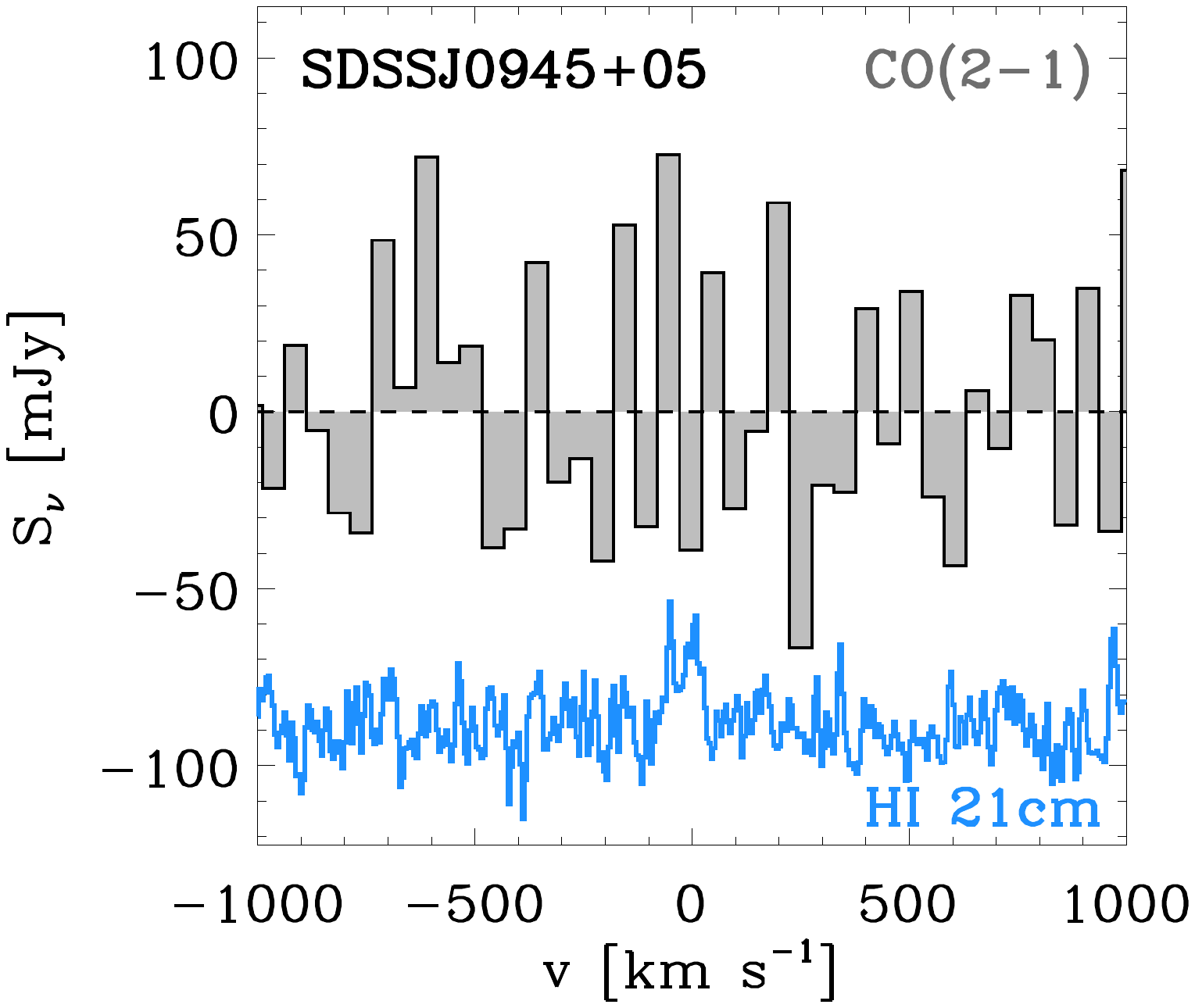}\quad
    \includegraphics[clip=true,trim=-0.4cm 0cm 0cm 0cm,width=0.18\textwidth,angle=90]{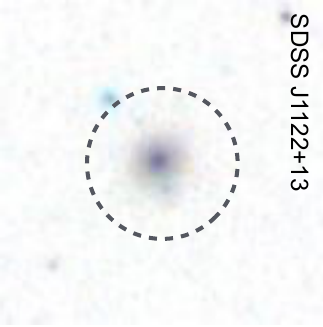}\quad
    \includegraphics[clip=true,trim=5.5cm 3.8cm 4cm 2cm,width=0.28\textwidth,angle=0]{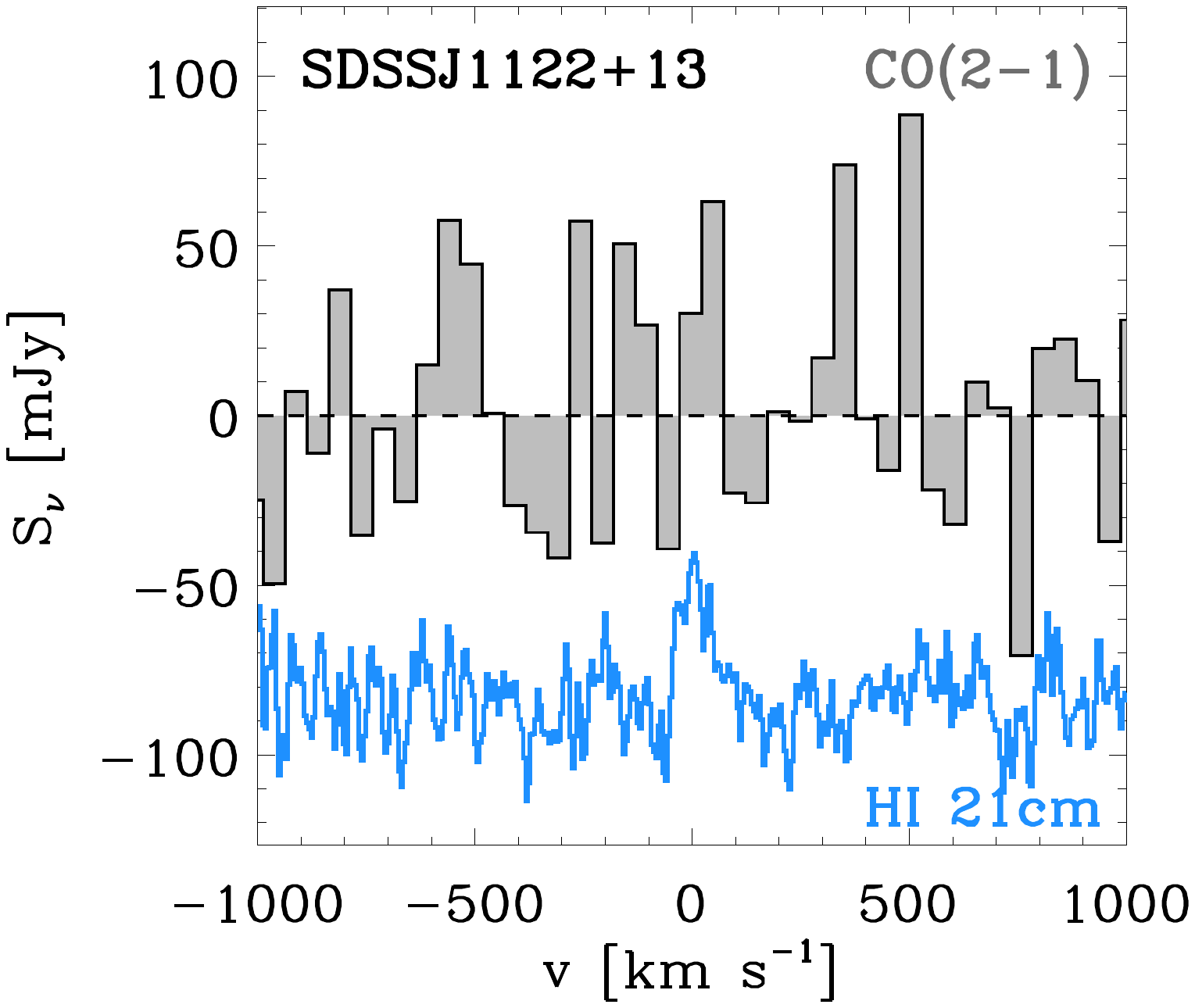}\\
      \includegraphics[clip=true,trim=-0.4cm 0cm 0cm 0cm,width=0.18\textwidth,angle=90]{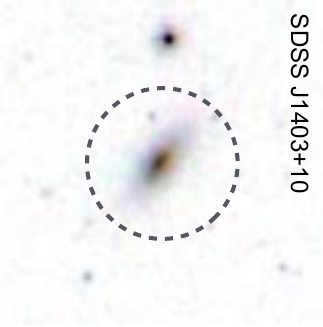}\quad
    \includegraphics[clip=true,trim=5.5cm 3.8cm 4cm 2cm,width=0.28\textwidth,angle=0]{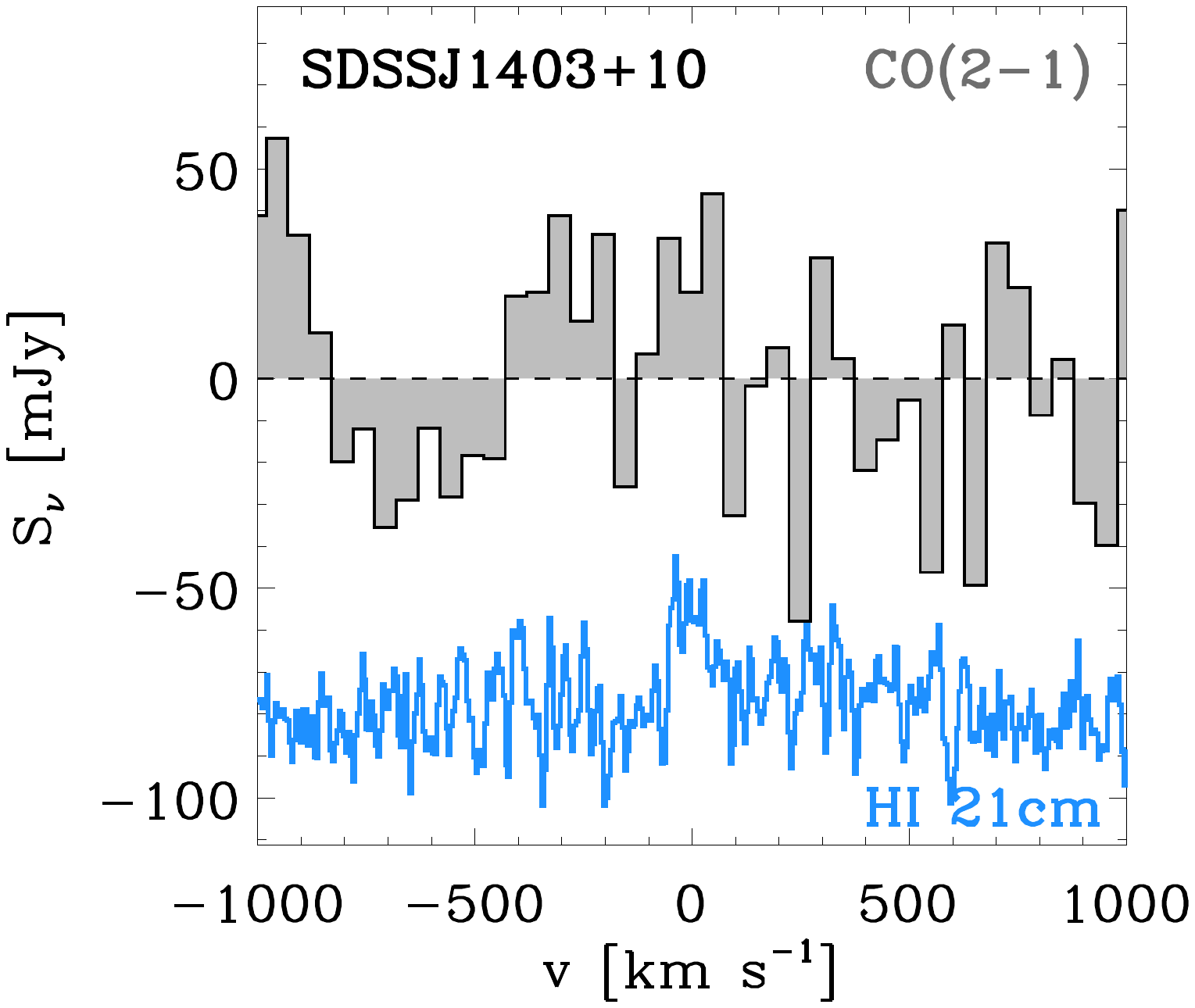}\quad
    \includegraphics[clip=true,trim=-0.4cm 0cm 0cm 0cm,width=0.18\textwidth,angle=90]{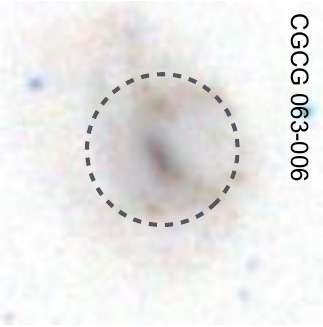}\quad
    \includegraphics[clip=true,trim=5.5cm 3.8cm 4cm 2cm,width=0.28\textwidth,angle=0]{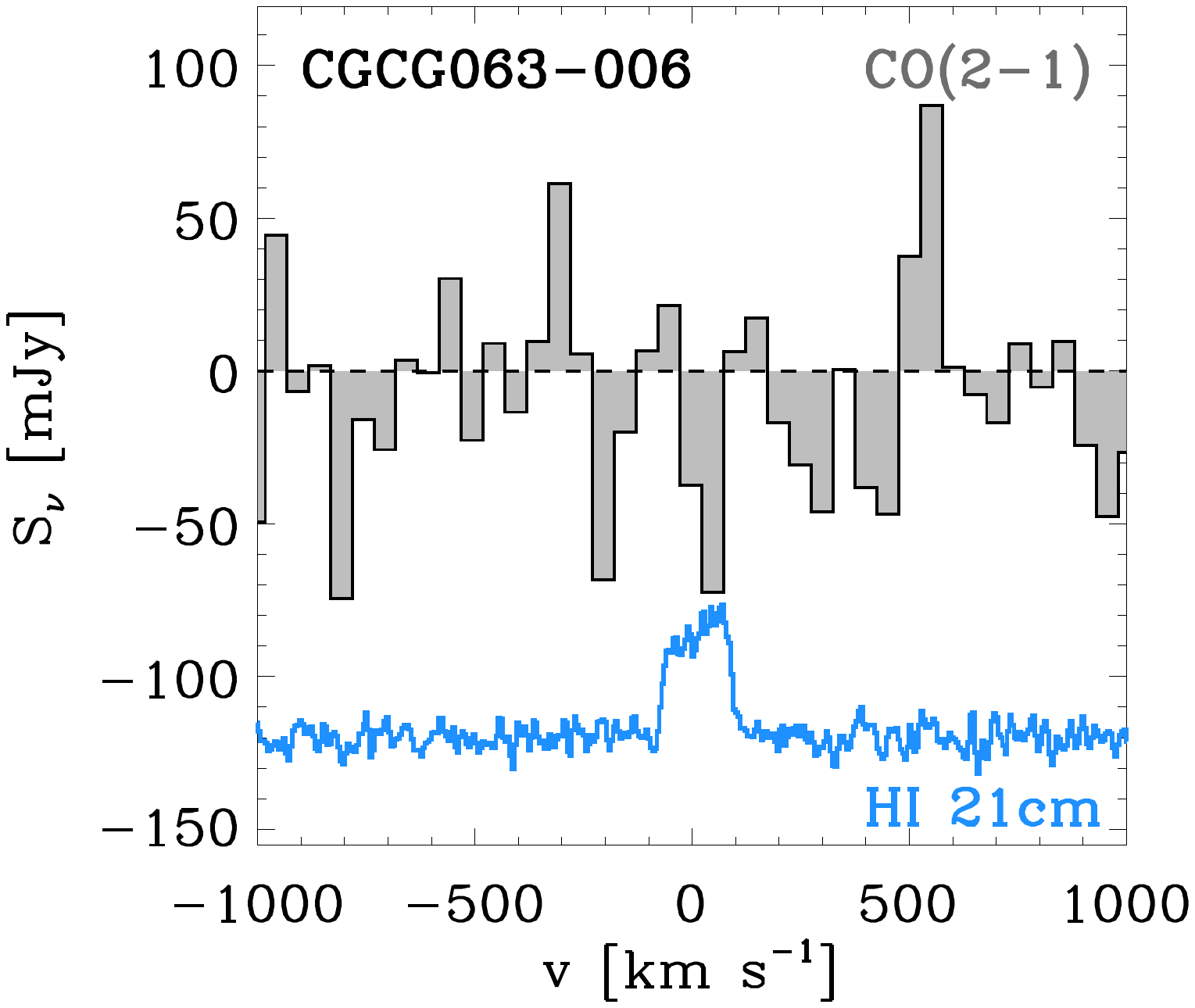}\\
     \caption{{\it Left panels:} SDSS cutout images ({\it g r i} composite, field of view = $60\arcsec\times60\arcsec$, scale = 0.5$\arcsec$/pixel, north is up and west is right) of ALLSMOG galaxies, showing the 27$''$ APEX beam at 230 GHz. {\it Right panels:} APEX CO(2-1) baseline-subtracted spectra, rebinned in bins of $\delta \varv=50$~\kms. The corresponding H{\sc i}~21cm spectra are also shown for comparison, after having been renormalised for visualisation purposes (H{\sc i} references are given in Table~\ref{table:HI_parameters}).}
   \label{fig:spectra5}
\end{figure*}
\begin{figure*}[tbp]
\centering
    \includegraphics[clip=true,trim=-0.4cm 0cm 0cm 0cm,width=0.18\textwidth,angle=90]{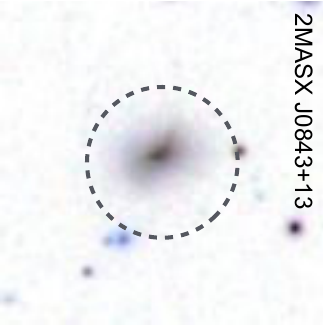}\quad
    \includegraphics[clip=true,trim=5.5cm 3.8cm 4cm 2cm,width=0.28\textwidth,angle=0]{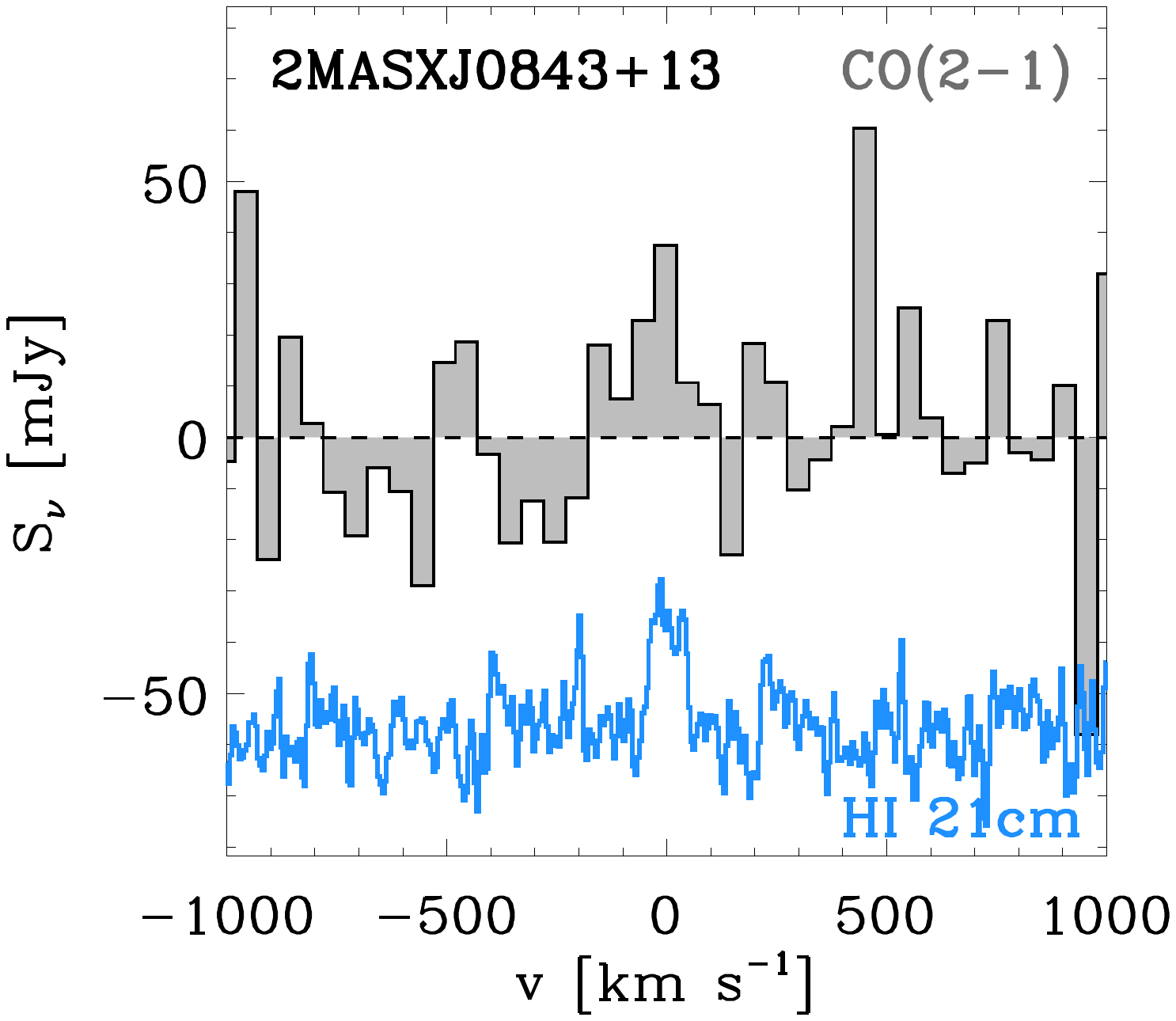}\quad
    \includegraphics[clip=true,trim=-0.4cm 0cm 0cm 0cm,width=0.18\textwidth,angle=90]{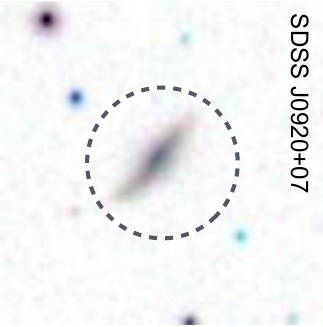}\quad
    \includegraphics[clip=true,trim=5.5cm 3.8cm 4cm 2cm,width=0.28\textwidth,angle=0]{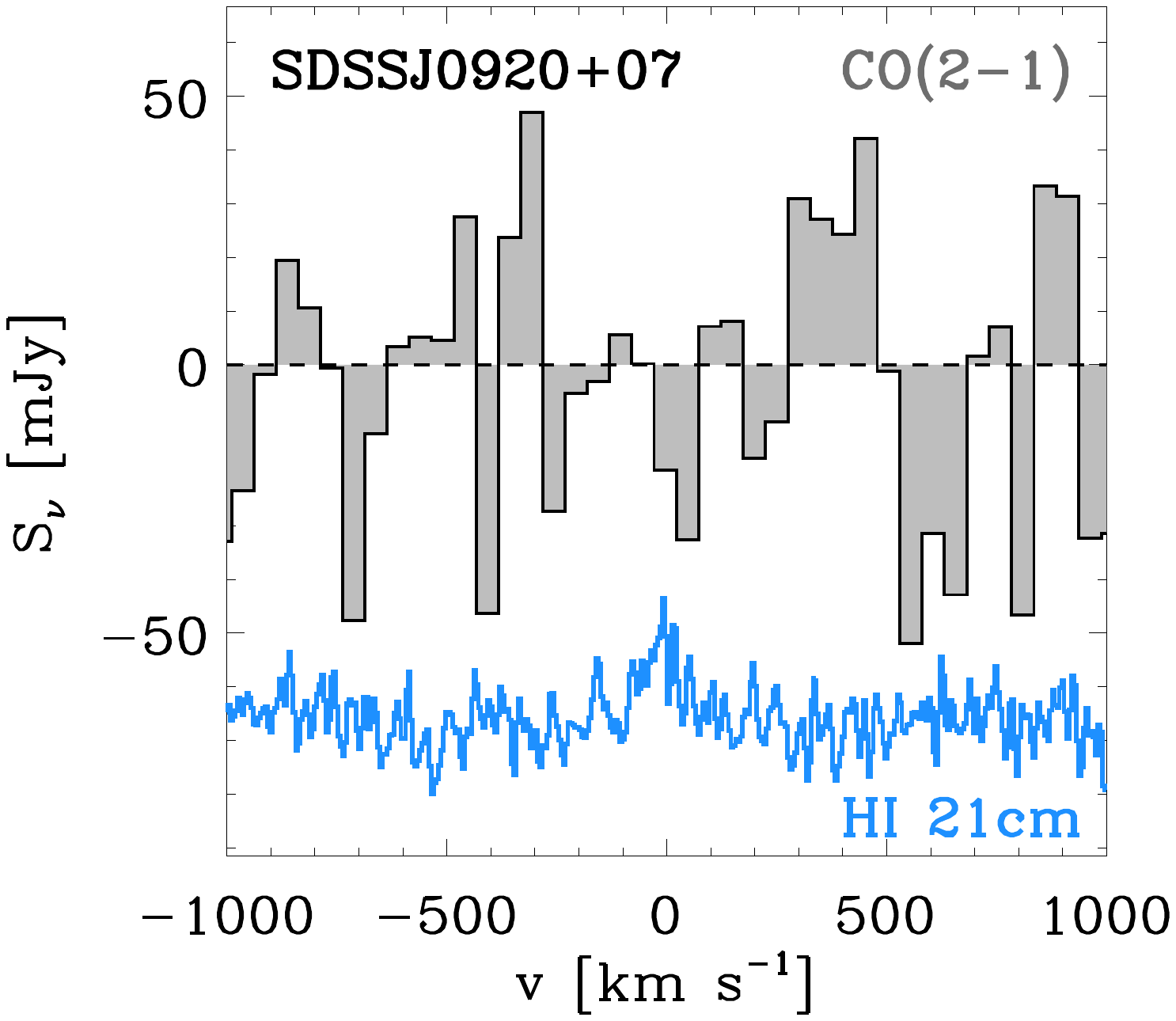}\\
    \includegraphics[clip=true,trim=-0.4cm 0cm 0cm 0cm,width=0.18\textwidth,angle=90]{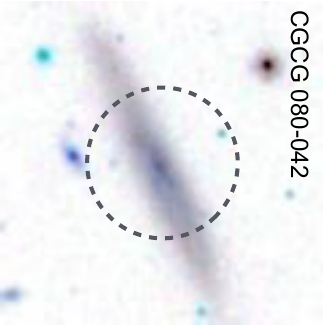}\quad
    \includegraphics[clip=true,trim=5.5cm 3.8cm 4cm 2cm,width=0.28\textwidth,angle=0]{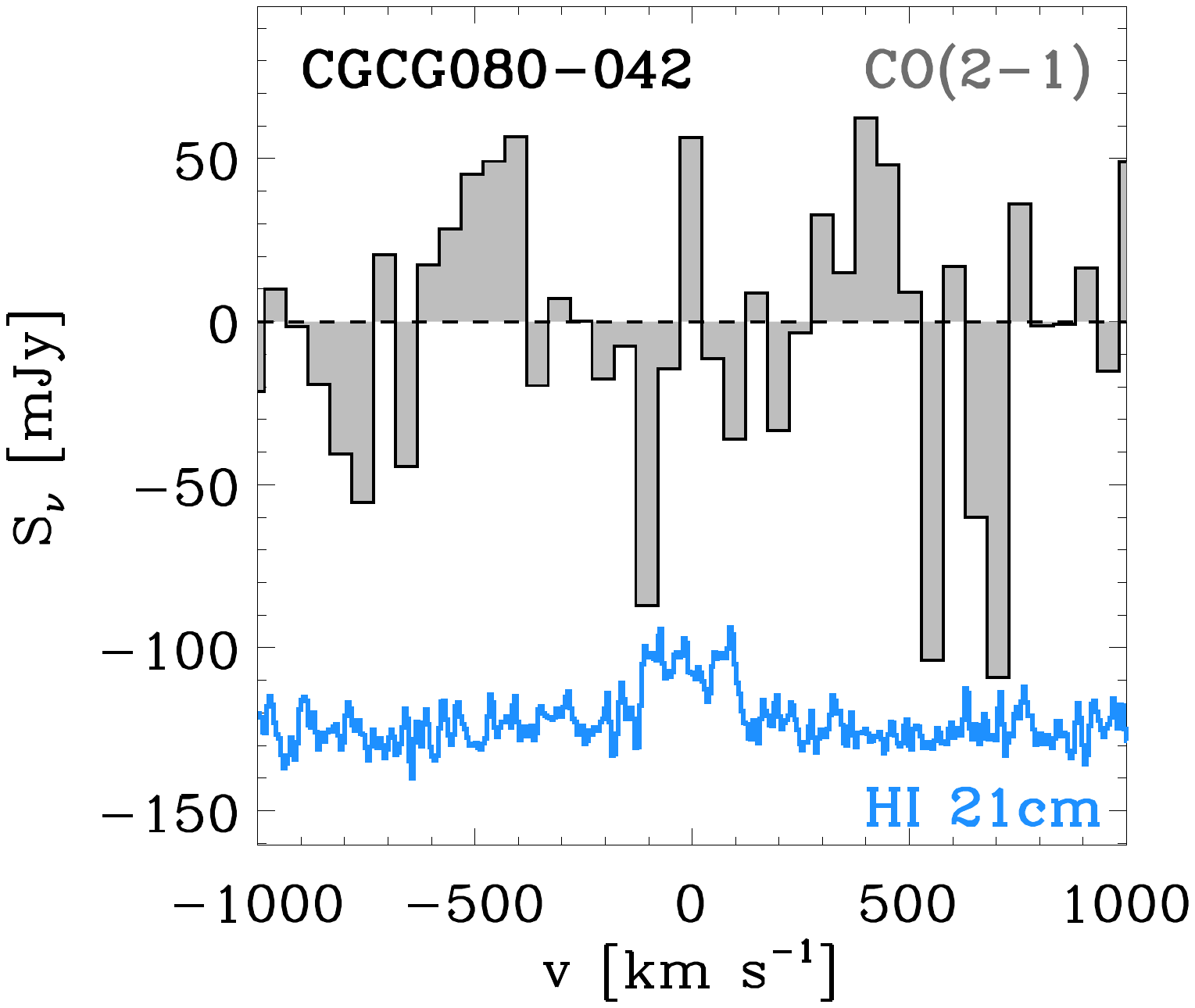}\quad
    \includegraphics[clip=true,trim=-0.4cm 0cm 0cm 0cm,width=0.18\textwidth,angle=90]{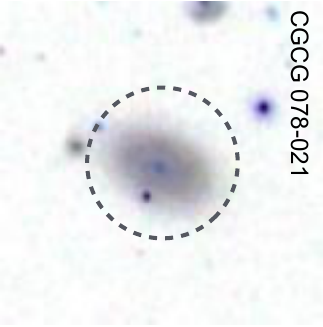}\quad
    \includegraphics[clip=true,trim=5.5cm 3.8cm 4cm 2cm,width=0.28\textwidth,angle=0]{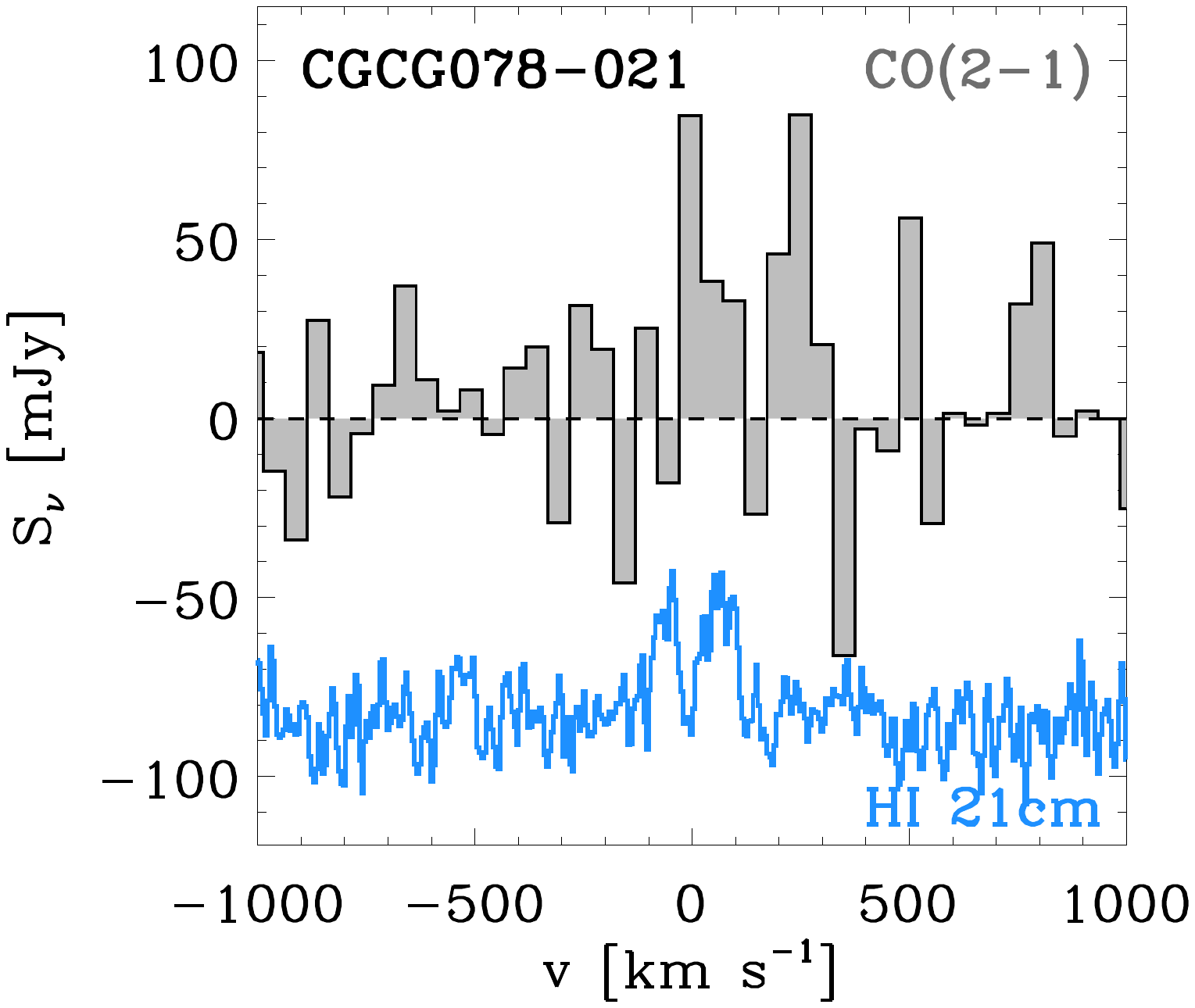}\\
      \includegraphics[clip=true,trim=-0.4cm 0cm 0cm 0cm,width=0.18\textwidth,angle=90]{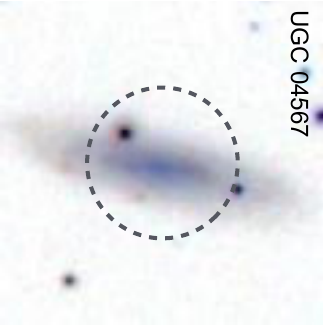}\quad
    \includegraphics[clip=true,trim=5.5cm 3.8cm 4cm 2cm,width=0.28\textwidth,angle=0]{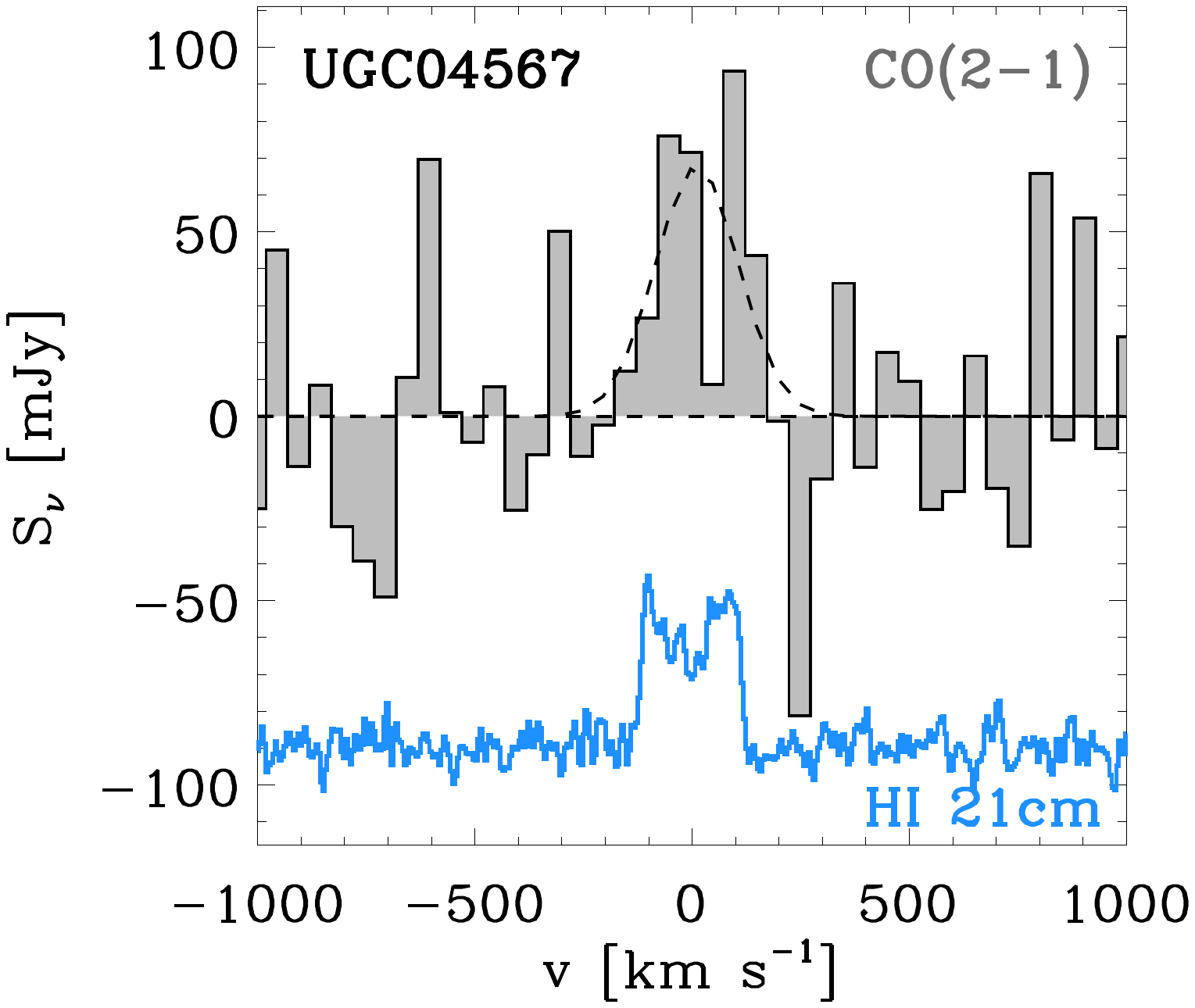}\quad
    \includegraphics[clip=true,trim=-0.4cm 0cm 0cm 0cm,width=0.18\textwidth,angle=90]{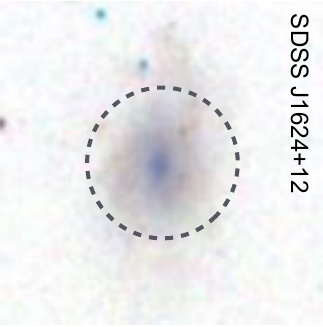}\quad
    \includegraphics[clip=true,trim=5.5cm 3.8cm 4cm 2cm,width=0.28\textwidth,angle=0]{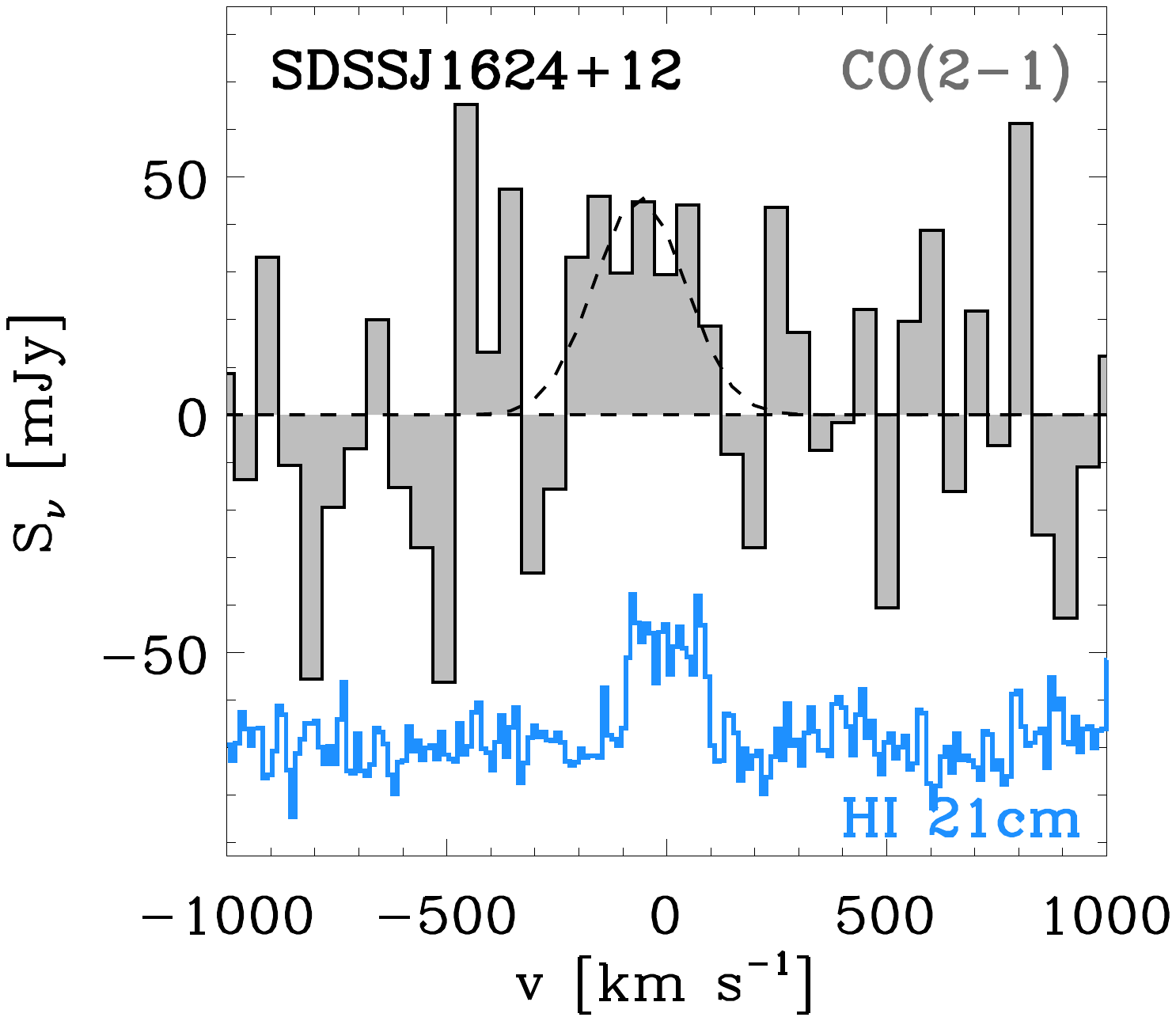}\\
      \includegraphics[clip=true,trim=-0.4cm 0cm 0cm 0cm,width=0.18\textwidth,angle=90]{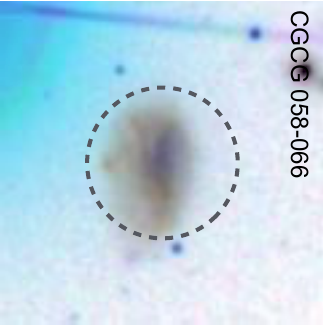}\quad
    \includegraphics[clip=true,trim=5.5cm 3.8cm 4cm 2cm,width=0.28\textwidth,angle=0]{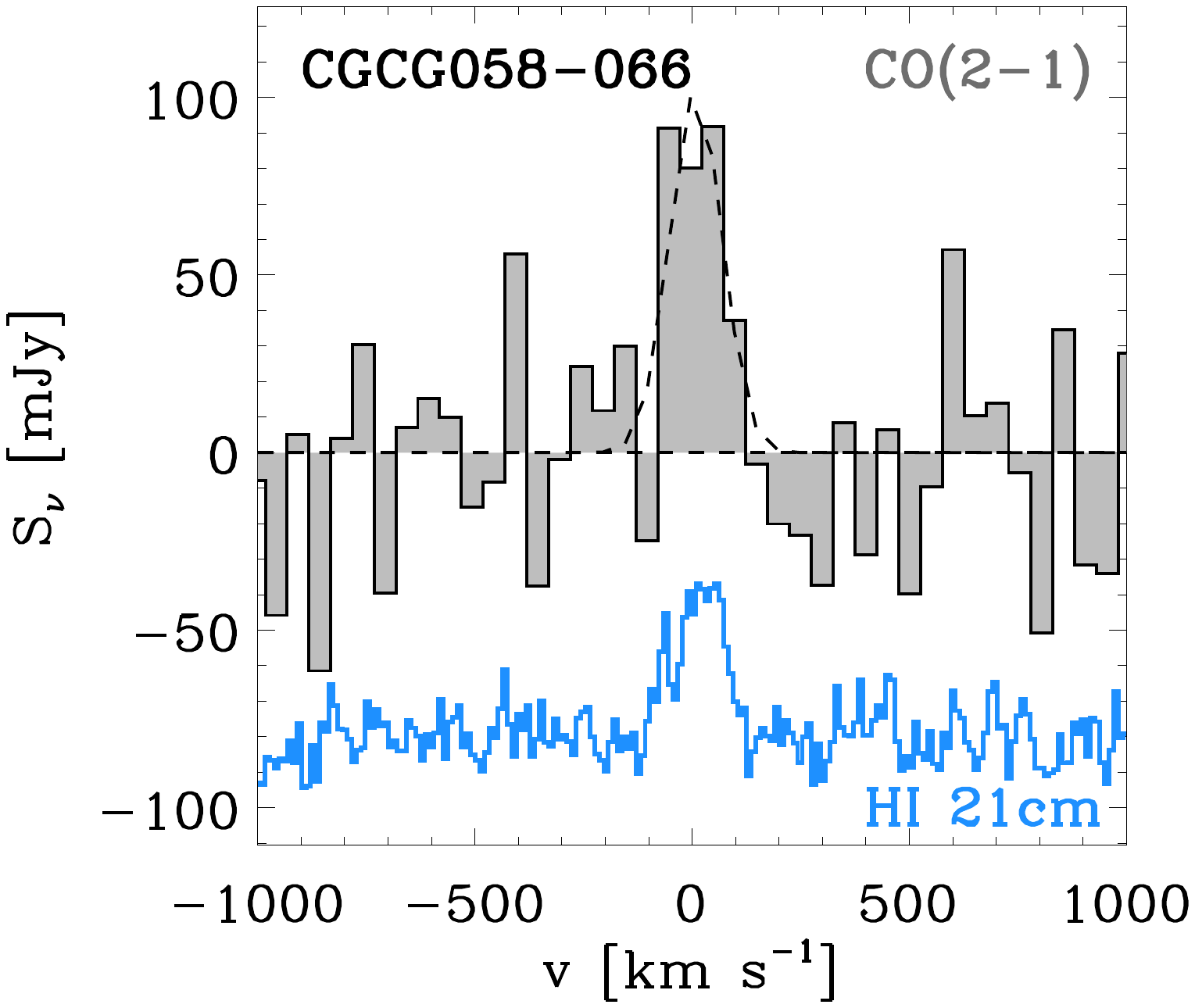}\quad
    \includegraphics[clip=true,trim=-0.4cm 0cm 0cm 0cm,width=0.18\textwidth,angle=90]{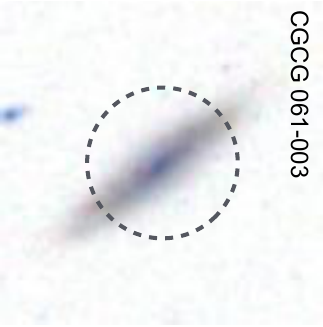}\quad
    \includegraphics[clip=true,trim=5.5cm 3.8cm 4cm 2cm,width=0.28\textwidth,angle=0]{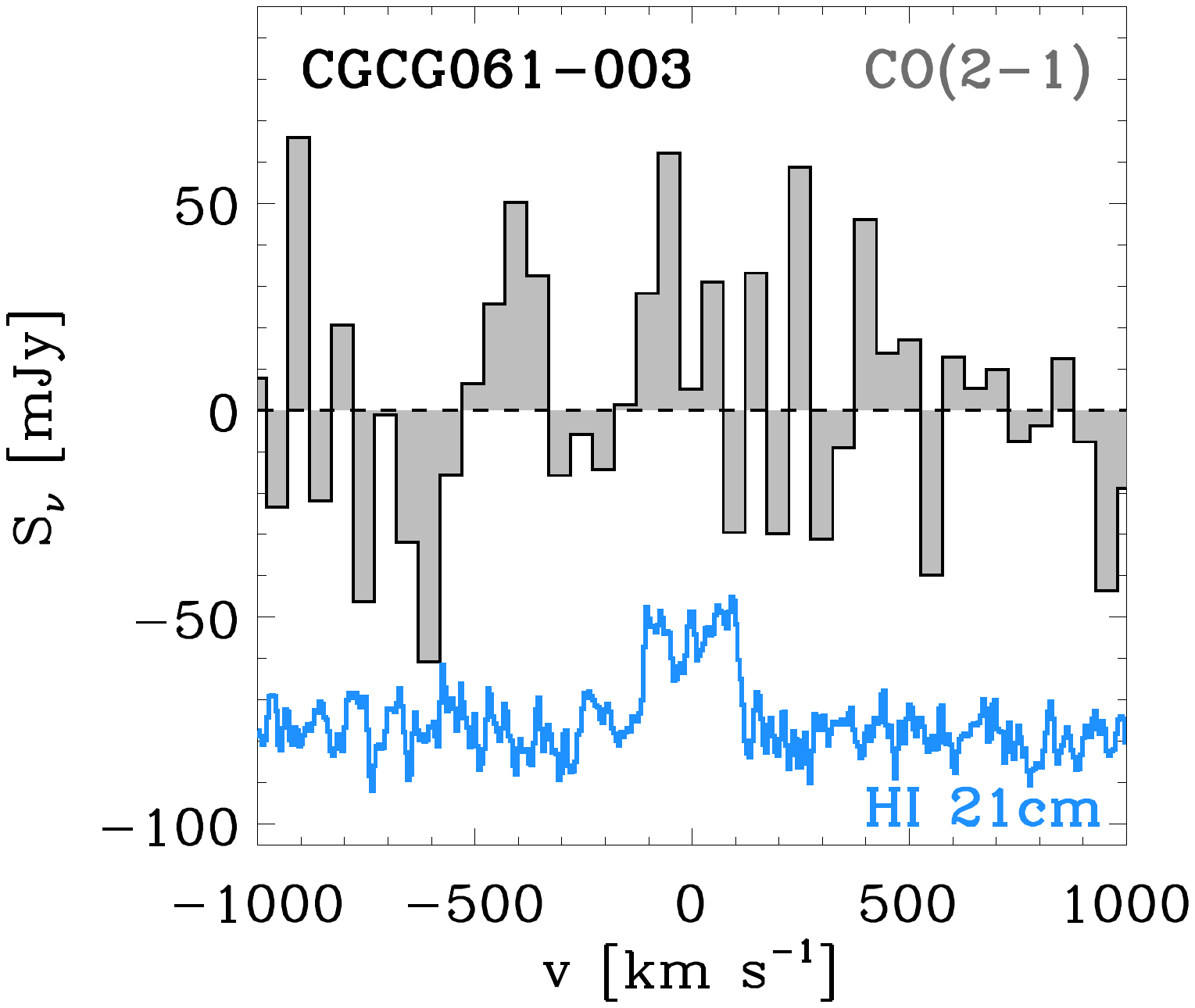}\\
      \includegraphics[clip=true,trim=-0.4cm 0cm 0cm 0cm,width=0.18\textwidth,angle=90]{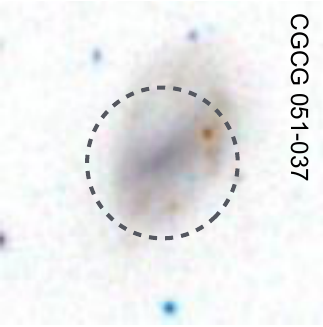}\quad
    \includegraphics[clip=true,trim=5.5cm 3.8cm 4cm 2cm,width=0.28\textwidth,angle=0]{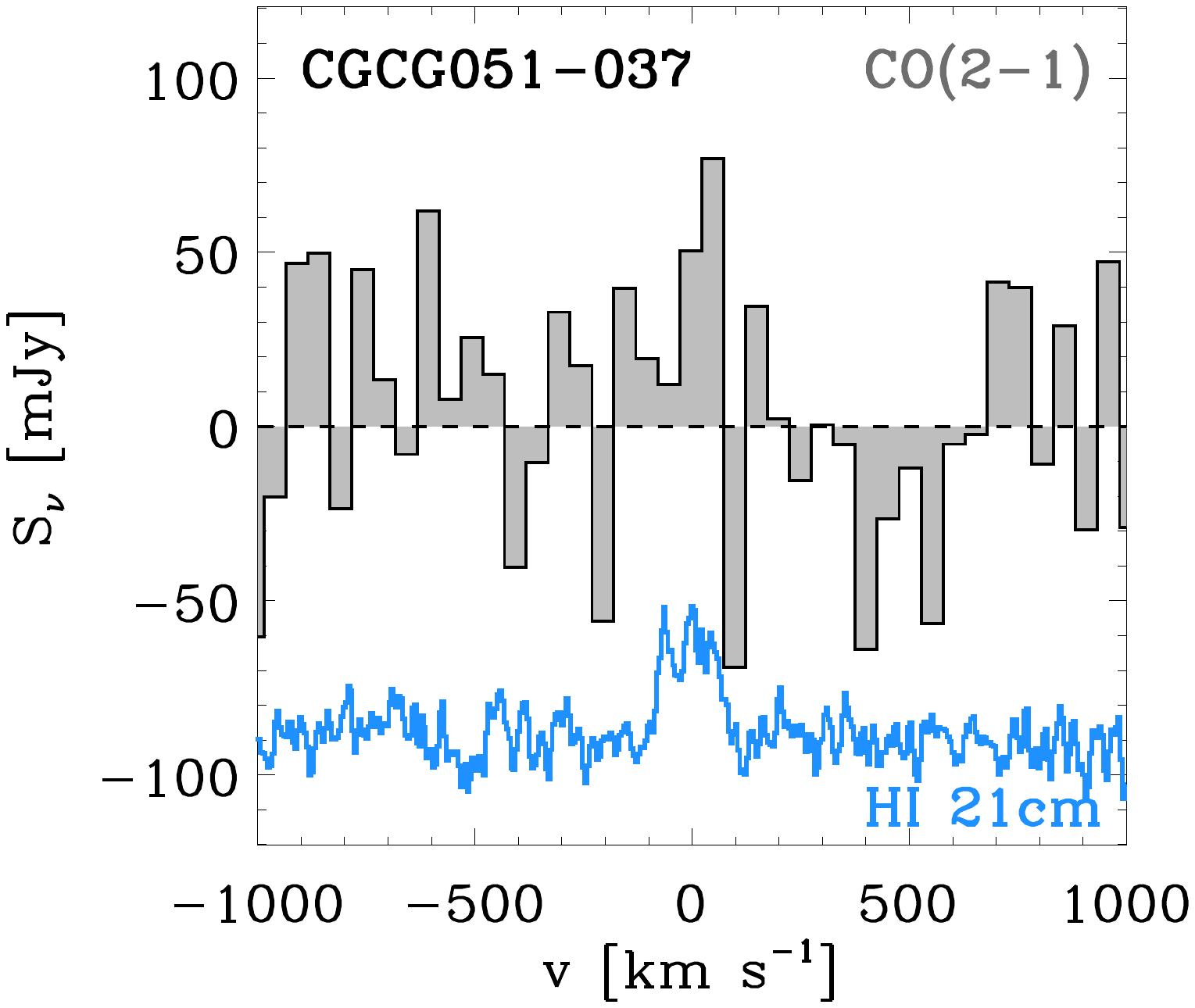}\quad
    \includegraphics[clip=true,trim=-0.4cm 0cm 0cm 0cm,width=0.18\textwidth,angle=90]{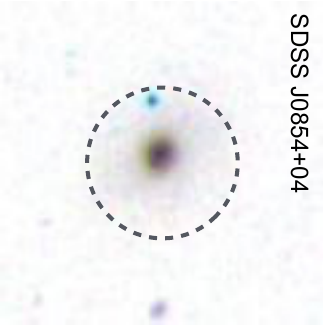}\quad
    \includegraphics[clip=true,trim=5.5cm 3.8cm 4cm 2cm,width=0.28\textwidth,angle=0]{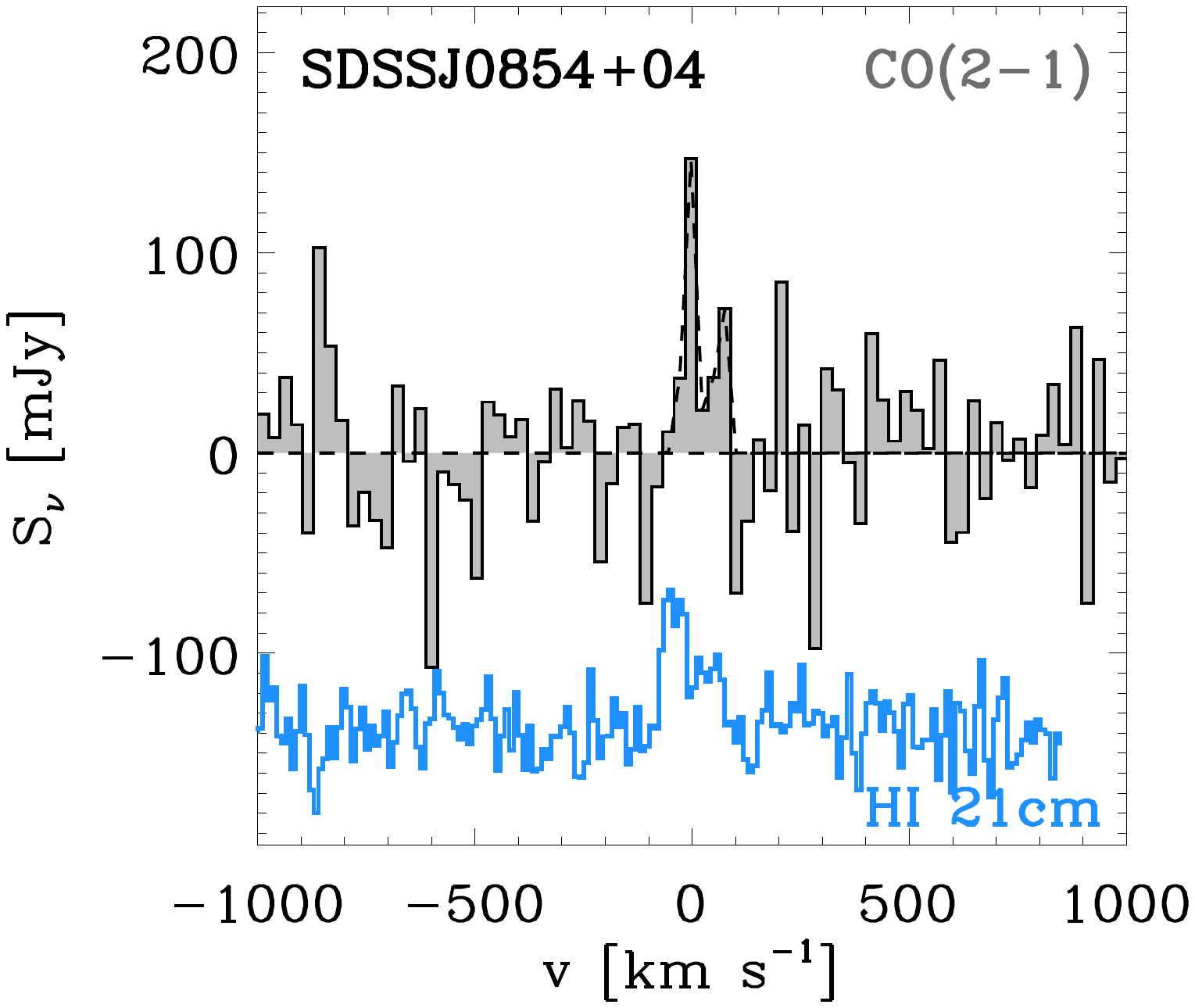}\\
     \caption{{\it Left panels:} SDSS cutout images ({\it g r i} composite, field of view = $60\arcsec\times60\arcsec$, scale = 0.5$\arcsec$/pixel, north is up and west is right) of ALLSMOG galaxies, showing the 27$''$ APEX beam at 230 GHz. {\it Right panels:} APEX CO(2-1) baseline-subtracted spectra, rebinned in bins of $\delta \varv=50$~\kms (2MASXJ0843+1303, SDSSJ0920+0759, CGCG080-042, CGCG078-021, UGC04567, SDSSJ1624+1251, CGCG058-066, CGCG061-003, CGCG051-037) or 25~\kms (SDSSJ0854+0418), depending on the width and S/N of the line. The corresponding H{\sc i}~21cm spectra are also shown for comparison, after having been renormalised for visualisation purposes (H{\sc i} references are given in Table~\ref{table:HI_parameters}).}
   \label{fig:spectra6}
\end{figure*}
\begin{figure*}[tbp]
\centering
    \includegraphics[clip=true,trim=-0.4cm 0cm 0cm 0cm,width=0.18\textwidth,angle=90]{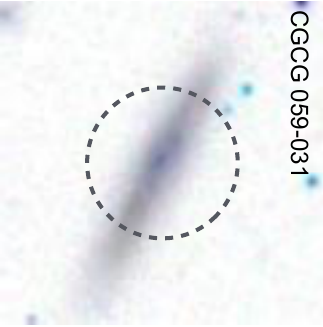}\quad
    \includegraphics[clip=true,trim=5.5cm 3.8cm 4cm 2cm,width=0.28\textwidth,angle=0]{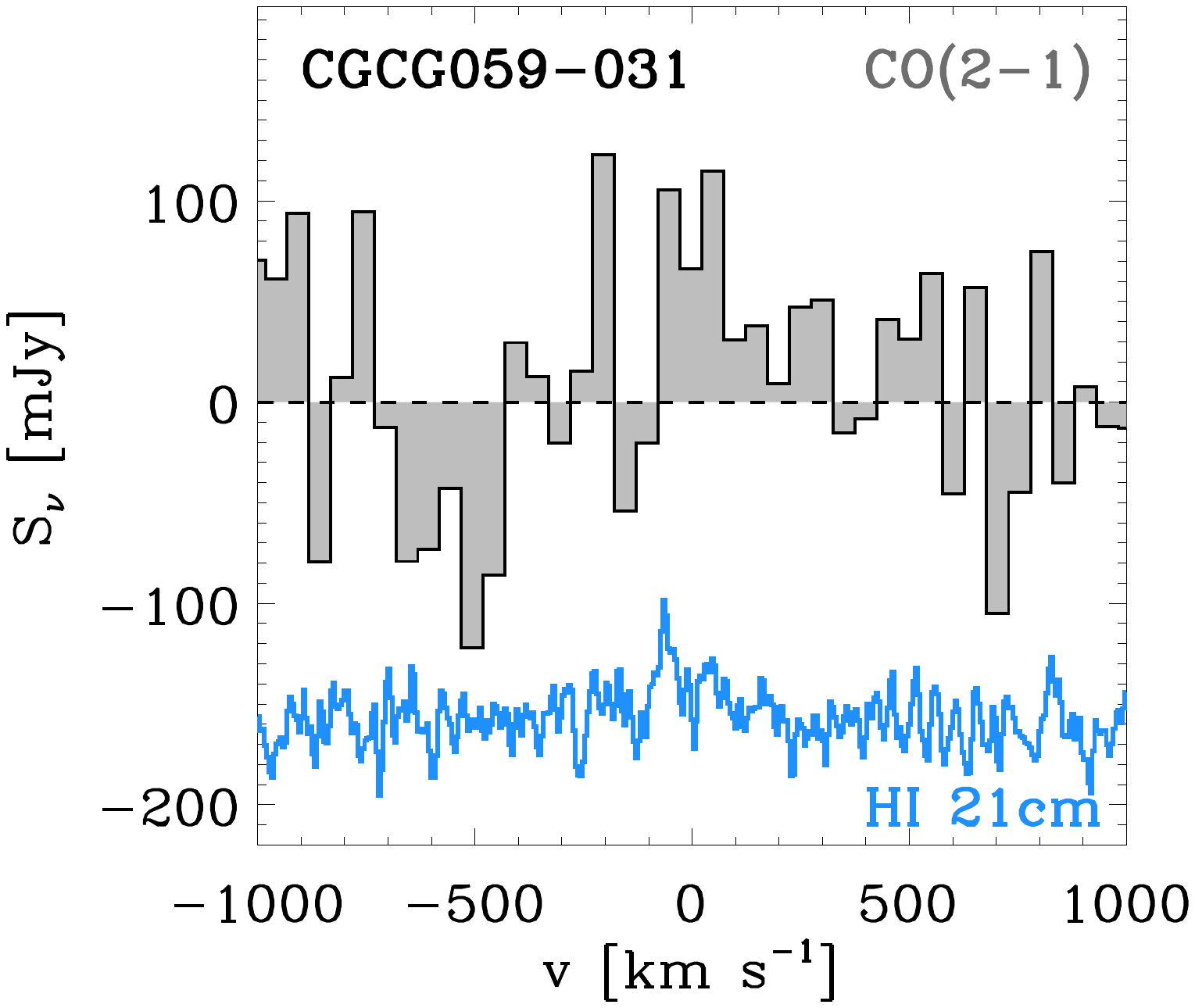}\quad
    \includegraphics[clip=true,trim=-0.4cm 0cm 0cm 0cm,width=0.18\textwidth,angle=90]{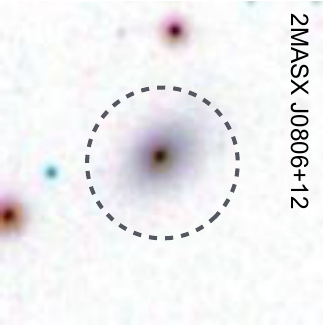}\quad
    \includegraphics[clip=true,trim=5.5cm 3.8cm 4cm 2cm,width=0.28\textwidth,angle=0]{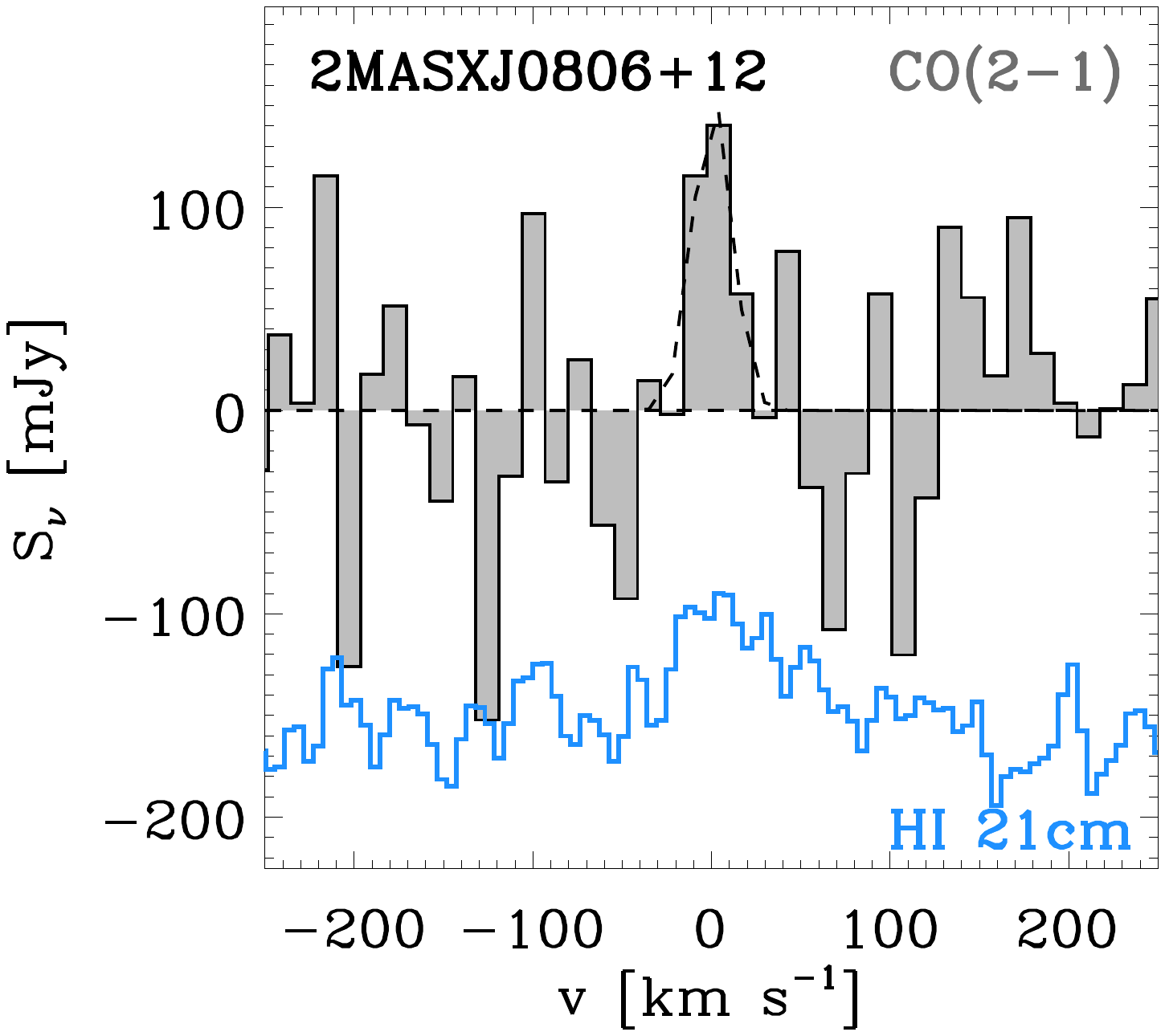}\\
    \includegraphics[clip=true,trim=-0.4cm 0cm 0cm 0cm,width=0.18\textwidth,angle=90]{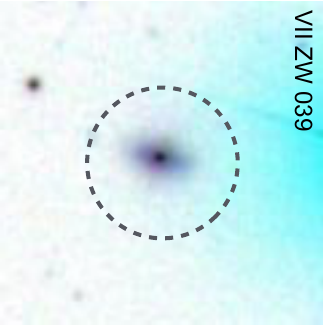}\quad
    \includegraphics[clip=true,trim=5.5cm 3.8cm 4cm 2cm,width=0.28\textwidth,angle=0]{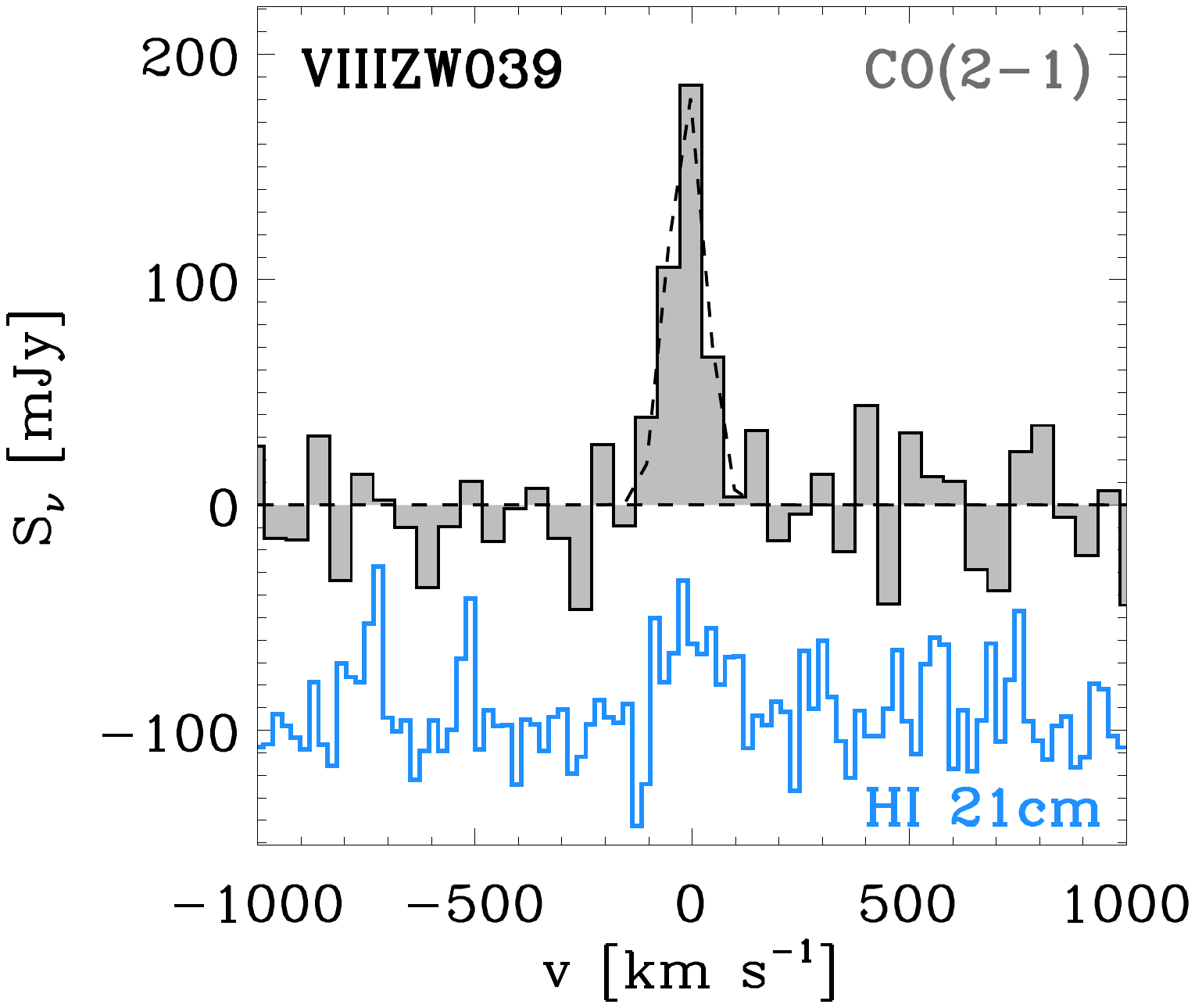}\quad
    \includegraphics[clip=true,trim=-0.4cm 0cm 0cm 0cm,width=0.18\textwidth,angle=90]{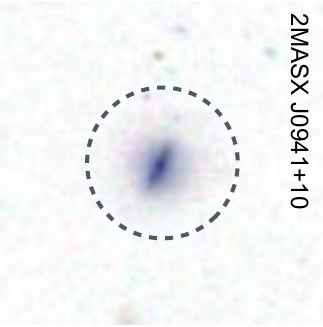}\quad
    \includegraphics[clip=true,trim=5.5cm 3.8cm 4cm 2cm,width=0.28\textwidth,angle=0]{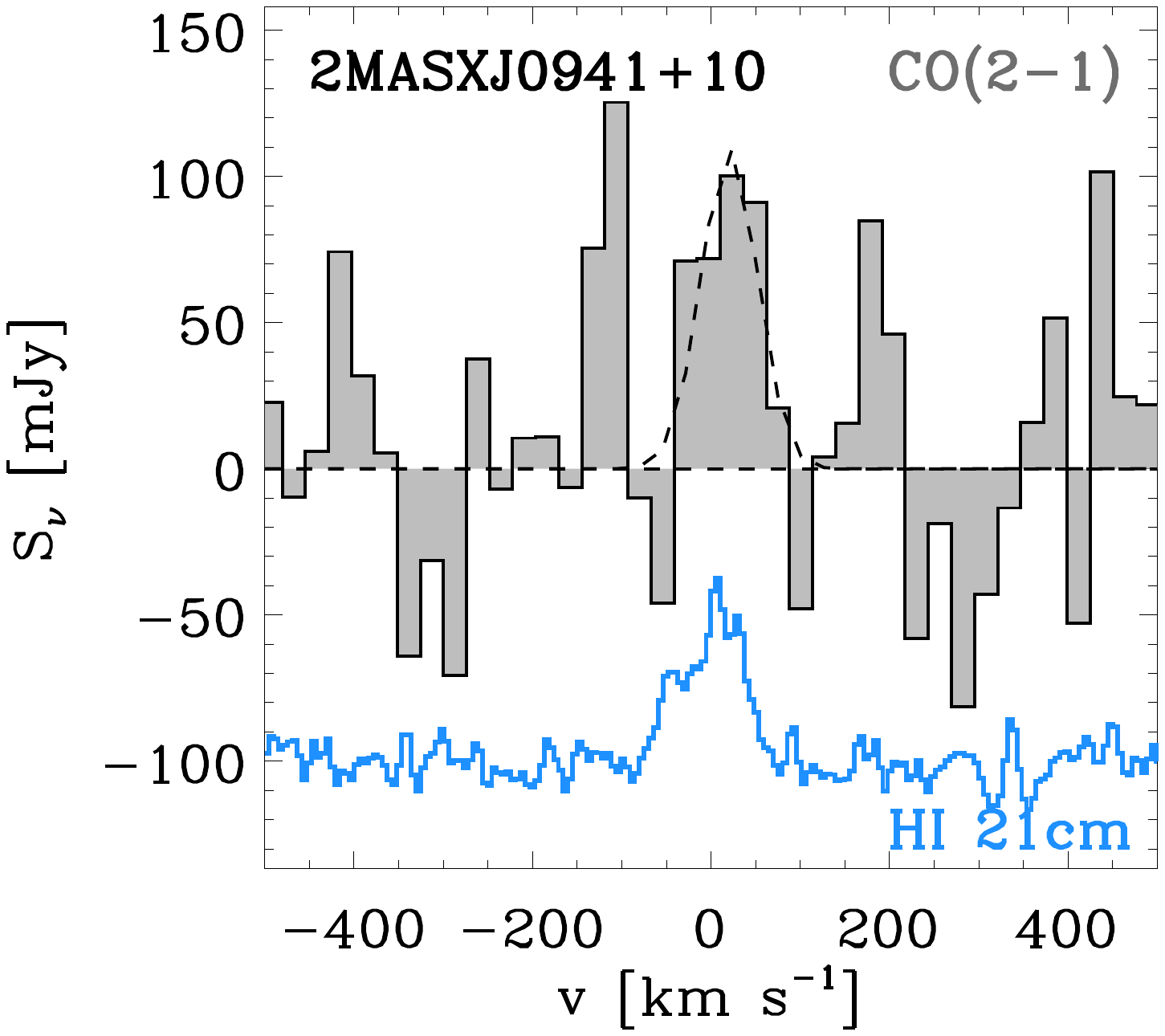}\\
      \includegraphics[clip=true,trim=-0.4cm 0cm 0cm 0cm,width=0.18\textwidth,angle=90]{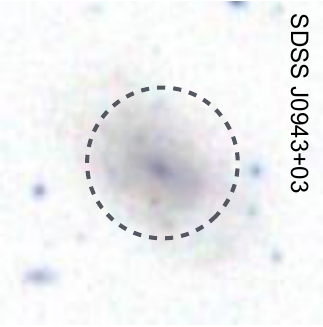}\quad
    \includegraphics[clip=true,trim=5.5cm 3.8cm 4cm 2cm,width=0.28\textwidth,angle=0]{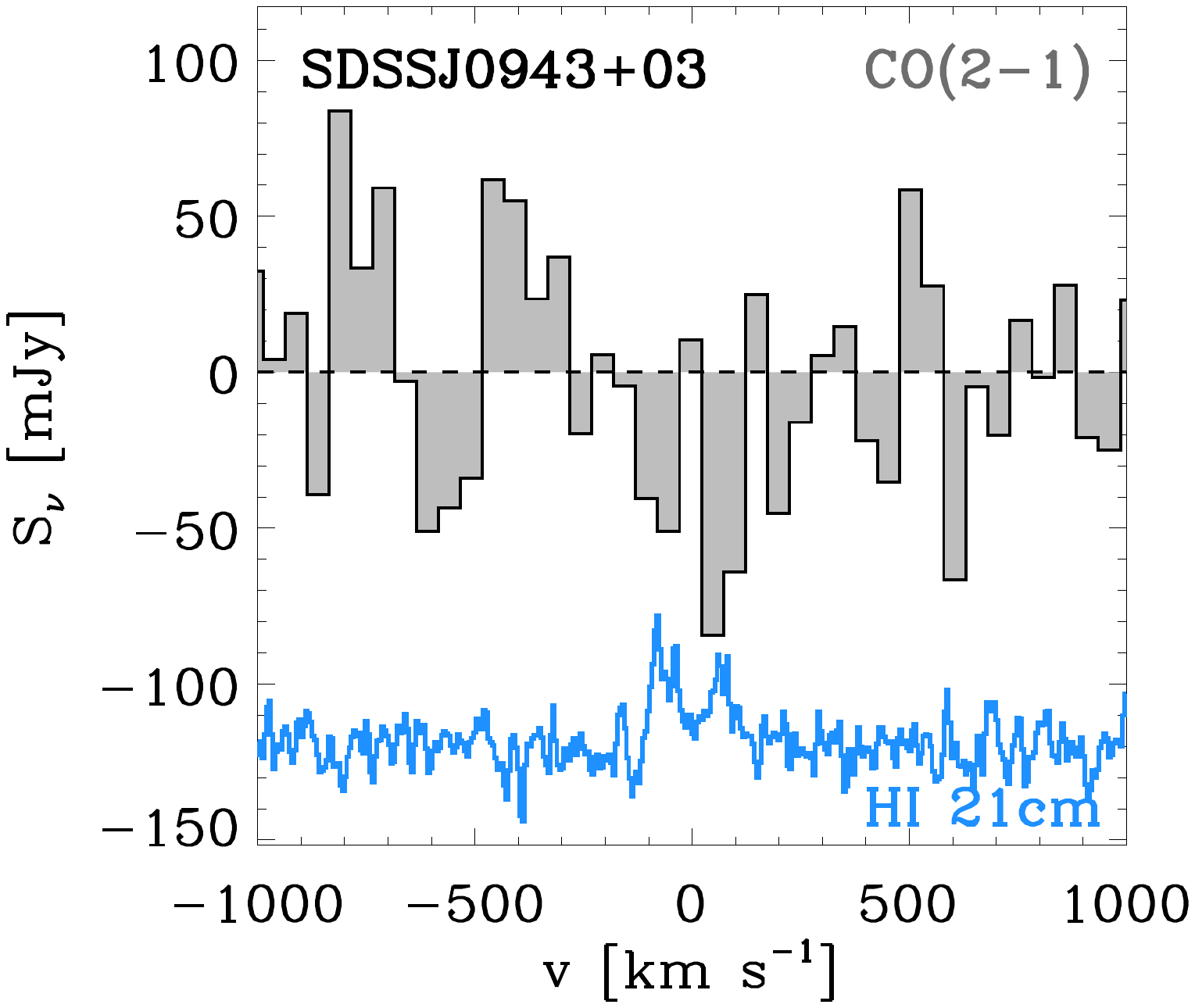}\quad
    \includegraphics[clip=true,trim=-0.4cm 0cm 0cm 0cm,width=0.18\textwidth,angle=90]{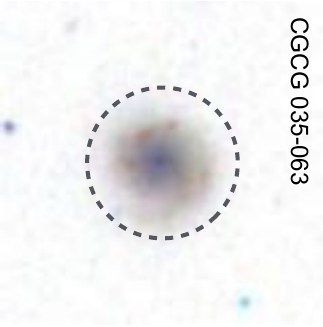}\quad
    \includegraphics[clip=true,trim=5.5cm 3.8cm 4cm 2cm,width=0.28\textwidth,angle=0]{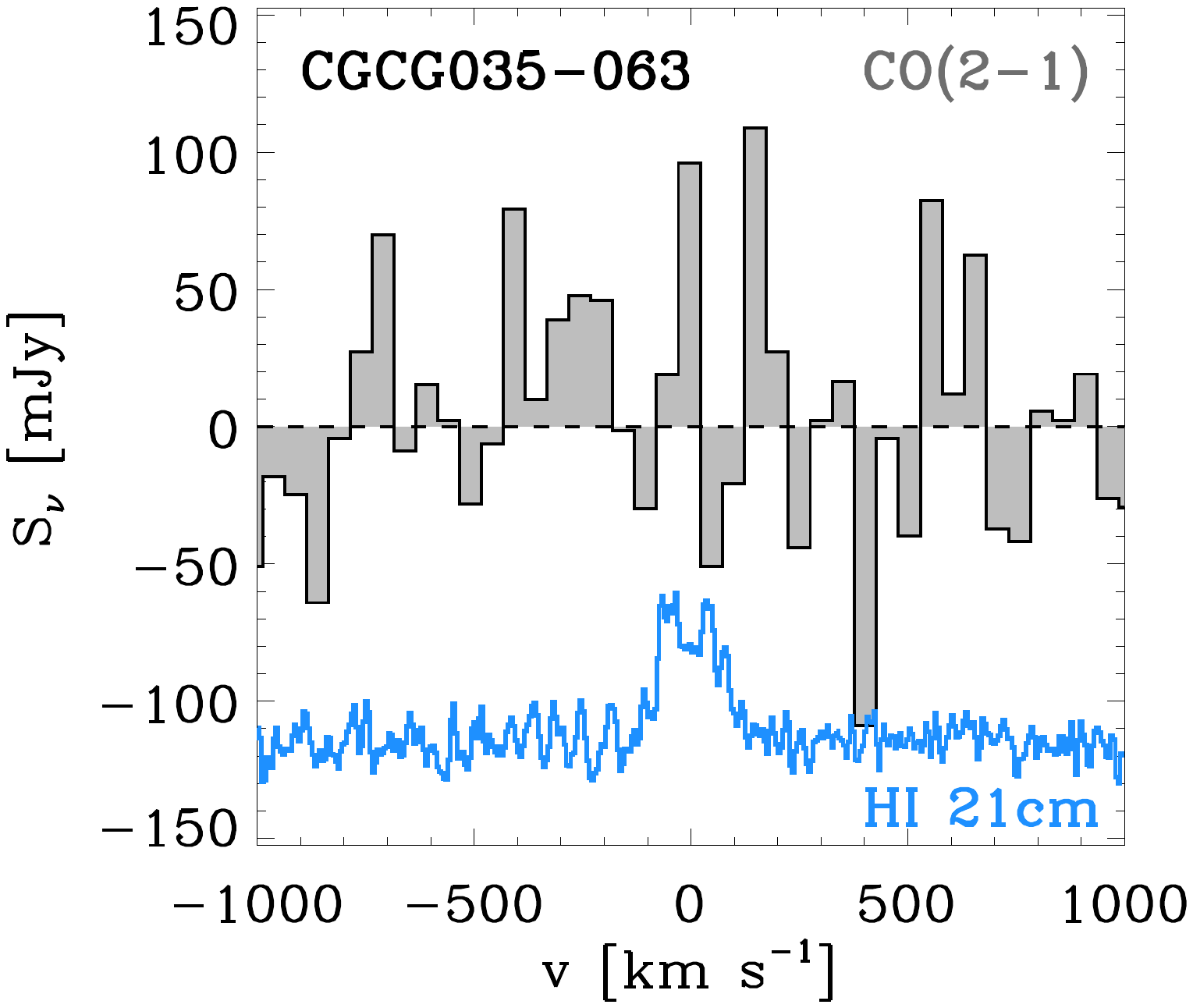}\\
      \includegraphics[clip=true,trim=-0.4cm 0cm 0cm 0cm,width=0.18\textwidth,angle=90]{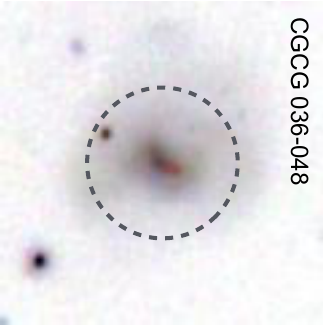}\quad
    \includegraphics[clip=true,trim=5.5cm 3.8cm 4cm 2cm,width=0.28\textwidth,angle=0]{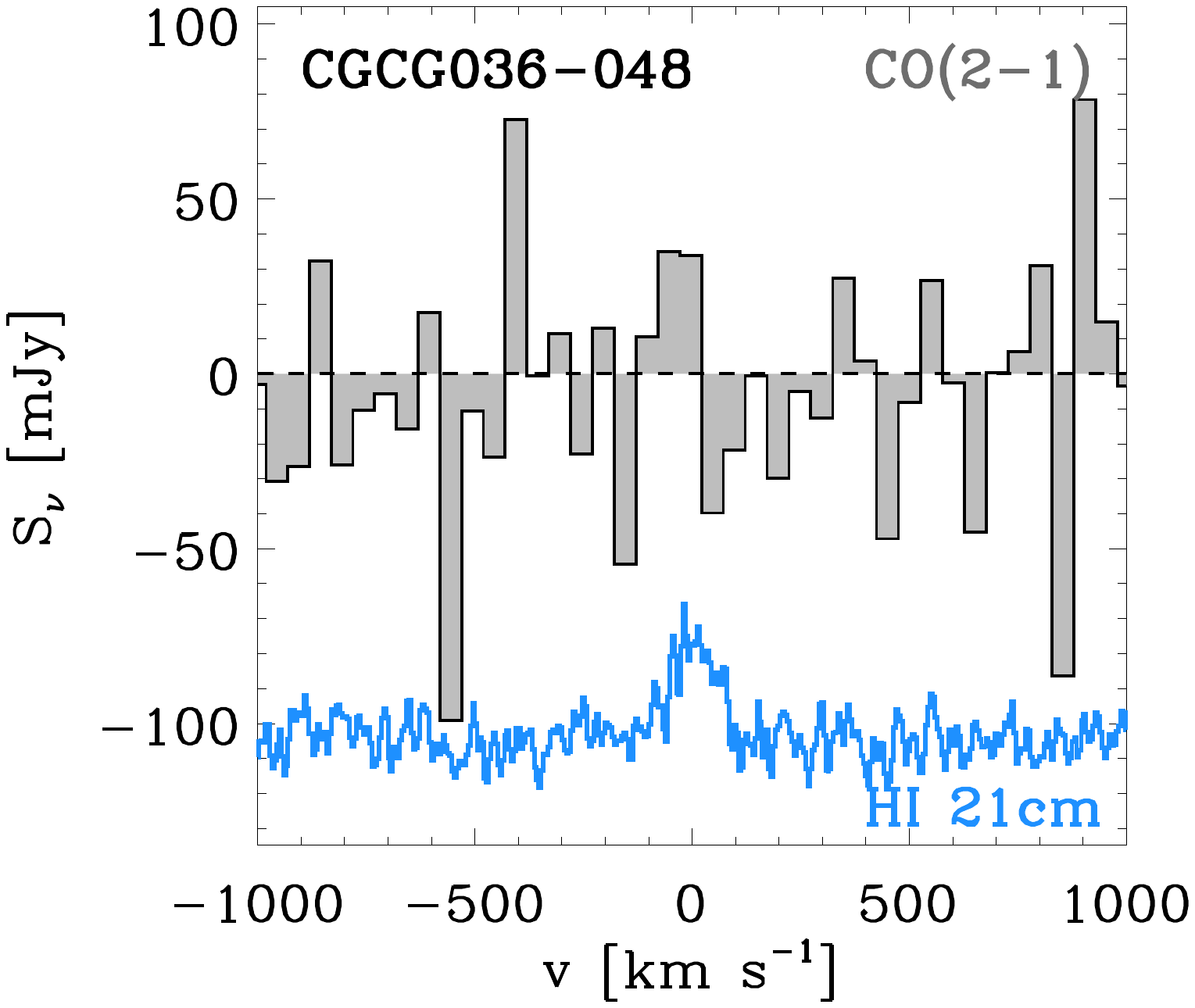}\quad
    \includegraphics[clip=true,trim=-0.4cm 0cm 0cm 0cm,width=0.18\textwidth,angle=90]{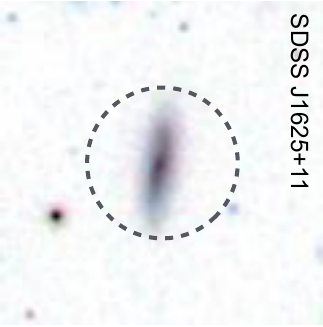}\quad
    \includegraphics[clip=true,trim=5.5cm 3.8cm 4cm 2cm,width=0.28\textwidth,angle=0]{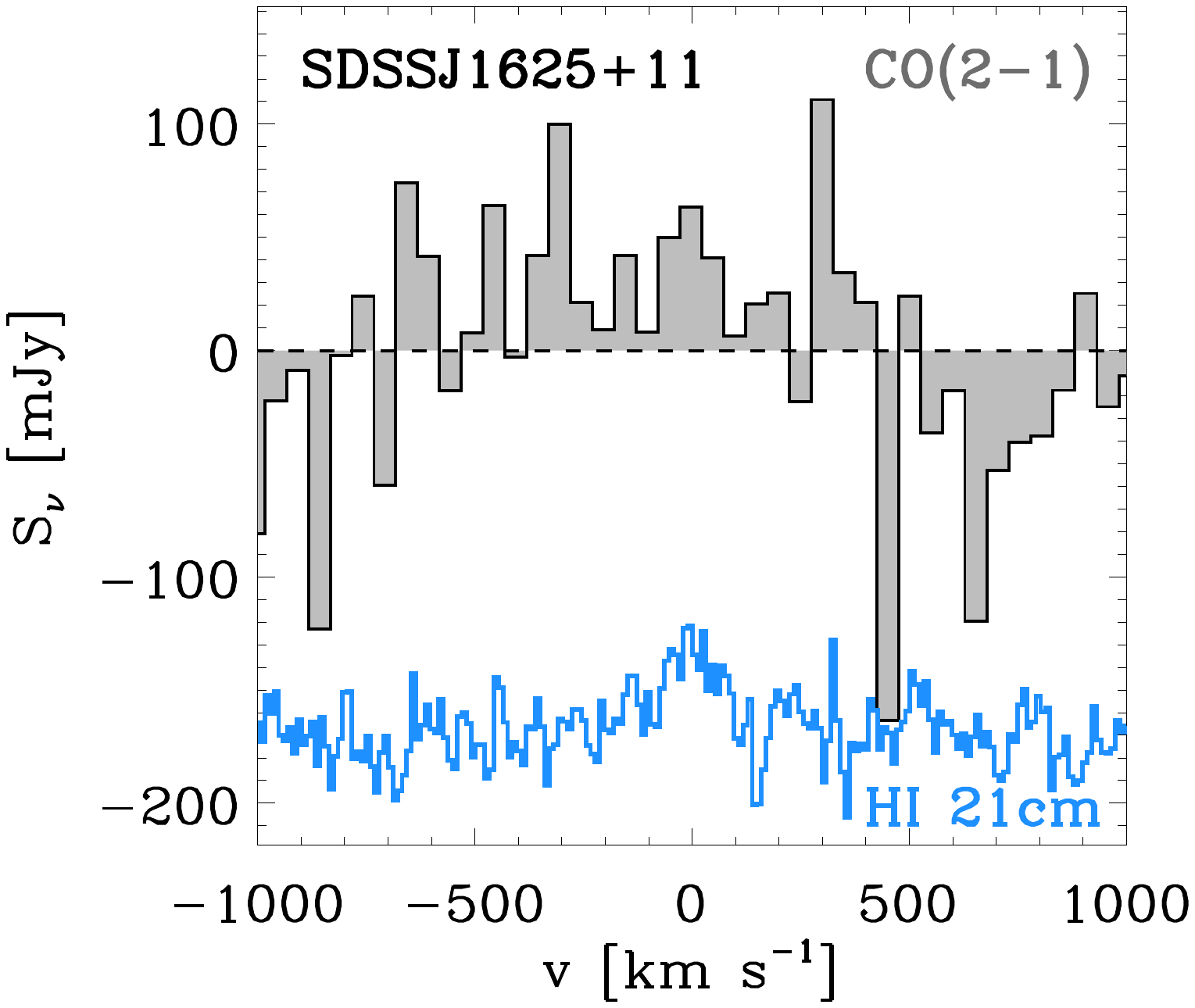}\\
      \includegraphics[clip=true,trim=-0.4cm 0cm 0cm 0cm,width=0.18\textwidth,angle=90]{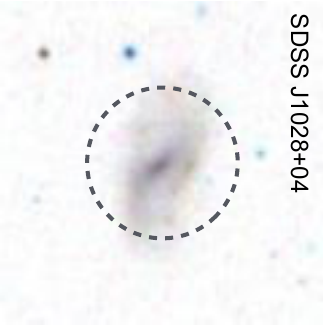}\quad
    \includegraphics[clip=true,trim=5.5cm 3.8cm 4cm 2cm,width=0.28\textwidth,angle=0]{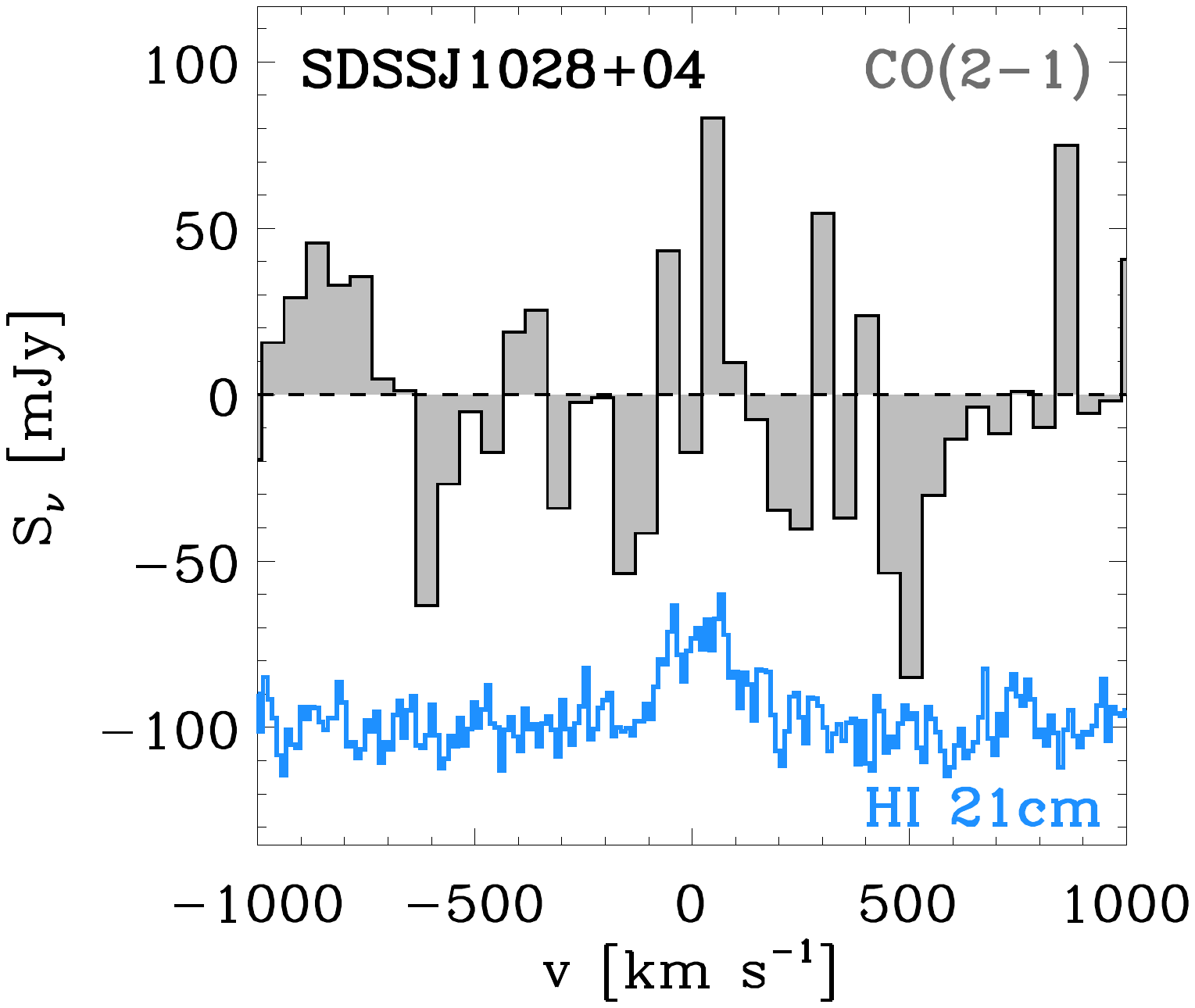}\quad
    \includegraphics[clip=true,trim=-0.4cm 0cm 0cm 0cm,width=0.18\textwidth,angle=90]{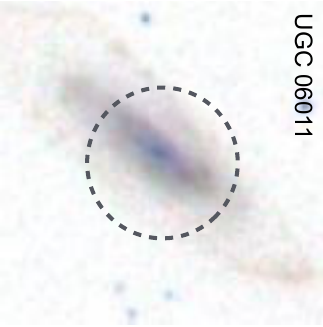}\quad
    \includegraphics[clip=true,trim=5.5cm 3.8cm 4cm 2cm,width=0.28\textwidth,angle=0]{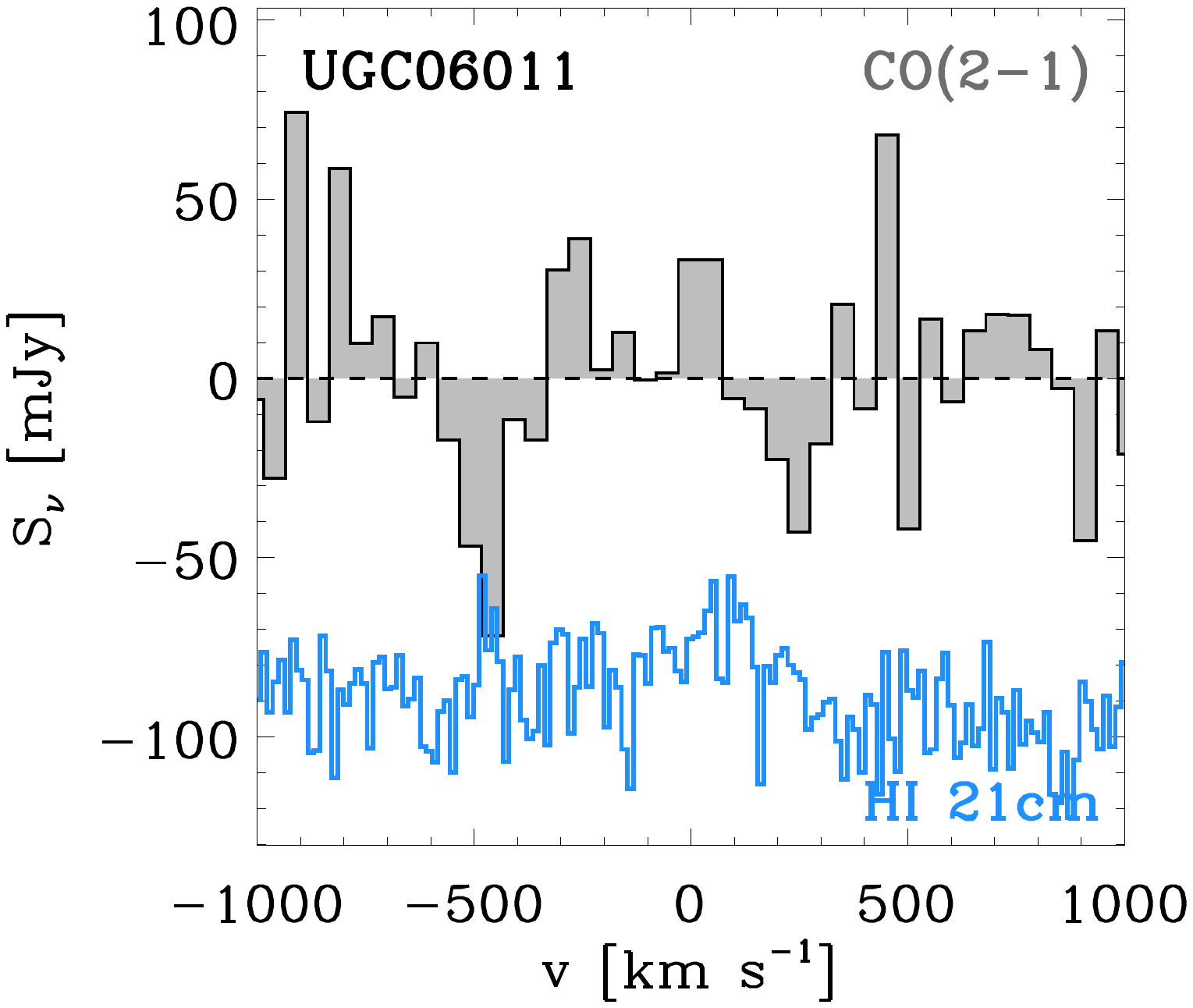}\\
     \caption{{\it Left panels:} SDSS cutout images ({\it g r i} composite, field of view = $60\arcsec\times60\arcsec$, scale = 0.5$\arcsec$/pixel, north is up and west is right) of ALLSMOG galaxies, showing the 27$''$ APEX beam at 230 GHz. {\it Right panels:} APEX CO(2-1) baseline-subtracted spectra, rebinned in bins of
     $\delta \varv=50$~\kms (CGCG059-031, VIIIZW039,  SDSSJ0943+0356, CGCG035-063, CGCG036-048, SDSSJ1625+1142, SDSSJ1028+0424, UGC06011),  25~\kms (2MASXJ0941+1056), 
     or 13~\kms (2MASXJ0806+1249), depending on the width and S/N of the line. The corresponding H{\sc i}~21cm spectra are also shown for comparison, after having been renormalised for visualisation purposes 
     (H{\sc i} references are given in Table~\ref{table:HI_parameters}).}
   \label{fig:spectra7}
\end{figure*}
\begin{figure*}[tbp]
\centering
    \includegraphics[clip=true,trim=-0.4cm 0cm 0cm 0cm,width=0.18\textwidth,angle=90]{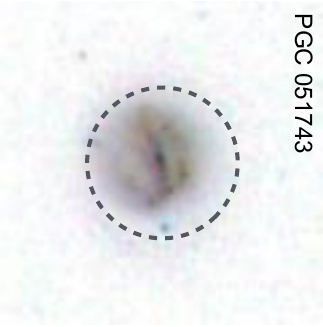}\quad
    \includegraphics[clip=true,trim=5.5cm 3.8cm 4cm 2cm,width=0.28\textwidth,angle=0]{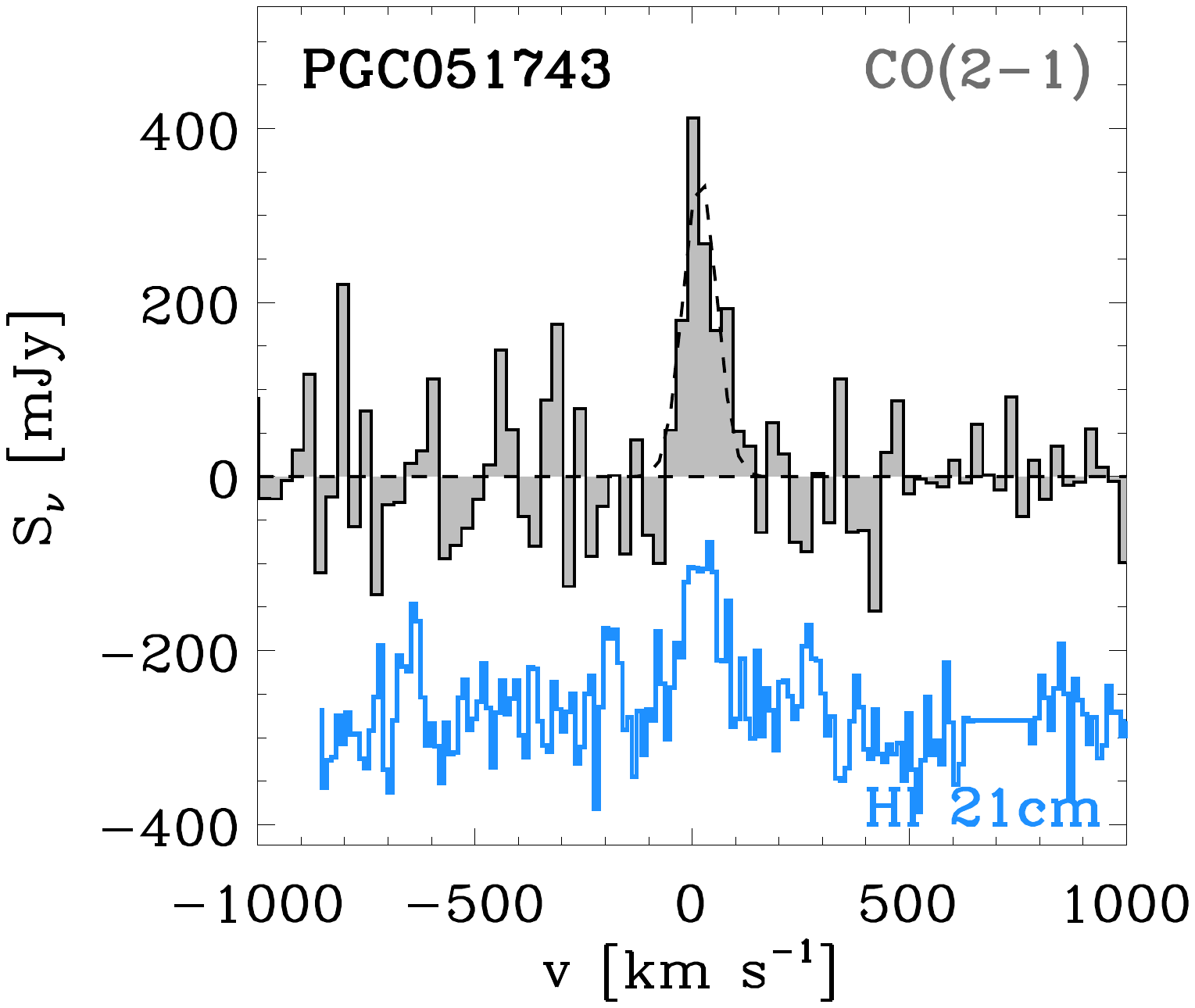}\quad
    \includegraphics[clip=true,trim=-0.4cm 0cm 0cm 0cm,width=0.18\textwidth,angle=90]{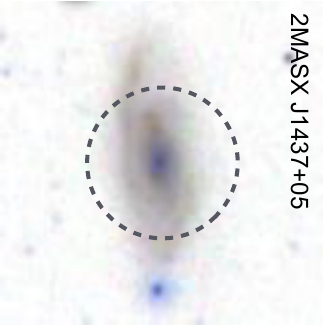}\quad
    \includegraphics[clip=true,trim=5.5cm 3.8cm 4cm 2cm,width=0.28\textwidth,angle=0]{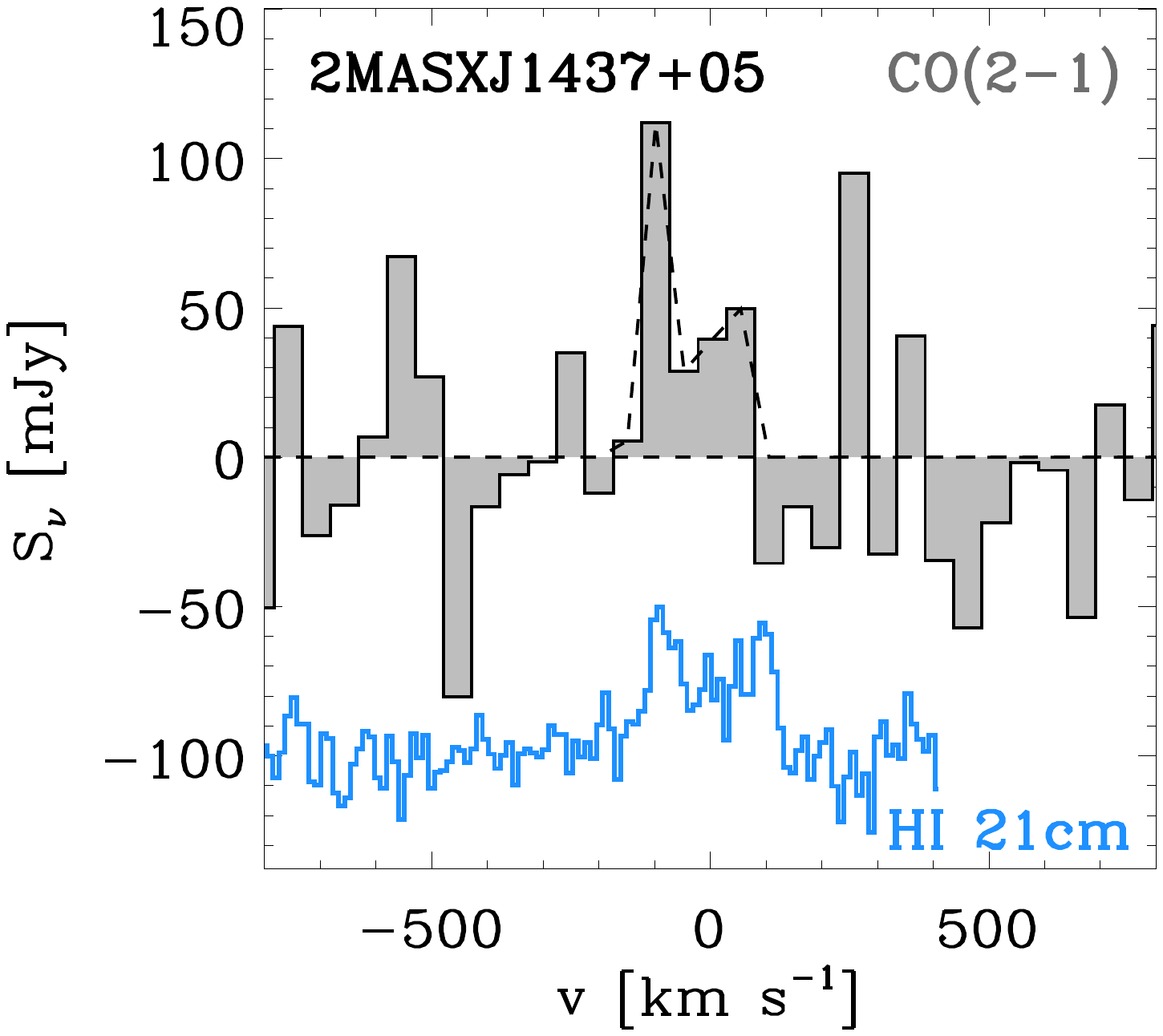}\\
    \includegraphics[clip=true,trim=-0.4cm 0cm 0cm 0cm,width=0.18\textwidth,angle=90]{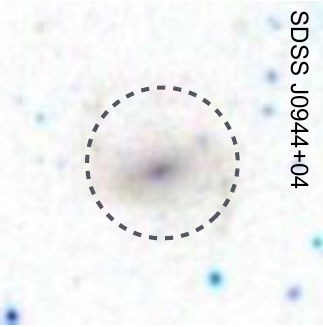}\quad
    \includegraphics[clip=true,trim=5.5cm 3.8cm 4cm 2cm,width=0.28\textwidth,angle=0]{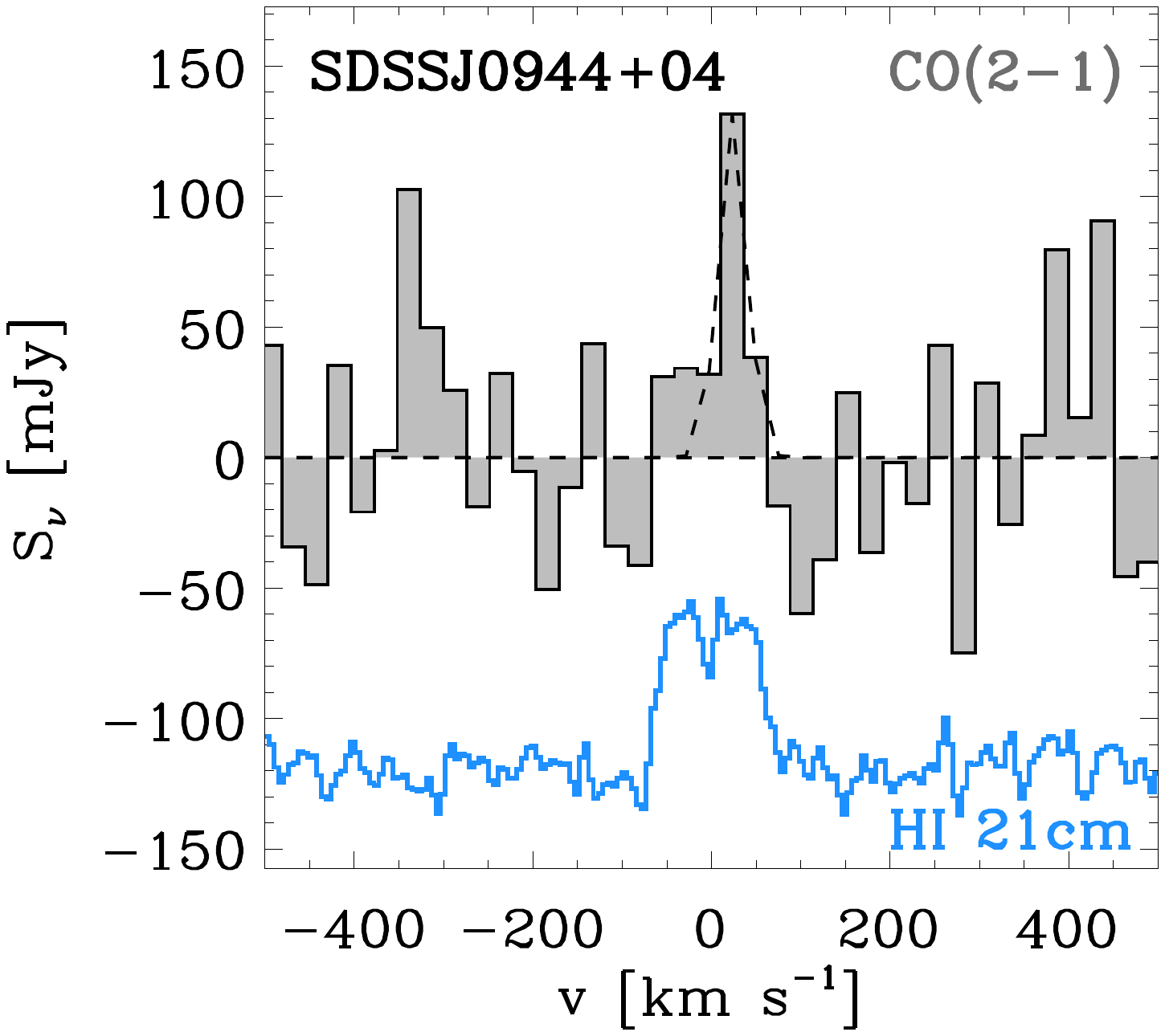}\quad
    \includegraphics[clip=true,trim=-0.4cm 0cm 0cm 0cm,width=0.18\textwidth,angle=90]{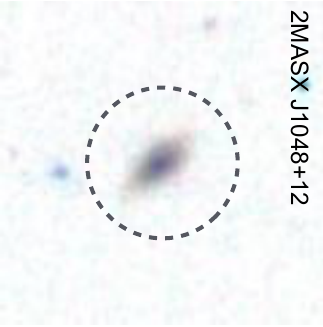}\quad
    \includegraphics[clip=true,trim=5.5cm 3.8cm 4cm 2cm,width=0.28\textwidth,angle=0]{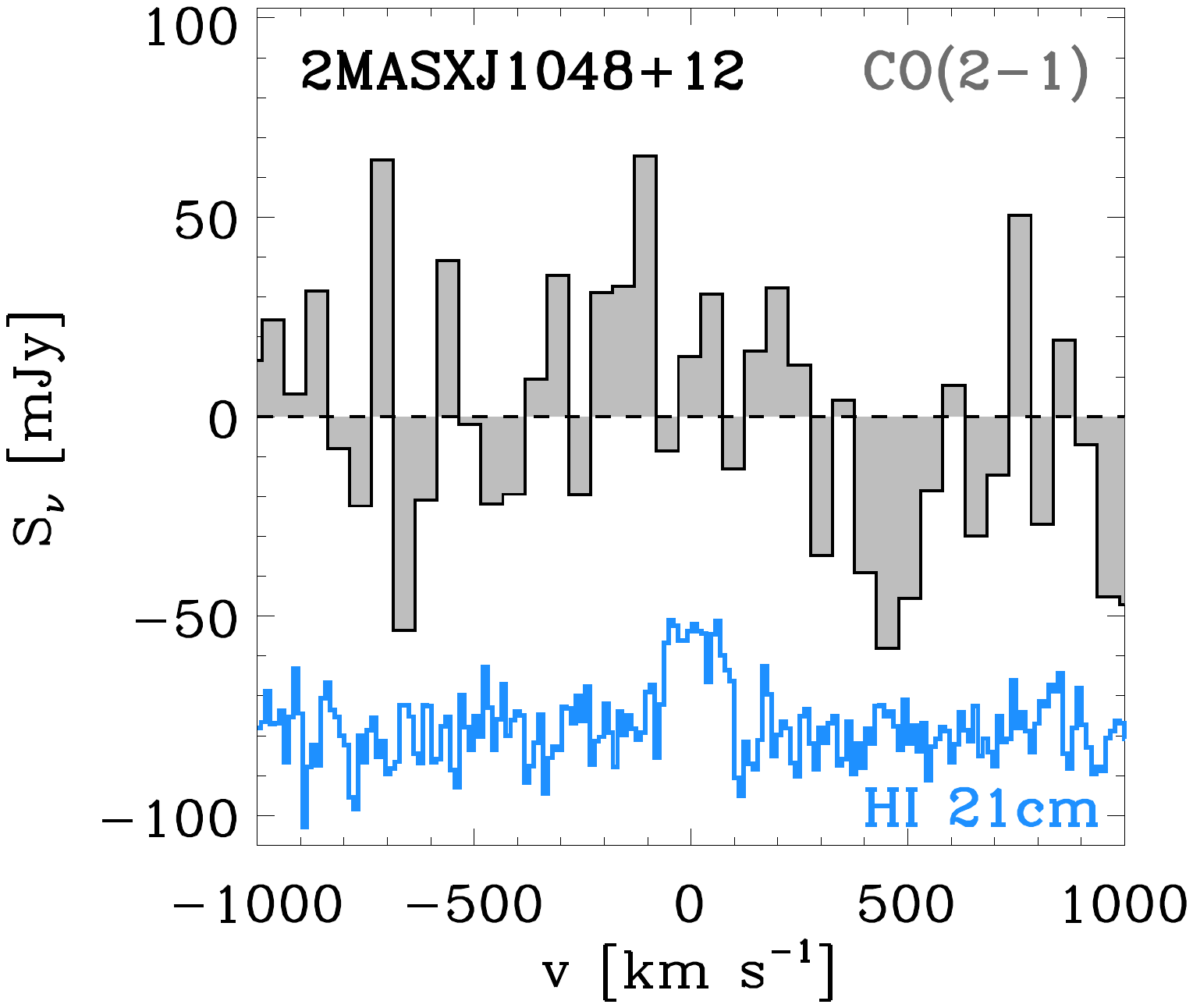}\\
      \includegraphics[clip=true,trim=-0.4cm 0cm 0cm 0cm,width=0.18\textwidth,angle=90]{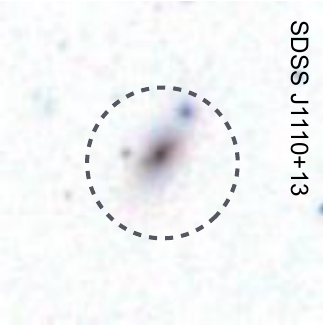}\quad
    \includegraphics[clip=true,trim=5.5cm 3.8cm 4cm 2cm,width=0.28\textwidth,angle=0]{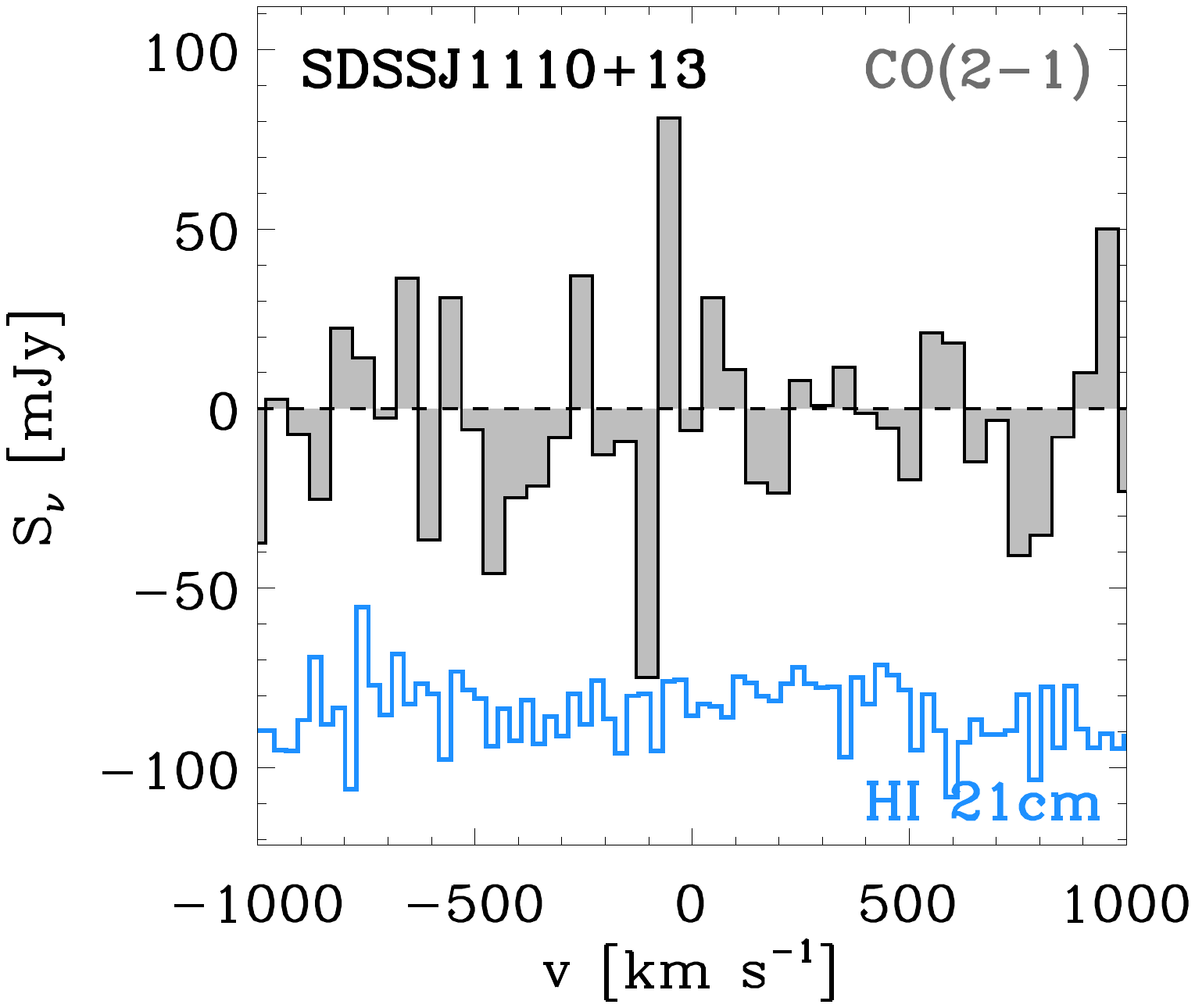}\quad
    \includegraphics[clip=true,trim=-0.4cm 0cm 0cm 0cm,width=0.18\textwidth,angle=90]{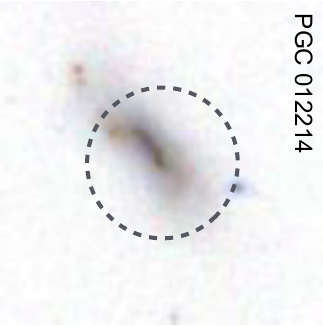}\quad
    \includegraphics[clip=true,trim=5.5cm 3.8cm 4cm 2cm,width=0.28\textwidth,angle=0]{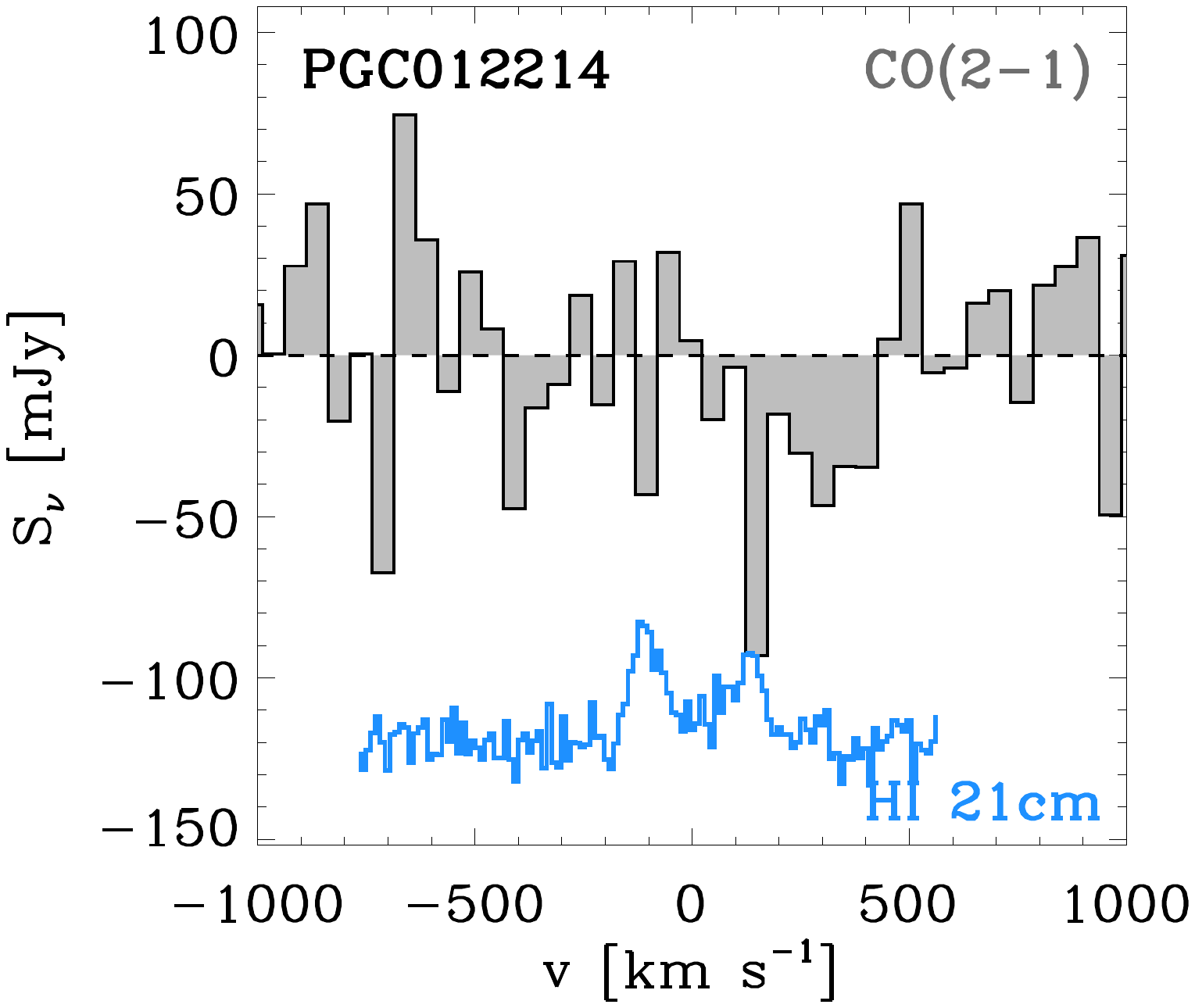}\\
      \includegraphics[clip=true,trim=-0.4cm 0cm 0cm 0cm,width=0.18\textwidth,angle=90]{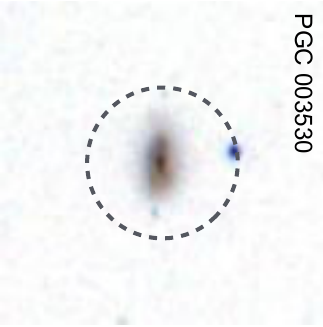}\quad
    \includegraphics[clip=true,trim=5.5cm 3.8cm 4cm 2cm,width=0.28\textwidth,angle=0]{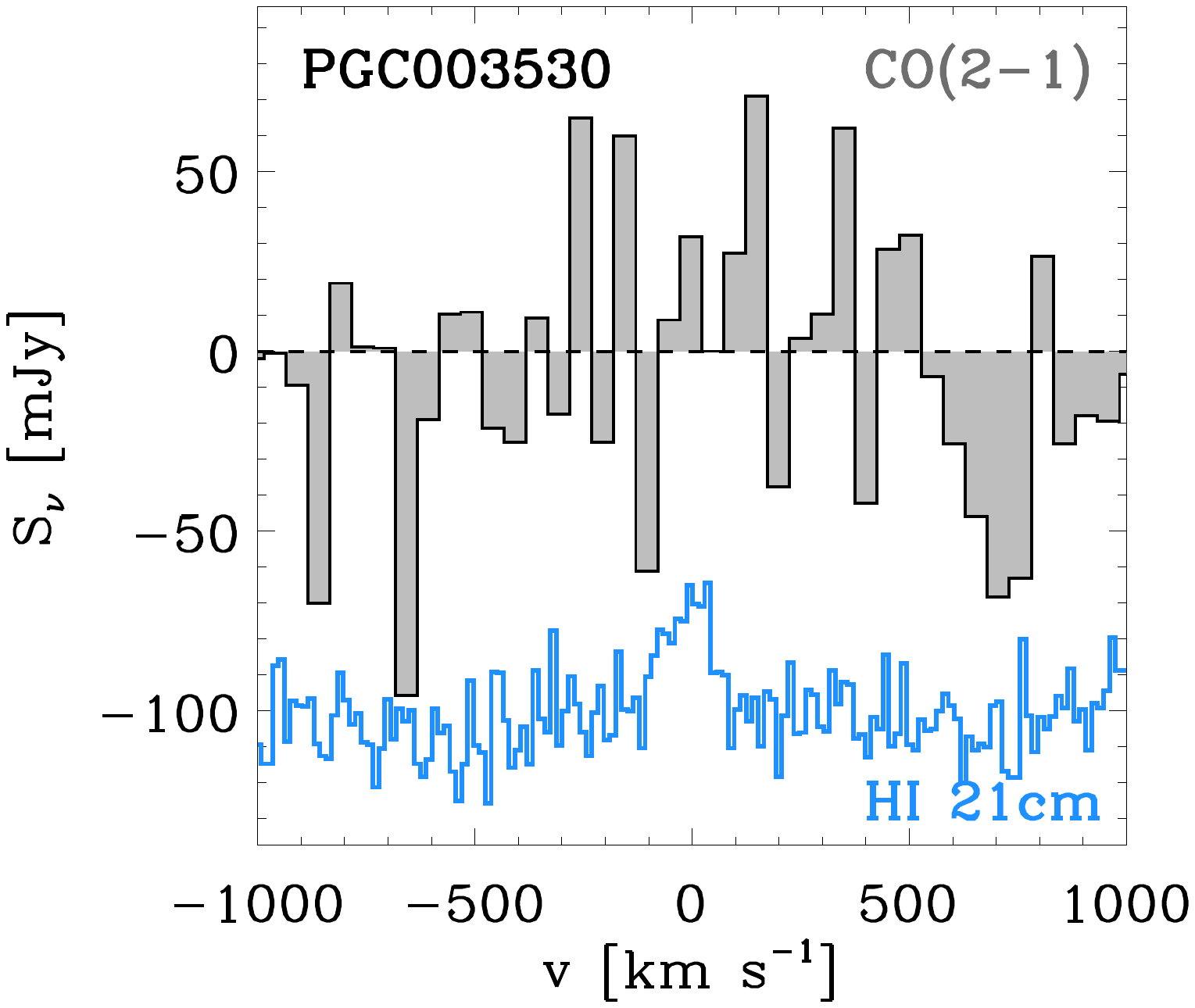}\quad
    \includegraphics[clip=true,trim=-0.4cm 0cm 0cm 0cm,width=0.18\textwidth,angle=90]{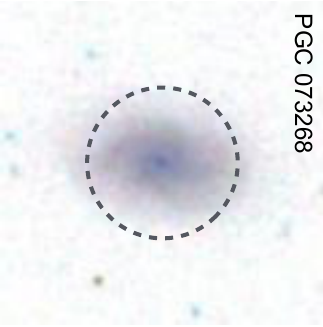}\quad
    \includegraphics[clip=true,trim=5.5cm 3.8cm 4cm 2cm,width=0.28\textwidth,angle=0]{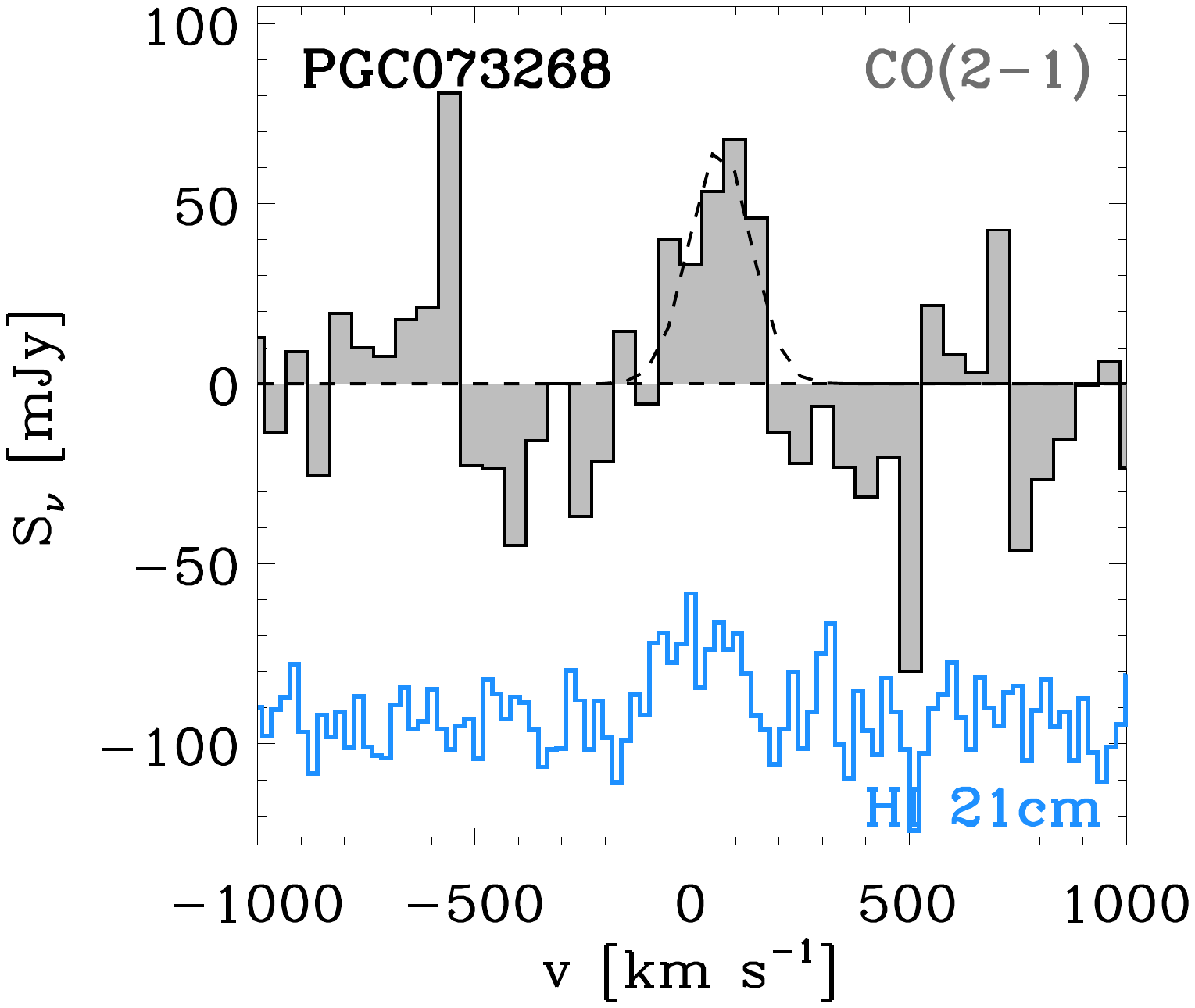}\\
      \includegraphics[clip=true,trim=-0.4cm 0cm 0cm 0cm,width=0.18\textwidth,angle=90]{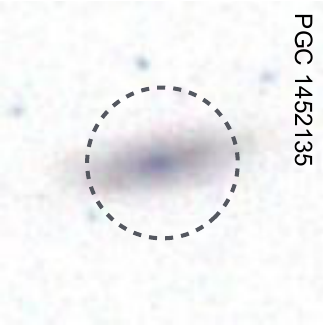}\quad
    \includegraphics[clip=true,trim=5.5cm 3.8cm 4cm 2cm,width=0.28\textwidth,angle=0]{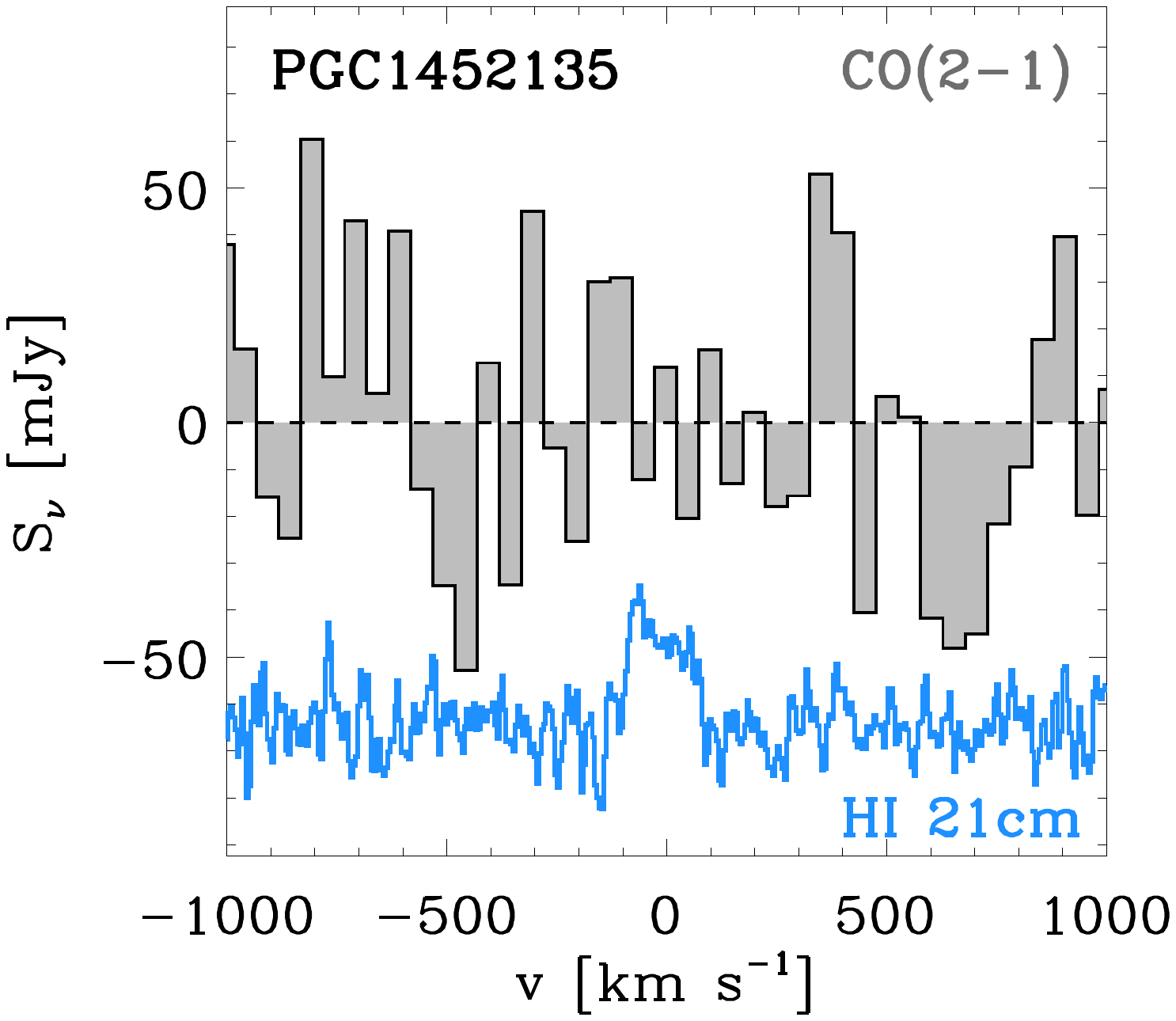}\quad
    \includegraphics[clip=true,trim=-0.4cm 0cm 0cm 0cm,width=0.18\textwidth,angle=90]{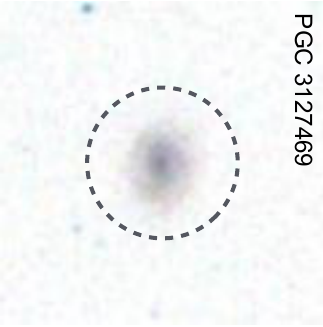}\quad
    \includegraphics[clip=true,trim=5.5cm 3.8cm 4cm 2cm,width=0.28\textwidth,angle=0]{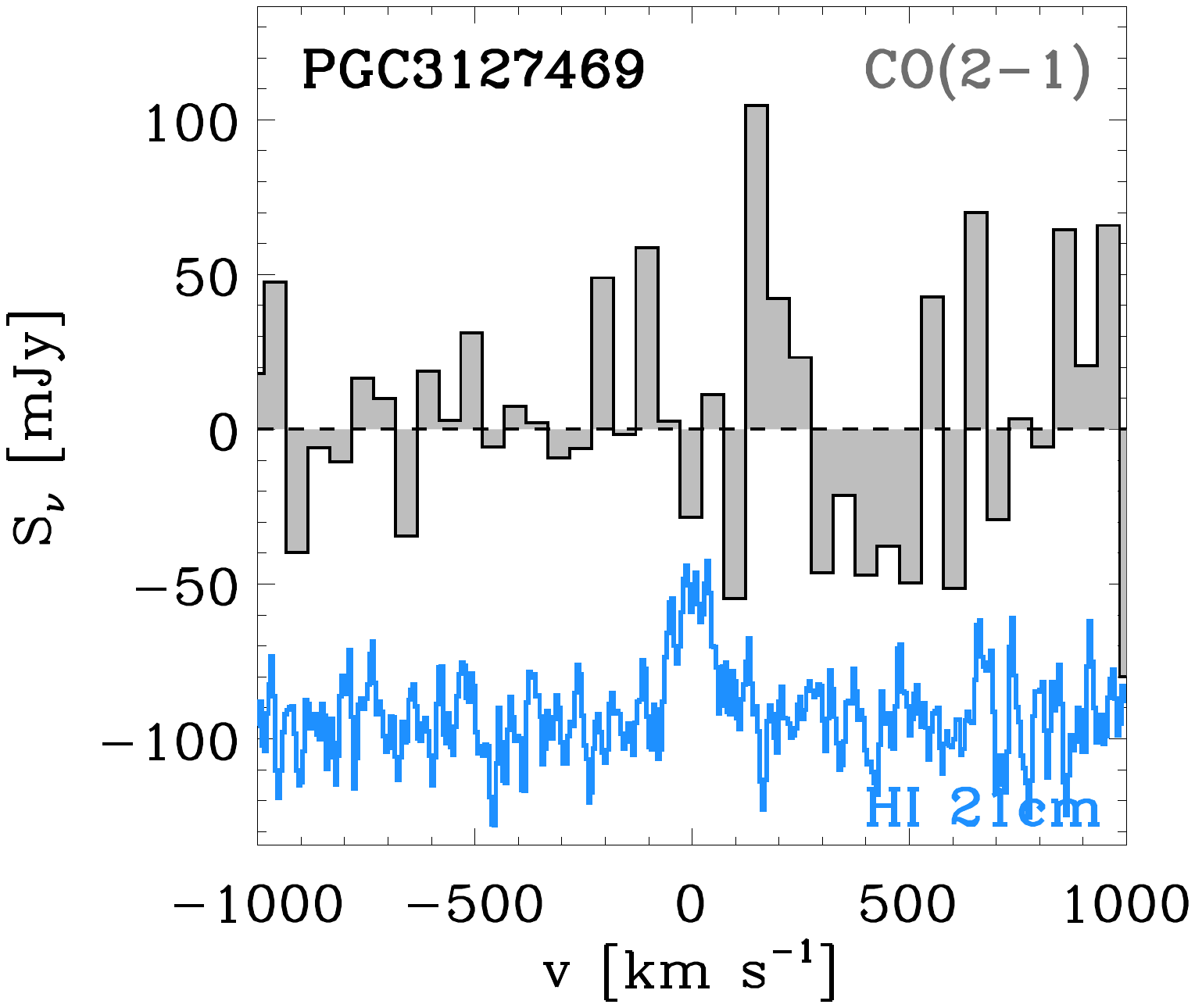}\\
     \caption{{\it Left panels:} SDSS cutout images ({\it g r i} composite, field of view = $60\arcsec\times60\arcsec$, scale = 0.5$\arcsec$/pixel, north is up and west is right) of ALLSMOG galaxies, showing the 27$''$ APEX beam at 230 GHz. {\it Right panels:} APEX CO(2-1) baseline-subtracted spectra, rebinned in bins of $\delta\varv=50$~\kms (2MASXJ1437+0500, 2MASXJ1048+1201, SDSSJ1110+1345, PGC012214, PGC003530, PGC073268, PGC1452135, PGC3127469), or 25~\kms (PGC051743, SDSSJ0944+0400), depending on the width and S/N of the line. The corresponding H{\sc i}~21cm spectra are also shown for comparison, after having been renormalised for visualisation purposes
     (H{\sc i} references are given in Table~\ref{table:HI_parameters}).}
   \label{fig:spectra8}
\end{figure*}
\begin{figure*}[tbp]
\centering
    \includegraphics[clip=true,trim=-0.4cm 0cm 0cm 0cm,width=0.18\textwidth,angle=90]{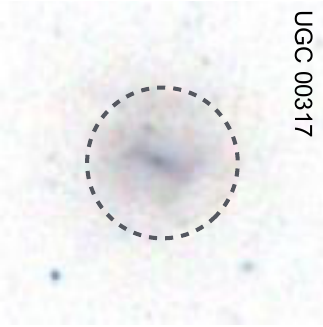}\quad
    \includegraphics[clip=true,trim=5.5cm 3.8cm 4cm 2cm,width=0.28\textwidth,angle=0]{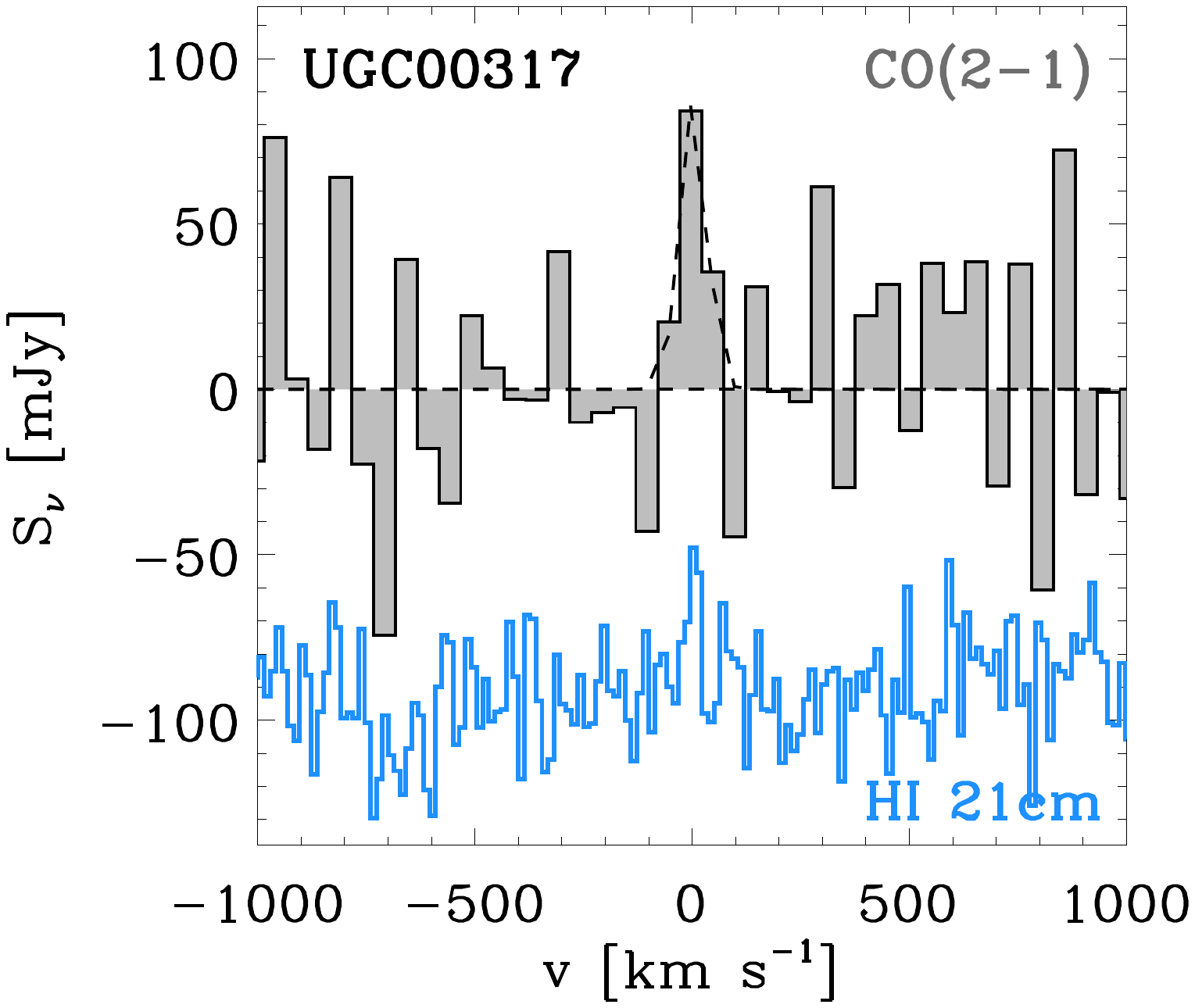}\quad
    \includegraphics[clip=true,trim=-0.4cm 0cm 0cm 0cm,width=0.18\textwidth,angle=90]{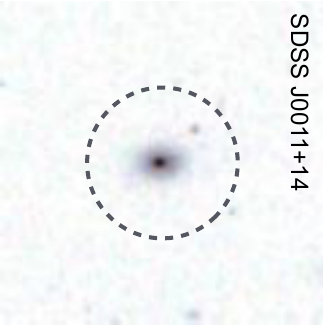}\quad
    \includegraphics[clip=true,trim=5.5cm 3.8cm 4cm 2cm,width=0.28\textwidth,angle=0]{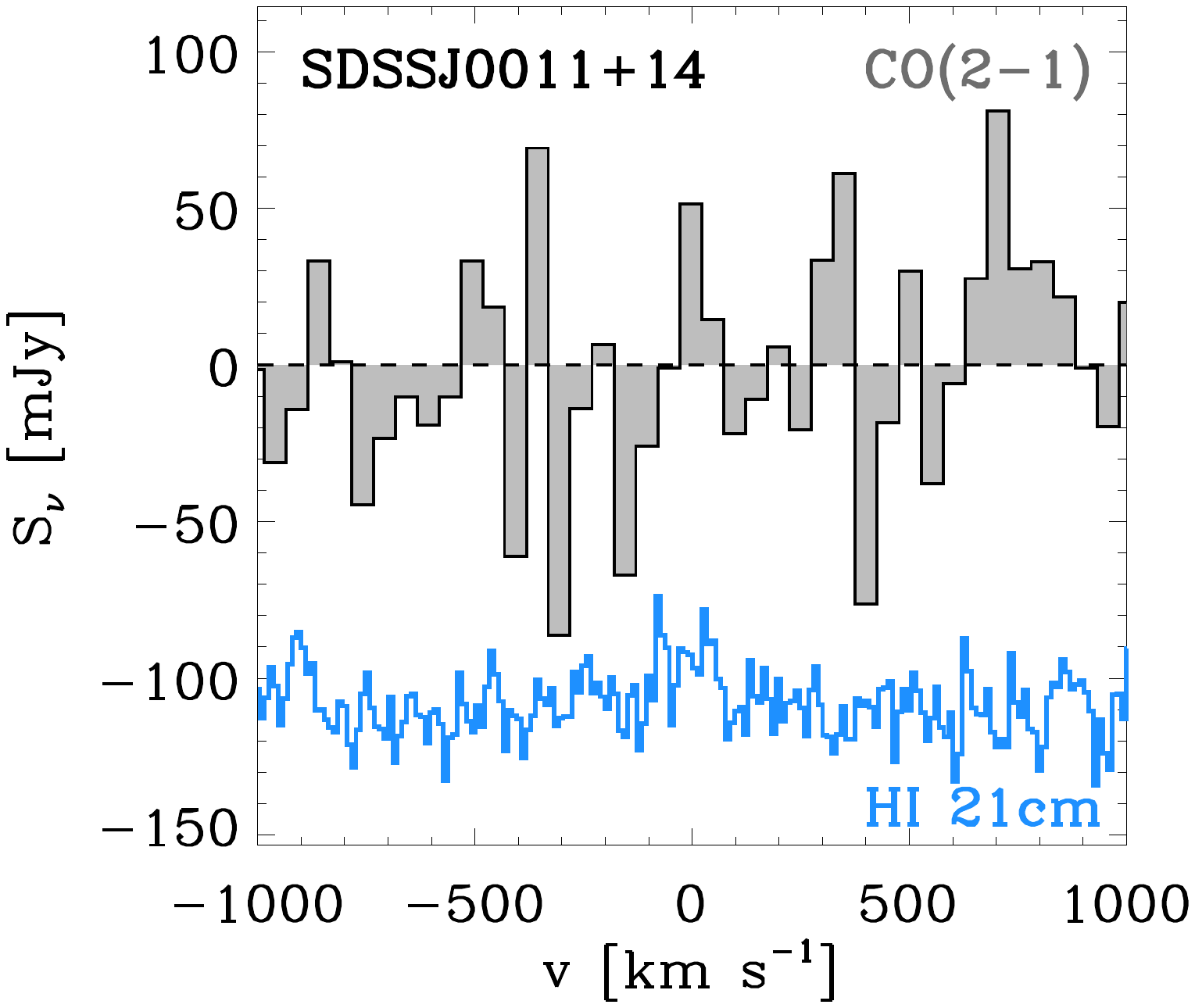}\\
    \includegraphics[clip=true,trim=-0.4cm 0cm 0cm 0cm,width=0.18\textwidth,angle=90]{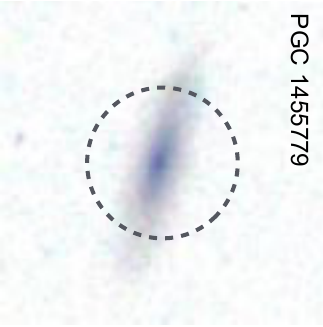}\quad
    \includegraphics[clip=true,trim=5.5cm 3.8cm 4cm 2cm,width=0.28\textwidth,angle=0]{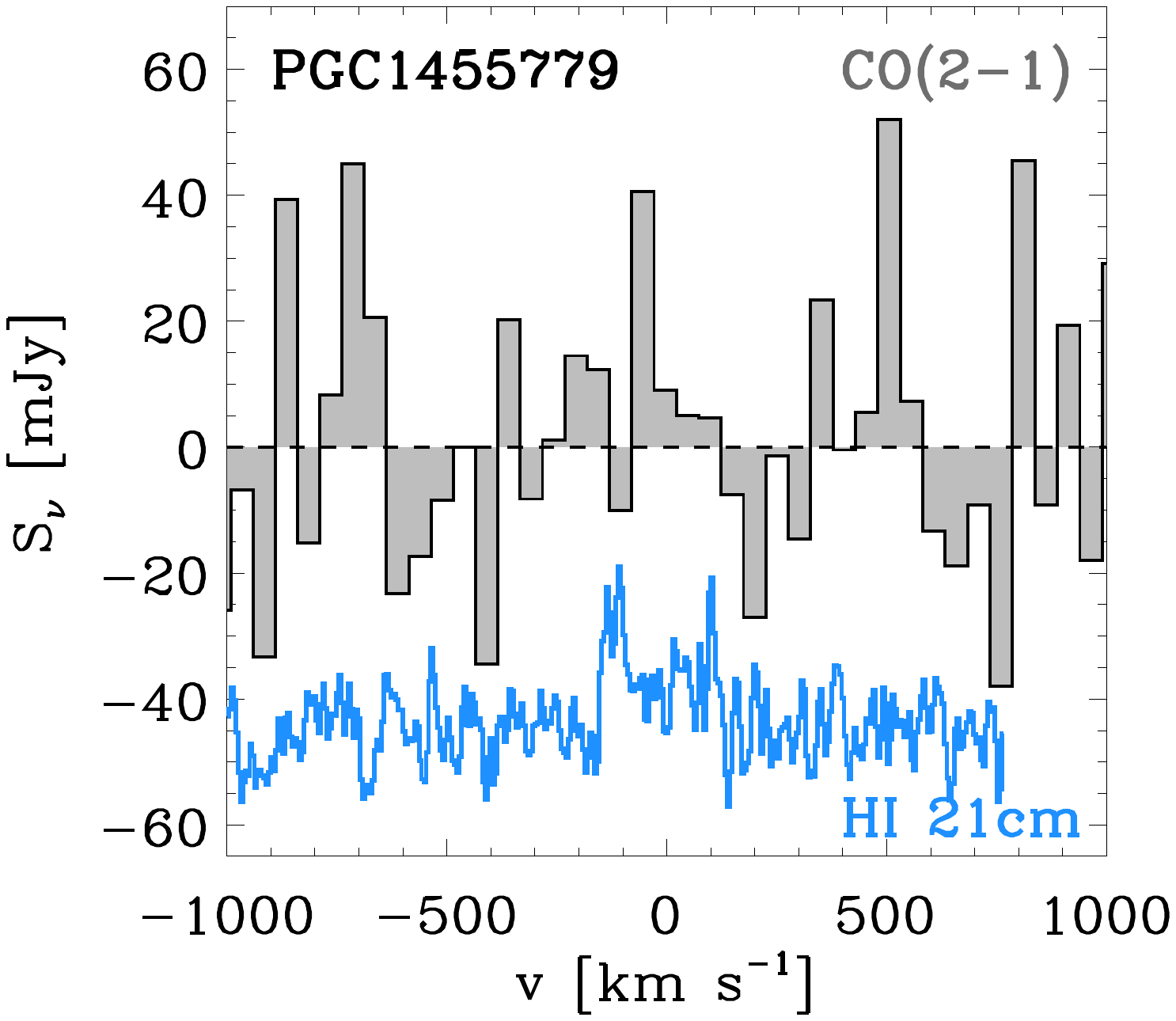}\quad
    \includegraphics[clip=true,trim=-0.4cm 0cm 0cm 0cm,width=0.18\textwidth,angle=90]{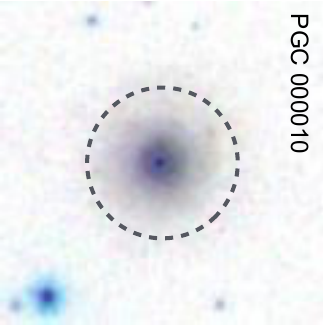}\quad
    \includegraphics[clip=true,trim=5.5cm 3.8cm 4cm 2cm,width=0.28\textwidth,angle=0]{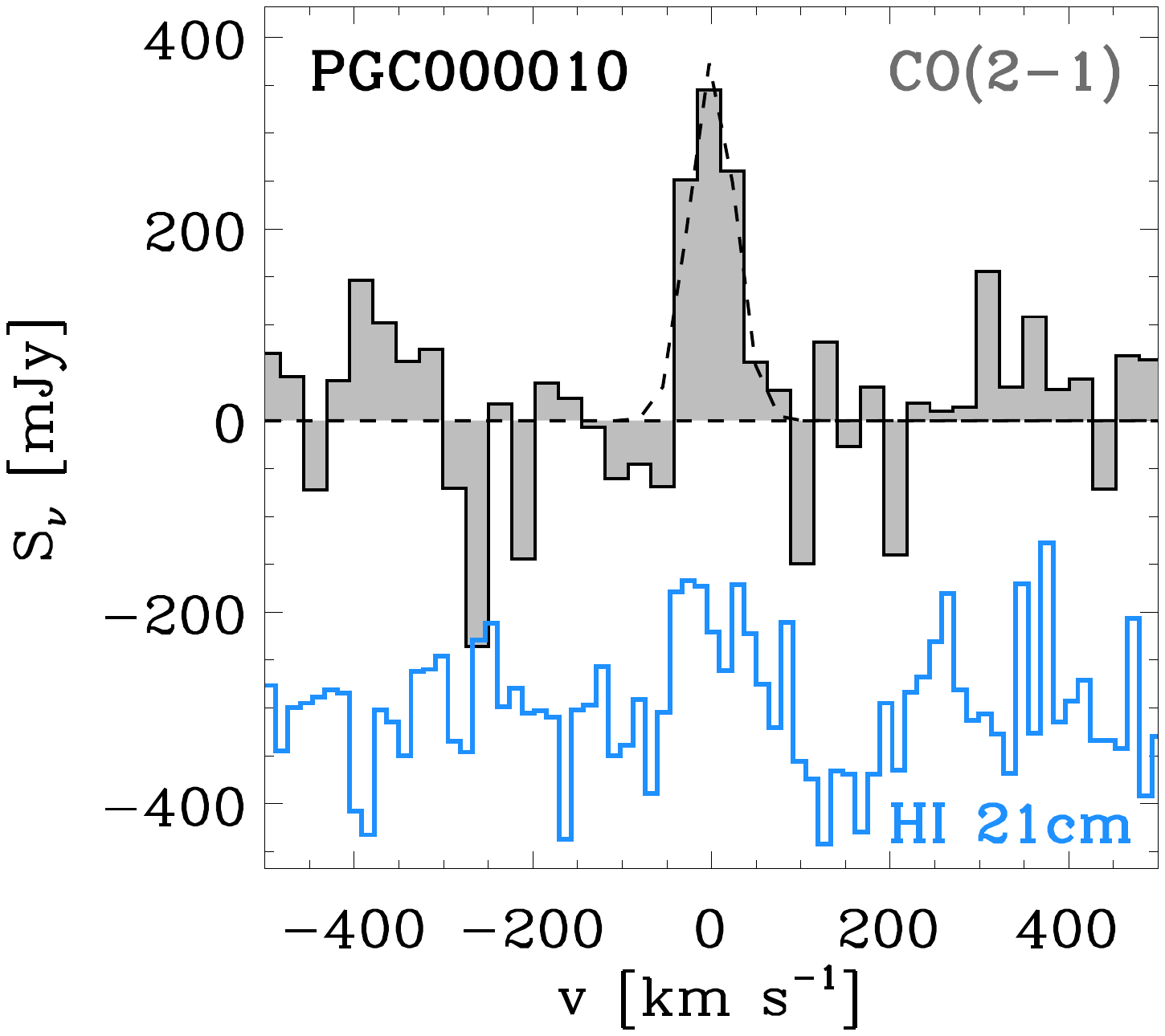}\\
      \includegraphics[clip=true,trim=-0.4cm 0cm 0cm 0cm,width=0.18\textwidth,angle=90]{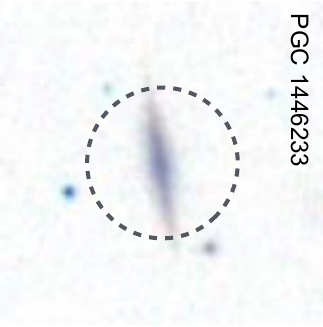}\quad
    \includegraphics[clip=true,trim=5.5cm 3.8cm 4cm 2cm,width=0.28\textwidth,angle=0]{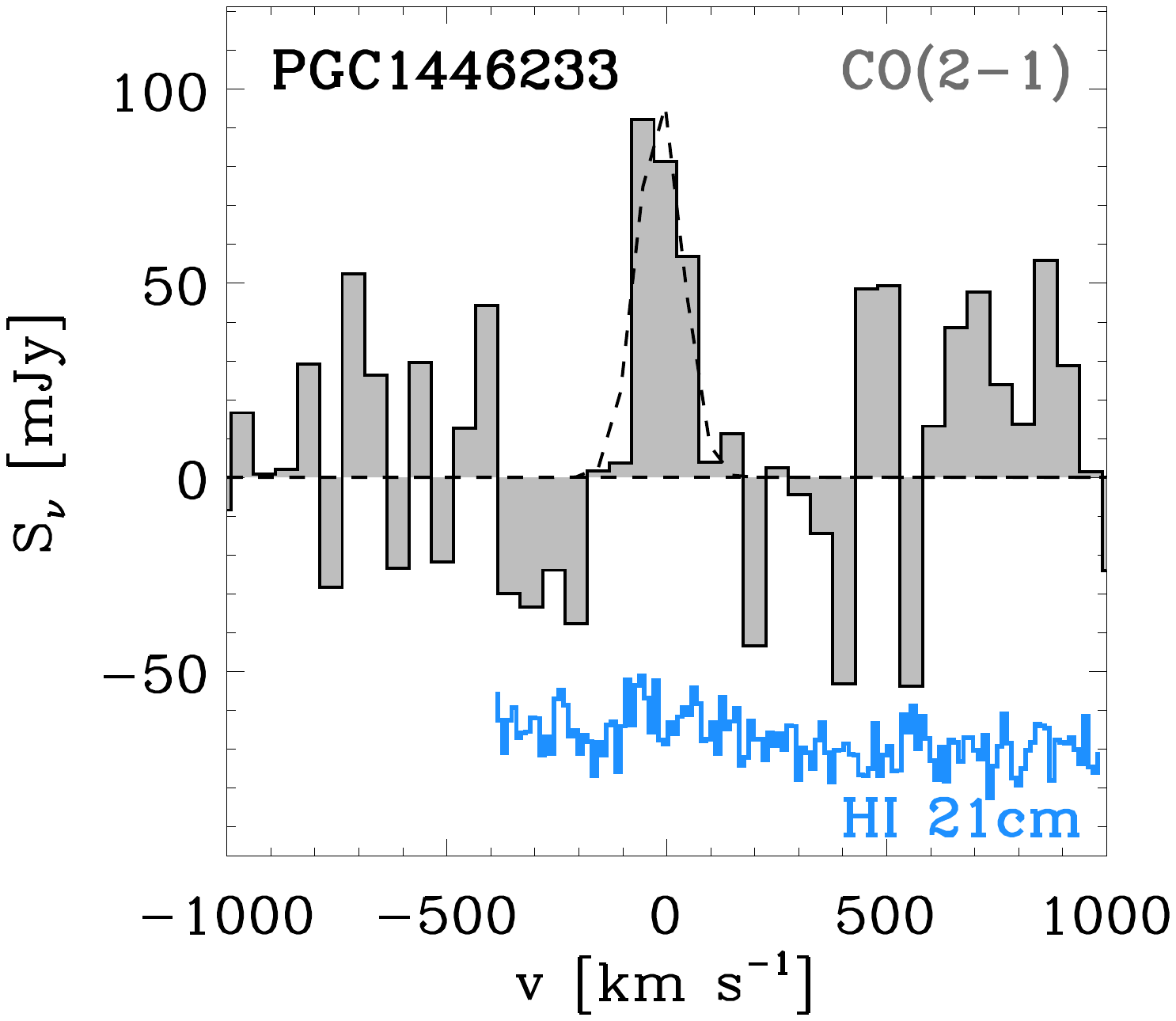}\quad
    \includegraphics[clip=true,trim=-0.4cm 0cm 0cm 0cm,width=0.18\textwidth,angle=90]{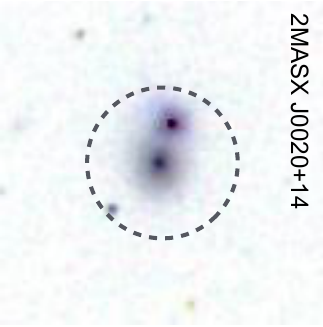}\quad
    \includegraphics[clip=true,trim=5.5cm 3.8cm 4cm 2cm,width=0.28\textwidth,angle=0]{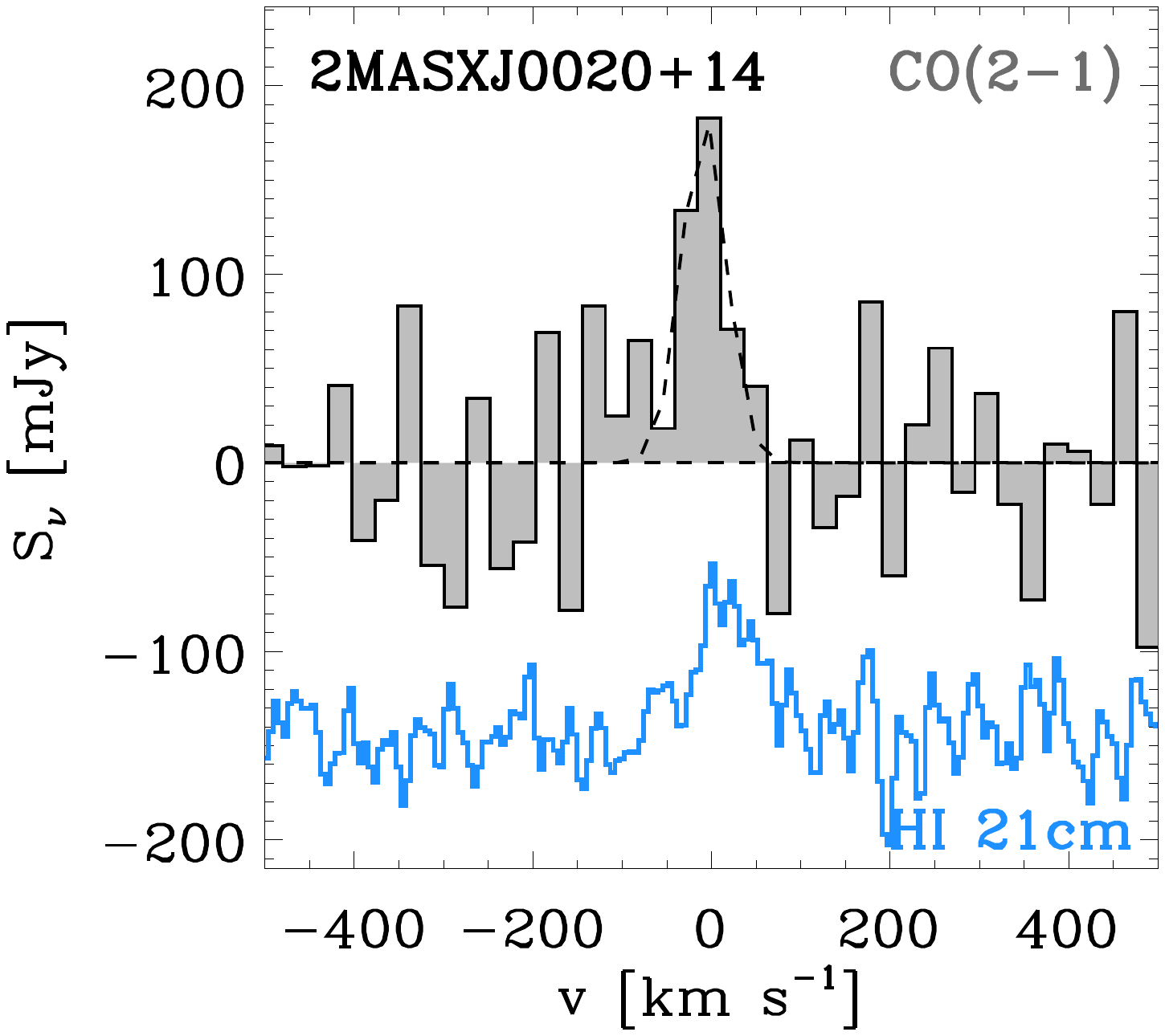}\\
      \includegraphics[clip=true,trim=-0.4cm 0cm 0cm 0cm,width=0.18\textwidth,angle=90]{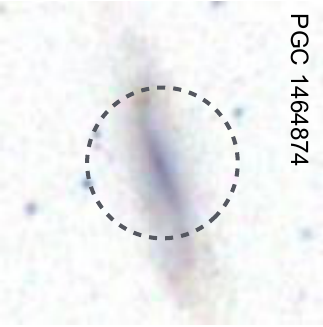}\quad
    \includegraphics[clip=true,trim=5.5cm 3.8cm 4cm 2cm,width=0.28\textwidth,angle=0]{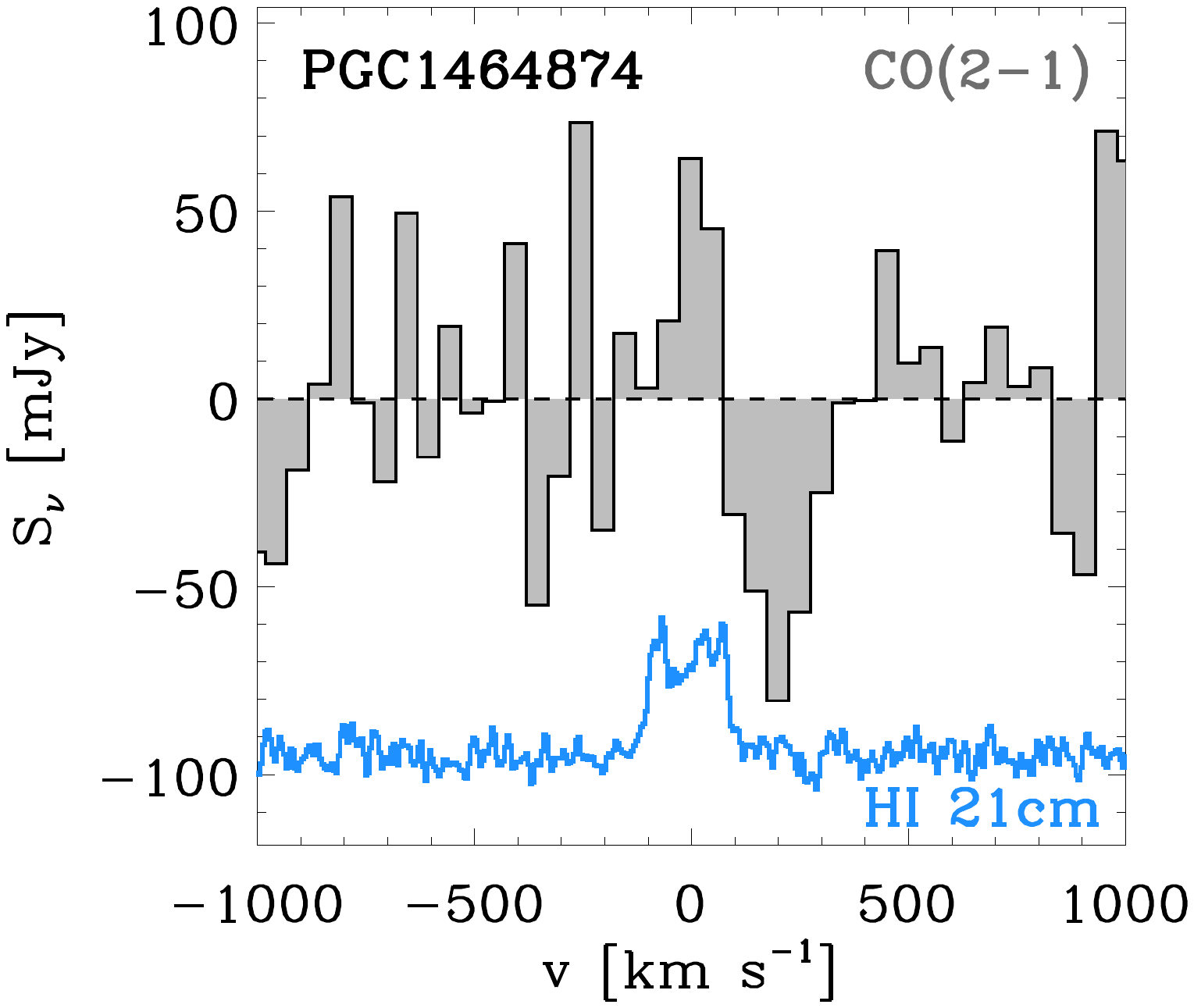}\quad
    \includegraphics[clip=true,trim=-0.4cm 0cm 0cm 0cm,width=0.18\textwidth,angle=90]{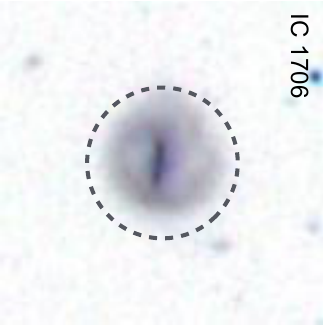}\quad
    \includegraphics[clip=true,trim=5.5cm 3.8cm 4cm 2cm,width=0.28\textwidth,angle=0]{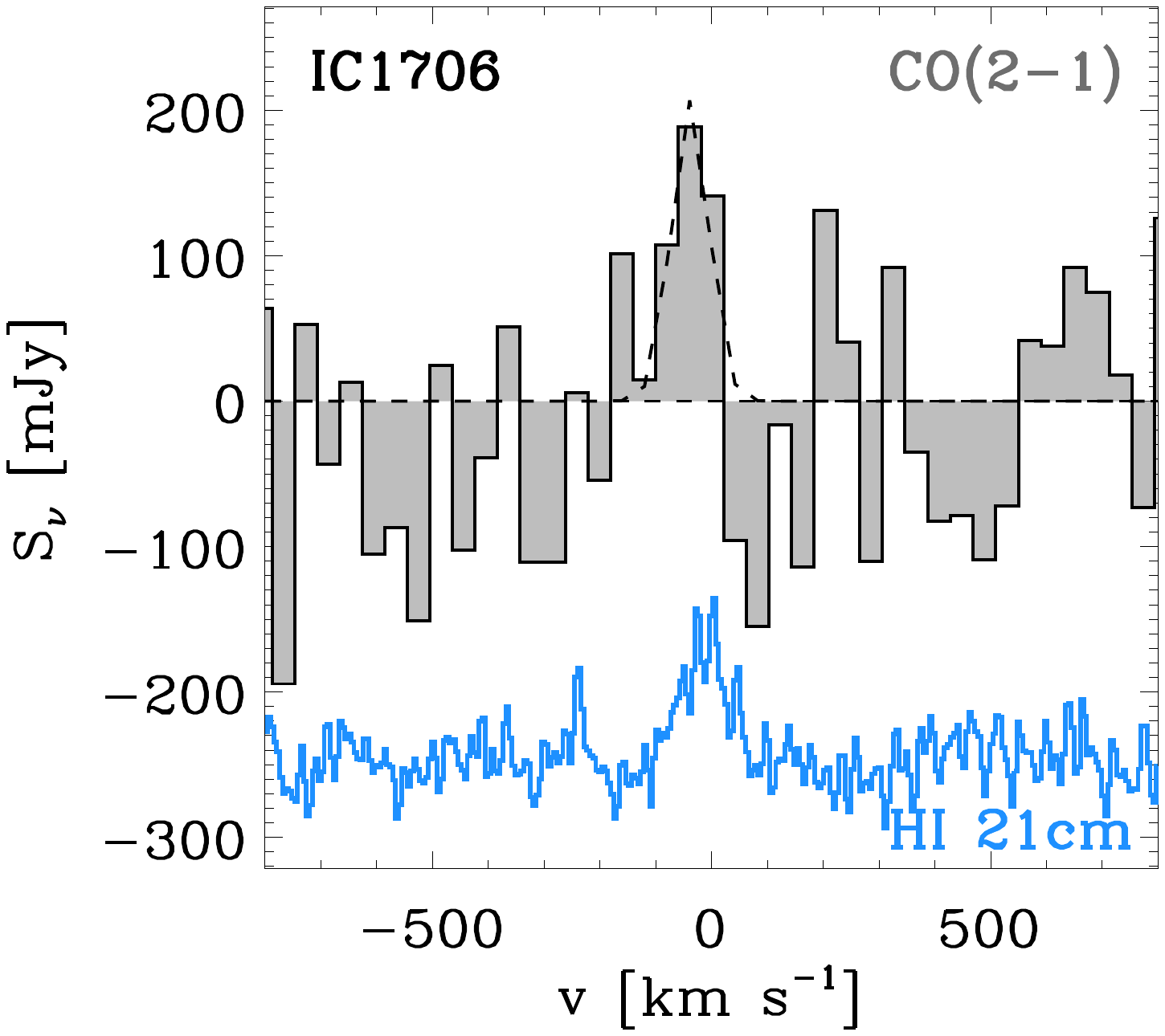}\\
     \caption{{\it Left panels:} SDSS cutout images ({\it g r i} composite, field of view = $60\arcsec\times60\arcsec$, scale = 0.5$\arcsec$/pixel, north is up and west is right) of ALLSMOG galaxies, showing the 27$''$ APEX beam at 230 GHz. {\it Right panels:} APEX CO(2-1) baseline-subtracted spectra, rebinned in bins of
     $\delta \varv=50$~\kms (UGC00317, SDSSJ0011+1428, PGC1455779, PGC1446233, PGC1464874), 40~\kms (IC1706), or 25~\kms (PGC000010, 2MASXJ0020+1413), depending on the width and S/N of the line. The corresponding H{\sc i}~21cm spectra are also shown for comparison, after having been renormalised for visualisation purposes 
     (H{\sc i} references are given in Table~\ref{table:HI_parameters}).}
   \label{fig:spectra9}
\end{figure*}
\begin{figure*}[tbp]
\centering
    \includegraphics[clip=true,trim=-0.4cm 0cm 0cm 0cm,width=0.18\textwidth,angle=90]{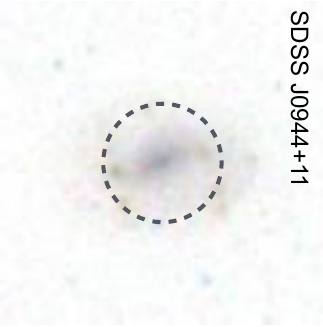}\quad
    \includegraphics[clip=true,trim=5.5cm 3.8cm 4cm 2cm,width=0.28\textwidth,angle=0]{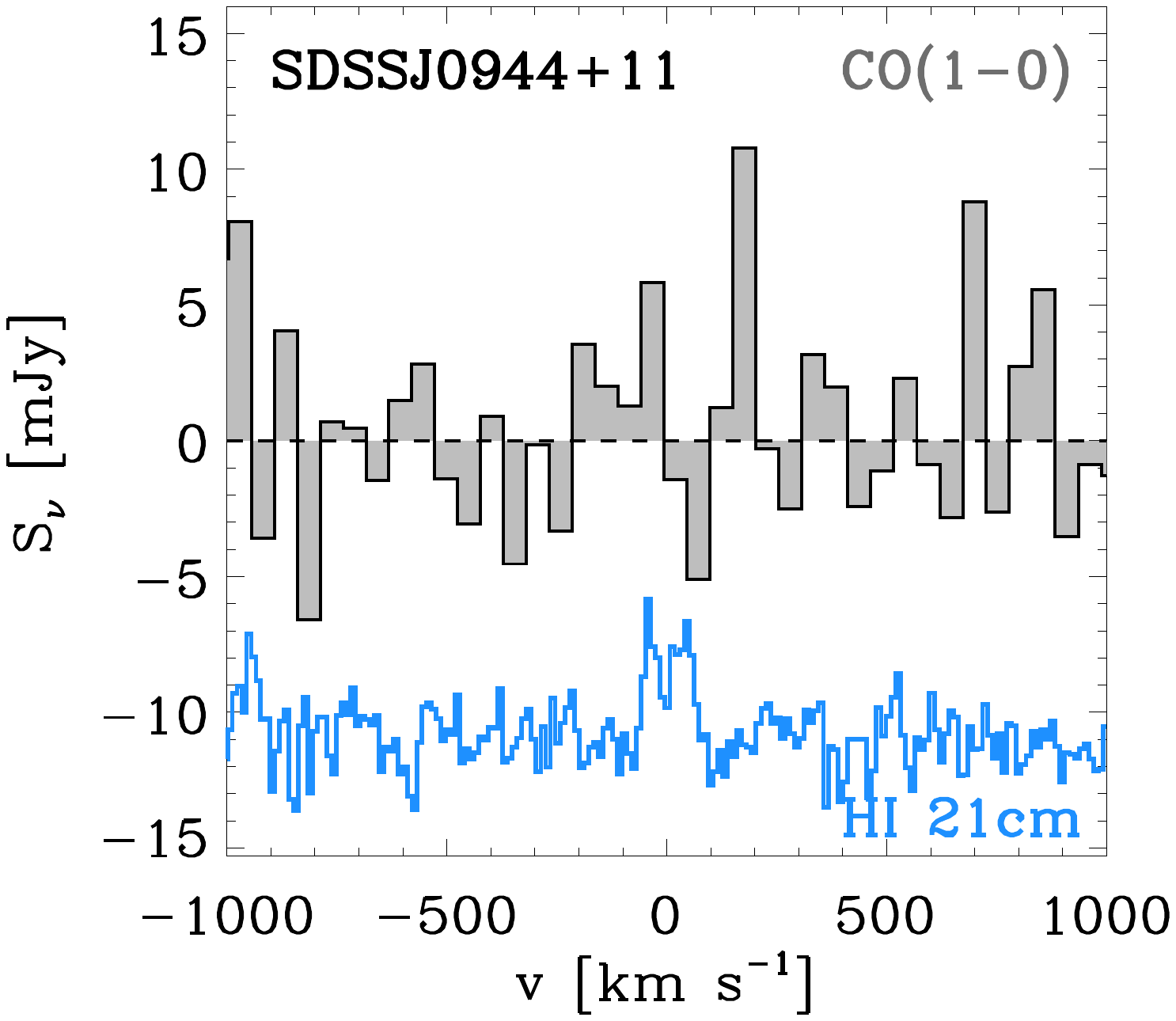}\quad
    \includegraphics[clip=true,trim=-0.4cm 0cm 0cm 0cm,width=0.18\textwidth,angle=90]{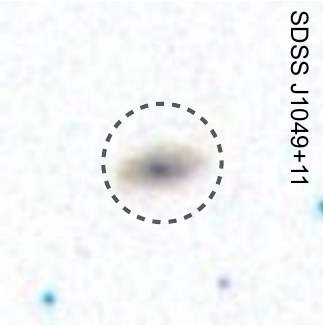}\quad
    \includegraphics[clip=true,trim=5.5cm 3.8cm 4cm 2cm,width=0.28\textwidth,angle=0]{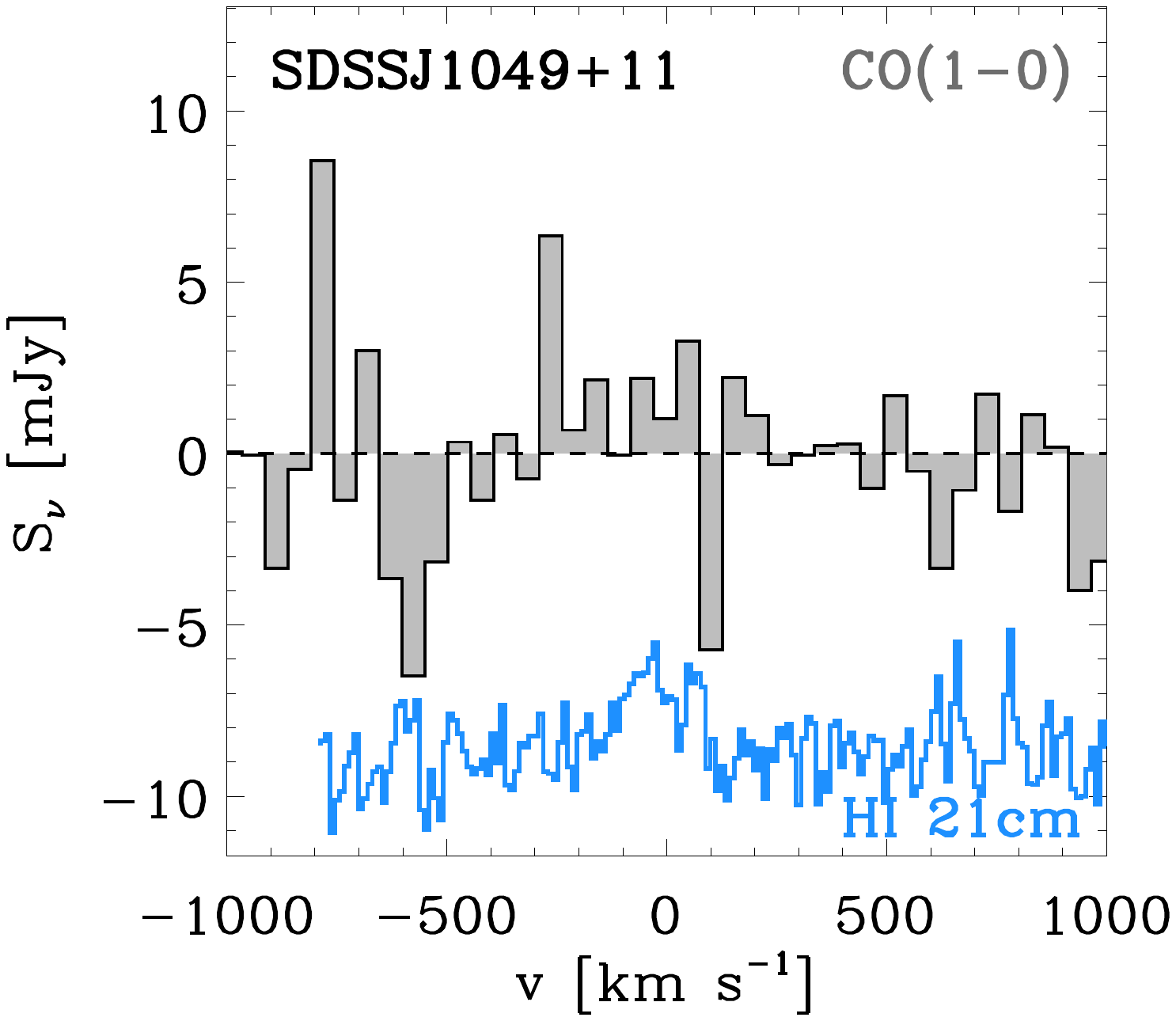}\\
    \includegraphics[clip=true,trim=-0.4cm 0cm 0cm 0cm,width=0.18\textwidth,angle=90]{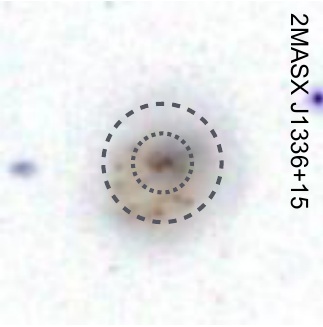}\quad
    \includegraphics[clip=true,trim=5.5cm 3.8cm 4cm 2cm,width=0.28\textwidth,angle=0]{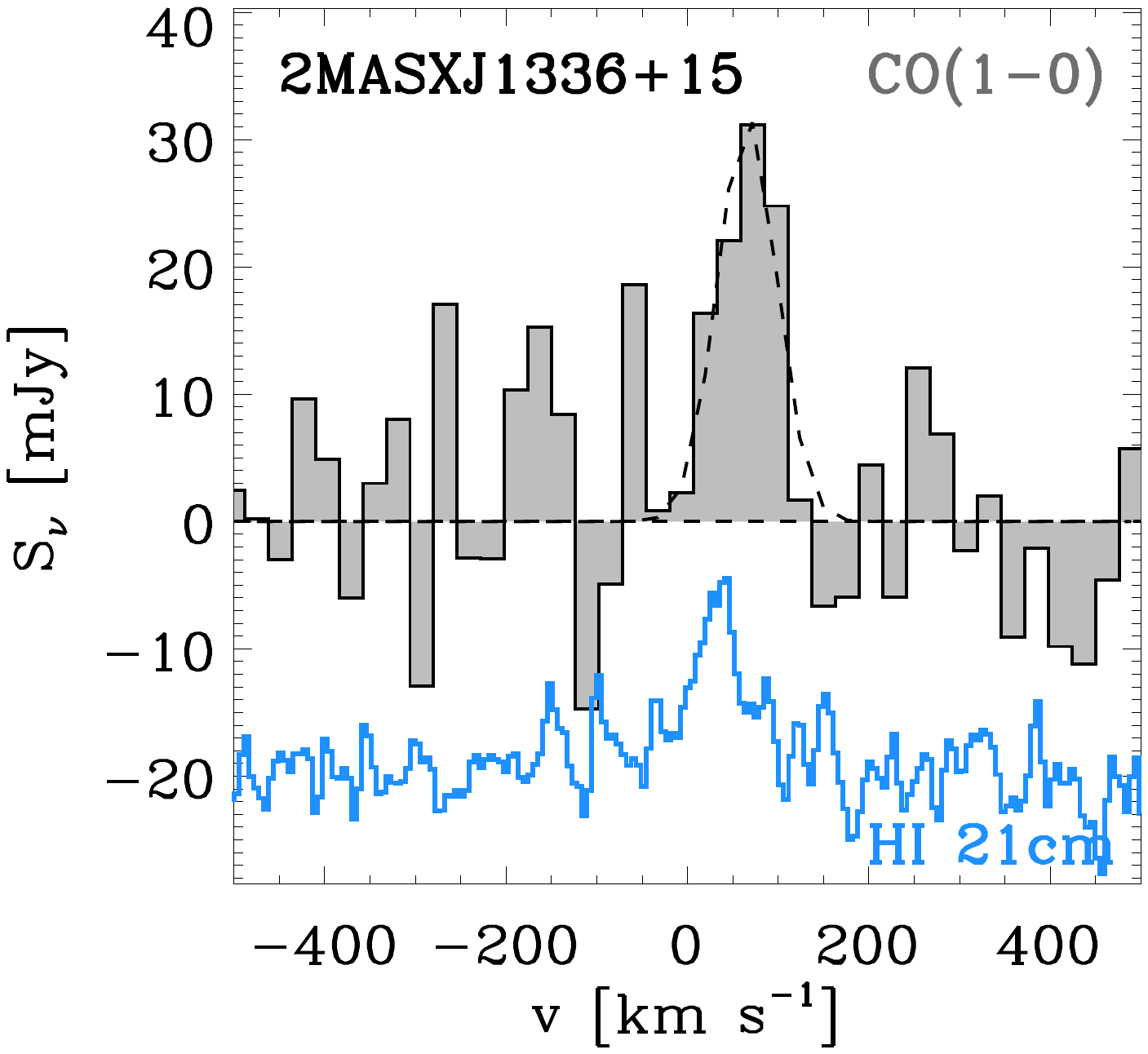}\quad
     \includegraphics[clip=true,trim=5.5cm 3.8cm 4cm 2cm,width=0.28\textwidth,angle=0]{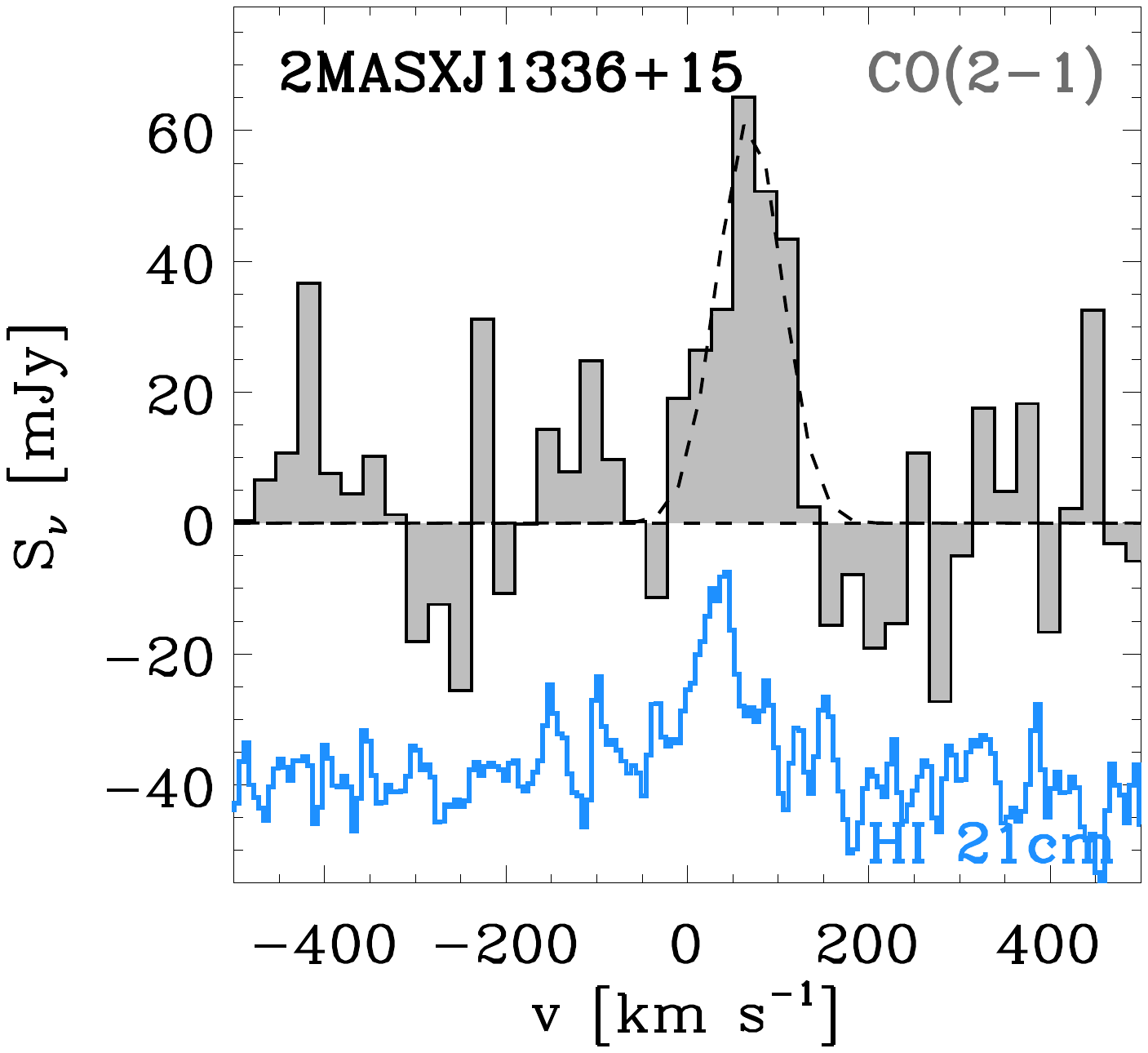}\\
    \includegraphics[clip=true,trim=-0.4cm 0cm 0cm 0cm,width=0.18\textwidth,angle=90]{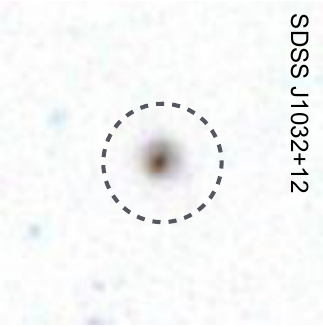}\quad
    \includegraphics[clip=true,trim=5.5cm 3.8cm 4cm 2cm,width=0.28\textwidth,angle=0]{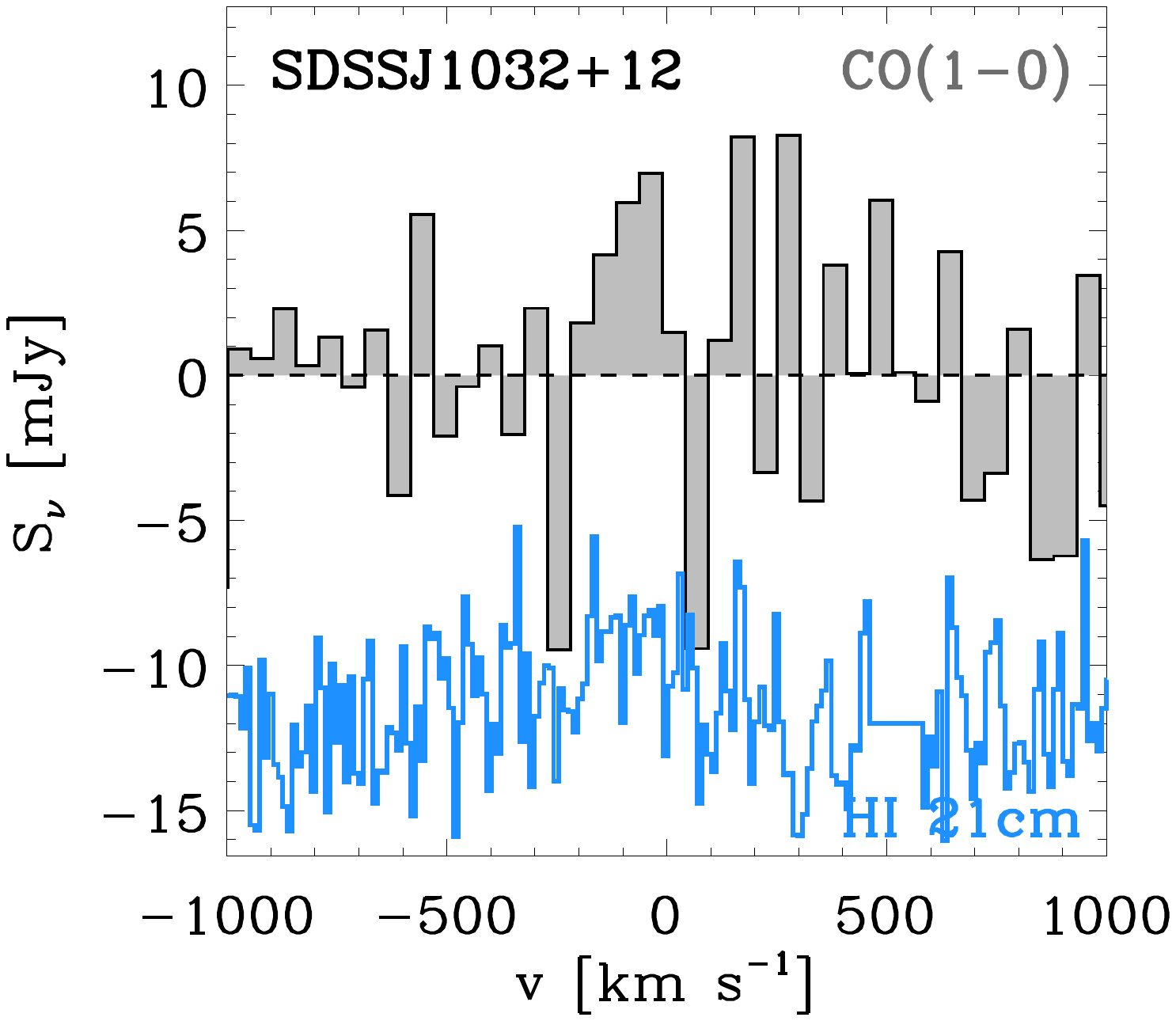}\quad
      \includegraphics[clip=true,trim=-0.4cm 0cm 0cm 0cm,width=0.18\textwidth,angle=90]{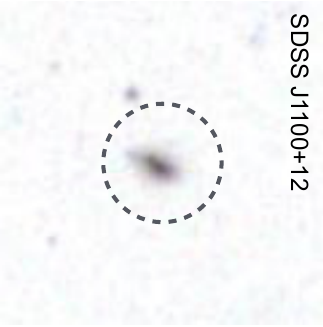}\quad
    \includegraphics[clip=true,trim=5.5cm 3.8cm 4cm 2cm,width=0.28\textwidth,angle=0]{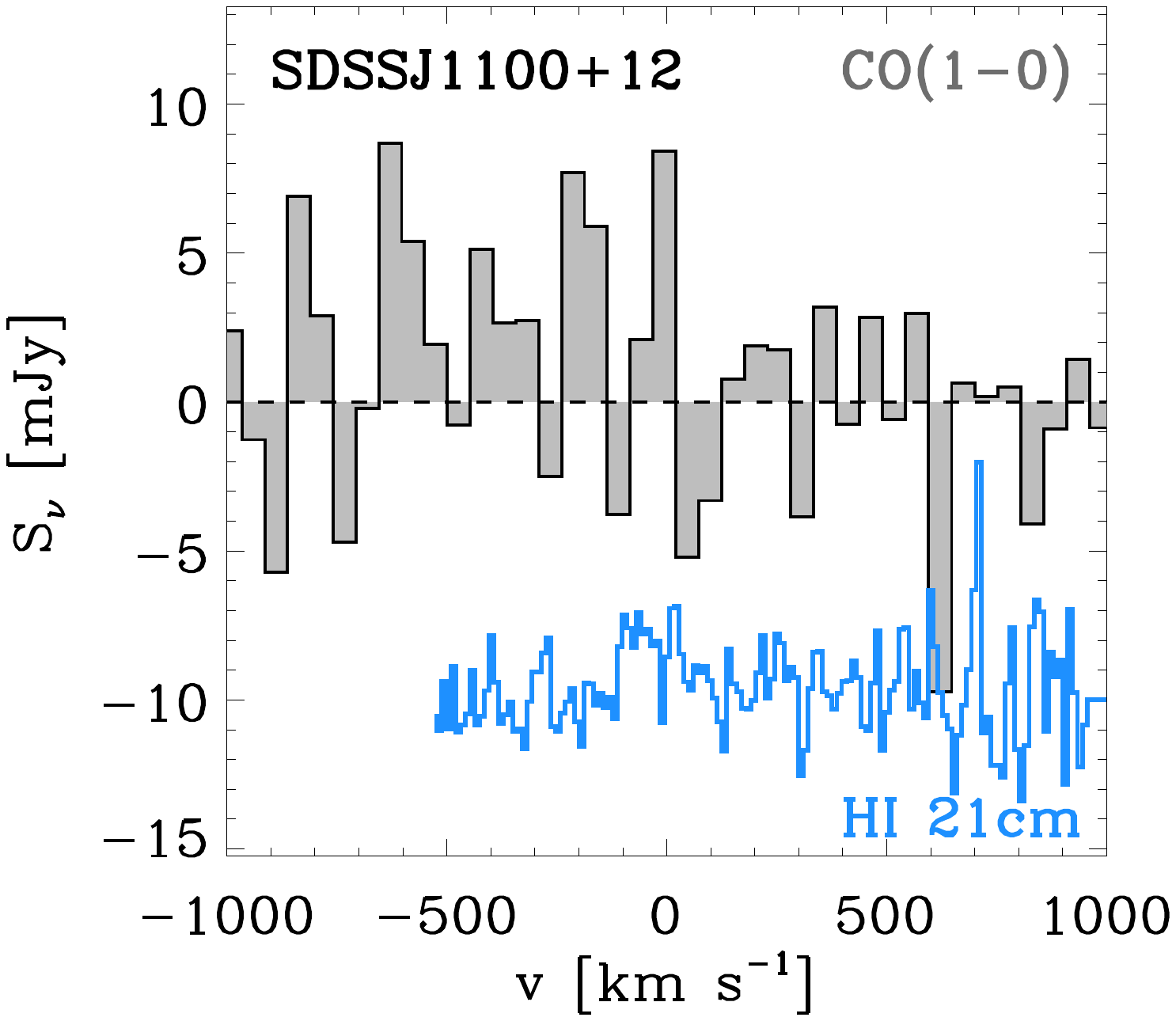}\\
    \includegraphics[clip=true,trim=-0.4cm 0cm 0cm 0cm,width=0.18\textwidth,angle=90]{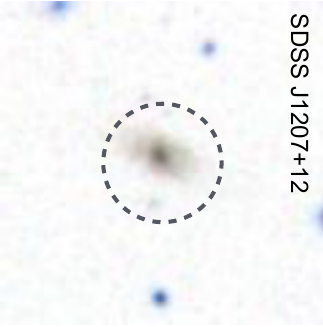}\quad
    \includegraphics[clip=true,trim=5.5cm 3.8cm 4cm 2cm,width=0.28\textwidth,angle=0]{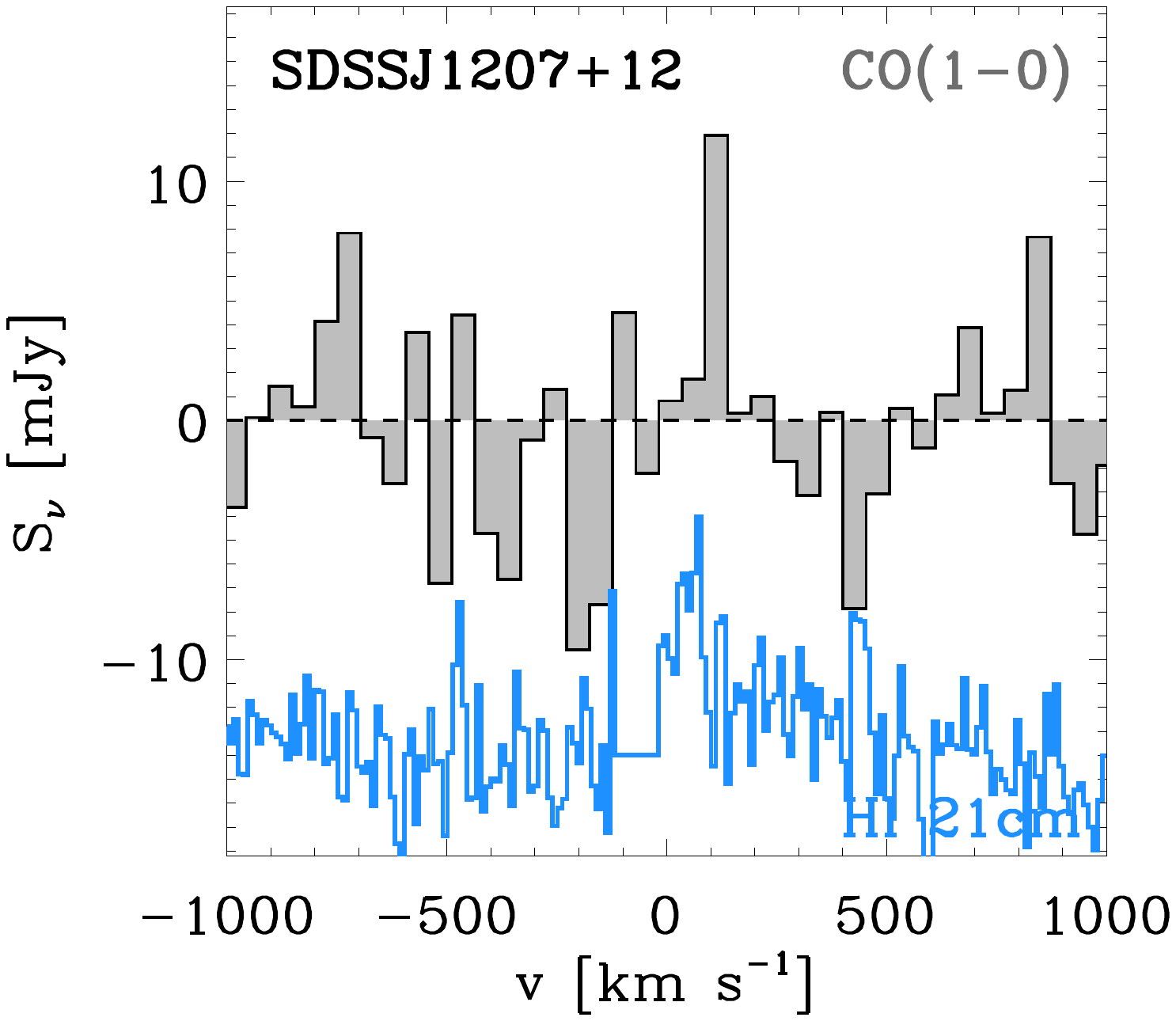}\quad
      \includegraphics[clip=true,trim=-0.4cm 0cm 0cm 0cm,width=0.18\textwidth,angle=90]{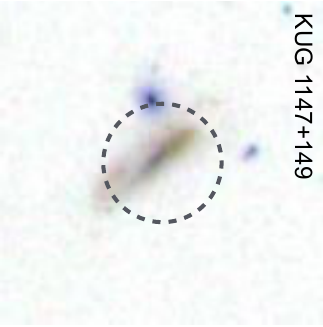}\quad
    \includegraphics[clip=true,trim=5.5cm 3.8cm 4cm 2cm,width=0.28\textwidth,angle=0]{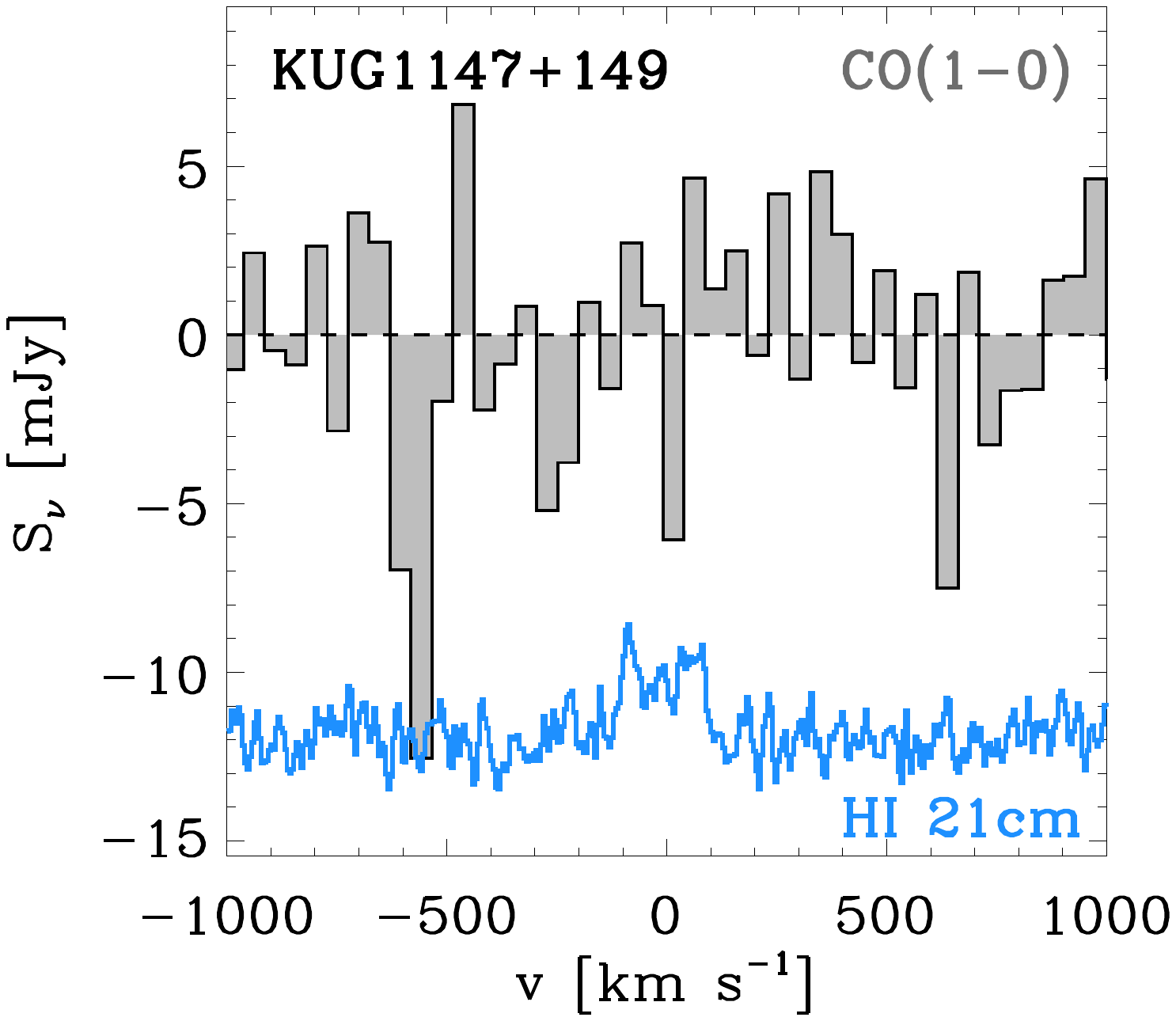}\\
    \includegraphics[clip=true,trim=-0.4cm 0cm 0cm 0cm,width=0.18\textwidth,angle=90]{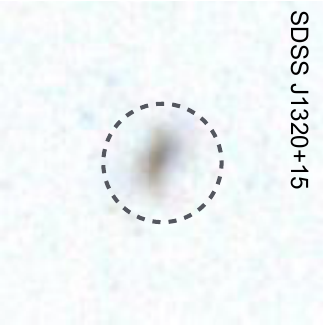}\quad
    \includegraphics[clip=true,trim=5.5cm 3.8cm 4cm 2cm,width=0.28\textwidth,angle=0]{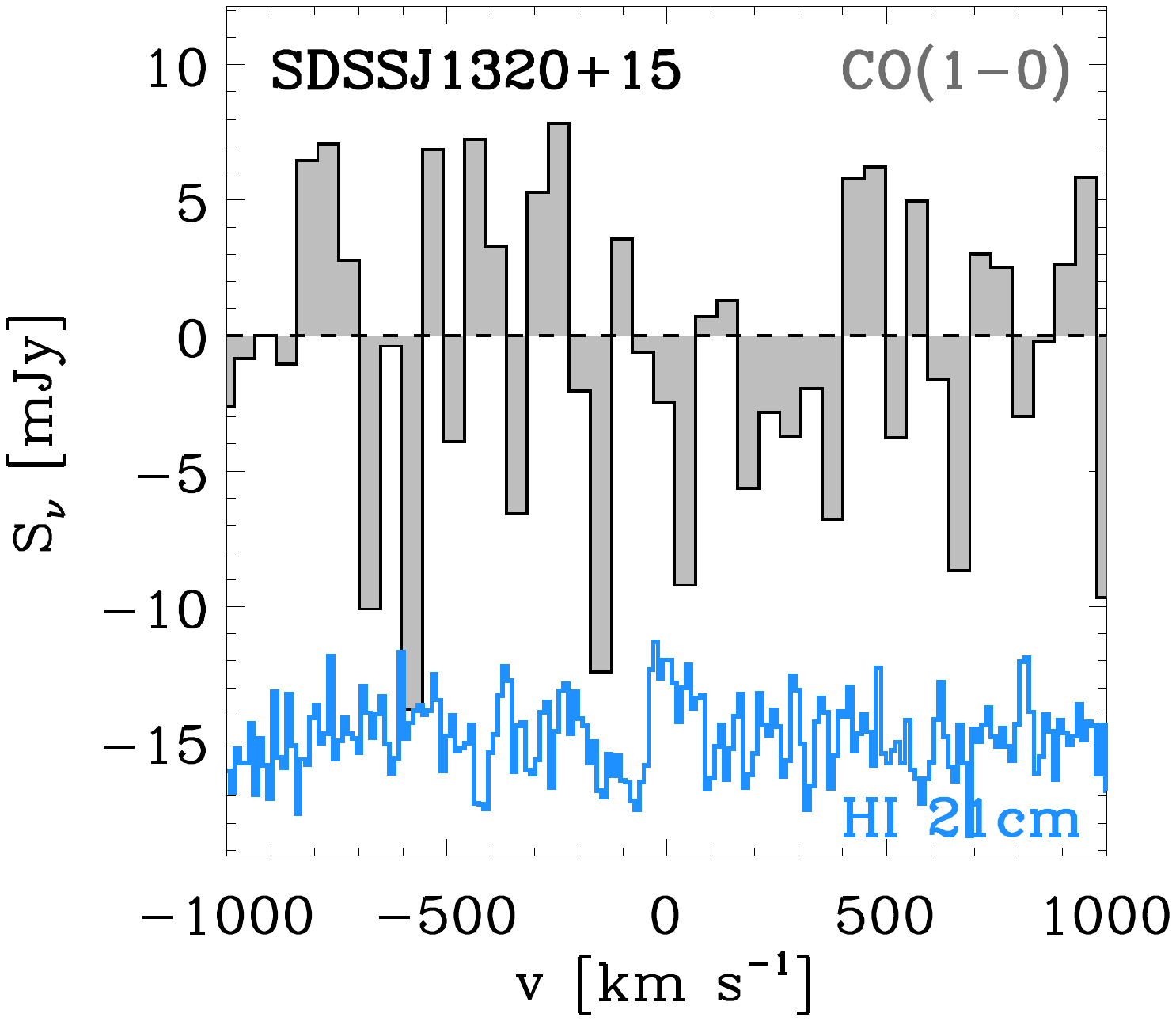}\quad
      \includegraphics[clip=true,trim=-0.4cm 0cm 0cm 0cm,width=0.18\textwidth,angle=90]{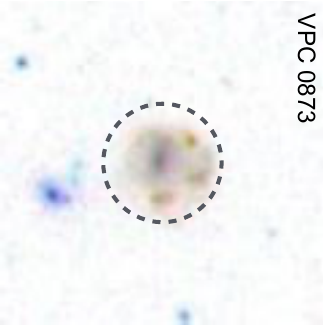}\quad
    \includegraphics[clip=true,trim=5.5cm 3.8cm 4cm 2cm,width=0.28\textwidth,angle=0]{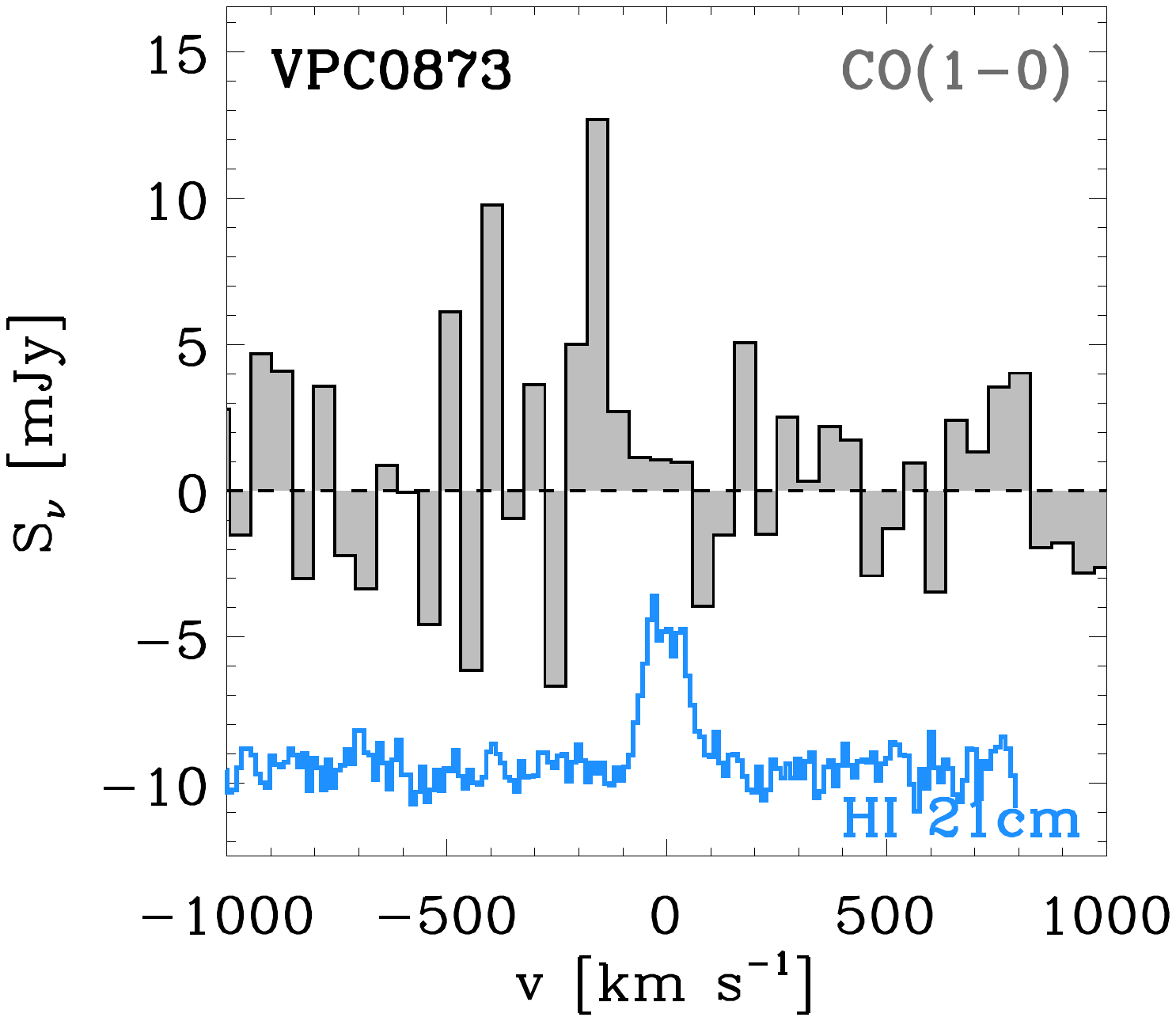}\\
     \caption{{\it Left panels:} SDSS cutout images ({\it g r i} composite, field of view = $60\arcsec\times60\arcsec$, scale = 0.5$\arcsec$/pixel, north is up and west is right) of the nine ALLSMOG galaxies observed with the IRAM~30m telescope, showing the 22$''$ IRAM 30m beam at 115 GHz. For 2MASXJ1336+1552 we also show the 11$''$ IRAM 30m beam at 230 GHz. {\it Right panels:} IRAM CO(1-0) baseline-subtracted spectra, rebinned in bins of
     $\delta \varv=50$~\kms (SDSSJ0944+1116, SDSSJ1049+1108, SDSSJ1032+1227, SDSSJ1100+1207, SDSSJ1207+1200, KUG1147+149,
     SDSSJ1320+1524, VPC0873), or 25~\kms (2MASXJ1336+1552, both the CO(1-0) and the CO(2-1) spectra), depending on the width and S/N of the line.
     The corresponding H{\sc i}~21cm spectra are also shown for comparison, after having been renormalised for visualisation purposes
     (H{\sc i} references are given in Table~\ref{table:HI_parameters}).}
   \label{fig:spectra10}
\end{figure*}
\end{appendix}
\clearpage
\newpage

\end{document}